\documentclass[12pt]{article}
\usepackage{float}
\usepackage{makeidx}
\usepackage{amssymb}
\usepackage{amsmath}
\usepackage{amsfonts}
\usepackage{lscape}
\usepackage{setspace}
\usepackage[letterpaper]{geometry}
\usepackage{chicago}
\usepackage{chbibref}
\usepackage[toc,page,title,titletoc,header]{appendix}
\usepackage{color}
\usepackage{epsfig}
\usepackage{hyperref}

\newtheorem{theorem}{Theorem}

\newtheorem{assumption}{Assumption}

\newtheorem{corollary}{Corollary}

\newtheorem{lemma}{Lemma}

\newtheorem{proposition}{Proposition}
\newtheorem{remark}{Remark}

\newenvironment{proof}[1][Proof]{\noindent \textbf{#1.} }{\  \rule{0.5em}{0.5em}}
\input{tcilatex}
\begin{document}

\title{\vspace*{-0.8in}{\Large Analysis of Multiple Long-Run Relations in
Panel Data Models}\thanks{{\footnotesize The views expressed in this paper
are those of the authors and do not necessarily reflect those of the Federal
Reserve Bank of Dallas or the Federal Reserve System. We gratefully
acknowledge our use of computational resources provided by the Big-Tex High
Performance Computing Group at the Federal Reserve Bank of Dallas.}}}
\author{{\normalsize Alexander Chudik} \\
{\normalsize Federal Reserve Bank of Dallas} \and {\normalsize M. Hashem
Pesaran} \\
{\normalsize Trinity College, Cambridge, UK and University of Southern
California, USA} \and {\normalsize Ron P. Smith} \\
{\normalsize Birkbeck, University of London, UK}}
\date{%
\normalsize%
\today }
\maketitle

\begin{abstract}
The literature on panel cointegration is extensive but does not cover data
sets where the cross section dimension, $n$, is larger than the time series
dimension $T$. This paper proposes a novel methodology that filters out the
short run dynamics using sub-sample time averages as deviations from their
full-sample counterpart, and estimates the number of long-run relations and
their coefficients using eigenvalues and eigenvectors of the pooled
covariance matrix of these sub-sample deviations. We refer to this procedure
as pooled minimum eigenvalue (PME). We show that PME estimator is consistent
and asymptotically normal as $n$ and $T$ $\rightarrow \infty $ jointly, such
that $T\approx n^{d}$, with $d>0$ for consistency and $d>1/2$ for asymptotic
normality. Extensive Monte Carlo studies show that the number of long-run
relations can be estimated with high precision, and the PME estimators have
good size and power properties. The utility of our approach is illustrated
by micro and macro applications using Compustat and Penn World Tables.
\end{abstract}



\thispagestyle{empty}\pagebreak

\pagenumbering{arabic}%

\section{Introduction}

\onehalfspacing%
\normalsize%

This paper provides a new methodology for the analysis of multiple long-run
relations in panel data models where the cross section dimension, $n$, is
large relative to the time series dimension, $T$. While there is an
extensive literature that considers multiple long-run (cointegrating)
relations for time series models, for panel data models with large $n$
researchers have mainly focussed on a single long-run relation with known
long-run causal links. The panel literature that does consider multiple
long-run relations assumes $n$ is fixed as $T\rightarrow \infty ,$ or adopt
sequential asymptotics whereby $T\rightarrow \infty $ first followed by $%
n\rightarrow \infty $, effectively requiring $T$ to be large relative to $n$
and do not cover many applications of interest in economics and finance that
involve many cross section units, such as firms and countries, observed over
relatively short time spans. One example is empirical corporate finance,
which investigates the stability of long-run relations, including financial
ratios, using accounting data, such as Compustat, where thousands of firms
are observed over relatively few time periods. 
\citeN{ColesLi2023}
provide examples from a number of sub-fields of corporate finance,
including: \ propensity to pay dividends, leverage, \ investment policy, and
firm performance. Another example is cross country empirical growth studies
that use data sets such as the Penn World Tables that provide annual data on
a range of macro variables for as many as $n=183$ countries over different
time periods, with a maximum time span of $T=70$. For both panel data sets
one would expect multiple long-run relations between the variables, some of
which may be the mean reverting ratios discussed in the macroeconomic and
finance literatures.

This paper proposes an estimation and testing strategy that applies to panel
data models with $n$ possibly much larger than $T$. We consider an $m\times
1 $ vector $\mathbf{w}_{it}$ for units $i=1,2,...,n$ over the time periods $%
t=1,2,...,T$, with an unknown number, $r_{0}\in \left\{ 0,1,...,m-1\right\} $%
, of linear combinations that are stationary. We refer to such linear
combinations as long-run relations. If it is known that all elements of $%
\mathbf{w}_{it}$ are $I\left( 1\right) $ then the stationary relations can
be viewed as cointegrating relations. The focus of our analysis is to
estimate $r_{0}$, the number of common long-run relations, and their
coefficients, when $r_{0}\geq 1$. We filter out the short run dynamics by
means of $q$ ($\geq 2$) non-overlapping sub-sample time averages, $\mathbf{%
\bar{w}}_{i\ell },$ $\ell =1,2,...,q,$ as deviations from their full-sample
counterpart, $\mathbf{\bar{w}}_{i\circ }$, namely $\mathbf{\bar{w}}_{i\ell }-%
\mathbf{\bar{w}}_{i\circ }$, and then construct a pooled sample covariance
matrix of these deviations which we denote by $\mathbf{Q}_{\bar{w}\bar{w}}$.
The number of long-run relations and their coefficients are estimated using
the eigenvalues and eigenvectors of $\mathbf{Q}_{\bar{w}\bar{w}}$. We refer
to this procedure as pooled minimum eigenvalue (PME), and note that it is
simple to implement, extends readily to unbalanced panels, and is shown to
be robust to stationary interactive time effects. It is semi-parametric
since it does not require modelling the short run dynamics and applies to
general linear process, thus allowing for moving average processes and is
not confined to vector autoregressions (VAR). Most importantly, the PME
approach does not require knowing long-run causal linkages that might exist
amongst the variables under consideration. To our knowledge, no other panel
estimation procedure exists for such a setting.

Denoting the first $r_{0}$ eigenvectors of $\mathbf{Q}_{\bar{w}\bar{w}}$ by $%
\mathbf{\hat{\beta}}_{j0},$ for $j=1,2,...,r_{0}$, we then consider
structural estimation of the long-run relations assuming they are subject to 
$r_{0}\times r_{0}$ exact identifying restrictions. Assuming $r_{0}$ is
known, we derive the asymptotic distribution of the exactly identified
long-run relations and propose consistent estimators for their covariance
matrices that does not require estimation of the dynamics of individual $%
\mathbf{w}_{it}$ processes; thus allowing us to test restrictions on the
elements of the exactly identified long-run relations with relatively short $%
T$. The identified long-run relations are shown to be consistent and
asymptotically normally distributed as $n$ and $T$ $\rightarrow \infty $
jointly such that $T\approx n^{d}$. For consistency only $d>0$ is required,
but for asymptotic normality a faster relative rate of $d>1/2$ is required.
Many panel time series estimation and inference procedures require $d>1$ ($%
n/T\rightarrow 0$). Our requirement $d>1/2$ indicates that the procedure
should work well with large $n$ and moderate $T$ allowing one to estimate
the coefficients of multiple long-run relations in such cases, without
estimating short-run dynamics.

We propose to estimate $r_{0}$ by the number of eigenvalues of $\mathbf{Q}_{%
\bar{w}\bar{w}}$ that fall below a given threshold $C_{T}=CT^{-\delta }$,
for some $C>0$ and $\delta >0$. One could use cross validation procedures to
set $C$ and $\delta $, but based on extensive Monte Carlo experiments we
have found that setting $C=1$ works well \textit{if} we base our selection
procedure on the eigenvalues of the correlation matrix, $\mathbf{R}_{_{\bar{w%
}\bar{w}}}=\left[ diag\left( \mathbf{Q}_{\bar{w}\bar{w}}\right) \right]
^{-1/2}\mathbf{Q}_{\bar{w}\bar{w}}\left[ diag\left( \mathbf{Q}_{\bar{w}\bar{w%
}}\right) \right] ^{-1/2}$. Such an estimator can be written conveniently as 
$\tilde{r}=\sum_{j=1}^{m}\mathcal{I}\left( \tilde{\lambda}_{j}<T^{-\delta
}\right) ,$ where $\tilde{\lambda}_{j},$ $j=1,2,...,m$ are the eigenvalues
of $\mathbf{R}_{_{\bar{w}\bar{w}}}$, and $\mathcal{I}\left( \mathcal{A}%
\right) =1$ if $\mathcal{A}$ is true and zero otherwise. In the Monte Carlo
experiments and the empirical applications we report results for $\delta
=(1/4,1/2)$, and find that overall setting $\delta =1/4$ works well.

Monte Carlo experiments show near-perfect performance of $\tilde{r}$ as an
estimator of $r_{0}$, when $\delta =1/4$, for all $\,n=50$, $500$, $1,000$, $%
3,000$ and $T=20$, $50$, $100$ sample size combinations and across a large
number of VAR and VARMA data generating processes, with and without
interactive time effects, non-Gaussian errors, generalized autoregressive
conditional heteroskedasticity (GARCH), threshold autoregressions (TAR), and
for different patterns of long-run causal ordering. $\tilde{r}$ performs
almost equally well for the smallest sample sizes of $T=20$ and $n=50$, as
well as for the largest $T=100$ and $n=3,000$. As an alternative approach we
considered Johansen's trace tests applied to each cross section unit
separately, and then estimated $r_{0}$ by the simple average of these
individual estimates. This was done purely for comparison since to the best
of our knowledge there are no other methods that apply to panels in the
literature.\ We found that this average type estimator performed reasonably
well when $T$ was large, but still fell short as compared to the
thresholding estimator.

The finite sample performance of PME estimator of the coefficients of the
long-run relations\ is found to be satisfactory with inference based on PME
estimator with $q=2$ (sub-sample time averages) generally more accurate in
terms of empirical size of the tests, compared with $q=4$, in line with
intuition that suggest a larger choice of $q$ is likely to result in a
larger finite-sample bias. We also considered a simple VAR(1) design with $%
m=2$, $r_{0}=1$ and one-way long-run causality to see how PME performs
compared to the many single equation estimators proposed in the literature
(and cited below). We found that in this simple case the PME\ estimator is
less efficient in terms of root mean square errors only when $T=100$.
However, PME with $q=2$ proved to be less biased and performed much better
in terms of size than the single equation approaches for all sample size
combinations.

To illustrate the utility of PME procedure we present one micro and one
macro application. The micro application considers a number of key financial
variables (in logs) and investigates if they are cointegrated, and whether
financial ratios can be regarded as stationary variables. To this end we
used accounting data for individual firms from CRSP/Compustat on their book
value (BV), market value (MV), short-term debt (SD), long-term debt (LD),
total assets (TA) and total debt outstanding (DO). The panels involving
these variables are unbalanced and cover the period $1950-2021$. We consider
firms with at least $20$ years of data, with $n$ varying between about $%
1,000 $ and $2,500$. The variables are grouped into three sets, where we
have prior expectations about possible cointegration and identification. The
first set considered has just two variables: the logarithm of total debt
outstanding and logarithm of total assets: \{$DO_{it}$, $TA_{it}$\}. The
ratio of total debt outstanding to total assets is often used as a measure
of leverage, which suggests a single hypothesized long-run relation. The
other two variable sets are: the logarithms of short and long term debt and
total assets, \{$SD_{it}$, $LD_{it}$, $TA_{it}$\}; and the logarithms of
total debt outstanding, book value and market value, \{$DO_{it}$, $BV_{it}$, 
$MV_{it}$\}. We expect two hypothesized long-run relations in these sets
with three variables. For each set of variables we provide estimates for the
full sample 1950-2021 as well as for a shorter sample that ends in 2010. The
estimates provide strong evidence of one long-run relation when we consider
two variables, and, with one exception, two long-run relations when we
consider panels with three variables. In the case of panels with $m=2$, we
illustrate that the PME estimates are invariant to normalization, which is
in contrast to the estimates obtained using panel regressions that depend on
which way the regression is run. For the relation between logarithms of debt
and total assets, we find the estimates of the long-run coefficients are
close to one in all cases, ranging from $1.113$ to $1.143$, and precisely
estimated. In the case of panels with $m=3$ we find $\tilde{r}=2$, and the
null hypothesis that long-run coefficients are equal to unity is not
rejected in about a quarter of the panel estimates. These results provide
partial support for use of logarithm of financial ratio in corporate
finance. In cases where the use of log ratio is not supported, one could use
the PME estimates of long-run relations in second stage regressions on
stationary variables that also include short run dynamics as well as other
stationary variables.

The macro application investigates long-run relations using unbalanced cross
country macroeconomic time series data from the Penn World Tables, featuring
up to $n=177$ countries over the years $1950-2019$. This dataset has a much
smaller cross-section dimension and a larger average time dimension compared
with the micro application. We focus on four key macro variables: per capita
real merchandise exports ($ex_{it}$) and imports ($im_{it}$), real labour
productivity per hour worked ($prod_{it}$), and real wages per hour worked ($%
wage_{it}$). The choice of these variables was motivated by two widely
maintained hypotheses. Firstly, real wages and productivity should balance
for steady state growth to be feasible. Secondly export and imports should
balance for international solvency, though the constraint may not be binding
for reserve-currency countries such as the US. These hypotheses are largely
confirmed for emerging economies, and, with notable departures from unit
long-run elasticities, also for advanced economies. In addition, when we
consider all the four variables together we uncover cross country evidence
on the long-run relation between exports and productivity without making any
assumption about the direction of causality between these variables.

\textbf{Related literature:} We first discuss the literature for a single
time series process, which could be viewed as the $p\times 1$ ($p=m$ $n$)
stacked vector, $\mathbf{w}_{t}=(\mathbf{w}_{1t}^{\prime },\mathbf{w}%
_{2t}^{\prime },...,\mathbf{w}_{nt}^{\prime })^{\prime }$. Our approach is
related to that of 
\citeANP{PhillipsOuliaris1990} (\citeyearNP{PhillipsOuliaris1988}, \citeyearNP{PhillipsOuliaris1990})
in that they also start from general linear processes. They propose testing
the null of no cointegration using the smallest eigenvalues of the spectral
density of $\Delta \mathbf{w}_{t}$ evaluated at zero frequency. However, it
is difficult to obtain reasonably precise estimates of the spectral density,
particularly in the presence of high persistence in first differences.
Attempting to eliminate the effects of the short run dynamics by using time
averages of sub-samples of time series data is also widely used. 
\citeN{MuellerWatson2018}
consider using sub-sample averages to estimate the long-run relation between
two variables $(y_{t}$ and $x_{t})$. Their estimated long-run coefficient
from regression of sub-sample averages of $y_{t}$\ on those of $x_{t}$ \ is
not the same as the reciprocal of the estimate that will be obtained from
the reverses regression. A panel version of their procedure can be
considered, but will be subject to the same limitations, namely it can
handle only one long-run relation and will require knowing the direction of
long-run causality. In not requiring any assumptions regarding the direction
of long-run causality our approach is comparable to the maximum likelihood
approach pioneered by 
\citeANP{Johansen1988} (\citeyearNP{Johansen1988}, \citeyearNP{Johansen1991a})
that allows for multiple long-run relations without assuming any long-run
causal orderings of the variables, but assumes a $VAR(s)$ specification in $%
\mathbf{w}_{t}$ where $p$ and $s$ are fixed (and quite small) relative to $T$%
. 
\citeANP{Onatski2018} (\citeyearNP{Onatski2018}, \citeyearNP{Onatski2019})%
\textbf{\ }investigate the asymptotic properties of Johansen test when $%
\mathbf{w}_{t}$ follows VAR(1) but allow $p,T\rightarrow \infty ,$ such that 
$p/T\rightarrow c\in (0,1]$. They provide theoretical arguments why
Johansen's test of cointegration rank is likely to be severely over-sized
even if $p$ takes moderate values. Extensions to higher order VARs are
provided by \cite{Bykhovskaya2022}. This is a promising approach which is
yet to be fully developed for the analysis of multiple cointegrations across
many units, which is the primary focus of this paper. Since the ordering of
the variables in the VAR does not affect the Johansen's tests of the
cointegration rank, without further restrictions the use of high-dimensional
VARs in $\mathbf{w}_{t}$ does not distinguish between cointegration across
units as compared to cointegration between the variables specific to the
cross section units. Also, the condition $p/T=nm/T\rightarrow c\in (0,1]$ is
unlikely to be met when $m>1$ and $n$ is of the same order of magnitude as $%
T $.

Turning to the panel cointegration literature, most studies consider $I(1)$
variables with a single cointegrating vector where the direction of long-run
causality is known. These estimators are typically generalizations of the
time series procedures such as the panel Fully Modified OLS of 
\citeANP{Pedroni1996} (\citeyearNP{Pedroni1996}, \citeyearNP{Pedroni2001}, \citeyearNP{Pedroni2001ReStat})%
, the Pooled Mean Group (PMG) estimator of 
\citeN{PesaranShinSmith1999}%
,or the panel Dynamic OLS of 
\citeN{MarkSul2003}%
. There are panel generalizations of Johansen's approach, such as 
\citeN{GroenKleibergen2003}%
, and 
\citeN{LarssonLyhagen2007}%
, which can be used to test for the number of cointegrating relations and
estimate their parameters. These are based on a vector error correction
model, VECM, which can deal with multiple cointegrating vectors, but require 
$T$ to be large relative to $n.$ In their applications 
\citeN{LarssonLyhagen2007}
have $m=3,$ $n=4.$ 
\citeN{Breitung2005}
proposes a systems estimator, but that requires that every cross section
unit cointegrate. 
\citeN{ChudikPesaranSmith2023}
suggest system pooled mean group estimator for a single common long-run
relation coefficient $\theta $, that can handle any long-run causal ordering
and allow some units to fail to cointegrate, but again requires $T$ to be
large relative to $n$. \ 

A large number of other topics have been examined within the context of a
panel with a single cointegrating relation. These include: estimation with
I(1) latent factors: 
\citeN{BaiKaoNg2009}
and 
\citeN{KapetaniosPesaranYamagata2011}%
; structural breaks: 
\citeN{BanerjeeCarrion2024}
and 
\citeN{DitzenKaraviasWesterlund2025}%
; and non-linear effects: 
\citeN{deJongWagner2025}%
. Further details can be found in the surveys by 
\citeN{Breitung2008}
and 
\citeN{Choi2015a}
that also cover testing for cointegration using residuals (%
\citeANP{Westerlund2005}, \citeyearNP{Westerlund2005}%
), and second generation panel unit root tests allowing for cross section
dependence (%
\citeANP{Pesaran2007}, \citeyearNP{Pesaran2007}%
). As this brief overview indicates, none of the methods advanced in the
literature consider multiple long-run relations when $n>>T$.

\textbf{Outline of the paper}: The rest of the paper is set out as follows:
Section \ref{Prelim} sets out the panel data model and introduces the PME
estimator. Section \ref{Assum} introduces the assumptions and discusses the
identification conditions. Section \ref{LongRR} gives a formal description
of the PME estimator and some of its asymptotic properties. Section \ref%
{distribution}\ considers identification and derives the asymptotic
distribution of the exactly identified PME estimator. Section \ref{r_sel}
shows how $r_{0}$ can be estimated by eigenvalue thresholding. Section \ref%
{IntEffects} allows for interactive time effects. Section \ref{qCTchoices}
discusses the choice of the number of sub-sample averages, $q$, and how to
set the parameters of the thresholding estimator of $r_{0}$. Section \ref{MC}
provides Monte Carlo evidence on the small sample properties of the PME
estimators of $r_{0}$ and $\mathbf{\beta }_{j0}$, $j=1,2,...,r_{0}$. Section %
\ref{EA} discusses the empirical applications, and Section \ref{CON}
provides some concluding remarks. The proofs of the propositions and
theorems are provided in an appendix, with related lemmas given in a
supplement. This supplement also includes sub-sections on extensions of PME
to panels with interactive time effects, on how to implement the proposed
estimator for unbalanced panels, and gives details of the data generating
processes used in the Monte Carlo experiments, plus additional information
on data sources and the construction of the variables used in the empirical
applications.

\textbf{Notations:} Matrices are denoted by bold upper case letters and
vectors are denoted by bold lower case letters. All vectors are column
vectors. $\left\Vert \mathbf{x}\right\Vert $ denotes the Euclidean norm of a
vector $\mathbf{x}$. $rank\left( \mathbf{A}\right) $ denotes the column rank
of $\mathbf{A}$. $vec\left( \mathbf{A}\right) $ denotes vectorization of $%
\mathbf{A}$. $tr(\mathbf{A})$ denotes the trace of a square matrix $\mathbf{A%
}$. Eigenvalues of $m\times m$ symmetric positive semi-definite real matrix $%
\mathbf{A}$ sorted in ascending order are $0\leq \lambda _{1}(\mathbf{A}%
)\leq \lambda _{2}(\mathbf{A})\leq ...\leq \lambda _{m}(\mathbf{A})$. $%
\lVert \mathbf{A}\rVert $ is the spectral norm of $\mathbf{A}$. Small and
large finite positive constants that do not depend on sample sizes $n$ and $%
T $ are denoted by $\epsilon $ and $K$, respectively. These constants can
take different values at different instances in the paper. $T_{n}\approx
n^{d}$ if {there exist $n_{0}\geq 1$ and positive constants }$\epsilon ${\
and }$K${, such that }$\inf_{n\geq n_{0}}\left( T_{n}/n^{d}\right) \geq
\epsilon ${\ and $\sup_{n\geq n_{0}}\left( T_{n}/n^{d}\right) \leq K$.} {For
simplicity of exposition we omit subscript }$n$ and write $T\approx n^{d}$.
Convergence in probability and distribution are denoted by $\rightarrow _{p}$
and $\rightarrow _{d}$, respectively. In this paper $o_{p}\left( 1\right) $
is short for sequence of random variables, random vectors or random matrices
that converge to zero in probability as $n\rightarrow \infty $ for all
values of $d>1/2$. $\mathbf{A}_{n}=O_{p}\left( 1\right) $ if sequence $%
\mathbf{A}_{n}$ is bounded in probability. $a_{n}=O(b_{n})$\emph{\ }denotes
the deterministic sequence $\left\{ a_{n}\right\} $\ is at most of order $%
b_{n}$\emph{. }Equivalence of asymptotic distributions is denoted by $%
\overset{a}{\thicksim }$.

\section{Preliminaries\label{Prelim}}

Consider the following general linear model for $\mathbf{w}_{it}$ 
\begin{equation}
\mathbf{w}_{it}=\mathbf{a}_{i}+\mathbf{G}_{i}\mathbf{f}_{t}+\mathbf{C}_{i}%
\mathbf{s}_{it}+\mathbf{v}_{it},\text{ for }t=1,2,...,T;i=1,2,...,n,
\label{Grep}
\end{equation}%
where $\mathbf{w}_{it}$ is an $m\times 1$ vector of outcomes, $\mathbf{a}%
_{i} $ is $m\times 1$ vector of fixed effects, $\mathbf{f}_{t}$ is a vector
of stationary latent factors with associated loading matrices, $\mathbf{G}%
_{i}$. $\mathbf{s}_{it}$ is the partial sum process defined by 
\begin{equation}
\mathbf{s}_{it}=\mathbf{u}_{i1}+\mathbf{u}_{i2}+...+\mathbf{u}_{it},\text{
for }t\geq 1\text{, and }\mathbf{s}_{it}=\mathbf{0}\text{, for }t<1,
\label{sit}
\end{equation}%
$\mathbf{u}_{it}$ is independently distributed over $i$ and $t$ with mean
zero and the $m\times m$ positive definite matrix, $\mathbf{\Sigma }_{i}$, $%
\mathbf{C}_{i}$ is an $m\times m$ matrix of fixed coefficients, and $\mathbf{%
v}_{it}=\mathbf{C}_{i}^{\ast }(L)\mathbf{u}_{it}$, where $\mathbf{C}%
_{i}^{\ast }(L)=\sum_{\ell =0}^{\infty }\mathbf{C}_{i\ell }^{\ast }L^{\ell }$%
. This model covers many specifications of interest such as vector
autoregressions, error correction models, as well as first-differenced
stationary models. It allows for interactive time effects which reduce to
time effects under the so-called parallel trends assumption, namely setting $%
\mathbf{G}_{i}=\mathbf{G}$ for all $i$. In stacked form the model for all $n$
units can be written as $\mathbf{w}_{t}=\mathbf{a+Gf}_{t}+\mathbf{Cs}_{t}+%
\mathbf{C}^{\ast }(L)\mathbf{u}_{t},$ where $\mathbf{w}_{t}=(\mathbf{w}%
_{1t}^{\prime },\mathbf{w}_{2t}^{\prime },...,\mathbf{w}_{nt}^{\prime
})^{\prime },$ $\mathbf{a=}\left( \mathbf{a}_{1}^{\prime },\mathbf{a}%
_{2}^{\prime },...,\mathbf{a}_{n}^{\prime }\right) ^{\prime }$, $\mathbf{G=(G%
}_{1}^{\prime },\mathbf{G}_{2}^{\prime },...,\mathbf{G}_{n}^{\prime
})^{\prime }$, and $\mathbf{u}_{t}=(\mathbf{u}_{1t}^{\prime },\mathbf{u}%
_{2t}^{\prime },...,\mathbf{u}_{nt}^{\prime })^{\prime }$. Under our
specification $\mathbf{C}$ and $\mathbf{C}_{i}^{\ast }(L)$ are assumed to be
block-diagonal matrices with $\mathbf{C}_{i}$ and $\mathbf{C}_{i}^{\ast }(L)$
as their $i^{th}$ block, respectively. Such restrictions seem inevitable
when $n$ is large relative to $T$, and seems plausible considering that we
allow for cross-sectional dependence through the common factors, $\mathbf{f}%
_{t}$.

Assuming that $\mathbf{C}_{i}$ has rank $m-r_{0}>0$ for all $i$, we are
interested in estimating $r_{0}$, and the associated stationary linear
combinations defined by $\mathbf{\beta }_{j0}^{\prime }\mathbf{w}_{it},$ $%
j=1,2,...,r_{0},$ where $\mathbf{B}_{0}=\left( \mathbf{\beta }_{10},\mathbf{%
\beta }_{20},...,\mathbf{\beta }_{r_{0}0}\right) $ is the $m\times r_{0}$
matrix of long-run relations that are common across all $i$, and satisfies $%
\mathbf{B}_{0}^{\prime }\mathbf{C}_{i}=0$. We also consider estimation of
long-run relations subject to the exactly identifying restrictions that are
motivated by the theory. The estimator we propose involves splitting the
data for each unit into $q\geq 2$ sub-samples; taking time averages of these
sub-samples and forming a pooled demeaned covariance matrix we label $%
\mathbf{Q}_{\bar{w}\bar{w}}$. The eigenvalues of this matrix allow us to
estimate $r_{0},$ and the eigenvectors corresponding to the first $r_{0}$
eigenvalues provide estimates of $\mathbf{\beta }_{j0}$. But to simplify the
exposition and focus on the main contribution of the paper, initially we
abstract from the interactive time effects, but return to this complication
in Section \ref{IntEffects}, where we show that our analysis remains valid
so long as the latent factors are stationary.

It is possible to allow for non-linear features, such as GARCH and threshold
autoregressions, so long as the effects of shocks to $\mathbf{u}_{it}$ decay
exponentially fast. But to keep the theoretical analyses relatively simple,
we only consider the robustness of our estimation and testing strategies to
such non-linear effects using Monte Carlo experiments.

\section{Assumptions and identification conditions\label{Assum}}

We directly work with (\ref{Grep}) and make the following assumptions:

\begin{assumption}
\textbf{\ }\label{ASS1} The error terms, $\mathbf{u}_{it}$, are distributed
independently over $i=1,2,...,n$ and $t=1,2,....,T$ with $E(\mathbf{u}_{it})=%
\mathbf{0,}$ and the covariance matrix $\,E(\mathbf{u}_{it}\mathbf{u}%
_{it}^{\prime })=\mathbf{\Sigma }_{i}$, where $\mathbf{\Sigma }_{i}$ is a
positive definite matrix, $\inf_{i}\mathbf{\lambda }_{1}\left( \mathbf{%
\Sigma }_{i}\right) >\epsilon $, $\sup_{i}\mathbf{\lambda }_{m}\left( 
\mathbf{\Sigma }_{i}\right) <K$, and $\sup_{it}E\left\Vert \mathbf{u}%
_{it}\right\Vert ^{4+\epsilon }<K$, for some $\epsilon >0$.
\end{assumption}

\begin{assumption}
\label{ASS2} The coefficient matrices $\mathbf{C}_{i}$ and $\mathbf{C}%
_{i\ell }^{\ast }$, are non-stochastic constants such that $%
\sup_{i}\left\Vert \mathbf{C}_{i}\right\Vert <K$ and $\sup_{i}\left\Vert 
\mathbf{C}_{i\ell }^{\ast }\right\Vert <K\rho ^{\ell }$, where $\rho $ lies
in the range $0<\rho <1$. The $m\times m$ matrix $\mathbf{C}_{i}$ has rank $%
m-r_{0}$, for $i=1,2,...,n.$
\end{assumption}

\begin{assumption}
\label{ASS3} (a) Let 
\begin{equation}
\mathbf{\Psi }_{n}=n^{-1}\sum_{i=1}^{n}\mathbf{C}_{i}\mathbf{\Sigma }_{i}%
\mathbf{C}_{i}^{\prime }\text{, and }\mathbf{\Psi }=lim_{n\rightarrow \infty
}\mathbf{\Psi }_{n}.  \label{epsin1}
\end{equation}%
Then there exists $n_{0}$ such that for all $n>n_{0}$, $rank(\mathbf{\Psi }%
_{n})=rank(\mathbf{\Psi })=m-r_{0}>0$. (b) The orthonormalized eigenvectors
associated with the $r_{0}$ zero eigenvalues of $\ \left( \frac{q-1}{6q}%
\right) \mathbf{\Psi }_{n}$\textbf{\ }are denoted by $\mathbf{\beta }_{j0}$,
for $j=1,2,...,r_{0}$, and the orthonormalized eigenvectors associated with
the ordered non-zero eigenvalues of $\left( \frac{q-1}{6q}\right) \mathbf{%
\Psi }_{n}$, namely $\lambda _{r_{0}+1}\leq \lambda _{r_{0}+2}\leq ...\leq
\lambda _{m}$, by $\mathbf{\beta }_{j}$, for $r_{0}+1,r_{0}+2,...,m$.
Specifically%
\begin{equation}
\left( \frac{q-1}{6q}\right) \mathbf{\Psi }_{n}\mathbf{\beta }_{j0}=0\text{,
for }j=1,2,...,r_{0}\text{,}  \label{coin0}
\end{equation}%
and 
\begin{equation}
\left( \frac{q-1}{6q}\right) \mathbf{\Psi }_{n}\mathbf{\beta }_{j}=\lambda
_{j}\mathbf{\beta }_{j}\text{, for }j=r_{0}+1,r_{0}+2,...,m\text{.}
\label{noncoin}
\end{equation}
\end{assumption}

\begin{remark}
Under Assumptions \ref{ASS1}-\ref{ASS2} $\left\{ \mathbf{v}_{it}\right\} $
has absolute summable autocovariances, $\sum_{h=0}^{\infty
}\sup_{i}\left\Vert \mathbf{\Gamma }_{i}(h)\right\Vert <K$, where $\mathbf{%
\Gamma }_{i}(h)=E\left( \mathbf{v}_{it}\mathbf{v}_{i,t-h}^{\prime }\right) $
is the autocovariance function of $\mathbf{v}_{it}$. See Lemma \ref{Lacf} in
the supplement, Section \ref{A2}.
\end{remark}

\begin{remark}
Under Assumptions \ref{ASS3}, $\mathbf{\beta }_{j0}^{\prime }\mathbf{\Psi }%
_{n}\mathbf{\beta }_{j0}=0$, for $j=1,2,...,r_{0}$ and $\lambda _{j}=\left( 
\frac{q-1}{6q}\right) \mathbf{\beta }_{j}^{\prime }\mathbf{\Psi }_{n}\mathbf{%
\beta }_{j}>0$, for $j=r_{0}+1,r_{0}+2,...,m$. Nonzero eigenvalues\textbf{\ }%
$\lambda _{j}$, for $j=r_{0}+1,r_{0}+2,...,m$, and the corresponding
eigenvectors $\mathbf{\beta }_{j}$, for $j=r_{0}+1,r_{0}+2,...,m$, depend on 
$n$, but to simplify the notations we avoid using the subscript $n$.
\end{remark}

\begin{remark}
\label{RemarkPP}Under part (b) of Assumption \ref{ASS3} $\mathbf{\Psi }_{n}$
and $\mathbf{\Psi }$ can be written as $\mathbf{\Psi }_{n}=\mathbf{P}%
_{n}^{\prime }\mathbf{P}_{n}$, and $\mathbf{\Psi =P}^{\prime }\mathbf{P,}$
where $\mathbf{P}_{n}$ and $\mathbf{P}$ are $(m-r_{0})\times m$ full rank
matrices, with $rank\left( \mathbf{P}_{n}\right) =rank\left( \mathbf{P}%
\right) =m-r_{0}$, for all $n>n_{0}$.
\end{remark}

\begin{remark}
Condition $rank\left( \mathbf{C}_{i}\right) =m-r_{0}$ for all $i=1,2,...,n$
in Assumption \ref{ASS2} can be relaxed to allow for no stochastic trends
for some cross section units, so long as the rank condition $rank\left( 
\mathbf{\Psi }_{n}\right) =rank\left( \mathbf{\Psi }\right) =m-r_{0}$ holds.
Specifically, suppose that $\mathbf{C}_{i}=\mathbf{0}$ for $i=1,2,...,n_{1}$%
, but the cointegration rank condition holds for the remaining units. Then 
\begin{equation*}
\mathbf{\Psi }_{n}=n^{-1}\sum_{i=1}^{n}\mathbf{C}_{i}\mathbf{\Sigma }_{i}%
\mathbf{C}_{i}^{\prime }\text{ }=(1-\pi _{n})\left( \frac{1}{n-n_{1}}%
\sum_{i=n_{1}+1}^{n}\mathbf{C}_{i}\mathbf{\Sigma }_{i}\mathbf{C}_{i}^{\prime
}\right) ,\text{ }
\end{equation*}%
where $\pi _{n}=n_{1}/n$ is the proportion of units without stochastic
trends. Then the rank requirement continues to hold if $\pi _{n}>0$, and $%
\frac{1}{n-n_{1}}\sum_{i=n_{1}+1}^{n}\mathbf{C}_{i}\mathbf{\Sigma }_{i}%
\mathbf{C}_{i}^{\prime }$ tends to a matrix having rank $m-r_{0}$. But for
clarity of exposition we maintain Assumption \ref{ASS2} without loss of
generality.
\end{remark}

\section{Estimation of long-run relations\label{LongRR}}

\subsection{Introducing sub-sample time averages}

We base our estimation procedure on non-overlapping sub-sample time averages
of $\mathbf{w}_{it}$. For the ease of exposition, suppose the panel data
under consideration is balanced, $T$ is divisible by $q$, and consider $q$ $%
(\geq 2)$ non-overlapping time averages of equal length $T_{q}$ defined by%
\begin{equation}
\mathbf{\bar{w}}_{i\ell }=\frac{1}{T_{q}}\sum_{t=(\ell -1)T_{q}+1}^{\ell
T_{q}}\mathbf{w}_{it},\text{ for }\ell =1,2,...,q\text{,}  \label{la}
\end{equation}%
where $T_{q}=T/q$. To simplify the exposition we abstract from interactive
effects and apply the above time average operator to (\ref{Grep}) to obtain%
\footnote{%
Sections \ref{Aunb} and \ref{TSIE} of the supplement consider unbalanced
panels and models with interactive effects.} 
\begin{equation}
\mathbf{\bar{w}}_{i\ell }=\mathbf{a}_{i}+\mathbf{C}_{i}\mathbf{\bar{s}}%
_{i\ell }+\mathbf{\bar{v}}_{i\ell }\mathbf{,}\text{ for }\ell =1,2,...,q,
\label{wbaris}
\end{equation}%
where $\mathbf{\bar{v}}_{i\ell }=T_{q}^{-1}\sum_{t=(\ell -1)T_{q}+1}^{\ell
T_{q}}\mathbf{v}_{it}$ and $\mathbf{\bar{s}}_{i\ell
}=T_{q}^{-1}\sum_{t=(\ell -1)T_{q}+1}^{\ell T_{q}}\mathbf{s}_{it}.$ We now
use standard de-meaning procedure and eliminate $\mathbf{a}_{i}$ from (\ref%
{wbaris}) to obtain 
\begin{equation}
\mathbf{\bar{w}}_{i\ell }-\mathbf{\bar{w}}_{i\circ }=\mathbf{C}_{i}\left( 
\mathbf{\bar{s}}_{i\ell }-\mathbf{\bar{s}}_{i\circ }\right) +\left( \mathbf{%
\bar{v}}_{i\ell }-\mathbf{\bar{v}}_{i\circ }\right) ,\text{ for }\ell
=1,2,...,q,  \label{wbar}
\end{equation}%
where $\mathbf{\bar{w}}_{i\circ }=q^{-1}\sum_{\ell =1}^{q}\mathbf{\bar{w}}%
_{i\ell }$, and similarly $\mathbf{\bar{s}}_{i\circ }=q^{-1}\sum_{\ell
=1}^{q}\mathbf{\bar{s}}_{i\ell }$ and $\mathbf{\bar{v}}_{i\circ
}=q^{-1}\sum_{\ell =1}^{q}\mathbf{\bar{v}}_{i\ell }$. Consider now the $%
m\times m$ pooled sample covariance matrix 
\begin{equation}
\mathbf{Q}_{\bar{w}\bar{w}}=n^{-1}\sum_{i=1}^{n}\mathbf{Q}_{\bar{w}_{i}\bar{w%
}_{i}}  \label{Qbar}
\end{equation}%
where 
\begin{equation}
\mathbf{Q}_{\bar{w}_{i}\bar{w}_{i}}=T^{-1}q^{-1}\sum_{\ell =1}^{q}\left( 
\mathbf{\bar{w}}_{i\ell }-\mathbf{\bar{w}}_{i\circ }\right) \left( \mathbf{%
\bar{w}}_{i\ell }-\mathbf{\bar{w}}_{i\circ }\right) ^{\prime }.
\label{Qibar}
\end{equation}%
The limiting value of $\mathbf{Q}_{\bar{w}\bar{w}}$, as $n,T\rightarrow
\infty $, plays a critical role in our approach to estimation of long-run
relations. Using (\ref{wbar}) in (\ref{Qibar}) we first note that 
\begin{equation}
\mathbf{Q}_{\bar{w}_{i}\bar{w}_{i}}=\mathbf{C}_{i}\mathbf{Q}_{\bar{s}_{i}%
\bar{s}_{i}}\mathbf{C}_{i}^{\prime }+\mathbf{C}_{i}\mathbf{Q}_{\bar{s}_{i}%
\bar{v}_{i}}+\mathbf{Q}_{\bar{v}_{i}\bar{s}_{i}}^{\prime }\mathbf{C}%
_{i}^{\prime }+\mathbf{Q}_{\bar{v}_{i}\bar{v}_{i}},  \label{Qwiwi}
\end{equation}%
where $T\mathbf{Q}_{\bar{s}_{i}\bar{s}_{i}}=q^{-1}\sum_{\ell =1}^{q}\left( 
\mathbf{\bar{s}}_{i\ell }-\mathbf{\bar{s}}_{i\circ }\right) \left( \mathbf{%
\bar{s}}_{i\ell }-\mathbf{\bar{s}}_{i\circ }\right) ^{\prime },$ $T\mathbf{Q}%
_{\bar{s}_{i}\bar{v}_{i}}=q^{-1}\sum_{\ell =1}^{q}\left( \mathbf{\bar{s}}%
_{i\ell }-\mathbf{\bar{s}}_{i\circ }\right) \left( \mathbf{\bar{v}}_{i\ell }-%
\mathbf{\bar{v}}_{i\circ }\right) ^{\prime }=T\mathbf{Q}_{\bar{v}_{i}\bar{s}%
_{i}}^{\prime },$ and $T\mathbf{Q}_{\bar{v}_{i}\bar{v}_{i}}=q^{-1}\sum_{\ell
=1}^{q}\left( \mathbf{\bar{v}}_{i\ell }-\mathbf{\bar{v}}_{i\circ }\right)
\left( \mathbf{\bar{v}}_{i\ell }-\mathbf{\bar{v}}_{i\circ }\right) ^{\prime
} $. Since $\left\{ \mathbf{v}_{it}\right\} $ is covariance stationary with
absolute summable autocovariances and $\left\{ \mathbf{s}_{it}\right\} $ is
a partial sum process it then follows that $\mathbf{\bar{v}}_{i\ell }-%
\mathbf{\bar{v}}_{i\circ }=$ $O_{p}\left( T^{-1/2}\right) $, and $\mathbf{%
\bar{s}}_{i\ell }-\mathbf{\bar{s}}_{i\circ }=O_{p}(T^{1/2})$. Moreover, as
established in Lemma \ref{LQB}, 
\begin{equation}
\sup_{i}E\left\Vert \mathbf{Q}_{\bar{v}_{i}\bar{v}_{i}}\right\Vert =O\left(
T^{-2}\right) \text{, and }\sup_{i}E\left\Vert \mathbf{Q}_{\bar{s}_{i}\bar{v}%
_{i}}\right\Vert =O\left( T^{-1}\right) \text{.}  \label{sup_vv_sv}
\end{equation}

\subsection{Pooled minimum eigenvalue (PME) estimator}

Our proposed estimation procedure is based on eigenvalues and eigenvectors
of $\mathbf{Q}_{\bar{w}\bar{w}}$, defined by (\ref{Qbar}). Averaging $%
\mathbf{Q}_{\bar{w}_{i}\bar{w}_{i}}$ in (\ref{Qwiwi}) over all cross section
units now yields:%
\begin{equation}
\mathbf{Q}_{\bar{w}\bar{w}}=n^{-1}\sum_{i=1}^{n}\mathbf{C}_{i}\mathbf{Q}_{%
\bar{s}_{i}\bar{s}_{i}}\mathbf{C}_{i}^{\prime }+n^{-1}\sum_{i=1}^{n}\mathbf{C%
}_{i}\mathbf{Q}_{\bar{s}_{i}\bar{v}_{i}}+n^{-1}\sum_{i=1}^{n}\mathbf{Q}_{%
\bar{v}_{i}\bar{s}_{i}}^{\prime }\mathbf{C}_{i}^{\prime
}+n^{-1}\sum_{i=1}^{n}\mathbf{Q}_{\bar{v}_{i}\bar{v}_{i}}\text{.}
\label{qwbd}
\end{equation}%
The pooled minimum eigenvalue (PME) estimator of $\mathbf{\beta }_{j0},$ $%
j=1,2,...,r_{0}$, is given by the $j^{th}$ orthonormalized eigenvector of $%
\mathbf{Q}_{\bar{w}\bar{w}}$ , $\mathbf{\hat{\beta}}_{j}$, associated with
its $r_{0}$ smallest eigenvalues, $\hat{\lambda}_{1}\leq \hat{\lambda}%
_{2}\leq ...\leq \hat{\lambda}_{r_{0}}$. Specifically, for $j=1,2,....,m$, $%
\mathbf{Q}_{\bar{w}\bar{w}}\mathbf{\hat{\beta}}_{j}\mathbf{=}\hat{\lambda}%
_{j}\mathbf{\hat{\beta}}_{j}$\textbf{, }such that $\mathbf{\hat{\beta}}%
_{j}^{\prime }\mathbf{\hat{\beta}}_{j}=1$, and $\hat{\lambda}_{j}=\mathbf{%
\hat{\beta}}_{j}^{^{\prime }}\mathbf{Q}_{\bar{w}\bar{w}}\mathbf{\hat{\beta}}%
_{j}$. In matrix notations we have 
\begin{equation}
\mathbf{\hat{B}}=\left( \mathbf{\hat{\beta}}_{1},\mathbf{\hat{\beta}}%
_{2},...,\mathbf{\hat{\beta}}_{m}\right) \text{, }\left\Vert \mathbf{\hat{B}}%
\right\Vert =1,  \label{PME}
\end{equation}%
and the PME estimator of $\mathbf{B}_{0}$ is given by $\mathbf{\hat{B}}%
_{0}=\left( \mathbf{\hat{\beta}}_{1},\mathbf{\hat{\beta}}_{2},...,\mathbf{%
\hat{\beta}}_{r_{0}}\right) .$

\subsection{Consistency of the PME estimator\label{consistency}}

Under Assumption \ref{ASS1},$\ \mathbf{Q}_{\bar{s}_{i}\bar{s}_{i}}$ has the
following exact moment (established in Lemma \ref{LEQss}) 
\begin{equation}
E\left( \mathbf{Q}_{\bar{s}_{i}\bar{s}_{i}}\right) =\frac{(q-1)}{6}\left( 
\frac{1}{q}+\frac{1}{T^{2}}\right) \mathbf{\Sigma }_{i}.  \label{EQsisi}
\end{equation}%
Hence $n^{-1}\sum_{i=1}^{n}\mathbf{C}_{i}E\left( \mathbf{Q}_{\bar{s}_{i}\bar{%
s}_{i}}\right) \mathbf{C}_{i}^{\prime }=\frac{(q-1)}{6}\left( \frac{1}{q}+%
\frac{1}{T^{2}}\right) \mathbf{\Psi }_{n}\mathbf{,}$ where $\mathbf{\Psi }%
_{n}$ is defined by (\ref{epsin1}), and by Assumption \ref{ASS3} is assumed
to have rank $m-r_{0}>0$. Furthermore, using $\sup_{i}\left\Vert \mathbf{C}%
_{i}\right\Vert <K$ and (\ref{sup_vv_sv}) then 
\begin{eqnarray}
n^{-1}\sum_{i=1}^{n}\mathbf{C}_{i}\mathbf{Q}_{\bar{s}_{i}\bar{v}_{i}}
&=&O_{p}\left( T^{-1}\right) ,\text{ }n^{-1}\sum_{i=1}^{n}\mathbf{Q}_{\bar{v}%
_{i}\bar{s}_{i}}^{\prime }\mathbf{C}_{i}^{\prime }=O_{p}\left( T^{-1}\right)
,  \label{supQsv} \\
\text{ and \ }n^{-1}\sum_{i=1}^{n}\mathbf{Q}_{\bar{v}_{i}\bar{v}_{i}}
&=&O_{p}\left( n^{-1/2}T^{-2}\right) .  \notag
\end{eqnarray}%
Using the above results in (\ref{qwbd}) we have 
\begin{equation}
\mathbf{Q}_{\bar{w}\bar{w}}-\frac{(q-1)}{6q}\mathbf{\Psi }%
_{n}=n^{-1}\sum_{i=1}^{n}\mathbf{C}_{i}\left[ \mathbf{Q}_{\bar{s}_{i}\bar{s}%
_{i}}-E\left( \mathbf{Q}_{\bar{s}_{i}\bar{s}_{i}}\right) \right] \mathbf{C}%
_{i}^{\prime }+O_{p}\left( T^{-1}\right) .  \label{gapQww}
\end{equation}%
Further $n^{-1}\sum_{i=1}^{n}\mathbf{C}_{i}\left[ \mathbf{Q}_{\bar{s}_{i}%
\bar{s}_{i}}-E\left( \mathbf{Q}_{\bar{s}_{i}\bar{s}_{i}}\right) \right] 
\mathbf{C}_{i}^{\prime }=q^{-1}\sum_{\ell =1}^{q}\mathbf{G}_{\ell }-\mathbf{G%
}_{0}$, where 
\begin{equation*}
\mathbf{G}_{\ell }=n^{-1}\sum_{i=1}^{n}\mathbf{C}_{i}\left\{ \left( \frac{%
\mathbf{\bar{s}}_{i\ell }}{\sqrt{T}}\right) \left( \frac{\mathbf{\bar{s}}%
_{i\ell }}{\sqrt{T}}\right) ^{\prime }-E\left[ \left( \frac{\mathbf{\bar{s}}%
_{i\ell }}{\sqrt{T}}\right) \left( \frac{\mathbf{\bar{s}}_{i\ell }}{\sqrt{T}}%
\right) ^{\prime }\right] \right\} \mathbf{C}_{i}^{\prime }\text{ , \ for }%
\ell =1,2,...,q,
\end{equation*}%
and%
\begin{equation*}
\mathbf{G}_{0}=n^{-1}\sum_{i=1}^{n}\mathbf{C}_{i}\left\{ \left( \frac{%
\mathbf{\bar{s}}_{i\circ }}{\sqrt{T}}\right) \left( \frac{\mathbf{\bar{s}}%
_{i\circ }}{\sqrt{T}}\right) ^{\prime }-E\left[ \left( \frac{\mathbf{\bar{s}}%
_{i\circ }}{\sqrt{T}}\right) \left( \frac{\mathbf{\bar{s}}_{i\circ }}{\sqrt{T%
}}\right) ^{\prime }\right] \right\} \mathbf{C}_{i}^{\prime }.
\end{equation*}%
Under Assumptions \ref{ASS1} and \ref{ASS2}, $\mathbf{\bar{s}}_{i\ell }$ and 
$\mathbf{\bar{s}}_{i\circ }$ are cross-sectionally independent random
variables and $\sup_{i}\left\Vert \mathbf{C}_{i}\right\Vert <K$. Further, $%
T_{q}^{-1/2}\mathbf{\bar{s}}_{i\ell }$ and $T^{-1/2}\mathbf{\bar{s}}_{i\circ
}$ are scaled partial sums of $u_{it}\,\ $and tend to bounded random
variables. See, for example, result (d) of Proposition 17.1 in 
\citeN{Hamilton1994}%
. Therefore, $\mathbf{G}_{\ell }$ and $\mathbf{G}_{0}$ both converge at the
rate of $n^{-1/2}$ to their means that are zero, by construction. Namely $%
\mathbf{G}_{\ell }=O_{p}\left( n^{-1/2}\right) $ and $\mathbf{G}%
_{0}=O_{p}\left( n^{-1/2}\right) $. Hence%
\begin{equation}
n^{-1}\sum_{i=1}^{n}\mathbf{C}_{i}\left[ \mathbf{Q}_{\bar{s}_{i}\bar{s}%
_{i}}-E\left( \mathbf{Q}_{\bar{s}_{i}\bar{s}_{i}}\right) \right] \mathbf{C}%
_{i}^{\prime }=O_{p}\left( n^{-1/2}\right) ,  \label{CQss}
\end{equation}%
and using this result in (\ref{gapQww}) yields%
\begin{equation*}
\mathbf{Q}_{\bar{w}\bar{w}}-\frac{(q-1)}{6q}\mathbf{\Psi }_{n}=O_{p}\left(
n^{-1/2}\right) +O_{p}\left( T^{-1}\right) .
\end{equation*}%
Therefore, for a fixed $q\left( \geq 2\right) $, $\mathbf{Q}_{\bar{w}\bar{w}%
}\rightarrow _{p}\frac{(q-1)}{6q}\mathbf{\Psi }$, as $n,T\rightarrow \infty $
jointly such that $T_{n}\approx n^{d}$ and $d>0$, where $\mathbf{\Psi }%
=lim_{n\rightarrow \infty }\mathbf{\Psi }_{n}$. This result is formally
established in the following proposition.

\begin{proposition}
\label{Qwbar}Consider the panel data model for $\mathbf{w}_{it}$ given by (%
\ref{Grep}) without the interactive time effects ($\mathbf{G}_{i}=0$), and
suppose Assumptions \ref{ASS1} to \ref{ASS3} hold. Consider the $m\times m$
pooled sample covariance matrix $\mathbf{Q}_{\bar{w}\bar{w}}$ defined by (%
\ref{Qbar}). Then$\ $ for $q\geq 2$ and a fixed $m$ we have%
\begin{equation}
\mathbf{Q}_{\bar{w}\bar{w}}=\frac{(q-1)}{6q}\mathbf{\Psi }_{n}+O_{p}\left(
n^{-1/2}\right) +O_{p}\left( T^{-1}\right) ,  \label{Qbar2}
\end{equation}%
\begin{equation}
\mathbf{Q}_{\bar{w}\bar{w}}\mathbf{\beta }_{j0}=O_{p}\left( n^{-1/2}\right)
+O_{p}\left( T^{-1}\right) \text{, for }j=1,2,...,r_{0},  \label{QbarBeta}
\end{equation}%
and $\mathbf{Q}_{\bar{w}\bar{w}}\rightarrow _{p}\frac{(q-1)}{6q}\mathbf{\Psi 
}$, as $n,T\rightarrow \infty $ jointly such that $T_{n}\approx n^{d}$ and $%
d>0$, where $\mathbf{\Psi }$ and $\mathbf{\Psi }_{n}$ are defined in
Assumption \ref{ASS3}.
\end{proposition}

For a known $r_{0}$ the PME estimator of $\mathbf{B}_{0}$, is given by $%
\mathbf{\hat{B}}_{0}=\left( \mathbf{\hat{\beta}}_{1},\mathbf{\hat{\beta}}%
_{2},...,\mathbf{\hat{\beta}}_{r_{0}}\right) $, where $\mathbf{\hat{\beta}}%
_{j}$ , $j=1,2,...,m$ are the orthonormalized eigenvectors of $\mathbf{Q}_{%
\bar{w}\bar{w}}$, as set out below equation (\ref{qwbd}). Then 
\begin{equation}
\mathbf{\hat{B}}_{0}^{\prime }\mathbf{Q}_{\bar{w}\bar{w}}\mathbf{\hat{B}}%
_{0}=\frac{(q-1)}{6q}\mathbf{\hat{B}}_{0}^{\prime }\mathbf{\Psi \hat{B}}%
_{0}+O_{p}\left( n^{-1/2}\right) +O_{p}\left( T^{-1}\right) .  \label{a1}
\end{equation}%
and $\mathbf{\hat{B}}_{0}^{\prime }\mathbf{Q}_{\bar{w}\bar{w}}\mathbf{\hat{B}%
}_{0}$ is asymptotically minimized when $\mathbf{\hat{B}}_{0}^{\prime }%
\mathbf{\Psi \mathbf{\hat{B}}}_{0}=\mathbf{0}$, and this occurs if $\mathbf{%
\hat{B}}_{0}$ lies in the space spanned by the $r_{0}$ eigenvectors of $%
\mathbf{\Psi }$ that are associated with its $r_{0}$ zero eigenvalues,
namely if and only if $\mathbf{\hat{B}}=\mathbf{\mathring{B}}_{0}\mathbf{H}$%
, for some $r_{0}\times r_{0}$ non-singular rotation matrix, $\mathbf{H}$.
Hence, we have $\mathbf{\hat{B}}_{r}\mathbf{H}\rightarrow _{p}\mathbf{%
\mathring{B}}_{0}$ as $n,T\rightarrow \infty $ jointly such that $T\approx
n^{d}$ and $d>0$. A formal statement is provided in the following theorem
with proofs in the Appendix.

\begin{theorem}
\label{Tcons}Consider the panel data model for the $m\times 1$ vector $%
\mathbf{w}_{it}$ given by (\ref{Grep}) without the interactive time effects (%
$\mathbf{G}_{i}=0$), and suppose that Assumptions \ref{ASS1} to \ref{ASS3}
hold and the number of long-run relations, $r_{0}$, is known. Let $\mathbf{%
\hat{B}}_{0}=\left( \mathbf{\hat{\beta}}_{1},\mathbf{\hat{\beta}}_{2},...,%
\mathbf{\hat{\beta}}_{r_{0}}\right) $ be the $m\times r_{0}$ matrix formed
from the orthonormalized eigenvectors of $\mathbf{Q}_{\bar{w}\bar{w}}$,
defined by (\ref{Qbar}), associated with its $r_{0}$ smallest eigenvalues.
Then for a fixed $m$ and $q$ $(\geq 2),$ $\mathbf{\hat{B}}_{0}\mathbf{%
H\rightarrow }_{p}\mathbf{\mathring{B}}_{0}$ as $n,T\rightarrow \infty $
jointly such that $T_{n}\approx n^{d}$ and $d>0$, for any $r_{0}\times r_{0}$
non-singular matrix, $\mathbf{H}$.
\end{theorem}

\section{Identification and asymptotic distribution of PME estimator\label%
{distribution}}

We focus on the case of exact identification of the long-run relations,
noting that the estimation of $r_{0}$ is invariant on how exact
identification is achieved.

\subsection{Exact identifying conditions\label{eic}}

We assume there exist $r_{0}^{2}$ a \textit{priori given and theoretically
meaningful }exact identifying restrictions on $\mathbf{B}_{0}$ given by 
\begin{equation}
\mathbf{R}\mathbf{\mathring{B}}_{0}=\mathbf{(R}_{1},\mathbf{R}_{2})\left( 
\begin{array}{c}
\mathbf{\mathring{B}}_{0,1} \\ 
\mathbf{\mathring{B}}_{0,2}%
\end{array}%
\right) \mathbf{=A,}  \label{ExactRes}
\end{equation}%
where $\mathbf{\mathring{B}}_{0}$ is the $m\times r_{0}$ matrix of
identified long-run relations, $\mathbf{R}_{1}$,$\mathbf{R}_{2}$ and $%
\mathbf{A}$ are $r_{0}\times r_{0},$ $r_{0}\times (m-r_{0})$ and $%
r_{0}\times r_{0}$ matrices of known fixed constants, with $rank\left( 
\mathbf{A}\right) =rank\left( \mathbf{R}_{1}\right) =r_{0}<m$. \ Then it
follows that $\mathbf{H}=\left( \mathbf{RB}_{0}\right) ^{-1}\mathbf{A}$, and
the PME estimator of $\mathbf{\mathbf{\mathring{B}}_{0}}$ is given by%
\begin{equation}
\widehat{\mathbf{\mathbf{\mathring{B}}}}_{0}\mathbf{=\hat{B}}_{0}\left( 
\mathbf{R\hat{B}}_{0}\right) ^{-1}\mathbf{A,}  \label{BETAdot_hat}
\end{equation}%
where $\mathbf{\hat{B}}_{0}=\left( \mathbf{\hat{\beta}}_{10},\mathbf{\hat{%
\beta}}_{20},...,\mathbf{\hat{\beta}}_{r_{0},0}\right) $. The exact
identifying restrictions, (\ref{ExactRes}), will often take the form $%
\mathbf{\mathring{B}}_{0}=\left( \mathbf{I}_{r_{0}},\mathbf{\mathring{B}}%
_{0,2}^{\prime }\right) ^{\prime }$, where $\mathbf{\mathring{B}}_{0,1}=%
\mathbf{I}_{r_{0}}$\ is an identity matrix of order $r_{0}$. Without loss of
generality, we consider this formulation and denote $\mathbf{\mathring{B}}%
_{0,2}$ by $\mathbf{\Theta }$. Once $\mathbf{\Theta }$ is estimated it is
possible to estimate $\mathbf{\mathring{B}}_{0,1}$ under more general
restrictions as $\mathbf{\mathring{B}}_{0,1}=\mathbf{R}_{1}^{-1}\left( 
\mathbf{A-R}_{2}\mathbf{\Theta }\right) .$

\begin{proposition}
\label{PropID} Consider the $r_{0}^{2}$ exact identifying restrictions given
by (\ref{ExactRes}), and suppose $m\times r$ matrix of long-run relations $%
\mathbf{\mathring{B}}_{0}$ is normalized as $\mathbf{\mathring{B}}%
_{0}=\left( \mathbf{I}_{r_{0}},\mathbf{\Theta }^{\prime }\right) ^{\prime }$%
, and Assumption \ref{ASS3} holds. Partition $\mathbf{\Psi }$ conformably
with $\mathbf{\mathring{B}}_{0}$ as $\mathbf{\Psi =}\left( 
\begin{array}{cc}
\mathbf{\Psi }_{11} & \mathbf{\Psi }_{21}^{\prime } \\ 
\mathbf{\Psi }_{21} & \mathbf{\Psi }_{22}%
\end{array}%
\right) $. Then $\mathbf{I}_{r_{0}}+\mathbf{\Psi }_{11}$ and $\mathbf{\Psi }%
_{22}$ are respectively $r_{0}\times r_{0}$ and $(m-r_{0})\times (m-r_{0})$
positive definite matrices and $\mathbf{\Theta }$ is uniquely determined by $%
\mathbf{\Theta =-\Psi }_{22}^{-1}\mathbf{\Psi }_{21},$ subject to the
restrictions $\mathbf{\Psi }_{11}=\mathbf{\Psi }_{21}^{\prime }\mathbf{\Psi }%
_{22}^{-1}\mathbf{\Psi }_{21}$. A proof is provided in the Appendix.
\end{proposition}

\subsection{Asymptotic distribution\label{AsyDis}}

To derive the asymptotic distribution of $\widehat{\mathbf{\mathring{B}}}%
_{0}-\mathbf{\mathring{B}}_{0}$, note from (\ref{QwwD}) that%
\begin{equation}
\mathbf{Q}_{\bar{w}\bar{w}}\sqrt{n}T\left( \widehat{\mathbf{\mathring{B}}_{0}%
}-\mathbf{\mathring{B}}_{0}\right) =-\left( n^{-1/2}\sum_{i=1}^{n}T\mathbf{C}%
_{i}\mathbf{Q}_{\bar{s}_{i}\bar{v}_{i}}\right) \mathbf{\mathring{B}}_{0}-%
\frac{\sqrt{n}}{T}\left( n^{-1}\sum_{i=1}^{n}T^{2}\mathbf{Q}_{\bar{v}_{i}%
\bar{v}_{i}}\right) \mathbf{\mathring{B}}_{0}+O_{p}\left( T^{1}\right) 
\mathbf{.}  \label{qwb}
\end{equation}%
By Lemma \ref{L_sqvvs}, $n^{-1}\sum_{i=1}^{n}T^{2}\mathbf{Q}_{\bar{v}_{i}%
\bar{v}_{i}}=O_{p}\left( n^{-1/2}\right) $ and since $\left\Vert \mathbf{%
\mathring{B}}_{0}\right\Vert <K$, then%
\begin{equation}
\mathbf{Q}_{\bar{w}\bar{w}}\sqrt{n}T\left( \widehat{\mathbf{\mathring{B}}}%
_{0}-\mathbf{\mathring{B}}_{0}\right) =-\left( n^{-1/2}\sum_{i=1}^{n}T%
\mathbf{C}_{i}\mathbf{Q}_{\bar{s}_{i}\bar{v}_{i}}\right) \mathbf{\mathring{B}%
}_{0}+O_{p}\left( T^{-1}\right) .  \label{DQww1}
\end{equation}%
Further, write the first term as%
\begin{eqnarray*}
\left( n^{-1/2}\sum_{i=1}^{n}T\mathbf{C}_{i}\mathbf{Q}_{\bar{s}_{i}\bar{v}%
_{i}}\right) \mathbf{\mathring{B}}_{0} &=&\left( n^{-1/2}\sum_{i=1}^{n}\left[
T\mathbf{C}_{i}\mathbf{Q}_{\bar{s}_{i}\bar{v}_{i}}-\mathbf{C}_{i}E\left( T%
\mathbf{Q}_{\bar{s}_{i}\bar{v}_{i}}\right) \right] \right) \mathbf{\mathring{%
B}}_{0} \\
&&+\frac{\sqrt{n}}{T}\left( n^{-1}\sum_{i=1}^{n}\mathbf{C}_{i}E\left( T^{2}%
\mathbf{Q}_{\bar{s}_{i}\bar{v}_{i}}\right) \right) \mathbf{\mathring{B}}_{0},
\end{eqnarray*}%
and using $\sup_{i}\left\Vert E\left( \mathbf{Q}_{\bar{s}_{i}\bar{v}%
_{i}}\right) \right\Vert =O\left( T^{-2}\right) $ (established in Lemma \ref%
{LQB}))%
\begin{equation}
n^{-1}\sum_{i=1}^{n}\mathbf{C}_{i}E\left( T^{2}\mathbf{Q}_{\bar{s}_{i}\bar{v}%
_{i}}\right) \mathbf{\mathring{B}}_{0}=O(1).  \label{T2qsvb}
\end{equation}%
Using the above results in (\ref{DQww1}) we have 
\begin{equation}
\mathbf{Q}_{\bar{w}\bar{w}}\sqrt{n}T\left( \widehat{\mathbf{\mathring{B}}}%
_{0}-\mathbf{\mathring{B}}_{0}\right) =-n^{-1/2}\sum_{i=1}^{n}\mathbf{Z}%
_{i}+O_{p}\left( \frac{\sqrt{n}}{T}\right) ,  \label{DQww2}
\end{equation}%
where $\mathbf{Z}_{i}=\mathbf{C}_{i}\left[ T\mathbf{Q}_{\bar{s}_{i}\bar{v}%
_{i}}-E\left( T\mathbf{Q}_{\bar{s}_{i}\bar{v}_{i}}\right) \right] \mathbf{%
\mathring{B}}_{0}$. Recalling that $T\mathbf{Q}_{\bar{s}_{i}\bar{v}%
_{i}}=q^{-1}\sum_{\ell =1}^{q}\left( \mathbf{\bar{s}}_{i\ell }-\mathbf{\bar{s%
}}_{i\circ }\right) \left( \mathbf{\bar{v}}_{i\ell }-\mathbf{\bar{v}}%
_{i\circ }\right) ^{\prime }$, then $\mathbf{Z}_{i}$ can be written as%
\begin{equation}
\mathbf{Z}_{i}=\mathbf{C}_{i}\left[ q^{-1}\sum_{\ell =1}^{q}\left( \mathbf{%
\bar{s}}_{i\ell }-\mathbf{\bar{s}}_{i\circ }\right) \left( \mathbf{\bar{v}}%
_{i\ell }-\mathbf{\bar{v}}_{i\circ }\right) ^{\prime }\right] \mathbf{%
\mathring{B}}_{0}-\mathbf{C}_{i}\left\{ q^{-1}\sum_{\ell =1}^{q}E\left[
\left( \mathbf{\bar{s}}_{i\ell }-\mathbf{\bar{s}}_{i\circ }\right) \left( 
\mathbf{\bar{v}}_{i\ell }-\mathbf{\bar{v}}_{i\circ }\right) ^{\prime }\right]
\right\} \mathbf{\mathring{B}}_{0}\text{.}  \label{zi}
\end{equation}%
Writing (\ref{DQww2}) in vec form%
\begin{equation*}
\left( \mathbf{I}_{r}\mathbf{\otimes Q}_{\bar{w}\bar{w}}\right) \sqrt{n}T%
\text{ $vec$}\left( \widehat{\mathbf{\mathring{B}}}_{0}-\mathbf{\mathring{B}}%
_{0}\right) =n^{-1/2}\sum_{i=1}^{n}\left( \mathbf{\mathring{B}}_{0}^{\prime }%
\mathbf{\otimes \mathbf{C}}_{i}\right) \left[ \mathbf{\bar{\xi}}%
_{iq}-E\left( \mathbf{\bar{\xi}}_{iq}\right) \right] +O_{p}\left( \frac{%
\sqrt{n}}{T}\right) ,
\end{equation*}%
where $\mathbf{\bar{\xi}}_{iq}=q^{-1}\sum_{\ell =1}^{q}\mathbf{\xi }_{i\ell
} $, and $\mathbf{\xi }_{i\ell }=vec\left[ \left( \mathbf{\bar{s}}_{i\ell }-%
\mathbf{\bar{s}}_{i\circ }\right) \left( \mathbf{\bar{v}}_{i\ell }-\mathbf{%
\bar{v}}_{i\circ }\right) ^{\prime }\right] =\left( \mathbf{\bar{v}}_{i\ell
}-\mathbf{\bar{v}}_{i\circ }\right) \mathbf{\otimes }\left( \mathbf{\bar{s}}%
_{i\ell }-\mathbf{\bar{s}}_{i\circ }\right) $. Under Assumption \ref{ASS1}, $%
\mathbf{\bar{v}}_{i\ell }$, $\mathbf{\bar{s}}_{i\ell }$, $\mathbf{\bar{v}}%
_{i\circ }$ and $\mathbf{\bar{s}}_{i\circ }$ are distributed independently
over $i$. Hence, $\mathbf{\bar{\xi}}_{iq}$ is distributed independently over 
$i$. In addition, $\sup_{i}\left\Vert \left( \mathbf{\mathring{B}}%
_{0}^{\prime }\mathbf{\otimes \mathbf{C}}_{i}\right) \right\Vert =\left\Vert 
\mathbf{\mathring{B}}_{0}\right\Vert \sup_{i}\left\Vert \mathbf{C}%
_{i}\right\Vert <K$ and using Lemma \ref{Ld}, it follows that for $%
(n,T)\rightarrow \infty \,,\ $jointly such that $T\approx n^{d}$ and $d>0$,
we have 
\begin{equation}
n^{-1/2}\sum_{i=1}^{n}\left( \mathbf{\mathring{B}}_{0}^{\prime }\mathbf{%
\otimes \mathbf{C}}_{i}\right) \left[ \mathbf{\bar{\xi}}_{iq}-E\left( 
\mathbf{\bar{\xi}}_{iq}\right) \right] \rightarrow _{d}N\left( \mathbf{%
0,\Omega }_{q}\right) ,  \label{egzd}
\end{equation}%
where 
\begin{equation}
\mathbf{\Omega }_{q}=lim_{n,T\rightarrow \infty }\left[ n^{-1}\sum_{i=1}^{n}%
\left( \mathbf{\mathring{B}}_{0}^{\prime }\mathbf{\otimes \mathbf{C}}%
_{i}\right) \mathbf{\Omega }_{\bar{\xi}_{iq}}\left( \mathbf{\mathring{B}}_{0}%
\mathbf{\otimes \mathbf{C}}_{i}^{\prime }\right) \right] ,  \label{oz}
\end{equation}%
and%
\begin{equation}
\mathbf{\Omega }_{\bar{\xi}_{iq}}=Var\left( \mathbf{\bar{\xi}}_{iq}\right)
=q^{-2}\sum_{\ell =1}^{q}\sum_{\ell ^{\prime }=1}^{q}Cov\left( \mathbf{\xi }%
_{i\ell },\mathbf{\xi }_{i\ell ^{\prime }}\right) .  \label{Omegaegzibar}
\end{equation}

Using (\ref{egzd}) in (\ref{DQww2}) yields asymptotic normality of $\widehat{%
\mathbf{\mathring{B}}}$, which is formally established in the following
theorem.

\begin{theorem}
\label{Tb}Consider the panel data model for the $m\times 1$ vector $\mathbf{w%
}_{it},$ given by (\ref{Grep}) without interactive time effects ($\mathbf{G}%
_{i}=\mathbf{0}$). Suppose that Assumptions \ref{ASS1} to \ref{ASS3} hold, $%
m $ and $q$ $(\geq 2)$ are fixed integers, and the number of long-run
relations, $r_{0}$ ($m>r_{0}>0$) is known. Suppose further that the long-run
relations, $\mathbf{\mathring{B}}_{0}$, of interest are subject to the exact
identifying restrictions, $\mathbf{R}\mathbf{\mathring{B}}_{0}\mathbf{=A}$%
\textbf{, }given by (\ref{ExactRes}), and consider the PME estimator of $%
\mathbf{\mathring{B}}_{0}$\textbf{\ }given by 
\begin{equation*}
\widehat{\mathbf{\mathbf{\mathring{B}}}}_{0}\mathbf{=\hat{B}}_{0}\left( 
\mathbf{R\hat{B}}_{0}\right) ^{-1}\mathbf{A,}
\end{equation*}%
where $\mathbf{\hat{B}}_{0}=\left( \mathbf{\hat{\beta}}_{10},\mathbf{\hat{%
\beta}}_{20},...,\mathbf{\hat{\beta}}_{r_{0},0}\right) $ are the first $%
r_{0} $ orthonormalized eigenvectors of $\mathbf{Q}_{\bar{w}\bar{w}}$
defined by (\ref{Qbar}). Then%
\begin{equation}
\sqrt{n}T\left( \mathbf{I}_{r_{0}}\mathbf{\otimes Q}_{\bar{w}\bar{w}}\right) 
\text{$vec$}\left( \widehat{\mathbf{\mathring{B}}}_{0}-\mathbf{\mathring{B}}%
_{0}\right) \rightarrow _{d}N\left( \mathbf{0,\Omega }_{q}\right) ,
\label{dist}
\end{equation}%
as $\left( n,T\right) \rightarrow \infty $, jointly such that $T\approx
n^{d} $ for $d>1/2$, where $\mathbf{\Omega }_{q}$ is defined by (\ref{oz}),
and $\mathbf{Q}_{\bar{w}\bar{w}}\rightarrow _{p}\frac{(q-1)}{6q}\mathbf{\Psi 
}$. (see (\ref{Qbar2})). A proof is provided in the Appendix.
\end{theorem}

Theorem \ref{Tb} can be readily used to obtain asymptotic distribution of
any linear combination of $\widehat{\mathbf{\mathring{B}}}_{0}$. One notable
case of interest is to consider the exact identifying restrictions $\mathbf{%
\mathring{B}}_{0,1}=\mathbf{I}_{r_{0}}$ discussed above. Under these
restrictions $\mathbf{\mathring{B}}_{0,1}=\widehat{\mathbf{\mathring{B}}}%
_{0,1}=\mathbf{I}_{r_{0}}$, and $\widehat{\mathbf{\mathring{B}}}_{0}-\mathbf{%
\mathring{B}}_{0}=\left( 
\begin{array}{cc}
\mathbf{0} & \mathbf{\hat{\Theta}}^{\prime }-\mathbf{\Theta }_{0}^{\prime }%
\end{array}%
\right) $. Partitioning $\mathbf{Q}_{\bar{w}\bar{w}}$ and $\mathbf{\Omega }%
_{z}$ accordingly, we have%
\begin{equation*}
\sqrt{n}T\mathbf{Q}_{\bar{w}\bar{w}}\left( \widehat{\mathbf{\mathring{B}}}%
_{0}-\mathbf{\mathring{B}}_{0}\right) =\left( 
\begin{array}{cc}
\mathbf{Q}_{11,\bar{w}\bar{w}} & \mathbf{Q}_{12,\bar{w}\bar{w}} \\ 
\mathbf{Q}_{21,\bar{w}\bar{w}} & \mathbf{Q}_{22,\bar{w}\bar{w}}%
\end{array}%
\right) \left( 
\begin{array}{c}
\mathbf{0} \\ 
\sqrt{n}T\left( \mathbf{\hat{\Theta}}-\mathbf{\Theta }_{0}\right)%
\end{array}%
\right) ,
\end{equation*}%
where $\mathbf{Q}_{22,\bar{w}\bar{w}}\rightarrow _{p}\frac{(q-1)}{6q}\mathbf{%
\Psi }_{22}$, and $\mathbf{\Psi }_{22}$ is the $(m-r_{0})\times (m-r_{0})$
lower right block of $\mathbf{\Psi }$ which is positive definite (see
Proposition \ref{PropID}). We obtain the following corollary.

\begin{corollary}
Suppose assumptions of Theorem \ref{Tb} hold and consider the exact
identifying restrictions $\mathbf{\mathring{B}}_{0,1}=\widehat{\mathbf{%
\mathring{B}}}_{0,1}=\mathbf{I}_{r_{0}}$. Suppose $r_{0}$ is known, $q\geq 2$%
, and let $\mathbf{\hat{\Theta}}$ and $\mathbf{\Theta }_{0}$ be the lower $%
\left( m-r_{0}\right) \times r_{0}$ block of $\widehat{\mathbf{\mathring{B}}}%
_{0}$ and $\mathbf{\mathring{B}}_{0}$, respectively. Then, 
\begin{equation}
\sqrt{n}T\text{ $vec$}\left( \mathbf{\hat{\Theta}}-\mathbf{\Theta }%
_{0}\right) \rightarrow _{d}N\left( \mathbf{0,\Omega }_{\theta q}\right) ,
\label{Distheta}
\end{equation}%
as $n,T\rightarrow \infty $, jointly such that $T\approx n^{d}$ for $d>1/2$
,where%
\begin{equation}
\mathbf{\Omega }_{\theta q}=\left( \frac{6q}{q-1}\right) ^{2}\left( \mathbf{I%
}_{r}\mathbf{\otimes \Psi }_{22}^{-1}\right) \mathbf{\Omega }_{q,22}\left( 
\mathbf{I}_{r}\mathbf{\otimes \Psi }_{22}^{-1}\right) \text{,}
\label{Omegatheta}
\end{equation}%
$\mathbf{\Omega }_{q,22}$ is the $\left( m-r_{0}\right) \times \left(
m-r_{0}\right) $ lower right block of $\mathbf{\Omega }_{q}$ given by (\ref%
{oz}), and $\mathbf{\Psi }_{22}$ is the $(m-r_{0})\times (m-r_{0})$ lower
right block of $\mathbf{\Psi }$.
\end{corollary}

\subsection{Estimation of the asymptotic covariance of $\mathbf{\hat{\Theta}}
$}

To consistently estimate $\mathbf{\Omega }_{\theta q}$ we need a consistent
estimator of $\mathbf{\Omega }_{q,22}$, since $\left[ (q-1)/6q\right] 
\mathbf{\Psi }_{22}$ can be consistently estimated by $\mathbf{Q}_{\bar{w}%
\bar{w},22}$. Consider $\mathbf{\Omega }_{q}$ given by (\ref{oz}). Using (%
\ref{wbar}) note that%
\begin{eqnarray*}
\left( \mathbf{\bar{w}}_{i\ell }-\mathbf{\bar{w}}_{i\circ }\right) \left( 
\mathbf{\bar{w}}_{i\ell }-\mathbf{\bar{w}}_{i\circ }\right) ^{\prime }%
\mathbf{\beta }_{0} &=&\left[ \mathbf{C}_{i}\left( \mathbf{\bar{s}}_{i\ell }-%
\mathbf{\bar{s}}_{i\circ }\right) +\left( \mathbf{\bar{v}}_{i\ell }-\mathbf{%
\bar{v}}_{i\circ }\right) \right] \left[ \left( \mathbf{\bar{s}}_{i\ell }-%
\mathbf{\bar{s}}_{i\circ }\right) ^{\prime }\mathbf{C}_{i}^{\prime }+\left( 
\mathbf{\bar{v}}_{i\ell }-\mathbf{\bar{v}}_{i\circ }\right) ^{\prime }\right]
\mathbf{\mathring{B}}_{0}\text{,} \\
&=&\mathbf{C}_{i}\left( \mathbf{\bar{s}}_{i\ell }-\mathbf{\bar{s}}_{i\circ
}\right) \left( \mathbf{\bar{v}}_{i\ell }-\mathbf{\bar{v}}_{i\circ }\right)
^{\prime }\mathbf{\mathring{B}}_{0}+\left( \mathbf{\bar{v}}_{i\ell }-\mathbf{%
\bar{v}}_{i\circ }\right) \left( \mathbf{\bar{v}}_{i\ell }-\mathbf{\bar{v}}%
_{i\circ }\right) ^{\prime }\mathbf{\mathring{B}}_{0}\text{.}
\end{eqnarray*}%
Let $\mathbf{\zeta }_{i\ell }=vec\left[ \left( \mathbf{\bar{w}}_{i\ell }-%
\mathbf{\bar{w}}_{i\circ }\right) \left( \mathbf{\bar{w}}_{i\ell }-\mathbf{%
\bar{w}}_{i\circ }\right) ^{\prime }\mathbf{\mathring{B}}_{0}\right] $, and $%
\mathbf{\eta }_{i\ell }=vec\left[ \left( \mathbf{\bar{v}}_{i\ell }-\mathbf{%
\bar{v}}_{i\circ }\right) \left( \mathbf{\bar{v}}_{i\ell }-\mathbf{\bar{v}}%
_{i\circ }\right) ^{\prime }\right] $, where recall that $\mathbf{\eta }%
_{i\ell }=O_{p}(T^{-1})$ uniformly in $i$ and $\ell $. Then $\mathbf{\zeta }%
_{i\ell }=\left( \mathbf{\mathring{B}}_{0}^{\prime }\mathbf{\otimes \mathbf{C%
}}_{i}\right) \mathbf{\xi }_{i\ell }+\left( \mathbf{\mathring{B}}%
_{0}^{\prime }\mathbf{\otimes I}_{m}\right) \mathbf{\eta }_{i\ell }$, and%
\begin{equation*}
\mathbf{\bar{\zeta}}_{iq}=q^{-1}\sum_{\ell =1}^{q}\mathbf{\zeta }_{i\ell
}=\left( \mathbf{\mathring{B}}_{0}^{\prime }\mathbf{\otimes \mathbf{C}}%
_{i}\right) \left( q^{-1}\sum_{\ell =1}^{q}\mathbf{\xi }_{i\ell }\right)
+\left( \mathbf{\mathring{B}}_{0}^{\prime }\mathbf{\otimes I}_{m}\right)
\left( q^{-1}\sum_{\ell =1}^{q}\mathbf{\eta }_{i\ell }\right) =\left( 
\mathbf{\mathring{B}}_{0}^{\prime }\mathbf{\otimes \mathbf{C}}_{i}\right) 
\mathbf{\bar{\xi}}_{iq}+O_{p}(T^{-1}),
\end{equation*}%
where $\mathbf{\bar{\xi}}_{iq}=q^{-1}\sum_{\ell =1}^{q}\mathbf{\xi }_{i\ell
} $, as before. It follows 
\begin{equation*}
n^{-1}\sum_{i=1}^{n}\mathbf{\bar{\zeta}}_{iq}\mathbf{\bar{\zeta}}%
_{iq}^{\prime }=n^{-1}\sum_{i=1}^{n}\left( \mathbf{\mathring{B}}_{0}^{\prime
}\mathbf{\otimes \mathbf{C}}_{i}\right) \mathbf{\bar{\xi}}_{iq}\mathbf{\bar{%
\xi}}_{iq}^{\prime }\left( \mathbf{\mathring{B}}_{0}\mathbf{\otimes \mathbf{C%
}}_{i}^{\prime }\right) +O_{p}(T^{-1}),
\end{equation*}%
and $E\left( \mathbf{\bar{\xi}}_{iq}\right) =O(T^{-1})$. Hence, as $%
n,T\rightarrow \infty $ jointly such that $T\approx n^{d}$ for $d>1/2$,%
\begin{equation*}
n^{-1}\sum_{i=1}^{n}\mathbf{\bar{\zeta}}_{iq}\mathbf{\bar{\zeta}}%
_{iq}^{\prime }\rightarrow _{p}\text{ }lim_{n,T}\left[ n^{-1}\sum_{i=1}^{n}%
\left( \mathbf{\mathring{B}}_{0}^{\prime }\mathbf{\otimes \mathbf{C}}%
_{i}\right) E\left( \mathbf{\bar{\xi}}_{iq}\mathbf{\bar{\xi}}_{iq}^{\prime
}\right) \left( \mathbf{\mathring{B}}_{0}\mathbf{\otimes \mathbf{C}}%
_{i}^{\prime }\right) \right] =\mathbf{\Omega }_{q}\text{.}
\end{equation*}%
Therefore, $\mathbf{\Omega }_{q}$ can be consistently estimated by the $%
m^{2}\times m^{2}$ matrix 
\begin{equation*}
\mathbf{\hat{\Omega}}_{q}=n^{-1}\sum_{i=1}^{n}\widehat{\mathbf{\bar{\zeta}}}%
_{iq}\widehat{\mathbf{\bar{\zeta}}}_{iq}^{\prime }=\left( 
\begin{array}{cc}
\mathbf{\hat{\Omega}}_{q,11} & \mathbf{\hat{\Omega}}_{q,12} \\ 
\mathbf{\hat{\Omega}}_{q,21} & \mathbf{\hat{\Omega}}_{q,22}%
\end{array}%
\right) ,
\end{equation*}%
where%
\begin{equation*}
\widehat{\mathbf{\bar{\zeta}}}_{iq}=q^{-1}\sum_{\ell =1}^{q}vec\left[ \left( 
\mathbf{\bar{w}}_{i\ell }-\mathbf{\bar{w}}_{i\circ }\right) \left( \mathbf{%
\bar{w}}_{i\ell }-\mathbf{\bar{w}}_{i\circ }\right) ^{\prime }\widehat{%
\mathbf{\mathring{B}}_{0}}\right] =q^{-1}\sum_{\ell =1}^{q}\left[ \widehat{%
\mathbf{\mathring{B}}}_{0}^{\prime }\left( \mathbf{\bar{w}}_{i\ell }-\mathbf{%
\bar{w}}_{i\circ }\right) \mathbf{\otimes I}_{m}\right] \left( \mathbf{\bar{w%
}}_{i\ell }-\mathbf{\bar{w}}_{i\circ }\right) \text{.}
\end{equation*}%
Let $\widehat{\mathbf{\mathring{B}}_{0}}^{\prime }\left( \mathbf{\bar{w}}%
_{i\ell }-\mathbf{\bar{w}}_{i\circ }\right) =\widehat{\mathbf{\bar{E}}}%
_{i\ell }$, the $r\times 1$ vector of error corrections for the sub-sample $%
\ell $, then 
\begin{equation}
\mathbf{\hat{\Omega}}_{q}=n^{-1}\sum_{i=1}^{n}q^{-2}\sum_{\ell
=1}^{q}\sum_{\ell ^{\prime }=1}^{q}\left( \widehat{\mathbf{\bar{E}}}_{i\ell }%
\mathbf{\otimes I}_{m}\right) \left( \mathbf{\bar{w}}_{i\ell }-\mathbf{\bar{w%
}}_{i\circ }\right) \left( \mathbf{\bar{w}}_{i\ell ^{\prime }}-\mathbf{\bar{w%
}}_{i\circ }\right) ^{\prime }\left( \widehat{\mathbf{\bar{E}}}_{i\ell
^{\prime }}^{\prime }\mathbf{\otimes I}_{m}\right) .  \label{Omegaz}
\end{equation}%
Using the above results we now have 
\begin{equation}
\widehat{Var\left[ vec\left( \mathbf{\hat{\Theta}}\right) \right] }=\frac{1}{%
nT^{2}}\mathbf{Q}_{\bar{w}\bar{w},22}^{-1}\mathbf{\hat{\Omega}}_{q,22}%
\mathbf{Q}_{\bar{w}\bar{w},22}^{-1}.  \label{VarHatTheta}
\end{equation}

When $r_{0}=1$, estimate of the error correction term $\widehat{\mathbf{%
\mathring{B}}}_{0}^{\prime }\left( \mathbf{\bar{w}}_{i\ell }-\mathbf{\bar{w}}%
_{i\circ }\right) =\widehat{\bar{e}}_{i\ell }$ is a scalar and we can write $%
\widehat{\mathbf{\bar{\zeta}}}_{iq}=q^{-1}\sum_{\ell =1}^{q}\left( \mathbf{%
\bar{w}}_{i\ell }-\mathbf{\bar{w}}_{i\circ }\right) \widehat{\bar{e}}_{i\ell
}$, and 
\begin{eqnarray*}
\mathbf{\hat{\Omega}}_{q} &=&n^{-1}\sum_{i=1}^{n}\left[ q^{-1}\sum_{\ell
=1}^{q}\left( \mathbf{\bar{w}}_{i\ell }-\mathbf{\bar{w}}_{i\circ }\right) 
\widehat{\bar{e}}_{i\ell }\right] \left[ q^{-1}\sum_{\ell =1}^{q}\left( 
\mathbf{\bar{w}}_{i\ell }-\mathbf{\bar{w}}_{i\circ }\right) ^{\prime }%
\widehat{\bar{e}}_{i\ell }\right] \text{,} \\
&=&n^{-1}\sum_{i=1}^{n}\left[ q^{-2}\sum_{\ell =1}^{q}\sum_{\ell ^{\prime
}=1}^{q}\left( \mathbf{\bar{w}}_{i\ell }-\mathbf{\bar{w}}_{i\circ }\right)
\left( \mathbf{\bar{w}}_{i\ell ^{\prime }}-\mathbf{\bar{w}}_{i\circ }\right)
^{\prime }\widehat{\bar{e}}_{i\ell }\widehat{\bar{e}}_{i\ell ^{\prime }}%
\right] \text{,}
\end{eqnarray*}%
which resembles the robust covariance matrix estimator that arises in
estimation of panel data models with short $T$ and large $n$. Here $q$ plays
the role of $T$.

\section{Estimation of $r_{0}$ by eigenvalue thresholding\label{r_sel}}

Under Assumption \ref{ASS3}, the true number of common long-run relations, $%
r_{0}$, is defined by $rank(\mathbf{\Psi }_{n})=m-r_{0}>0$, where $\mathbf{%
\Psi }_{n}=n^{-1}\sum_{i=1}^{n}\mathbf{C}_{i}\mathbf{\Sigma }_{i}\mathbf{C}%
_{i}^{\prime }$. Subject to this condition, $\mathbf{\Psi }_{n}\mathbf{\beta 
}_{j,0}=0$, for $j=1,2,...,r_{0}$, where $\mathbf{\beta }_{j,0}$ is the $%
j^{th}$ long-run relation ($j\leq r_{0}$). The $m\times r_{0}$ matrix of
long-run relations is denoted by $\mathbf{B}_{0}$. It is also worth bearing
in mind that under Assumption \ref{ASS3}, $\mathbf{\Psi }_{n}\mathbf{\beta }%
_{j}\neq \mathbf{0}$, for $j=r_{0}+1,r_{0}+2,...m$, namely cannot be spanned
by the $r_{0}$ columns of $\mathbf{B}_{0}$. See (\ref{coin0}) and (\ref%
{noncoin}). There is a clear shift in the ordered eigenvalues of $\mathbf{%
\Psi }_{n}$ from $\lambda _{r_{0}}=$ $0$ to $\lambda _{r_{0}+1}>0$, which
allows us to propose a thresholding estimator of $r_{0}$ applied to the
eigenvalues of $\mathbf{Q}_{\bar{w}\bar{w}}$, noting that under our
assumptions $\mathbf{Q}_{\bar{w}\bar{w}}$ tends to $\left( \frac{q-1}{6q}%
\right) \mathbf{\Psi }$ as $n$ and $T\rightarrow \infty $. See result (\ref%
{Qbar2}) of Proposition \ref{Qwbar}. Such an estimator can be written
conveniently as 
\begin{equation}
\hat{r}=\sum_{j=1}^{m}\mathcal{I}\left( \hat{\lambda}_{j}<C_{T}\right) ,
\label{rhat}
\end{equation}%
where $\hat{\lambda}_{1}\leq \hat{\lambda}_{2}\leq ....\leq \hat{\lambda}%
_{m} $ are ordered eigenvalues of $\mathbf{Q}_{\bar{w}\bar{w}}$, and $%
\mathcal{I}\left( \mathcal{A}\right) =1$ if $\mathcal{A}$ is true or zero
otherwise, and $C_{T}=KT^{-\delta }$, for some $\delta >0$. This estimator
is invariant to the ordering of the eigenvalues, but using the ordering
helps with the exposition and the rationale behind the proofs.

To establish the consistency of $\hat{r}$ as an estimator of $r_{0}$, we
first note that (\ref{rhat}) can be written equivalently as%
\begin{equation}
\hat{r}-r_{0}=-\sum_{j=1}^{r_{0}}\mathcal{I}\left( \hat{\lambda}_{j}\geq
C_{T}\right) +\sum_{j=r_{0}+1}^{m}\mathcal{I}\left( \hat{\lambda}%
_{j}<C_{T}\right) ,  \label{rhatgap}
\end{equation}%
which in turn yields: 
\begin{equation}
E\left\vert \hat{r}-r_{0}\right\vert \leq \sum_{j=1}^{r_{0}}\Pr \left( \hat{%
\lambda}_{j}\geq C_{T}\right) +\sum_{j=r_{0}+1}^{m}\Pr \left( \hat{\lambda}%
_{j}<C_{T}\right) .  \label{egap}
\end{equation}%
Again noting the ordering of the eigenvalues, $\Pr \left( \hat{\lambda}%
_{j}\geq C_{T}\right) \leq \Pr \left( \hat{\lambda}_{1}\geq C_{T}\right) $,
for $j=2,3,...,r_{0},$ and $\Pr \left( \hat{\lambda}_{r_{0}+1}<C_{T}\right)
\geq \Pr \left( \hat{\lambda}_{j}<C_{T}\right) $, for $%
j=r_{0}+2,r_{0}+3,...,m$. Using these results in (\ref{egap}) we have 
\begin{equation}
E\left\vert \hat{r}-r_{0}\right\vert \leq r_{0}\Pr \left( \hat{\lambda}%
_{1}\geq C_{T}\right) +(m-r_{0})\Pr \left( \hat{\lambda}_{r_{0}+1}<C_{T}%
\right) .  \label{Eup}
\end{equation}%
Similarly,%
\begin{eqnarray}
E\left( \hat{r}-r_{0}\right) ^{2} &\leq &r_{0}^{2}\Pr \left( \hat{\lambda}%
_{1}\geq C_{T}\right) +(m-r_{0})^{2}\Pr \left( \hat{\lambda}%
_{r_{0}+1}<C_{T}\right)  \label{MSErhat} \\
&&+2r_{0}(m-r_{0})\sqrt{\Pr \left( \hat{\lambda}_{1}\geq C_{T}\right) \Pr
\left( \hat{\lambda}_{r_{0}+1}<C_{T}\right) }\text{.}  \notag
\end{eqnarray}%
Also, by Markov inequality there exists $\epsilon >0$ such that 
\begin{equation}
\Pr \left( \left\vert \hat{r}-r_{0}\right\vert >\epsilon \right) \leq
(r_{0}/\epsilon )\Pr \left( \hat{\lambda}_{1}\geq C_{T}\right) +\left[
(m-r_{0})/\epsilon \right] \Pr \left( \hat{\lambda}_{r_{0}+1}<C_{T}\right) ,
\label{Prrhat}
\end{equation}%
Again by Markov inequality $\Pr \left( \hat{\lambda}_{1}\geq C_{T}\right)
\leq C_{T}^{-1}E\left( \hat{\lambda}_{1}\right) $, and using result (\ref{el}%
) established in Lemma \ref{LQb}, we have 
\begin{equation}
\Pr \left( \hat{\lambda}_{1}\geq C_{T}\right) =O\left(
C_{T}^{-1}n^{-1/2}T^{-2}\right) .  \label{PrL1}
\end{equation}%
Consider now $\Pr \left( \hat{\lambda}_{r_{0}+1}<C_{T}\right) $, and recall
that 
\begin{equation}
\mathbf{Q}_{\bar{w}\bar{w}}\mathbf{\hat{\beta}}_{j}=\hat{\lambda}_{j}\mathbf{%
\hat{\beta}}_{j}\text{, and }\hat{\lambda}_{j}=\mathbf{\hat{\beta}}%
_{j}^{\prime }\mathbf{Q}_{\bar{w}\bar{w}}\mathbf{\hat{\beta}}_{j}\text{, for 
}j=r_{0}+1,r_{0}+2,...,m,  \label{ljhat}
\end{equation}%
with associated population values given by (see Assumption \ref{ASS3}).%
\begin{equation}
\frac{(q-1)}{6q}\mathbf{\Psi }_{n}\mathbf{\beta }_{j}=\lambda _{j}\mathbf{%
\beta }_{j}\text{, and }\frac{(q-1)}{6q}\mathbf{\beta }_{j}^{\prime }\mathbf{%
\Psi }_{n}\mathbf{\beta }_{j}=\lambda _{j}\text{\textbf{, }for }%
j=r_{0}+1,r_{0}+2,...,m.  \label{lj}
\end{equation}%
where $\mathbf{\Psi }_{n}=n^{-1}\sum_{i=1}^{n}\mathbf{C}_{i}\mathbf{\Sigma }%
_{i}\mathbf{C}_{i}^{\prime }\succeq 0$, and $\mathbf{\beta }_{j}^{\prime }%
\mathbf{\beta }_{j}=1$. Furthermore, $\lambda _{j}>0$ and $\mathbf{\Psi }_{n}%
\mathbf{\beta }_{j}\neq 0,$ for $j>r_{0}$. But using (\ref{qwbd}) and
results established in Lemma \ref{LQB} and Proposition \ref{Qwbar} we have $%
E\left( \mathbf{Q}_{\bar{w}\bar{w}}\right) =\frac{(q-1)}{6q}\mathbf{\Psi }%
_{n}+O\left( T^{-2}\right) $, and 
\begin{equation*}
\mathbf{\tilde{Q}}_{\bar{w}\bar{w}}=\mathbf{Q}_{\bar{w}\bar{w}}-E\left( 
\mathbf{Q}_{\bar{w}\bar{w}}\right) =n^{-1}\sum_{i=1}^{n}\mathbf{C}_{i}\left[ 
\mathbf{Q}_{\bar{s}_{i}\bar{s}_{i}}-E\left( \mathbf{Q}_{\bar{s}_{i}\bar{s}%
_{i}}\right) \right] \mathbf{C}_{i}^{\prime }+O_{p}\left( T^{-1}\right)
=O_{p}\left( n^{-1/2}\right) +O_{p}\left( T^{-1}\right) .
\end{equation*}%
Using the above results then (\ref{lj}) can be written equivalently as $%
E\left( \mathbf{Q}_{\bar{w}\bar{w}}\right) \mathbf{\beta }_{j}=\lambda _{j}%
\mathbf{\beta }_{j}+O\left( T^{-2}\right) $,$\,$\ and $\lambda _{j}=\mathbf{%
\beta }_{j}^{\prime }E\left( \mathbf{Q}_{\bar{w}\bar{w}}\right) \mathbf{%
\beta }_{j}+O\left( T^{-2}\right) $. Therefore, together with (\ref{ljhat}),
we have%
\begin{equation*}
\hat{\lambda}_{j}-\lambda _{j}=\mathbf{\hat{\beta}}_{j}^{\prime }\mathbf{Q}_{%
\bar{w}\bar{w}}\mathbf{\hat{\beta}}_{j}-\mathbf{\beta }_{j}^{\prime }E\left( 
\mathbf{Q}_{\bar{w}\bar{w}}\right) \mathbf{\beta }_{j}+O\left( T^{-2}\right)
\end{equation*}%
and rearranged as 
\begin{eqnarray}
&&\left( \left[ E\left( \mathbf{Q}_{\bar{w}\bar{w}}\right) -\mathbf{I}%
_{m}\lambda _{j}\right] \right) \left( \mathbf{\hat{\beta}}_{j}-\mathbf{%
\beta }_{j}\right) -\left( \hat{\lambda}_{j}-\lambda _{j}\right) \mathbf{%
\beta }_{j}  \label{betaja2} \\
&=&-\mathbf{\tilde{Q}}_{\bar{w}\bar{w}}\mathbf{\beta }_{j}+\left( \hat{%
\lambda}_{j}-\lambda _{j}\right) \left( \mathbf{\hat{\beta}}_{j}-\mathbf{%
\beta }_{j}\right) -\mathbf{\tilde{Q}}_{\bar{w}\bar{w}}\left( \mathbf{\hat{%
\beta}}_{j}-\mathbf{\beta }_{j}\right) +O\left( T^{-2}\right) .  \notag
\end{eqnarray}%
Also, since $\mathbf{\beta }_{j}^{\prime }E\left( \mathbf{Q}_{\bar{w}\bar{w}%
}\right) =\lambda _{j}\mathbf{\beta }_{j}^{\prime }+O\left( T^{-2}\right) ,$
then \textbf{\ }%
\begin{equation}
\hat{\lambda}_{j}-\lambda _{j}=\mathbf{\beta }_{j}^{\prime }\mathbf{\tilde{Q}%
}_{\bar{w}\bar{w}}\mathbf{\beta }_{j}+2\lambda _{j}\mathbf{\beta }%
_{j}^{\prime }\left( \mathbf{\hat{\beta}}_{j}-\mathbf{\beta }_{j}\right) +2%
\mathbf{\beta }_{j}^{\prime }\mathbf{\tilde{Q}}_{\bar{w}\bar{w}}\left( 
\mathbf{\hat{\beta}}_{j}-\mathbf{\beta }_{j}\right) +\left( \mathbf{\hat{%
\beta}}_{j}-\mathbf{\beta }_{j}\right) ^{\prime }\mathbf{Q}_{\bar{w}\bar{w}%
}\left( \mathbf{\hat{\beta}}_{j}-\mathbf{\beta }_{j}\right) +O\left(
T^{-2}\right) .  \label{lambdaja2}
\end{equation}%
where $\mathbf{\tilde{Q}}_{\bar{w}\bar{w}}=O_{p}\left( n^{-1/2}\right)
+O_{p}\left( T^{-1}\right) $. Pre-multiplying both sides of the above
equations by $\sqrt{n}$ and stacking them in matrix notation we will have 
\begin{equation*}
\left( 
\begin{array}{cc}
1 & -2\lambda _{j}\mathbf{\beta }_{j}^{\prime } \\ 
-\mathbf{\beta }_{j} & E\left( \mathbf{Q}_{\bar{w}\bar{w}}\right) -\lambda
_{j}\mathbf{I}_{m}%
\end{array}%
\right) \left( 
\begin{array}{c}
\sqrt{n}\left( \hat{\lambda}_{j}-\lambda _{j}\right) \\ 
\sqrt{n}\left( \mathbf{\hat{\beta}}_{j}-\mathbf{\beta }_{j}\right)%
\end{array}%
\right) =\left( 
\begin{array}{c}
\sqrt{n}\mathbf{\beta }_{j}^{\prime }\mathbf{\tilde{Q}}_{\bar{w}\bar{w}}%
\mathbf{\beta }_{j} \\ 
-\sqrt{n}\mathbf{\tilde{Q}}_{\bar{w}\bar{w}}\mathbf{\beta }_{j}%
\end{array}%
\right) +\mathbf{p}_{nT},
\end{equation*}%
where%
\begin{equation*}
\mathbf{p}_{nT}=n^{-1/2}\left( 
\begin{array}{c}
2\mathbf{\beta }_{j}^{\prime }\sqrt{n}\mathbf{\tilde{Q}}_{\bar{w}\bar{w}}%
\sqrt{n}\left( \mathbf{\hat{\beta}}_{j}-\mathbf{\beta }_{j}\right) +\sqrt{n}%
\left( \mathbf{\hat{\beta}}_{j}-\mathbf{\beta }_{j}\right) ^{\prime }\mathbf{%
Q}_{\bar{w}\bar{w}}\sqrt{n}\left( \mathbf{\hat{\beta}}_{j}-\mathbf{\beta }%
_{j}\right) \\ 
\sqrt{n}\left( \hat{\lambda}_{j}-\lambda _{j}\right) \sqrt{n}\left( \mathbf{%
\hat{\beta}}_{j}-\mathbf{\beta }_{j}\right) -\sqrt{n}\mathbf{\tilde{Q}}_{%
\bar{w}\bar{w}}\sqrt{n}\left( \mathbf{\hat{\beta}}_{j}-\mathbf{\beta }%
_{j}\right)%
\end{array}%
\right) +O(n^{1/2}T^{-2})\text{.}
\end{equation*}%
It then follows that $\mathbf{p}_{nT}$ is of lower order as compared to$%
\sqrt{n}\left( \hat{\lambda}_{j}-\lambda _{j}\right) $ and $\sqrt{n}\left( 
\mathbf{\hat{\beta}}_{j}-\mathbf{\beta }_{j}\right) $, and as $%
n,T\rightarrow \infty $ jointly, such that $T\approx n^{d}$ for $d>1/2$, we
have%
\begin{equation*}
\mathbf{\Omega }_{j}\left( 
\begin{array}{c}
\sqrt{n}\left( \hat{\lambda}_{j}-\lambda _{j}\right) \\ 
\sqrt{n}\left( \mathbf{\hat{\beta}}_{j}-\mathbf{\beta }_{j}\right)%
\end{array}%
\right) =\left( 
\begin{array}{c}
\sqrt{n}\mathbf{\beta }_{j}^{\prime }\mathbf{\tilde{Q}}_{\bar{w}\bar{w}}%
\mathbf{\beta }_{j} \\ 
-\sqrt{n}\mathbf{\tilde{Q}}_{\bar{w}\bar{w}}\mathbf{\beta }_{j}%
\end{array}%
\right) +o_{p}(1),
\end{equation*}%
where $\mathbf{\Omega }_{j}=\left( 
\begin{array}{cc}
1 & -2\lambda _{j}\mathbf{\beta }_{j}^{\prime } \\ 
-\mathbf{\beta }_{j} & E\left( \mathbf{Q}_{\bar{w}\bar{w}}\right) -\lambda
_{j}\mathbf{I}_{m}%
\end{array}%
\right) $. Using partitioned inverse it is easily seen that $\mathbf{\Omega }%
_{j}$ has an inverse if $\mathbf{\Upsilon }_{j}=E\left( \mathbf{Q}_{\bar{w}%
\bar{w}}\right) -\lambda _{j}\mathbf{I}_{m}-2\lambda _{j}\mathbf{\beta }_{j}%
\mathbf{\beta }_{j}^{\prime }$ is invertible. To check the invertibility of $%
\mathbf{\Upsilon }_{j}$ we note that $\mathbf{\Upsilon }_{j}\mathbf{\beta }%
_{j}=-2\lambda _{j}\mathbf{\beta }_{j}+O\left( T^{-2}\right) $, and since $%
\lambda _{j}>0$ for $j>r_{0}$ it then follows that $\mathbf{\Upsilon }_{j}$
must be invertible. Therefore, we can now solve for $\sqrt{n}\left( \hat{%
\lambda}_{j}-\lambda _{j}\right) $, in terms of a linear combination of$%
\sqrt{n}\mathbf{\beta }_{j}^{\prime }\mathbf{\tilde{Q}}_{\bar{w}\bar{w}}%
\mathbf{\beta }_{j}$ and $\sqrt{n}\mathbf{\tilde{Q}}_{\bar{w}\bar{w}}\mathbf{%
\beta }_{j}$, and its asymptotic distribution can be derived accordingly.
Consider the asymptotic distribution of these two terms, and note that since 
$\mathbf{Q}_{\bar{s}_{i}\bar{s}_{i}}=T^{-1}q^{-1}\sum_{\ell =1}^{q}\left( 
\mathbf{\bar{s}}_{i\ell }-\mathbf{\bar{s}}_{i\circ }\right) \left( \mathbf{%
\bar{s}}_{i\ell }-\mathbf{\bar{s}}_{i\circ }\right) ^{\prime }$, then%
\begin{equation*}
\sqrt{n}\mathbf{\beta }_{j}^{\prime }\mathbf{\tilde{Q}}_{\bar{w}\bar{w}}%
\mathbf{\beta }_{j}=n^{-1/2}\sum_{i=1}^{n}\mathbf{\beta }_{j}^{\prime }%
\mathbf{C}_{i}\left[ \mathbf{Q}_{\bar{s}_{i}\bar{s}_{i}}-E\left( \mathbf{Q}_{%
\bar{s}_{i}\bar{s}_{i}}\right) \right] \mathbf{C}_{i}^{\prime }\mathbf{\beta 
}_{j}=n^{-1/2}\sum_{i=1}^{n}q^{-1}\sum_{\ell =1}^{q}\left[ \zeta _{ij,\ell
}^{2}-E\left( \zeta _{ij,\ell }^{2}\right) \right] ,
\end{equation*}%
where$\ \zeta _{ij,\ell }=T^{-1/2}\mathbf{\beta }_{j}^{\prime }\mathbf{C}%
_{i}\left( \mathbf{\bar{s}}_{i\ell }-\mathbf{\bar{s}}_{i\circ }\right) $.
Also by Minkowski inequality 
\begin{equation*}
\left( E\left\vert \zeta _{ij,\ell }\right\vert ^{4+\epsilon }\right)
^{1/4+\epsilon }\leq \left\Vert \mathbf{C}_{i}\right\Vert \left( \left[
E\left\Vert T^{-1/2}\mathbf{\bar{s}}_{i\ell }\right\Vert ^{4+\epsilon }%
\right] ^{1/4+\epsilon }+\left[ E\left\Vert T^{-1/2}\mathbf{\bar{s}}_{i\ell
}\right\Vert ^{4+\epsilon }\right] ^{1/4+\epsilon }\right) \text{.}
\end{equation*}%
Using Lemma \ref{Llc} in the supplement we have $\sup_{i,\ell }\left\Vert
T^{-1/2}\mathbf{\bar{s}}_{i\ell }\right\Vert ^{4+\epsilon }<K$, and hence $%
\left( E\left\vert \zeta _{ij,\ell }\right\vert ^{4+\epsilon }\right)
^{1/4+\epsilon }<K$ and $\sqrt{n}\mathbf{\beta }_{j}^{\prime }\mathbf{\tilde{%
Q}}_{\bar{w}\bar{w}}\mathbf{\beta }_{j}\rightarrow _{d}N(0,\varpi _{j}^{2})$%
, for some $\varpi _{j}^{2}>0$. Similarly, it follows that all $m$
elements\thinspace of $\sqrt{n}\mathbf{\tilde{Q}}_{\bar{w}\bar{w}}\mathbf{%
\beta }_{j}=n^{-1/2}\sum_{i=1}^{n}\mathbf{C}_{i}\left[ \mathbf{Q}_{\bar{s}%
_{i}\bar{s}_{i}}-E\left( \mathbf{Q}_{\bar{s}_{i}\bar{s}_{i}}\right) \right] 
\mathbf{C}_{i}^{\prime }\mathbf{\beta }_{j}$ are asymptotically normally
distributed with zero means and finite variances. Therefore, it also follows
that $\sqrt{n}\left( \hat{\lambda}_{j}-\lambda _{j}\right) \overset{a}{%
\thicksim }N(0,\varpi _{\lambda _{j}}^{2})$, for some $\varpi _{\lambda
_{j}}^{2}>0$. Using this result and setting $j=r_{0}+1$ we have 
\begin{equation*}
\Pr \left( \hat{\lambda}_{r_{0}+1}<C_{T}\right) \overset{a}{\thicksim }%
\,\Phi \left[ \frac{\sqrt{n}}{\varpi _{\lambda _{r_{0}+1}}}\left(
C_{T}-\lambda _{r_{0}+1}\right) \right] =\Phi \left[ \frac{-\lambda
_{r_{0}+1}\sqrt{n}}{\varpi _{\lambda _{r_{0}+1}}}\left( 1-\frac{C_{T}}{%
\lambda _{r_{0}+1}}\right) \right] .
\end{equation*}%
Since $\lambda _{r_{0}+1}>0$, then $\Pr \left( \hat{\lambda}%
_{r_{0}+1}<C_{T}\right) \rightarrow 0$, as $n$ $\rightarrow \infty $ if $%
C_{T}<\lambda _{r_{0}+1}$. Recall also from (\ref{PrL1}) that $\Pr \left( 
\hat{\lambda}_{1}\geq C_{T}\right) =O\left( C_{T}^{-1}n^{-1/2}T^{-2}\right) $%
, and $\Pr \left( \hat{\lambda}_{1}\geq C_{T}\right) \rightarrow 0$ if $%
C_{T}^{-1}$ does not rise too fast with $T$. Setting $T=KT^{-\delta }$, and
recalling that $n\approx T^{1/d}$, these two conditions on $C_{T}$ are met
if $0<\delta <2+(1/2d)$. Using this result in (\ref{MSErhat}) and (\ref%
{Prrhat}), it follows that $\hat{r}\rightarrow _{p}r_{0}$ and $E\left( \hat{r%
}-r_{0}\right) ^{2}\rightarrow 0$, as $n$ and $T\rightarrow \infty $, if $%
\delta $ is set close to zero such that $KT^{-\delta }<$ $\lambda _{r_{0}+1}$%
. \ These results also suggest that the probability of selecting too many
long-run relations is more affected by $n$ than $T$, and the probability of
selecting too few long-run relations tends to zero much faster with $T$ than
with $n$. We need large $n$ for not selecting more than $r_{0}$ long-run
relations. This latter probability, $\Pr \left( \hat{\lambda}%
_{j}<C_{T}\right) $ for $j>r_{0}$, is also affected by the size of $\lambda
_{r_{0}+1}$ which measures the degree to which there is a transition from a
stationary linear combination under\ which $\lambda _{j}=0$ for $j\leq r_{0}$%
, to $\lambda _{j}>0$ for ~$j>r_{0}$.

Our theoretical derivations also provide some insight on how to set $K$ and $%
\delta $ when choosing $C_{T}$. It is clear that $\delta $ need not be too
large, so long as $K\thickapprox \lambda _{r_{0}+1}$. In practice, this can
be achieved approximately, by appropriate scaling of the observations, $%
\mathbf{w}_{it}$, as discussed below.

\begin{remark}
The above derivations also suggest that our proposed selection/estimation
procedure would be valid even if there were near stationary relations,
namely if $\lambda _{r_{0}+1}\thickapprox n^{-b}$ for some $b>0$. Then $\Pr
\left( \hat{\lambda}_{r_{0}+1}<C_{T}\right) \rightarrow 0,$ so long as $%
b<1/2 $, which could be viewed as local-to-zero eigenvalue. Such cases will
not be pursued in this paper, where we require $\lambda _{r_{0}+1}>0$.
\end{remark}

\section{Allowing for interactive time effects\label{IntEffects}}

The model with interactive time effects is given by (\ref{Grep}), which we
reproduce here for convenience: 
\begin{equation}
\mathbf{w}_{it}=\mathbf{a}_{i}+\mathbf{G}_{i}\mathbf{f}_{t}+\mathbf{C}_{i}%
\mathbf{s}_{it}+\mathbf{v}_{it},\text{ for }t=1,2,...,T;i=1,2,...,n,
\label{gmf}
\end{equation}%
where $\mathbf{f}_{t}$ is an $m_{f}\times 1$ vector of latent factors with $%
\mathbf{G}_{i}$ the associated $m\times m_{f}$ matrix of factor loadings. We
assume $\mathbf{f}_{t}$ is covariance stationary, and treat the factor
loadings, $\mathbf{G}_{i}$,\ as nonstochastic, without placing any
restrictions on them, besides being uniformly bounded. Hence, latent factors
could be strong, semi-strong or weak. However, we do not allow for the
possibility of unit roots in latent factors, and therefore do not consider
the case of cointegration between $\mathbf{w}_{it}$ and latent factors.
Specifically, we assume:

\begin{assumption}
\label{ASSfactors}(i) The $m_{f}\times 1$ vector of latent factors, $\mathbf{%
f}_{t}$, is given by $\mathbf{f}_{t}=\mathbf{\Phi }_{f}(L)\mathbf{%
\varepsilon }_{ft}=\sum_{j=0}^{\infty }\mathbf{\Phi }_{f\ell }L^{\ell }%
\mathbf{\varepsilon }_{f,t-\ell }$, for $t=...0,1,2,...,T$, where $\mathbf{%
\varepsilon }_{ft}$ is an $m_{f}\times 1$ vector of errors distributed
independently over $t$ with $E(\mathbf{\varepsilon }_{ft})=\mathbf{0,}$ and $%
\sup_{it}E\left\Vert \mathbf{\varepsilon }_{ft}\right\Vert ^{4+\epsilon }<K$%
, for some $\epsilon >0$. $\mathbf{\varepsilon }_{ft}$ is independently
distributed of $\mathbf{u}_{it^{\prime }}$, for all $i,t,t^{\prime }$. (ii)
The $m\times m_{f}$ coefficient matrices $\mathbf{\Phi }_{f\ell }$ are
non-stochastic constants such that $\left\Vert \mathbf{\Phi }_{f\ell
}\right\Vert <K\rho ^{\ell }$, where $\rho $ lies in the range $0<\rho <1$.
(iii) $\mathbf{G}_{i}\,\ $are nonstochastic constants such that $%
\sup_{i}\left\Vert \mathbf{G}_{i}\right\Vert <K$.
\end{assumption}

Under Assumption \ref{ASSfactors}, $E\left( \mathbf{G}_{i}\mathbf{f}%
_{t}\right) $ is time invariant, and together with Assumption \ref{ASS1}, $%
E\left( \mathbf{w}_{it}\right) $ continues to be time invariant. Subtracting
sub-sample time averages from the full sample time average, $\mathbf{\bar{w}}%
_{i\ell }-\mathbf{\bar{w}}_{i\circ }$, will therefore continue to remove
unit-specific means.\ More specifically, under (\ref{gmf}) $\mathbf{Q}_{\bar{%
w}_{i}\bar{w}_{i}}$ given by (\ref{Qwiwi}) has the following expension 
\begin{equation}
\mathbf{Q}_{\bar{w}_{i}\bar{w}_{i}}=\mathbf{C}_{i}\mathbf{Q}_{\bar{s}_{i}%
\bar{s}_{i}}\mathbf{C}_{i}^{\prime }+\mathbf{C}_{i}\mathbf{Q}_{\bar{s}_{i}%
\bar{v}_{i}}+\mathbf{C}_{i}\mathbf{Q}_{\bar{s}_{i}\bar{f}_{i}}+\mathbf{Q}_{%
\bar{v}_{i}\bar{s}_{i}}^{\prime }\mathbf{C}_{i}^{\prime }+\mathbf{Q}_{\bar{f}%
_{i}\bar{s}_{i}}^{\prime }\mathbf{C}_{i}+\mathbf{Q}_{\bar{f}_{i}\bar{f}_{i}}+%
\mathbf{Q}_{\bar{f}_{i}\bar{v}_{i}}+\mathbf{Q}_{\bar{v}_{i}\bar{f}_{i}}+%
\mathbf{Q}_{\bar{v}_{i}\bar{v}_{i}},  \label{Qwiwi_f}
\end{equation}%
where $\mathbf{Q}_{\bar{f}_{i}\bar{s}_{i}}=\left( Tq\right) ^{-1}\sum_{\ell
=1}^{q}\mathbf{G}_{i}\left( \mathbf{\bar{f}}_{\ell }-\mathbf{\bar{f}}_{\circ
}\right) \left( \mathbf{\bar{s}}_{i\ell }-\mathbf{\bar{s}}_{i\circ }\right)
^{\prime }=\mathbf{Q}_{\bar{s}_{i}\bar{f}_{i}}^{\prime }$,%
\begin{equation*}
\mathbf{Q}_{\bar{f}_{i}\bar{v}_{i}}=\left( Tq\right) ^{-1}\sum_{\ell =1}^{q}%
\mathbf{G}_{i}\left( \mathbf{\bar{f}}_{\ell }-\mathbf{\bar{f}}_{\circ
}\right) \left( \mathbf{\bar{v}}_{i\ell }-\mathbf{\bar{v}}_{i\circ }\right)
^{\prime }=\mathbf{Q}_{\bar{v}_{i}\bar{f}_{i}}^{\prime }
\end{equation*}
$\mathbf{Q}_{\bar{f}_{i}\bar{f}_{i}}=\left( Tq\right) ^{-1}\sum_{\ell =1}^{q}%
\mathbf{G}_{i}\left( \mathbf{\bar{f}}_{\ell }-\mathbf{\bar{f}}_{\circ
}\right) \left( \mathbf{\bar{f}}_{\ell }-\mathbf{\bar{f}}_{\circ }\right)
^{\prime }\mathbf{G}_{i}^{\prime }$, and the terms not involving the latent
factor are as before. By Lemma \ref{LQB} in the supplement we have%
\begin{equation}
\sup_{i}E\left\Vert \mathbf{Q}_{\bar{f}_{i}\bar{f}_{i}}\right\Vert =O\left(
T^{-2}\right) \text{, }\sup_{i}E\left\Vert \mathbf{Q}_{\bar{f}_{i}\bar{v}%
_{i}}\right\Vert =O\left( T^{-2}\right) \text{, and }\sup_{i}E\left\Vert 
\mathbf{Q}_{\bar{f}_{i}\bar{s}_{i}}\right\Vert =O\left( T^{-1}\right) \text{.%
}  \label{Order_fterms}
\end{equation}%
Pooling over $i$, $\mathbf{Q}_{\bar{w}\bar{w}}=n^{-1}\sum_{i=1}^{n}\mathbf{Q}%
_{\bar{w}_{i}\bar{w}_{i}}$, and using the above results together with those
already established in (\ref{supQsv}) and (\ref{CQss}) now yields%
\begin{equation}
\mathbf{Q}_{\bar{w}\bar{w}}=\frac{(q-1)}{6q}\mathbf{\Psi }_{n}+O_{p}\left(
n^{-1/2}\right) +O_{p}\left( T^{-1}\right) ,\text{ and }\mathbf{Q}_{\bar{w}%
\bar{w}}\mathbf{\beta }_{j0}=O_{p}\left( n^{-1/2}\right) +O_{p}\left(
T^{-1}\right) \text{,}  \label{Qwwfo}
\end{equation}%
which is identical to results (\ref{Qbar2})-(\ref{QbarBeta}) in Proposition %
\ref{Qwbar} for models without interactive time effects.\textbf{\ }%
Similarly, inclusion of interactive time effects does not alter the
convergence rate of the exactly identified PME estimator $\widehat{\mathbf{%
\mathring{B}}}_{0}$, given by (\ref{BETAdot_hat}), and its asymptotic
distribution will remain correctly centered at zero. To see this consider
the following expression for $\mathbf{Q}_{\bar{w}\bar{w}}\sqrt{n}T\left( 
\widehat{\mathbf{\mathring{B}}}_{0}-\mathbf{\mathring{B}}_{0}\right) $,
which extends (\ref{qwb}) to panel data models with interactive time
effects, 
\begin{eqnarray*}
\mathbf{Q}_{\bar{w}\bar{w}}\sqrt{n}T\left( \widehat{\mathbf{\mathring{B}}}%
_{0}-\mathbf{\mathring{B}}_{0}\right) &=&-\left( n^{-1/2}\sum_{i=1}^{n}T%
\mathbf{C}_{i}\mathbf{Q}_{\bar{s}_{i}\bar{v}_{i}}\right) \mathbf{\mathring{B}%
}_{0}-\left( n^{-1/2}\sum_{i=1}^{n}T\mathbf{C}_{i}\mathbf{Q}_{\bar{s}_{i}%
\bar{f}_{i}}\right) \mathbf{\mathring{B}}_{0} \\
&&-\frac{\sqrt{n}}{T}\left( n^{-1}\sum_{i=1}^{n}T^{2}\mathbf{Q}_{\bar{f}_{i}%
\bar{f}_{i}}\right) \mathbf{\mathring{B}}_{0}-\frac{\sqrt{n}}{T}\left[
n^{-1}\sum_{i=1}^{n}T^{2}\left( \mathbf{Q}_{\bar{f}_{i}\bar{v}_{i}}+\mathbf{Q%
}_{\bar{v}_{i}\bar{f}_{i}}\right) \right] \mathbf{\mathring{B}}_{0} \\
&&-\frac{\sqrt{n}}{T}\left( n^{-1}\sum_{i=1}^{n}T^{2}\mathbf{Q}_{\bar{v}_{i}%
\bar{v}_{i}}\right) \mathbf{\mathring{B}}_{0}+O_{p}\left( T^{-1}\right) 
\mathbf{.}
\end{eqnarray*}%
The new terms involve matrices $\mathbf{Q}_{\bar{s}_{i}\bar{f}_{i}}$, $%
\mathbf{Q}_{\bar{f}_{i}\bar{f}_{i}}$ and $\mathbf{Q}_{\bar{f}_{i}\bar{v}%
_{i}}=\mathbf{Q}_{\bar{v}_{i}\bar{f}_{i}}^{\prime }$. Using the bounds in (%
\ref{Order_fterms}), we obtain $E\left\Vert n^{-1}\sum_{i=1}^{n}T^{2}\mathbf{%
Q}_{\bar{f}_{i}\bar{f}_{i}}\right\Vert \leq
n^{-1}\sum_{i=1}^{n}T^{2}E\left\Vert \mathbf{Q}_{\bar{f}_{i}\bar{f}%
_{i}}\right\Vert =O\left( T^{-2}\right) $, and similarly \newline
$E\left\Vert n^{-1}\sum_{i=1}^{n}T^{2}\left( \mathbf{Q}_{\bar{f}_{i}\bar{v}%
_{i}}+\mathbf{Q}_{\bar{v}_{i}\bar{f}_{i}{}_{i}}\right) \right\Vert <K$. In
addition, by Lemma \ref{L_sqvvs} recall that $n^{-1}\sum_{i=1}^{n}T^{2}%
\mathbf{Q}_{\bar{v}_{i}\bar{v}_{i}}=O_{p}\left( n^{-1/2}\right) $. Using
these results and noting that $\left\Vert \mathbf{\mathring{B}}%
_{0}\right\Vert <K$, we have%
\begin{equation}
\mathbf{Q}_{\bar{w}\bar{w}}\sqrt{n}T\left( \widehat{\mathbf{\mathring{B}}}%
_{0}-\mathbf{\mathring{B}}_{0}\right) =-\left( n^{-1/2}\sum_{i=1}^{n}T%
\mathbf{C}_{i}\mathbf{Q}_{\bar{s}_{i}\bar{v}_{i}}\right) \mathbf{\mathring{B}%
}_{0}-\left( n^{-1/2}\sum_{i=1}^{n}T\mathbf{C}_{i}\mathbf{Q}_{\bar{s}_{i}%
\bar{f}_{i}}\right) \mathbf{\mathring{B}}_{0}+O_{p}\left( \frac{\sqrt{n}}{T}%
\right) \text{.}  \label{qbf}
\end{equation}%
To simplify the notations we set $\mathbf{\omega }_{it}=\mathbf{f}_{it}+%
\mathbf{v}_{it}$, and note that 
\begin{equation*}
\mathbf{Q}_{\bar{s}_{i}\bar{\omega}_{i}}=\mathbf{Q}_{\bar{s}_{i}\bar{v}_{i}}+%
\mathbf{Q}_{\bar{s}_{i}\bar{f}_{i}}=T^{-1}q^{-1}\sum_{\ell =1}^{q}\left( 
\mathbf{\bar{s}}_{i\ell }-\mathbf{\bar{s}}_{i\circ }\right) \left( \mathbf{%
\bar{\omega}}_{i\ell }-\mathbf{\bar{\omega}}_{i\circ }\right) ^{\prime }%
\text{.}
\end{equation*}%
Then (\ref{qbf}) can be written as 
\begin{equation*}
\mathbf{Q}_{\bar{w}\bar{w}}\sqrt{n}T\left( \widehat{\mathbf{\mathring{B}}}%
_{0}-\mathbf{\mathring{B}}_{0}\right) =-\left( n^{-1/2}\sum_{i=1}^{n}T%
\mathbf{C}_{i}\mathbf{Q}_{\bar{s}_{i}\bar{\omega}_{i}}\right) \mathbf{%
\mathring{B}}_{0}+O_{p}\left( \frac{\sqrt{n}}{T}\right) .
\end{equation*}%
Let $\mathbf{Z}_{i}^{\ast }=\mathbf{C}_{i}\left[ T\mathbf{Q}_{\bar{s}_{i}%
\bar{\omega}_{i}}-E\left( T\mathbf{Q}_{\bar{s}_{i}\bar{\omega}_{i}}\right) %
\right] \mathbf{\mathring{B}}_{0}$. Since $\mathbf{s}_{it}$ is independent
of $\mathbf{f}_{t}$ and $E\left( \mathbf{s}_{it}\right) =\mathbf{0}$, we
have $E\left( \mathbf{Q}_{\bar{s}_{i}\bar{f}_{i}}\right) =\mathbf{0}$, and
using (\ref{T2qsvb}) we obtain $\left\Vert n^{-1}\sum_{i=1}^{n}\mathbf{C}%
_{i}E\left( T^{2}\mathbf{Q}_{\bar{s}_{i}\bar{\omega}_{i}}\right) \mathbf{%
\beta }_{0}\right\Vert <K$. It now follows that 
\begin{equation}
\mathbf{Q}_{\bar{w}\bar{w}}\sqrt{n}T\left( \widehat{\mathbf{\mathring{B}}}%
_{0}-\mathbf{\mathring{B}}_{0}\right) =-n^{-1/2}\sum_{i=1}^{n}\mathbf{Z}%
_{i}^{\ast }+O_{p}\left( \frac{\sqrt{n}}{T}\right) ,  \label{qbz}
\end{equation}%
where $\mathbf{Z}_{i}^{\ast }=\mathbf{C}_{i}\left[ T\mathbf{Q}_{\bar{s}_{i}%
\bar{\omega}_{i}}-E\left( T\mathbf{Q}_{\bar{s}_{i}\bar{\omega}_{i}}\right) %
\right] \mathbf{\beta }_{0}$. Writing (\ref{qbz}) in vec form%
\begin{equation*}
\left( \mathbf{I}_{r}\mathbf{\otimes Q}_{\bar{w}\bar{w}}\right) \sqrt{n}T%
\text{ }vec\left( \widehat{\mathbf{\mathring{B}}}_{0}-\mathbf{\mathring{B}}%
_{0}\right) =n^{-1/2}\sum_{i=1}^{n}\left( \mathbf{\mathring{B}}_{0}^{\prime }%
\mathbf{\otimes \mathbf{C}}_{i}\right) \left[ \mathbf{\xi }_{iq}^{\ast
}-E\left( \mathbf{\xi }_{iq}^{\ast }\right) \right] +O_{p}\left( \frac{\sqrt{%
n}}{T}\right) ,
\end{equation*}%
where $\mathbf{\xi }_{iq}^{\ast }$ is the $m^{2}\times 1$ vector given by $%
\mathbf{\xi }_{iq}^{\ast }=q^{-1}\sum_{s=1}^{q}\left( \mathbf{\bar{\omega}}%
_{is}-\mathbf{\bar{\omega}}_{i}\right) \mathbf{\otimes }\left( \mathbf{\bar{s%
}}_{is}-\mathbf{\bar{s}}_{i}\right) $. Although $\mathbf{\xi }_{iq}^{\ast }$
is not independent over $i$, it is a martingale difference sequence, and 
\newline
$n^{-1/2}\sum_{i=1}^{n}\left( \mathbf{\mathring{B}}_{0}^{\prime }\mathbf{%
\otimes \mathbf{C}}_{i}\right) \left[ \mathbf{\xi }_{iq}^{\ast }-E\left( 
\mathbf{\xi }_{iq}^{\ast }\right) \right] $ converges to a normal
distribution as $n,T\rightarrow \infty $, jointly. Therefore, $\widehat{%
\mathbf{\mathring{B}}}$ will continue to be asymptotically normally
distributed, with its asymptotic distribution correctly centered at zero.
Even though the variance of the asymptotic distribution of PME estimator in
general depends on the interactive time effects, the estimator of the
asymptotic variance given by (\ref{VarHatTheta}) will continue to be
consistent under Assumption \ref{ASSfactors}, and inference can be carried
out in the same way as in panel data models without interactive time effects.

Overall, we find that the PME estimator is robust to interactive time
effects so long as Assumption \ref{ASSfactors} holds. Theorems and \ref%
{Tconsf} and \ref{Tbf} in the supplement provide formal statements regarding
consistency and the asymptotic normality of the PME estimators in presence
of interactive time effects.

\section{How to choose $q$ and $C_{T}$\label{qCTchoices}}

To implement the estimation of $r_{0}$ and the associated long-run relations
we need to decide on $q$, and $C_{T}=CT^{-\delta }$ that enter the
thresholding estimation of $r_{0}$.

\subsection{Choice of $q$\label{qchoice}}

To ensure that $\mathbf{Q}_{\bar{w}\bar{w}}=n^{-1}T^{-1}q^{-1}\sum_{i=1}^{n}%
\sum_{\ell =1}^{q}\left( \mathbf{\bar{w}}_{i\ell }-\mathbf{\bar{w}}_{i\circ
}\right) \left( \mathbf{\bar{w}}_{i\ell }-\mathbf{\bar{w}}_{i\circ }\right)
^{\prime }$ does not depend on the fixed effects, given by $E(\mathbf{w}%
_{it})=\mathbf{a}_{i}$, we need at least two sub-samples, namely $q\geq 2$.
To reduce the variance of time series dependence of $\mathbf{\bar{w}}_{i\ell
}-\mathbf{\bar{w}}_{i\circ }$ over the sub-samples we need $T/q$ to be
reasonably large. An optimum choice of $q$ most likely depends on $T$, and
not so much on $n$. To ensure that the mathematical derivations are
manageable and transparent, so far we have assumed that $q$ is fixed as $%
T\rightarrow \infty $. But to select $q$ we must allow $q$ to depend on $T$,
denoted as $q_{T}$, and consider values of $q_{T}$ that rise with $T$, but
at a slower rate such that $q_{T}/T\rightarrow 0$. Analogous to the problem
of selecting the lag order in time series literature, we conjecture setting $%
q_{T}$ to rise at the rate of $T^{1/3}$, and set $q_{T}$ to the lower
integer part of $\max (2,T^{1/3})$. This would suggest that values of $q_{T}$
equal to $2$, $3$ and $4$ for values of $T=20,50$ and $100$, respectively,
considered in our Monte Carlo simulations reported in Section \ref{MC}
below, where we consider the values of $2$ and $4$, to save space. For
estimation of $r_{0}$, the choice of $q=2$ works perfectly well for all
values of $T$ considered. But the higher value of $q=4$ does seem to perform
slightly better than $q=2$ for estimation of long-run coefficients when $%
T=100$.

\subsection{Choices of $C$ and $\protect\delta $ for estimation of $r_{0}$}

It is clear that eigenvalues of $\mathbf{Q}_{\bar{w}\bar{w}}$ depend on the
scale of the observations, $\mathbf{w}_{it}$, and some form of scaling of
data is required to reduce the sensitivity of the eigenvalues to scale. One
could resort to cross validation procedures to set $C$ and $\delta $, but
based on extensive Monte Carlo experiments, we have found that setting $C=1$
works well if we base our selection procedure on the eigenvalues of the
following correlation matrix 
\begin{equation}
\mathbf{R}_{_{\bar{w}\bar{w}}}=\left[ diag\left( \mathbf{Q}_{\bar{w}\bar{w}%
}\right) \right] ^{-1/2}\mathbf{Q}_{\bar{w}\bar{w}}\left[ diag\left( \mathbf{%
Q}_{\bar{w}\bar{w}}\right) \right] ^{-1/2}.  \label{Rww}
\end{equation}%
Accordingly, our proposed estimator of $r_{0}$ is given by 
\begin{equation}
\tilde{r}=\sum_{j=1}^{m}\mathcal{I}\left( \tilde{\lambda}_{j}<T^{-\delta
}\right) ,  \label{r_est}
\end{equation}%
where $\tilde{\lambda}_{j}$, for $j=1,2,...,m$ are the eigenvalues of $%
\mathbf{R}_{_{\bar{w}\bar{w}}}$.\footnote{%
Using the correlation matrix, $\mathbf{R}_{_{\bar{w}\bar{w}}}$, has the
advantage that it is unaffected by scaling, so long as the same scaling is
used across all cross section units. It is not invariant if the scaling
varies across units as well as across the variables. This issues is
addressed in Section \ref{Sup_MCs} of the supplement where we investigate
the small sample sensitivity of $\tilde{r}$ to differential scaling of the
variables across the units. We find that the small sample performance of $%
\tilde{r}$ as an estimator of $r_{0}$, is hardly affected by such
differential scaling.}

Also, based on our theoretical derivations any value of $\delta $ close to $%
zero$ should work. In the Monte Carlo experiments we consider the values of $%
\delta =1/4$ and $1/2$ and conclude that $\delta =1/4$ is a good overall
choice and cross-validation is not necessary for the implementation of our
estimation strategy.

\section{Monte Carlo Evidence\label{MC}}

We investigate small sample properties of the proposed PME estimator with
Monte Carlo experiments using both VARMA(1,1) and VAR(1) designs, with and
without interactive time effects. The designs we consider are all special
cases of the general linear model (\ref{Grep}). We set $m=3$, and generate
the $3\times 1$ vector $\mathbf{w}_{it}$ as $I(1)$ variables under three
scenarios: non-cointegration, $r_{0}=0$, and $r_{0}=1$ and $r_{0}=2$
cointegrating relations. We consider sample size combinations, $%
T=(20,50,100) $ and $n=(50,500,1000,3000)$, and report results for $q=2$ and 
$q=4$ (sub-samples), which are in line with our conjecture of setting $q$ in
line with the $\max (2,T^{1/3})$ rule. See Section \ref{qchoice}.

When $r_{0}=1$, there are a range of alternative estimators of the
cointegrating relation in the literature that we can use for comparison,
most of which assume the direction of long-run causality is known. To
accommodate existing estimators, we distinguish between experiments based on
long-run causal ordering. The PME estimator does not require the direction
of long-run causality to be known. When $r_{0}>1$, to the best of our
knowledge, there are no obvious alternative estimators of $r_{0}$ and the
associated cointegrating relations in the panel cointegration literature
that we can use. Accordingly, for the purpose of comparison, we report
results using a mean group version of Johansen's maximum likelihood
procedure, whereby we estimate $r_{0}$ and associated cointegrating vectors
(if any), for all individual units in the panel separately, and report the
frequency with which $r_{0}$ is selected by Johansen procedure across the $n$
units, and the mean group estimates of the cointegrating coefficients and
their standard errors.

Subsection \ref{DGPsubsection} outlines the data generating processes
(DGPs). Subsection \ref{number} gives the results for estimates of $r_{0}$.
Our results show near perfect performance for our proposed estimator of $%
r_{0}$, even for samples as small as $T=20$ and $n=50$. This is in line with
the theory developed in Section \ref{r_sel}. Subsection \ref{coefficients}
reports the results for the coefficients of the long-run relations assuming $%
r_{0}$ is known, which is justified considering the near perfect performance
of our estimator of $r_{0}$. The MC results provide simulation evidence that
the PME estimator performs well in panels with $n$ as large as $1,000$ and $%
T $ as small as $20$.

\subsection{Data generating processes\label{DGPsubsection}}

We consider experiments with and without long-run relations. In the
experiments with long-run relations, $\mathbf{w}_{it}$ is generated as%
\begin{equation}
\Delta \mathbf{w}_{it}=\mathbf{d}_{i}-\mathbf{\Pi }_{i}\mathbf{w}_{i,t-1}+%
\mathbf{u}_{it}-\mathbf{\Theta }_{i}\mathbf{u}_{i,t-1},  \label{varma}
\end{equation}%
for $i=1,2,...,n$, $t=1,2,...,T$, where $\mathbf{\Pi }_{i}=\mathbf{A}_{i}%
\mathbf{B}_{0}^{\prime },$ $\mathbf{A}_{i}$ is $m\times r_{0},$ $\mathbf{B}%
_{0}$ is $m\times r_{0}.$ We set $\mathbf{d}_{i}=\mathbf{\Pi }_{i}\mathbf{%
\mu }_{iw}$ to ensure no linear trends in data. See, for example, Section
5.7 in 
\citeN{Johansen1995}%
.\textbf{\ }The elements of $\mathbf{\mu }_{iw}$ are generated as $IIDN(0,1)$%
. We consider both VAR(1) and VARMA(1,1) designs. For VAR(1) we set $\mathbf{%
\Theta }_{i}=\mathbf{0}$, and for VARMA(1,1) with set $\mathbf{\Theta }%
_{i}=diag(\theta _{ij},j=1,2,...,m)$, and generate $\theta _{ij}$ as $IIDU%
\left[ -0.5,0.5\right] $. The errors $\mathbf{u}_{it}$, are generated
following both Gaussian and chi-squared distributions.

We initially consider $m=3$ variables in $\mathbf{w}_{it}=\left(
w_{it,1},w_{it,2},w_{it,3}\right) ^{\prime }$, with both one and two
long-run relations. For $r_{0}=1$, we set $\mathbf{B}_{0}=\mathbf{\beta }%
_{1,0}=\left( 1,0,-1\right) ^{\prime }$ and $\mathbf{A}%
_{i}=(a_{i,11},a_{i,21},...,a_{i,31})^{\prime }$. In this case, the long-run
relation is given by%
\begin{equation}
\mathbf{\beta }_{1,0}^{\prime }\mathbf{w}_{it}=w_{it,1}-w_{it,3}=\beta
_{11,0}w_{it,1}+\beta _{12,0}w_{it,2}+\beta _{13,0}w_{it,3}\text{,}
\label{r1}
\end{equation}%
with $\beta _{11,0}=1$, $\beta _{12,0}=0$ and $\beta _{13,0}=-1$. We
identify the long-run relation by imposing $\beta _{11,0}=1$ and estimate $%
\beta _{12,0}$ and $\beta _{13,0}$. When $r_{0}=2$, we set%
\begin{equation*}
\mathbf{B}_{0}=\left( \mathbf{\beta }_{1,0},\mathbf{\beta }_{2,0}\right)
=\left( 
\begin{array}{cc}
1 & 0 \\ 
0 & 1 \\ 
-1 & -1%
\end{array}%
\right) ,\text{ and }\mathbf{A}_{i}=\left( 
\begin{array}{cc}
a_{i,11} & a_{i,12} \\ 
a_{i,21} & a_{i22} \\ 
a_{i,31} & a_{i,32}%
\end{array}%
\right) \text{.}
\end{equation*}%
In this case, the two long-run relations are given by 
\begin{eqnarray}
\mathbf{\beta }_{1,0}^{\prime }\mathbf{w}_{it} &=&w_{it,1}-w_{it,3}=\beta
_{11,0}w_{it,1}+\beta _{12,0}w_{it,2}+\beta _{13,0}w_{it,3}\text{,}
\label{r2a} \\
\mathbf{\beta }_{2,0}^{\prime }\mathbf{w}_{it} &=&w_{it,2}-w_{it,3}=\beta
_{21,0}w_{it,1}+\beta _{22,0}w_{it,2}+\beta _{23,0}w_{it,3}\text{,}
\label{r2b}
\end{eqnarray}%
where $\beta _{11,0}=\beta _{22,0}=1$, $\beta _{12,0}=\beta _{21,0}=0$, and $%
\beta _{13,0}=\beta _{23,0}=-1$. We identify these long-run relations by
imposing $\beta _{11,0}=\beta _{22,0}=1$ and $\beta _{12,0}=\beta _{21,0}=0$%
, and we estimate $\beta _{13,0}$ and $\beta _{23,0}$.

We set the values of $\mathbf{A}_{i}$ to control the average speed of
convergence towards the long-run relations. For example, in the case where $%
r_{0}=1$ then $\mathbf{B}_{0}^{\prime }\mathbf{A}_{i}=\rho
_{i}=a_{i,11}-a_{i,31}$ and $\rho _{i}$, for $i=1,2,...,n$ are generated as $%
IIDU[0.1,0.2]$ representing slow convergence, and as $IIDU[0.1,0.3]$
representing moderate convergence. We then set $a_{i,21}=0$, which leaves us
with one free parameter in $\mathbf{A}_{i}$ which we use to set the system
measures of the fit, $PR_{nT}^{2}=0.2$ and $0.3,$ defined as a pooled $R^{2}$%
, given by equation (\ref{SysPR2}) in the supplement. Since $a_{i,11}$ and $%
a_{i,31}$ are both nonzero, the long-run causality runs from $\left(
w_{it,2},w_{it,3}\right) $ to $w_{it,1}$ as well as from $w_{it,1}$ to $%
w_{it,3}$. For $r_{0}=2$ the rate of convergence will depend on the
eigenvalues of $\mathbf{I}_{2}-\mathbf{B}_{0}^{\prime }\mathbf{A}_{i}$ and
the details of how we generate the elements of $\mathbf{A}_{i}$ are given in
the supplement.

We also consider experiments with interactive time effects, which are
obtained by augmenting the solution of (\ref{varma}) with $\mathbf{G}_{i}%
\mathbf{f}_{t}$, where $\mathbf{f}_{t}$ is and $m_{f}\times 1$ vector of
latent factors. Each of these factors are generated as AR(1) process with a
break in the AR coefficient, and the individual elements of the $m\times
m_{f}$ matrix of factor loadings $\mathbf{G}_{i}$ are generated as $IIDU%
\left[ 0.0.4\right] $. We set $m_{f}=4$. Specifically, we augment the
general linear process versions of the above VARMA(1,1) and VAR(1)
specifications with $\mathbf{G}_{i}\mathbf{f}_{t}$\thinspace , namely%
\begin{equation*}
\mathbf{w}_{it}=\mathbf{w}_{i0}+\mathbf{G}_{i}\mathbf{f}_{t}+\mathbf{C}_{i}%
\mathbf{s}_{it}+\mathbf{C}_{i}^{\ast }(L)\mathbf{u}_{it},
\end{equation*}%
where $\mathbf{C}_{i}$ and $\mathbf{C}_{i}^{\ast }(L)$ are obtained from 
\begin{equation*}
(\mathbf{I}_{3}-\mathbf{\Psi }_{i}L)\left[ \mathbf{C}_{i}+\mathbf{C}%
_{i}^{\ast }(L)(1-L)\right] =(\mathbf{I}_{3}-\mathbf{\Theta }_{i}L)(1-L)%
\text{,}
\end{equation*}%
and $\mathbf{\Psi }_{i}=\mathbf{I}_{3}-\mathbf{A}_{i}\mathbf{B}_{0}^{\prime
} $\textbf{. }See the supplement for further details.

For the experiments with no long-run relations ($r_{0}=0$) we generate $%
\Delta \mathbf{w}_{it}$ using the VAR(1) model in first-differences:%
\begin{equation}
\Delta \mathbf{w}_{it}=\mathbf{\Phi }_{i}\Delta \mathbf{w}_{i,t-1}+\mathbf{u}%
_{it},  \label{VARdif}
\end{equation}%
for $i=1,2,...,n$, $t=1,2,...,T$, where $\mathbf{u}_{it}\thicksim IIDN(%
\mathbf{0},\mathbf{\Sigma }_{ui})$. The elements of the covariance matrix $%
\mathbf{\Sigma }_{ui}=\left( \sigma _{i,\ell \ell ^{\prime }}\right) $ are
generated as $\sigma _{i,\ell \ell }=1$ for $i=1,2,...,n$ and $\ell =1,2,3$,
and $\sigma _{i,\ell \ell ^{\prime }}\thicksim IIDU(0,0.5)$, for $\ell \neq
\ell ^{\prime }\,$, and $i=1,2,...,n$. We use a diagonal matrix for $\mathbf{%
\Phi }_{i}=(\phi _{i,\ell \ell ^{\prime }})$, with $\phi _{i,\ell \ell }$
elements on its diagonal, for $r=1,2,...,m$. We consider three options for $%
\phi _{i,\ell \ell }$: ($i$) low values $\phi _{i,\ell \ell }\sim U[0,0.8]$,
($ii$) moderate values $\phi _{i,\ell \ell }\sim U[0.7,0.9]$, and ($iii$)
high values $\phi _{i,\ell \ell }\sim U[0.80,0.95]$. $\mathbf{w}_{it}$ is
then obtained by cumulating $\Delta \mathbf{w}_{it}$ from the initial value $%
\mathbf{w}_{i,0}=0$. Similarly to the experiment with long-run relations
given by (\ref{varma}), the model (\ref{VARdif}) is a special case of (\ref%
{Grep}). Specifically, (\ref{VARdif}) leads to $\mathbf{w}_{it}=\mathbf{w}%
_{i0}+\mathbf{G}_{i}\mathbf{f}_{t}+\mathbf{C}_{i}\mathbf{s}_{it}+\mathbf{C}%
_{i}^{\ast }(L)\mathbf{u}_{it},$ where $\mathbf{C}_{i}=\left( \mathbf{I}_{m}-%
\mathbf{\Phi }_{i}\right) ^{-1}$, and $\mathbf{C}_{i}^{\ast }(L)=-\mathbf{%
\Phi }_{i}\left( \mathbf{I}_{m}-\mathbf{\Phi }_{i}\right) ^{-1}\left( 
\mathbf{I}_{m}-\mathbf{\Phi }_{i}L\right) ^{-1}$. For experiments without
interactive time effects we set $\mathbf{G}_{i}=\mathbf{0}$, for all $i$.

In addition to the designs described above, we also consider a data
generating process taken from Section 3.1 of 
\citeN{ChudikPesaranSmith2021PB}%
. For this $m=2$, and $\mathbf{w}_{it}=\left( w_{1,it},w_{2,it}\right)
^{\prime }$ is generated as 
\begin{eqnarray*}
\Delta w_{1,it} &=&c_{i}-a_{i,11}\left( w_{1,i,t-1}-w_{2,i,t-1}\right)
+u_{1,it}\text{,} \\
\Delta w_{2,it} &=&u_{2,it}\text{,}
\end{eqnarray*}%
where $a_{i,11}\sim IIDU\left[ 0.2,0.3\right] $, $\mathbf{u}_{it}=\left(
u_{1,it},u_{2,it}\right) ^{\prime }$ is heteroskedastic and
cross-sectionally independent, and $u_{1,it}$ is correlated with $u_{2,it}$.%
\footnote{$u_{1,it}$ $=\sigma _{1i}e_{1,it}$, $u_{2,it}=\sigma _{2i}e_{2,it}$%
, $\sigma _{1,i}^{2},\sigma _{2,i}^{2}\sim IIDU\left[ 0.8,1.2\right] $,%
\begin{equation*}
\left( 
\begin{array}{c}
e_{1,it} \\ 
e_{2,it}%
\end{array}%
\right) \sim IIDN\left( \mathbf{0}_{2},\mathbf{\Sigma }_{e}\right) \text{, }%
\mathbf{\Sigma }_{e}\sim \left( 
\begin{array}{cc}
1 & \rho _{ei} \\ 
\rho _{ei} & 1%
\end{array}%
\right) \text{, and }\rho _{ei}\sim IIDU\left[ 0.3,0.7\right] \text{.}
\end{equation*}%
See Section 3.1 of 
\citeN{ChudikPesaranSmith2021PB}
for details.} We use this design to see how PME compares to single-equation
estimators that correctly assume long-run causality runs from $w_{2,it}$ to $%
w_{1,it}$. We would expect such estimators to perform reasonably well,
particularly when $T$ is large relative to $n$, and provide a good baseline
to evaluate the performance of PME in settings favorable to the single
equation techniques advanced in the literature.

In short, we have 71 experiments. 6 experiments with $m=3$ variables and no
long-run relations ($r_{0}=0$), given by 3 choices for the distribution of $%
\phi _{ij}$, and interactive time effects are included or not. $64=2^{5}$
experiments feature $m=3$ variables with long-run relations, given by
combinations of the choice of model (VAR(1) or VARMA(1,1)), $r_{0}=1$ or $2$%
, Gaussian or chi-square error distributions, moderate or slow speed of
convergence, system measure of fit $PR_{nT}^{2}$ $=0.2$ or $0.3$, and
interactive time effects are included or not. In addition, we have one
experiment with $m=2$ variables, $r_{0}=1$ long-run relation and one one-way
long-run causality taken directly from 
\citeN{ChudikPesaranSmith2021PB}%
.

To save space, we report only summary results that are averages across a
number of selected experiments, with the results for the individual
experiments available from the authors upon request. Section \ref{Sup_MCd}
of the supplement also provides further details of the Monte Carlo designs,
how the processes are initialized, and the rationale behind the
parameterization adopted. Additionally, Section \ref{Sup_MCr}\ of the
supplement shows that the results for estimation of $r_{0}$ the associated
long-run relations are robust to GARCH and threshold autoregressive effects.

\subsection{Small sample evidence on estimation of $r_{0}$ \label{number}}

We summarize the Monte Carlo findings for the estimation of $r_{0}$ by the
eigenvalue thresholding estimator $\tilde{r}$ given by (\ref{r_est}), using $%
\delta =1/4$ and $1/2$ in Tables 1-2. Table 1 reports selection frequencies
of $\tilde{r}=0$, $1$, $2$, $3$ using VAR(1) based DGPs without interactive
time effects, for the three cases of $r_{0}=0,1$ or $2$.\footnote{%
These results are averaged over 3 VAR(1) experiments in first differences\ ($%
r_{0}=0$), differeing in terms of autoregressive coefficients (low, medium
and high values), and over 8 VAR(1) experiments in levels featuring $r_{0}=1$
and $2$ long-run relationships, differing in terms of the error
distributions (Gaussian or chi-squared), fit (high or low), and speed of
convergence towards long run (moderate or low).} The theory indicated very
fast convergence of $\tilde{r}$ to $r_{0}$, and this is confirmed by the
Monte Carlo simulations. The eigenvalue thresholding estimator $\tilde{r}$
correctly identifies $r_{0}$ in $100$ per cent of cases, except for the very
small sample sizes considered. For $n=50$ and $T=20$, we see $5\%$
probability of $\tilde{r}$ overestimating the true number of long-run
relations when the smaller exponent $\delta =1/4$ is used in experiments
with $r_{0}=0$ in Table1. For comparison, we also report selection frequency
for the Johansen procedure using the trace statistic and the conventional
nominal level of 5 percent.\footnote{%
We assume the true lag order of the VAR design is known.} Whereas there is a
single estimate for the whole panel per replication using $\tilde{r},$ there
are $n$ such estimates per replication using the Johansen procedure ($\hat{r}%
_{i}$, $i=1,2,...,n$). We simply use them all in calculating the selection
frequency.\footnote{%
There are a number of ways that one might choose $r$ for the panel from
these $n$ tests, based on the modal selection or the average value of the
test statistic for instance. We do not explore these avenues here.} Hence $n$
will not influence these results, but increasing $T$ does improve the
frequency with which the correct value is selected using the trace
statistic. For $T=100$, frequency of correctly estimating the number of
long-run relations using the Johansen trace statistics is $69$ percent for $%
r_{0}=0$, $82$ percent for $r_{0}=1$, and only $31$ percent for $r_{0}=2$ in
Table 1.

Simulation results for the performance of $\tilde{r}$ as an estimator of $%
r_{0}$ in the case of experiments with interactive time effects are
presented in Section \ref{Sup_MCs} of the supplement, to save space. These
results continue to show near perfect performance of $\tilde{r}$ similarly
to the experiments summarized above for the case of panels without
interactive time effects.

Overall, our findings are in line with the theoretical insights of a very
fast convergence of $\tilde{r}$ to $r_{0}$. For this reason, we report next
on the small sample performance of the identified PME estimator assuming $%
r_{0}$ is known.\pagebreak

\begin{center}
\singlespacing%
TABLE 1: Selection frequencies, averaged across experiments, for the
estimation of $r_{0}=0,1,2,3$ by eigenvalue thresholding estimator, $\tilde{r%
}$, given by (\ref{r_est}) with $\delta =1/4$ and $1/2$ and by Johansen's
trace statistics using $VAR(1)$ as the DGP with $r_{0}=0,1,2,$ and without
interactive time effects.\smallskip

\renewcommand{\arraystretch}{1.0}\setlength{\tabcolsep}{4pt}%
\scriptsize%
\begin{tabular}{rrrrrrrrrrrrrrrrr}
\hline\hline
&  & \multicolumn{3}{c}{Frequency $\tilde{r}=0$} &  & \multicolumn{3}{c}{
Frequency $\tilde{r}=1$} &  & \multicolumn{3}{c}{Frequency $\tilde{r}=2$} & 
& \multicolumn{3}{c}{Frequency $\tilde{r}=3$} \\ 
\cline{3-5}\cline{7-9}\cline{11-13}\cline{15-17}
$n$ $\backslash $ $T$ &  & \textbf{20} & \textbf{50} & \textbf{100} &  & 
\textbf{20} & \textbf{50} & \textbf{100} &  & \textbf{20} & \textbf{50} & 
\textbf{100} &  & \textbf{20} & \textbf{50} & \textbf{100} \\ \hline
\multicolumn{10}{l}{\textbf{A. Experiments with }$r_{0}=0$} &  &  &  &  &  & 
&  \\ \hline
&  & \multicolumn{15}{l}{Correlation matrix eigenvalue thresholding
estimator $\tilde{r}$, with $\delta =1/4$} \\ \hline
\textbf{50} &  & 0.95 & 1.00 & 1.00 &  & 0.05 & 0.00 & 0.00 &  & 0.00 & 0.00
& 0.00 &  & 0.00 & 0.00 & 0.00 \\ 
\textbf{500} &  & 1.00 & 1.00 & 1.00 &  & 0.00 & 0.00 & 0.00 &  & 0.00 & 0.00
& 0.00 &  & 0.00 & 0.00 & 0.00 \\ 
\textbf{1,000} &  & 1.00 & 1.00 & 1.00 &  & 0.00 & 0.00 & 0.00 &  & 0.00 & 
0.00 & 0.00 &  & 0.00 & 0.00 & 0.00 \\ 
\textbf{3,000} &  & 1.00 & 1.00 & 1.00 &  & 0.00 & 0.00 & 0.00 &  & 0.00 & 
0.00 & 0.00 &  & 0.00 & 0.00 & 0.00 \\ \hline
&  & \multicolumn{15}{l}{Correlation matrix eigenvalue thresholding
estimator $\tilde{r}$, with $\delta =1/2$} \\ \hline
\textbf{50} &  & 1.00 & 1.00 & 1.00 &  & 0.00 & 0.00 & 0.00 &  & 0.00 & 0.00
& 0.00 &  & 0.00 & 0.00 & 0.00 \\ 
\textbf{500} &  & 1.00 & 1.00 & 1.00 &  & 0.00 & 0.00 & 0.00 &  & 0.00 & 0.00
& 0.00 &  & 0.00 & 0.00 & 0.00 \\ 
\textbf{1,000} &  & 1.00 & 1.00 & 1.00 &  & 0.00 & 0.00 & 0.00 &  & 0.00 & 
0.00 & 0.00 &  & 0.00 & 0.00 & 0.00 \\ 
\textbf{3,000} &  & 1.00 & 1.00 & 1.00 &  & 0.00 & 0.00 & 0.00 &  & 0.00 & 
0.00 & 0.00 &  & 0.00 & 0.00 & 0.00 \\ \hline
&  & \multicolumn{15}{l}{Selection of $r$ based on Johansen Trace statistics
($p=0.05$)} \\ \hline
\textbf{50} &  & 0.28 & 0.55 & 0.69 &  & 0.41 & 0.29 & 0.22 &  & 0.14 & 0.06
& 0.03 &  & 0.17 & 0.10 & 0.06 \\ 
\textbf{500} &  & 0.28 & 0.54 & 0.69 &  & 0.41 & 0.29 & 0.22 &  & 0.14 & 0.06
& 0.03 &  & 0.17 & 0.10 & 0.06 \\ 
\textbf{1,000} &  & 0.28 & 0.54 & 0.69 &  & 0.41 & 0.29 & 0.22 &  & 0.14 & 
0.06 & 0.03 &  & 0.17 & 0.10 & 0.06 \\ 
\textbf{3,000} &  & 0.28 & 0.54 & 0.69 &  & 0.41 & 0.29 & 0.22 &  & 0.14 & 
0.06 & 0.03 &  & 0.17 & 0.10 & 0.06 \\ \hline
\multicolumn{9}{l}{\textbf{B. Experiments with }$r_{0}=1$} &  &  &  &  &  & 
&  &  \\ \hline
&  & \multicolumn{15}{l}{Correlation matrix eigenvalue thresholding
estimator $\tilde{r}$, with $\delta =1/4$} \\ \hline
\textbf{50} &  & 0.00 & 0.00 & 0.00 &  & 1.00 & 1.00 & 1.00 &  & 0.00 & 0.00
& 0.00 &  & 0.00 & 0.00 & 0.00 \\ 
\textbf{500} &  & 0.00 & 0.00 & 0.00 &  & 1.00 & 1.00 & 1.00 &  & 0.00 & 0.00
& 0.00 &  & 0.00 & 0.00 & 0.00 \\ 
\textbf{1,000} &  & 0.00 & 0.00 & 0.00 &  & 1.00 & 1.00 & 1.00 &  & 0.00 & 
0.00 & 0.00 &  & 0.00 & 0.00 & 0.00 \\ 
\textbf{3,000} &  & 0.00 & 0.00 & 0.00 &  & 1.00 & 1.00 & 1.00 &  & 0.00 & 
0.00 & 0.00 &  & 0.00 & 0.00 & 0.00 \\ \hline
&  & \multicolumn{15}{l}{Correlation matrix eigenvalue thresholding
estimator $\tilde{r}$, with $\delta =1/2$} \\ \hline
\textbf{50} &  & 0.00 & 0.00 & 0.00 &  & 1.00 & 1.00 & 1.00 &  & 0.00 & 0.00
& 0.00 &  & 0.00 & 0.00 & 0.00 \\ 
\textbf{500} &  & 0.00 & 0.00 & 0.00 &  & 1.00 & 1.00 & 1.00 &  & 0.00 & 0.00
& 0.00 &  & 0.00 & 0.00 & 0.00 \\ 
\textbf{1,000} &  & 0.00 & 0.00 & 0.00 &  & 1.00 & 1.00 & 1.00 &  & 0.00 & 
0.00 & 0.00 &  & 0.00 & 0.00 & 0.00 \\ 
\textbf{3,000} &  & 0.00 & 0.00 & 0.00 &  & 1.00 & 1.00 & 1.00 &  & 0.00 & 
0.00 & 0.00 &  & 0.00 & 0.00 & 0.00 \\ \hline
&  & \multicolumn{15}{l}{Selection of $r$ based on Johansen Trace statistics
($p=0.05$)} \\ \hline
\textbf{50} &  & 0.46 & 0.26 & 0.01 &  & 0.36 & 0.57 & 0.82 &  & 0.07 & 0.08
& 0.08 &  & 0.11 & 0.09 & 0.08 \\ 
\textbf{500} &  & 0.46 & 0.26 & 0.02 &  & 0.36 & 0.57 & 0.82 &  & 0.07 & 0.08
& 0.08 &  & 0.11 & 0.09 & 0.08 \\ 
\textbf{1,000} &  & 0.46 & 0.26 & 0.01 &  & 0.36 & 0.57 & 0.82 &  & 0.07 & 
0.08 & 0.08 &  & 0.11 & 0.09 & 0.08 \\ 
\textbf{3,000} &  & 0.46 & 0.26 & 0.01 &  & 0.36 & 0.57 & 0.82 &  & 0.07 & 
0.08 & 0.08 &  & 0.11 & 0.09 & 0.08 \\ \hline
\multicolumn{9}{l}{\textbf{C. Experiments with }$r_{0}=2$} &  &  &  &  &  & 
&  &  \\ \hline
&  & \multicolumn{15}{l}{Correlation matrix eigenvalue thresholding
estimator $\tilde{r}$, with $\delta =1/4$} \\ \hline
\textbf{50} &  & 0.00 & 0.00 & 0.00 &  & 0.00 & 0.00 & 0.00 &  & 1.00 & 1.00
& 1.00 &  & 0.00 & 0.00 & 0.00 \\ 
\textbf{500} &  & 0.00 & 0.00 & 0.00 &  & 0.00 & 0.00 & 0.00 &  & 1.00 & 1.00
& 1.00 &  & 0.00 & 0.00 & 0.00 \\ 
\textbf{1,000} &  & 0.00 & 0.00 & 0.00 &  & 0.00 & 0.00 & 0.00 &  & 1.00 & 
1.00 & 1.00 &  & 0.00 & 0.00 & 0.00 \\ 
\textbf{3,000} &  & 0.00 & 0.00 & 0.00 &  & 0.00 & 0.00 & 0.00 &  & 1.00 & 
1.00 & 1.00 &  & 0.00 & 0.00 & 0.00 \\ \hline
&  & \multicolumn{15}{l}{Correlation matrix eigenvalue thresholding
estimator $\tilde{r}$, with $\delta =1/2$} \\ \hline
\textbf{50} &  & 0.00 & 0.00 & 0.00 &  & 0.00 & 0.00 & 0.00 &  & 1.00 & 1.00
& 1.00 &  & 0.00 & 0.00 & 0.00 \\ 
\textbf{500} &  & 0.00 & 0.00 & 0.00 &  & 0.00 & 0.00 & 0.00 &  & 1.00 & 1.00
& 1.00 &  & 0.00 & 0.00 & 0.00 \\ 
\textbf{1,000} &  & 0.00 & 0.00 & 0.00 &  & 0.00 & 0.00 & 0.00 &  & 1.00 & 
1.00 & 1.00 &  & 0.00 & 0.00 & 0.00 \\ 
\textbf{3,000} &  & 0.00 & 0.00 & 0.00 &  & 0.00 & 0.00 & 0.00 &  & 1.00 & 
1.00 & 1.00 &  & 0.00 & 0.00 & 0.00 \\ \hline
&  & \multicolumn{15}{l}{Selection of $r$ based on Johansen Trace statistics
($p=0.05$)} \\ \hline
\textbf{50} &  & 0.36 & 0.06 & 0.00 &  & 0.39 & 0.61 & 0.43 &  & 0.09 & 0.15
& 0.31 &  & 0.16 & 0.18 & 0.25 \\ 
\textbf{500} &  & 0.36 & 0.06 & 0.00 &  & 0.39 & 0.61 & 0.43 &  & 0.09 & 0.15
& 0.31 &  & 0.16 & 0.18 & 0.25 \\ 
\textbf{1,000} &  & 0.36 & 0.06 & 0.00 &  & 0.39 & 0.61 & 0.43 &  & 0.09 & 
0.15 & 0.31 &  & 0.16 & 0.18 & 0.25 \\ 
\textbf{3,000} &  & 0.36 & 0.06 & 0.00 &  & 0.39 & 0.61 & 0.43 &  & 0.09 & 
0.15 & 0.31 &  & 0.16 & 0.18 & 0.25 \\ \hline\hline
\end{tabular}%
\vspace*{-0.45in}
\end{center}

\begin{flushleft}
\singlespacing%
\scriptsize%
\singlespacing%
Notes: For $r_{0}=0,$ (panel A) there are 3 experiments: with low, medium
and high serial correlation in first-difference VAR(1)\ model. For $r_{0}=1,$
(panel B) and $r_{0}=2$ (panel C) there are 8 VAR(1) experiments differing
in terms of the error distributions (Gaussian or chi-square), fit (high or
low), and speed of convergence towards long run (moderate or low). Lag order
for the computation of Johansen's trace statistics is set equal to the true
lag order for the VAR. All experiments based on $R=2,000$ MC replications.
Results for individual Monte Carlo experiments are available from the
authors upon request. A summary of the different MC designs is given in
Subsection \ref{DGPsubsection}, with a detailed account of the data
generating processes provided in Section \ref{Sup_MCd} of the
supplement.\pagebreak
\end{flushleft}

\begin{center}
\normalsize%
\singlespacing%
TABLE 2: Selection frequencies, averaged across experiments, for the
estimation of $r_{0}=0,1,2,3$ by eigenvalue thresholding estimator with $%
\delta =1/4$ and $1/2$ and by Johansen's trace statistics using VARMA(1,1)
as the DGP with $r_{0}=1,2,$ and without interactive time effects.\smallskip

\renewcommand{\arraystretch}{1.0}\setlength{\tabcolsep}{4pt}%
\scriptsize%
\begin{tabular}{rrrrrrrrrrrrrrrrr}
\hline\hline
&  & \multicolumn{3}{c}{Frequency $\tilde{r}=0$} &  & \multicolumn{3}{c}{
Frequency $\tilde{r}=1$} &  & \multicolumn{3}{c}{Frequency $\tilde{r}=2$} & 
& \multicolumn{3}{c}{Frequency $\tilde{r}=3$} \\ 
\cline{3-5}\cline{7-9}\cline{11-13}\cline{15-17}
$n$ $\backslash $ $T$ &  & \textbf{20} & \textbf{50} & \textbf{100} &  & 
\textbf{20} & \textbf{50} & \textbf{100} &  & \textbf{20} & \textbf{50} & 
\textbf{100} &  & \textbf{20} & \textbf{50} & \textbf{100} \\ \hline
\multicolumn{9}{l}{\textbf{A. Experiments with }$r_{0}=1$} &  &  &  &  &  & 
&  &  \\ \hline
&  & \multicolumn{15}{l}{Correlation matrix eigenvalue thresholding
estimator $\tilde{r}$, with $\delta =1/4$} \\ \hline
\textbf{50} &  & 0.00 & 0.00 & 0.00 &  & 1.00 & 1.00 & 1.00 &  & 0.00 & 0.00
& 0.00 &  & 0.00 & 0.00 & 0.00 \\ 
\textbf{500} &  & 0.00 & 0.00 & 0.00 &  & 1.00 & 1.00 & 1.00 &  & 0.00 & 0.00
& 0.00 &  & 0.00 & 0.00 & 0.00 \\ 
\textbf{1,000} &  & 0.00 & 0.00 & 0.00 &  & 1.00 & 1.00 & 1.00 &  & 0.00 & 
0.00 & 0.00 &  & 0.00 & 0.00 & 0.00 \\ 
\textbf{3,000} &  & 0.00 & 0.00 & 0.00 &  & 1.00 & 1.00 & 1.00 &  & 0.00 & 
0.00 & 0.00 &  & 0.00 & 0.00 & 0.00 \\ \hline
&  & \multicolumn{15}{l}{Correlation matrix eigenvalue thresholding
estimator $\tilde{r}$, with $\delta =1/2$} \\ \hline
\textbf{50} &  & 0.00 & 0.00 & 0.00 &  & 1.00 & 1.00 & 1.00 &  & 0.00 & 0.00
& 0.00 &  & 0.00 & 0.00 & 0.00 \\ 
\textbf{500} &  & 0.00 & 0.00 & 0.00 &  & 1.00 & 1.00 & 1.00 &  & 0.00 & 0.00
& 0.00 &  & 0.00 & 0.00 & 0.00 \\ 
\textbf{1,000} &  & 0.00 & 0.00 & 0.00 &  & 1.00 & 1.00 & 1.00 &  & 0.00 & 
0.00 & 0.00 &  & 0.00 & 0.00 & 0.00 \\ 
\textbf{3,000} &  & 0.00 & 0.00 & 0.00 &  & 1.00 & 1.00 & 1.00 &  & 0.00 & 
0.00 & 0.00 &  & 0.00 & 0.00 & 0.00 \\ \hline
&  & \multicolumn{15}{l}{Selection of $r$ based on Johansen Trace statistics
($p=0.05$)} \\ \hline
\textbf{50} &  & 0.40 & 0.33 & 0.14 &  & 0.38 & 0.48 & 0.69 &  & 0.09 & 0.09
& 0.09 &  & 0.13 & 0.11 & 0.09 \\ 
\textbf{500} &  & 0.40 & 0.33 & 0.14 &  & 0.38 & 0.48 & 0.69 &  & 0.09 & 0.09
& 0.09 &  & 0.13 & 0.11 & 0.09 \\ 
\textbf{1,000} &  & 0.40 & 0.33 & 0.14 &  & 0.38 & 0.48 & 0.69 &  & 0.09 & 
0.09 & 0.09 &  & 0.13 & 0.11 & 0.09 \\ 
\textbf{3,000} &  & 0.40 & 0.33 & 0.14 &  & 0.38 & 0.48 & 0.69 &  & 0.09 & 
0.09 & 0.09 &  & 0.13 & 0.11 & 0.09 \\ \hline
\multicolumn{9}{l}{\textbf{B. Experiments with }$r_{0}=2$} &  &  &  &  &  & 
&  &  \\ \hline
&  & \multicolumn{15}{l}{Correlation matrix eigenvalue thresholding
estimator $\tilde{r}$, with $\delta =1/4$} \\ \hline
\textbf{50} &  & 0.00 & 0.00 & 0.00 &  & 0.00 & 0.00 & 0.00 &  & 1.00 & 1.00
& 1.00 &  & 0.00 & 0.00 & 0.00 \\ 
\textbf{500} &  & 0.00 & 0.00 & 0.00 &  & 0.00 & 0.00 & 0.00 &  & 1.00 & 1.00
& 1.00 &  & 0.00 & 0.00 & 0.00 \\ 
\textbf{1,000} &  & 0.00 & 0.00 & 0.00 &  & 0.00 & 0.00 & 0.00 &  & 1.00 & 
1.00 & 1.00 &  & 0.00 & 0.00 & 0.00 \\ 
\textbf{3,000} &  & 0.00 & 0.00 & 0.00 &  & 0.00 & 0.00 & 0.00 &  & 1.00 & 
1.00 & 1.00 &  & 0.00 & 0.00 & 0.00 \\ \hline
&  & \multicolumn{15}{l}{Correlation matrix eigenvalue thresholding
estimator $\tilde{r}$, with $\delta =1/2$} \\ \hline
\textbf{50} &  & 0.00 & 0.00 & 0.00 &  & 0.02 & 0.00 & 0.00 &  & 0.98 & 1.00
& 1.00 &  & 0.00 & 0.00 & 0.00 \\ 
\textbf{500} &  & 0.00 & 0.00 & 0.00 &  & 0.00 & 0.00 & 0.00 &  & 1.00 & 1.00
& 1.00 &  & 0.00 & 0.00 & 0.00 \\ 
\textbf{1,000} &  & 0.00 & 0.00 & 0.00 &  & 0.00 & 0.00 & 0.00 &  & 1.00 & 
1.00 & 1.00 &  & 0.00 & 0.00 & 0.00 \\ 
\textbf{3,000} &  & 0.00 & 0.00 & 0.00 &  & 0.00 & 0.00 & 0.00 &  & 1.00 & 
1.00 & 1.00 &  & 0.00 & 0.00 & 0.00 \\ \hline
&  & \multicolumn{15}{l}{Selection of $r$ based on Johansen Trace statistics
($p=0.05$)} \\ \hline
\textbf{50} &  & 0.42 & 0.34 & 0.10 &  & 0.37 & 0.42 & 0.47 &  & 0.08 & 0.09
& 0.21 &  & 0.13 & 0.15 & 0.21 \\ 
\textbf{500} &  & 0.42 & 0.34 & 0.10 &  & 0.37 & 0.42 & 0.47 &  & 0.08 & 0.10
& 0.21 &  & 0.13 & 0.15 & 0.22 \\ 
\textbf{1,000} &  & 0.42 & 0.34 & 0.10 &  & 0.37 & 0.42 & 0.47 &  & 0.08 & 
0.09 & 0.21 &  & 0.13 & 0.15 & 0.22 \\ 
\textbf{3,000} &  & 0.42 & 0.34 & 0.10 &  & 0.37 & 0.42 & 0.47 &  & 0.08 & 
0.09 & 0.21 &  & 0.13 & 0.15 & 0.22 \\ \hline\hline
\end{tabular}%
\vspace{-0.2in}
\end{center}

\begin{flushleft}
\singlespacing%
\scriptsize%
Notes: For $r_{0}=1,$ (panel A) and $r_{0}=2$ (panel B) there are 8
VARMA(1,1) experiments differing in terms of the error distributions
(Gaussian or chi-squared), fit (high or low), and speed of convergence
towards long run (moderate or low). Lag order for the computation of
Johansen's trace statistics is set equal to the integer part of $T^{-1/3}$.
All experiments are based on $R=2,000$ MC replications.\ Results for
individual Monte Carlo experiments are available from the authors upon
request. A summary of the different MC designs is given in Subsection \ref%
{DGPsubsection}, with a detailed account of the data generating processes
provided in Section \ref{Sup_MCd} of the supplement.
\end{flushleft}

\normalsize%
\onehalfspacing%

\pagebreak

\begin{center}
\singlespacing%
TABLE 3: Simulated bias, RMSE, size and power, averaged across 8
experiments, for PME and MG-Johansen estimators of long-run relations, using
a three variable VAR(1) with $r_{0}=2$ and without interactive time
effects\smallskip

\renewcommand{\arraystretch}{1.0}\setlength{\tabcolsep}{4pt}%
\scriptsize%
\begin{tabular}{rrrrrr|rrrr|rrrrrrrr}
\hline\hline
&  & \multicolumn{3}{c}{\textbf{Bias (}$\times 100$\textbf{)}} &  & 
\multicolumn{3}{|c}{\textbf{RMSE (}$\times 100$\textbf{)}} &  & 
\multicolumn{3}{|c}{\textbf{Size}($\times 100$)} & \multicolumn{1}{r|}{} & 
\multicolumn{3}{|c}{\textbf{Power (}$\times 100$\textbf{)}} &  \\ 
\cline{2-18}
$n\backslash T$ &  & \textbf{20} & \textbf{50} & \textbf{100} &  & \textbf{20%
} & \textbf{50} & \textbf{100} &  & \textbf{20} & \textbf{50} & \textbf{100}
&  & \multicolumn{1}{|r}{\textbf{20}} & \textbf{50} & \textbf{100} &  \\ 
\hline
\multicolumn{13}{l}{\textbf{A. Results for }$\beta _{13,0}$} & 
\multicolumn{1}{l}{} & \multicolumn{1}{l}{} & \multicolumn{1}{l}{} & 
\multicolumn{1}{l}{} &  \\ \hline
\multicolumn{1}{l}{} & \multicolumn{1}{l}{} & \multicolumn{15}{l}{PME
estimator with $q=2$ sub-samples} &  \\ \hline
\textbf{50} &  & -0.15 & -0.13 & -0.07 &  & 4.06 & 1.96 & 1.10 &  & 7.66 & 
7.10 & 7.39 &  & \multicolumn{1}{|r}{15.21} & 42.18 & 81.30 &  \\ 
\textbf{500} &  & -0.02 & -0.15 & -0.06 &  & 1.34 & 0.64 & 0.35 &  & 7.13 & 
5.83 & 5.17 &  & \multicolumn{1}{|r}{67.73} & 99.63 & 100.00 &  \\ 
\textbf{1,000} &  & 0.01 & -0.15 & -0.07 &  & 0.97 & 0.46 & 0.25 &  & 7.64 & 
6.73 & 6.48 &  & \multicolumn{1}{|r}{90.24} & 100.00 & 100.00 &  \\ 
\textbf{3,000} &  & -0.05 & -0.14 & -0.06 &  & 0.66 & 0.29 & 0.15 &  & 12.84
& 8.16 & 7.76 &  & \multicolumn{1}{|r}{99.96} & 100.00 & 100.00 &  \\ \hline
\multicolumn{1}{l}{} & \multicolumn{1}{l}{} & \multicolumn{15}{l}{PME
estimator with $q=4$ sub-samples} &  \\ \hline
\textbf{50} &  & -0.62 & -0.39 & -0.12 &  & 3.29 & 1.59 & 0.79 &  & 7.48 & 
8.33 & 7.63 &  & \multicolumn{1}{|r}{23.13} & 64.41 & 96.21 &  \\ 
\textbf{500} &  & -0.49 & -0.39 & -0.11 &  & 1.22 & 0.62 & 0.27 &  & 10.58 & 
13.04 & 7.58 &  & \multicolumn{1}{|r}{91.19} & 100.00 & 100.00 &  \\ 
\textbf{1,000} &  & -0.47 & -0.38 & -0.12 &  & 0.95 & 0.51 & 0.21 &  & 14.99
& 20.15 & 11.16 &  & \multicolumn{1}{|r}{99.31} & 100.00 & 100.00 &  \\ 
\textbf{3,000} &  & -0.53 & -0.38 & -0.12 &  & 0.76 & 0.43 & 0.15 &  & 35.96
& 47.89 & 20.91 &  & \multicolumn{1}{|r}{100.00} & 100.00 & 100.00 &  \\ 
\hline
\multicolumn{1}{l}{} & \multicolumn{1}{l}{} & \multicolumn{15}{l}{
MG-Johansen estimator} &  \\ \hline
\textbf{50} &  & 15.78 & -3.14 & -0.39 &  & \TEXTsymbol{>}100 & \TEXTsymbol{>%
}100 & 65.53 &  & 2.66 & 2.96 & 3.26 &  & \multicolumn{1}{|r}{2.67} & 10.45
& 56.48 &  \\ 
\textbf{500} &  & -12.34 & 1.92 & -0.86 &  & \TEXTsymbol{>}100 & \TEXTsymbol{%
>}100 & 85.82 &  & 1.98 & 1.89 & 2.66 &  & \multicolumn{1}{|r}{2.27} & 10.77
& 65.62 &  \\ 
\textbf{1,000} &  & -4.83 & -1.07 & 0.18 &  & \TEXTsymbol{>}100 & 
\TEXTsymbol{>}100 & 36.39 &  & 2.16 & 2.24 & 2.31 &  & \multicolumn{1}{|r}{
2.38} & 11.44 & 65.38 &  \\ 
\textbf{3,000} &  & -46.82 & -40.69 & -0.17 &  & \TEXTsymbol{>}100 & 
\TEXTsymbol{>}100 & 65.14 &  & 2.21 & 2.01 & 2.09 &  & \multicolumn{1}{|r}{
2.34} & 12.24 & 65.83 &  \\ \hline
\multicolumn{13}{l}{\textbf{B. Results for }$\beta _{23,0}$} & 
\multicolumn{1}{l}{} & \multicolumn{1}{|l}{} & \multicolumn{1}{l}{} & 
\multicolumn{1}{l}{} &  \\ \hline
\multicolumn{1}{l}{} & \multicolumn{1}{l}{} & \multicolumn{15}{l}{PME
estimator with $q=2$ sub-samples} &  \\ \hline
\textbf{50} &  & 0.02 & -0.15 & -0.07 &  & 4.11 & 1.96 & 1.09 &  & 8.26 & 
7.43 & 7.31 &  & \multicolumn{1}{|r}{15.78} & 41.90 & 81.03 &  \\ 
\textbf{500} &  & -0.21 & -0.16 & -0.07 &  & 1.34 & 0.65 & 0.35 &  & 6.99 & 
5.88 & 5.63 &  & \multicolumn{1}{|r}{67.41} & 99.65 & 100.00 &  \\ 
\textbf{1,000} &  & -0.07 & -0.14 & -0.07 &  & 0.98 & 0.46 & 0.25 &  & 7.53
& 6.32 & 6.61 &  & \multicolumn{1}{|r}{90.25} & 100.00 & 100.00 &  \\ 
\textbf{3,000} &  & -0.08 & -0.14 & -0.07 &  & 0.66 & 0.29 & 0.16 &  & 13.13
& 8.56 & 7.98 &  & \multicolumn{1}{|r}{99.97} & 100.00 & 100.00 &  \\ \hline
\multicolumn{1}{l}{} & \multicolumn{1}{l}{} & \multicolumn{15}{l}{PME
estimator with $q=4$ sub-samples} &  \\ \hline
\textbf{50} &  & -0.42 & -0.38 & -0.12 &  & 3.30 & 1.57 & 0.77 &  & 7.89 & 
8.05 & 6.72 &  & \multicolumn{1}{|r}{23.71} & 64.31 & 96.66 &  \\ 
\textbf{500} &  & -0.71 & -0.40 & -0.12 &  & 1.31 & 0.63 & 0.28 &  & 13.23 & 
13.69 & 8.26 &  & \multicolumn{1}{|r}{91.11} & 100.00 & 100.00 &  \\ 
\textbf{1,000} &  & -0.56 & -0.38 & -0.12 &  & 0.99 & 0.51 & 0.21 &  & 16.85
& 19.71 & 10.71 &  & \multicolumn{1}{|r}{99.25} & 100.00 & 100.00 &  \\ 
\textbf{3,000} &  & -0.57 & -0.38 & -0.12 &  & 0.78 & 0.43 & 0.16 &  & 37.17
& 48.16 & 21.74 &  & \multicolumn{1}{|r}{100.00} & 100.00 & 100.00 &  \\ 
\hline
\multicolumn{1}{l}{} & \multicolumn{1}{l}{} & \multicolumn{15}{l}{
MG-Johansen estimator} &  \\ \hline
\textbf{50} &  & -21.74 & 3.58 & 0.52 &  & \TEXTsymbol{>}100 & \TEXTsymbol{>}%
100 & 65.06 &  & 2.56 & 2.74 & 3.43 &  & \multicolumn{1}{|r}{14.82} & 15.76
& 56.87 &  \\ 
\textbf{500} &  & -6.62 & -1.29 & 0.65 &  & \TEXTsymbol{>}100 & \TEXTsymbol{>%
}100 & 79.29 &  & 2.23 & 1.94 & 2.58 &  & \multicolumn{1}{|r}{15.19} & 15.74
& 65.54 &  \\ 
\textbf{1,000} &  & 13.31 & 0.70 & -0.18 &  & \TEXTsymbol{>}100 & 
\TEXTsymbol{>}100 & 40.82 &  & 2.34 & 2.14 & 2.56 &  & \multicolumn{1}{|r}{
15.74} & 15.99 & 65.71 &  \\ 
\textbf{3,000} &  & 13.99 & 55.84 & 0.04 &  & \TEXTsymbol{>}100 & 
\TEXTsymbol{>}100 & 56.04 &  & 2.09 & 2.17 & 2.23 &  & \multicolumn{1}{|r}{
15.49} & 17.06 & 65.86 &  \\ \hline
\end{tabular}%
\vspace{-0.2in}
\end{center}

\begin{flushleft}
\scriptsize%
\singlespacing%
Notes: The long-run relations are given by $\mathbf{\beta }_{1,0}^{\prime }%
\mathbf{w}_{it}=w_{it,1}-w_{it,3}$ and $\mathbf{\beta }_{2,0}^{\prime }%
\mathbf{w}_{it}=w_{it,2}-w_{it,3}$, and identified using $\beta
_{11,0}=\beta _{22,0}=1$, and $\beta _{12,0}=\beta _{21,0}=0$. The 8
experiments differ with respect to distribution (Gaussian or chi-squared),
fit (high or low), and speed of convergence toward long run (moderate or
low). Size and power are computed at the five percent nominal level.
Reported results are based on $R=2,000$ Monte Carlo replications. Simulated
power are computed under $H_{1}:$ $\beta _{13}=-0.97$, $\beta _{23}=-0.97$,
as alternatives to $-1$ for both coefficients under the null. Results for
individual Monte Carlo experiments are available from the authors upon
request. A summary of the different MC designs is given in Subsection \ref%
{DGPsubsection}, with a detailed account of the data generating processes
provided in Section \ref{Sup_MCd} of the supplement.
\end{flushleft}

\normalsize%
\onehalfspacing%

\subsection{Estimation of coefficients of long-run relations\label%
{coefficients}}

\subsubsection{Comparison of PME with MG-Johansen in VAR(1) designs with
multiple long-run relations}

We first investigate how PME estimators of multiple long-run relations
compare with the Mean Group estimator based on Johansen's estimator of
individual cointegrating vectors (MG-Johansen). To this end, we focus on the
VAR(1) design with multiple long-run relations ($r_{0}=2$ long-run relations
among $m=3$ variables) and no interactive time effects. We assume the true
lag order is known in the case of MG-Johansen. This is a set up which we
believe is most favorable to Johansen procedure when applied to the
individual units. But we note that the PME estimator does not require the
knowledge of the true lag order. These long-run relations are identified
according to (\ref{r2a})-(\ref{r2b}) for both PME, and MG-Johansen's
estimators. While moments of Johansen's estimator do not exist, we expect
the simulated size and power of MG-Johansen to be good for a sufficiently
large $T$ relative to $n$.

Table 3 provides a summary of the results for estimation of $\beta _{13,0}$
(\thinspace $=$ $-1)$ in part A, and for $\beta _{23,0}$ ($=-1$) in part B.
The table report bias, root mean square error (RMSE), size of the tests at
5\% nominal level, and power of the tests ($H_{1}:$ $\beta _{13}=-0.97$, $%
\beta _{23,0}=-0.97$). All entries in this table are multiplied by $100$,
and averaged across $8$ experiments, defined in terms of error distributions
(Gaussian or chi-squared), fit (high or low), and speed of convergence
towards the long run (moderate or low).

We first note that the PME estimator works quite well, with relatively small
bias and RMSE that decline with $n$ and $T$. The choice of $q=2$ works well
for most $n$ and $T$ combinations, and $q=4$ performs better in terms of
RMSE only when $T=100$. In terms of size, PME with $q=2$ does better than
with $q=4$ for all $n$ and $T$ combinations, and has size close to the
nominal $5$ per cent level, together with satisfactory power that rapidly
rises to $100$ per cent as $n$ and $T$ are increased. However, even when $%
q=2 $ there is some evidence of moderate size distortions, particularly when 
$T=20$ and $n=3,000$. The size distortion of PME when $q=4$ seems to be
largely due to its bias that does not decline with $T$, despite the fall we
observe in its RMSE.

The results for the MG-Johansen estimator, also reported in Table 3, show
size below 5 percent and a very low power in comparison to PME. The very
large RMSE and bias entries could be due to lack of moments for the
unit-specific estimators obtained using the Johansen procedure. Clearly,
further research is needed for adapting the use of Johansen maximum
likelihood approach for use with large panels. We are using MG-Johansen
estimators here only for the purpose of comparisons when $r_{0}>1$. Below we
do consider other panel cointegration procedures when $r_{0}=1$, and we will
no longer report results for MG-Johansen as they are very similar to those
summarized in Table 3.

The above summary results do not depend much on which of the two
coefficients is considered.\pagebreak

\begin{center}
\singlespacing%
TABLE 4: Simulated bias, RMSE, size and power, averaged across 8
experiments, for PME estimators of long-run relations in the case of a three
variables VARMA(1,1) with $r_{0}=2$, and without interactive time
effects\smallskip

\renewcommand{\arraystretch}{0.8}\setlength{\tabcolsep}{5pt}%
\scriptsize%
\begin{tabular}{rrrrrr|rrrr|rrrr|rrrr}
\hline\hline
&  & \multicolumn{3}{c}{\textbf{Bias (}$\times 100$\textbf{)}} &  & 
\multicolumn{3}{|c}{\textbf{RMSE (}$\times 100$\textbf{)}} &  & 
\multicolumn{3}{|c}{\textbf{Size}($\times 100$)} &  & \multicolumn{3}{|c}{%
\textbf{Power (}$\times 100$\textbf{)}} &  \\ \cline{2-18}
$n\backslash T$ &  & \textbf{20} & \textbf{50} & \textbf{100} &  & \textbf{20%
} & \textbf{50} & \textbf{100} &  & \textbf{20} & \textbf{50} & \textbf{100}
&  & \textbf{20} & \textbf{50} & \textbf{100} &  \\ \hline
\multicolumn{15}{l}{\textbf{A. Results for }$\beta _{13,0}$} & 
\multicolumn{1}{l}{} & \multicolumn{1}{l}{} &  \\ \hline
\multicolumn{1}{l}{} & \multicolumn{1}{l}{} & \multicolumn{15}{l}{PME
estimator with $q=2$ sub-samples} &  \\ \hline
\textbf{50} &  & -0.20 & -0.27 & -0.14 &  & 6.92 & 4.05 & 2.44 &  & 8.49 & 
7.43 & 7.99 &  & 10.44 & 16.84 & 33.66 &  \\ 
\textbf{500} &  & 0.16 & -0.22 & -0.13 &  & 2.24 & 1.31 & 0.77 &  & 7.25 & 
5.80 & 5.81 &  & 29.62 & 72.41 & 96.88 &  \\ 
\textbf{1,000} &  & 0.23 & -0.20 & -0.13 &  & 1.66 & 0.93 & 0.55 &  & 8.40 & 
5.98 & 5.79 &  & 48.41 & 92.58 & 99.92 &  \\ 
\textbf{3,000} &  & 0.16 & -0.21 & -0.11 &  & 1.08 & 0.56 & 0.32 &  & 12.96
& 6.78 & 6.73 &  & 86.03 & 99.99 & 100.00 &  \\ \hline
\multicolumn{1}{l}{} & \multicolumn{1}{l}{} & \multicolumn{15}{l}{PME
estimator with $q=4$ sub-samples} &  \\ \hline
\textbf{50} &  & -0.83 & -0.73 & -0.24 &  & 5.60 & 3.30 & 1.80 &  & 7.13 & 
7.39 & 7.52 &  & 12.72 & 25.58 & 52.34 &  \\ 
\textbf{500} &  & -0.51 & -0.64 & -0.20 &  & 1.94 & 1.20 & 0.60 &  & 8.33 & 
9.94 & 7.17 &  & 53.05 & 92.56 & 99.86 &  \\ 
\textbf{1,000} &  & -0.46 & -0.63 & -0.21 &  & 1.50 & 0.96 & 0.44 &  & 10.73
& 13.52 & 8.18 &  & 76.78 & 99.49 & 100.00 &  \\ 
\textbf{3,000} &  & -0.53 & -0.63 & -0.20 &  & 1.11 & 0.75 & 0.30 &  & 24.57
& 32.18 & 14.53 &  & 98.21 & 100.00 & 100.00 &  \\ \hline
\multicolumn{15}{l}{\textbf{B. Results for }$\beta _{23,0}$} & 
\multicolumn{1}{l}{} & \multicolumn{1}{l}{} &  \\ \hline
\multicolumn{1}{l}{} & \multicolumn{1}{l}{} & \multicolumn{15}{l}{PME
estimator with $q=2$ sub-samples} &  \\ \hline
\textbf{50} &  & 0.06 & -0.23 & -0.12 &  & 6.84 & 4.07 & 2.40 &  & 8.50 & 
7.71 & 7.47 &  & 10.87 & 17.45 & 33.51 &  \\ 
\textbf{500} &  & -0.06 & -0.25 & -0.13 &  & 2.25 & 1.31 & 0.77 &  & 6.94 & 
5.81 & 6.01 &  & 29.10 & 72.41 & 96.82 &  \\ 
\textbf{1,000} &  & 0.11 & -0.21 & -0.13 &  & 1.64 & 0.93 & 0.55 &  & 8.03 & 
5.81 & 5.99 &  & 48.22 & 92.34 & 99.93 &  \\ 
\textbf{3,000} &  & 0.12 & -0.20 & -0.11 &  & 1.09 & 0.56 & 0.33 &  & 12.66
& 6.92 & 7.06 &  & 86.06 & 99.98 & 100.00 &  \\ \hline
\multicolumn{1}{l}{} & \multicolumn{1}{l}{} & \multicolumn{15}{l}{PME
estimator with $q=4$ sub-samples} &  \\ \hline
\textbf{50} &  & -0.51 & -0.65 & -0.22 &  & 5.48 & 3.28 & 1.74 &  & 7.36 & 
7.20 & 6.68 &  & 13.36 & 26.04 & 51.75 &  \\ 
\textbf{500} &  & -0.77 & -0.67 & -0.21 &  & 2.04 & 1.21 & 0.60 &  & 9.54 & 
10.01 & 7.12 &  & 52.29 & 92.70 & 99.88 &  \\ 
\textbf{1,000} &  & -0.59 & -0.63 & -0.21 &  & 1.53 & 0.96 & 0.44 &  & 11.66
& 14.24 & 8.08 &  & 76.76 & 99.49 & 100.00 &  \\ 
\textbf{3,000} &  & -0.57 & -0.63 & -0.20 &  & 1.13 & 0.75 & 0.31 &  & 24.94
& 32.18 & 14.91 &  & 98.21 & 100.00 & 100.00 &  \\ \hline\hline
\end{tabular}%
\vspace{-0.2in}
\end{center}

\begin{flushleft}
\scriptsize%
\singlespacing%
Notes: The long-run relations are given by $\mathbf{\beta }_{1,0}^{\prime }%
\mathbf{w}_{it}=w_{it,1}-w_{it,3}$ and $\mathbf{\beta }_{2,0}^{\prime }%
\mathbf{w}_{it}=w_{it,2}-w_{it,3}$, and identified using $\beta
_{11,0}=\beta _{22,0}=1$, and $\beta _{12,0}=\beta _{21,0}=0$. The 8
experiments differ with respect to distribution (Gaussian or chi-squared),
fit (high or low), and speed of convergence toward long run (moderate or
low). Reported results are based on $R=2,000$ Monte Carlo replications.
Simulated power are computed under $H_{1}:$ $\beta _{13}=-0.97$ and $\beta
_{23}=-0.97$, as alternatives to $-1$ for both coefficients under the null.
Results for individual Monte Carlo experiments are available from the
authors upon request. A summary of the different MC designs is given in
Subsection \ref{DGPsubsection}, with a detailed account of the data
generating processes provided in Section \ref{Sup_MCd} of the supplement.
Size and Power are computed at 5 percent nominal level. \vspace{-0.3in}
\end{flushleft}

\begin{center}
\normalsize%
\singlespacing%
TABLE 5: Simulated bias, RMSE, size and power, averaged across 8
experiments, for PME estimators of long-run relations in the case of three
variables VARMA(1,1) with $r_{0}=2$, and with interactive time
effects\smallskip

\renewcommand{\arraystretch}{0.8}\setlength{\tabcolsep}{5pt}%
\scriptsize%
\begin{tabular}{rrrrrr|rrrr|rrrr|rrrr}
\hline\hline
&  & \multicolumn{3}{c}{\textbf{Bias (}$\times 100$\textbf{)}} &  & 
\multicolumn{3}{|c}{\textbf{RMSE (}$\times 100$\textbf{)}} &  & 
\multicolumn{3}{|c}{\textbf{Size}($\times 100$)} &  & \multicolumn{3}{|c}{%
\textbf{Power (}$\times 100$\textbf{)}} &  \\ \cline{2-18}
$n\backslash T$ &  & \textbf{20} & \textbf{50} & \textbf{100} &  & \textbf{20%
} & \textbf{50} & \textbf{100} &  & \textbf{20} & \textbf{50} & \textbf{100}
&  & \textbf{20} & \textbf{50} & \textbf{100} &  \\ \hline
\multicolumn{15}{l}{\textbf{A. Results for }$\beta _{13,0}$} & 
\multicolumn{1}{l}{} & \multicolumn{1}{l}{} &  \\ \hline
\multicolumn{1}{l}{} & \multicolumn{1}{l}{} & \multicolumn{15}{l}{PME
estimator with $q=2$ sub-samples} &  \\ \hline
\textbf{50} &  & -0.20 & -0.26 & -0.14 &  & 6.89 & 4.05 & 2.44 &  & 8.56 & 
7.46 & 8.00 &  & 10.59 & 16.85 & 33.76 &  \\ 
\textbf{500} &  & 0.16 & -0.21 & -0.13 &  & 2.23 & 1.31 & 0.77 &  & 7.15 & 
5.83 & 5.79 &  & 29.86 & 72.38 & 96.91 &  \\ 
\textbf{1,000} &  & 0.24 & -0.20 & -0.13 &  & 1.65 & 0.93 & 0.55 &  & 8.38 & 
5.94 & 5.76 &  & 48.74 & 92.63 & 99.92 &  \\ 
\textbf{3,000} &  & 0.16 & -0.21 & -0.11 &  & 1.08 & 0.56 & 0.32 &  & 12.91
& 6.78 & 6.71 &  & 86.23 & 99.99 & 100.00 &  \\ \hline
\multicolumn{1}{l}{} & \multicolumn{1}{l}{} & \multicolumn{15}{l}{PME
estimator with $q=4$ sub-samples} &  \\ \hline
\textbf{50} &  & -0.81 & -0.72 & -0.23 &  & 5.54 & 3.29 & 1.80 &  & 7.04 & 
7.41 & 7.53 &  & 12.85 & 25.58 & 52.38 &  \\ 
\textbf{500} &  & -0.50 & -0.63 & -0.20 &  & 1.92 & 1.20 & 0.60 &  & 8.21 & 
9.96 & 7.13 &  & 53.69 & 92.61 & 99.85 &  \\ 
\textbf{1,000} &  & -0.44 & -0.63 & -0.20 &  & 1.47 & 0.95 & 0.44 &  & 10.59
& 13.56 & 8.22 &  & 77.28 & 99.50 & 100.00 &  \\ 
\textbf{3,000} &  & -0.52 & -0.63 & -0.20 &  & 1.09 & 0.75 & 0.30 &  & 24.09
& 32.06 & 14.57 &  & 98.36 & 100.00 & 100.00 &  \\ \hline
\multicolumn{15}{l}{\textbf{B. Results for }$\beta _{23,0}$} & 
\multicolumn{1}{l}{} & \multicolumn{1}{l}{} &  \\ \hline
\multicolumn{1}{l}{} & \multicolumn{1}{l}{} & \multicolumn{15}{l}{PME
estimator with $q=2$ sub-samples} &  \\ \hline
\textbf{50} &  & 0.05 & -0.22 & -0.12 &  & 6.81 & 4.06 & 2.40 &  & 8.41 & 
7.76 & 7.54 &  & 11.00 & 17.61 & 33.53 &  \\ 
\textbf{500} &  & -0.06 & -0.24 & -0.13 &  & 2.23 & 1.31 & 0.77 &  & 6.90 & 
5.79 & 6.00 &  & 29.31 & 72.36 & 96.86 &  \\ 
\textbf{1,000} &  & 0.11 & -0.21 & -0.13 &  & 1.64 & 0.93 & 0.55 &  & 7.99 & 
5.81 & 6.06 &  & 48.69 & 92.39 & 99.93 &  \\ 
\textbf{3,000} &  & 0.12 & -0.20 & -0.11 &  & 1.08 & 0.56 & 0.33 &  & 12.46
& 6.91 & 7.06 &  & 86.13 & 99.98 & 100.00 &  \\ \hline
\multicolumn{1}{l}{} & \multicolumn{1}{l}{} & \multicolumn{15}{l}{PME
estimator with $q=4$ sub-samples} &  \\ \hline
\textbf{50} &  & -0.49 & -0.64 & -0.22 &  & 5.43 & 3.27 & 1.74 &  & 7.45 & 
7.23 & 6.63 &  & 13.33 & 25.83 & 51.71 &  \\ 
\textbf{500} &  & -0.75 & -0.67 & -0.21 &  & 2.01 & 1.21 & 0.60 &  & 9.43 & 
9.98 & 7.16 &  & 52.87 & 92.71 & 99.88 &  \\ 
\textbf{1,000} &  & -0.57 & -0.63 & -0.21 &  & 1.51 & 0.96 & 0.44 &  & 11.54
& 14.21 & 8.06 &  & 77.26 & 99.50 & 100.00 &  \\ 
\textbf{3,000} &  & -0.55 & -0.62 & -0.20 &  & 1.11 & 0.75 & 0.31 &  & 24.68
& 32.24 & 14.89 &  & 98.32 & 100.00 & 100.00 &  \\ \hline
\end{tabular}%
\vspace{-0.2in}
\end{center}

\begin{flushleft}
\scriptsize%
\singlespacing%
Notes: See the notes to Table 4.\pagebreak
\end{flushleft}

\begin{center}
\begin{tabular}{cc}
\multicolumn{2}{c}{Long run parameters} \\ \hline\hline
-$\beta _{13,0}$ & -$\beta _{23,0}$ \\ 
\multicolumn{1}{l}{A. $T=20$} & \multicolumn{1}{l}{} \\ 
\multicolumn{1}{l}{\includegraphics[width=0.3\textwidth]{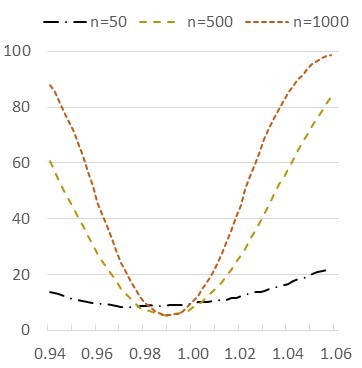}} & \multicolumn{1}{l}{\includegraphics[width=0.3\textwidth]{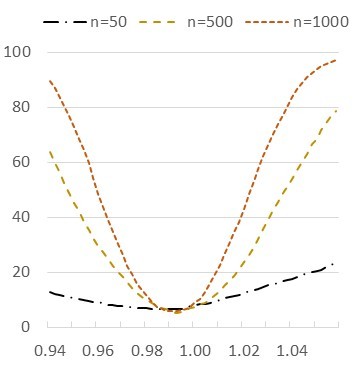}} \\ 
\multicolumn{1}{l}{B. $T=50$} & \multicolumn{1}{l}{} \\ 
\multicolumn{1}{l}{\includegraphics[width=0.3\textwidth]{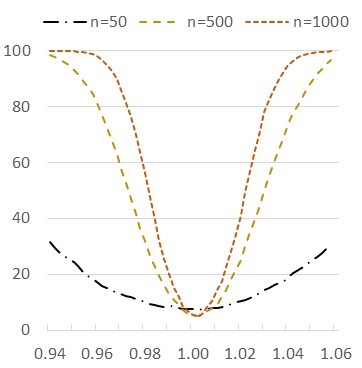}} & \multicolumn{1}{l}{\includegraphics[width=0.3\textwidth]{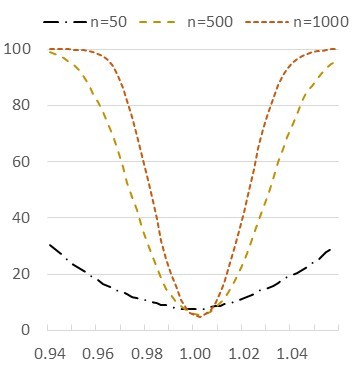}} \\ 
\multicolumn{1}{l}{C. $T=100$} & \multicolumn{1}{l}{} \\ 
\multicolumn{1}{l}{\includegraphics[width=0.3\textwidth]{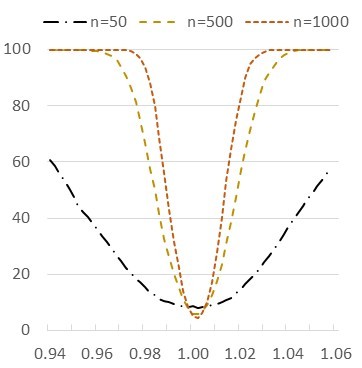}} & \multicolumn{1}{l}{\includegraphics[width=0.3\textwidth]{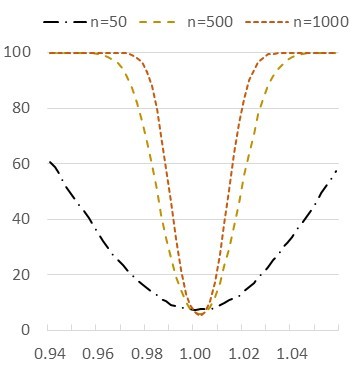}}%
\end{tabular}%
\bigskip \bigskip \bigskip

FIGURE 1:\ Empirical power curves for the tests based on PME\ estimators of $%
\beta _{13,0}$ and $\beta _{23,0}$ parameters of $r_{0}=2$ long-run
relations, using VARMA(1,1) without interactive time effects, slow speed of
convergence, $PR_{nT}^{2}=0.2$, and Gaussian errors. PME estimators use $q=2$
sub-samples. \pagebreak

\begin{tabular}{cc}
\multicolumn{2}{c}{Long run parameters} \\ \hline\hline
-$\beta _{13,0}$ & -$\beta _{23,0}$ \\ 
\multicolumn{1}{l}{A. $T=20$} & \multicolumn{1}{l}{} \\ 
\multicolumn{1}{l}{\includegraphics[width=0.3\textwidth]{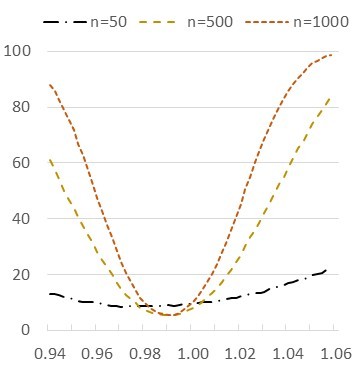}} & \multicolumn{1}{l}{\includegraphics[width=0.3\textwidth]{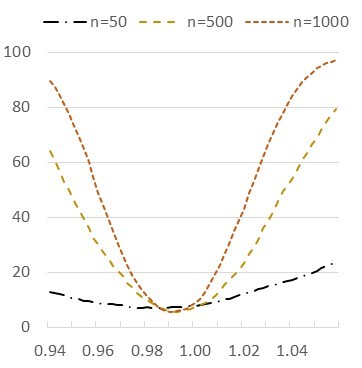}} \\ 
\multicolumn{1}{l}{B. $T=50$} & \multicolumn{1}{l}{} \\ 
\multicolumn{1}{l}{\includegraphics[width=0.3\textwidth]{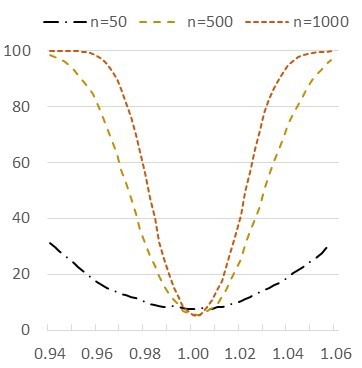}} & \multicolumn{1}{l}{\includegraphics[width=0.3\textwidth]{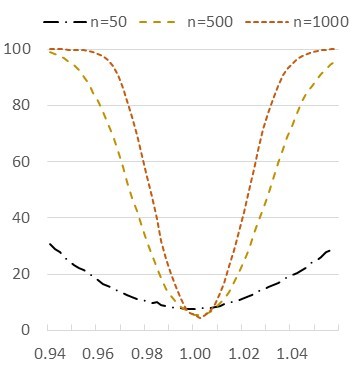}} \\ 
\multicolumn{1}{l}{C. $T=100$} & \multicolumn{1}{l}{} \\ 
\multicolumn{1}{l}{\includegraphics[width=0.3\textwidth]{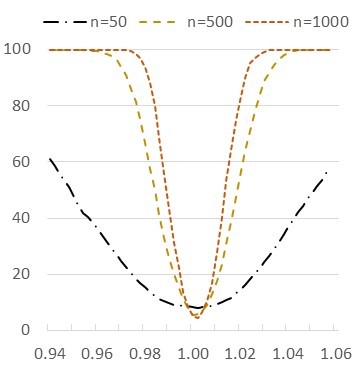}} & \multicolumn{1}{l}{\includegraphics[width=0.3\textwidth]{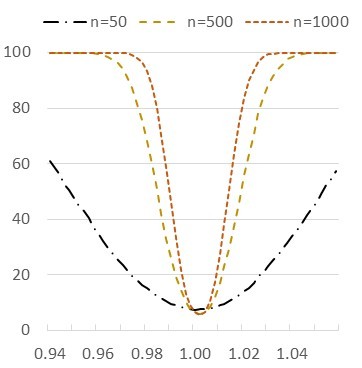}}%
\end{tabular}%
\bigskip \bigskip \bigskip

FIGURE 2:\ Empirical power curves for the tests based on PME\ estimators of $%
\beta _{13,0}$ and $\beta _{23,0}$ parameters of $r_{0}=2$ long-run
relations, using VARMA(1,1) with interactive time effects, slow speed of
convergence, $PR_{nT}^{2}=0.2$, and Gaussian errors. PME estimators use $q=2$
sub-samples. \pagebreak

\singlespacing%
TABLE 6: Simulated bias, RMSE, size and power for estimation of long-run
relation using VAR(1) with $m=2$ variables, $r_{0}=1$, and one-way long-run
causality\smallskip

\renewcommand{\arraystretch}{1.0}\setlength{\tabcolsep}{4pt}%
\scriptsize%
\begin{tabular}{rrrrr|rrrr|rrrr|rrrr}
\hline\hline
& \multicolumn{4}{c|}{\textbf{Bias (}$\times 100$\textbf{)}} & 
\multicolumn{4}{|c|}{\textbf{RMSE (}$\times 100$\textbf{)}} & 
\multicolumn{4}{|c|}{\textbf{Size }($\times 100$)} & \multicolumn{4}{|c}{%
\textbf{Power (}$\times 100$\textbf{)}} \\ \cline{2-17}
$n\backslash T$ & \textbf{20} & \textbf{50} & \textbf{100} &  & \textbf{20}
& \textbf{50} & \textbf{100} &  & \textbf{20} & \textbf{50} & \textbf{100} & 
& \textbf{20} & \textbf{50} & \textbf{100} &  \\ \hline
& \multicolumn{16}{l}{PME estimator with $q=2$ sub-samples} \\ \hline
\textbf{50} & -0.98 & -0.19 & -0.03 &  & 4.92 & 2.11 & 1.05 &  & 7.95 & 6.90
& 6.30 &  & 15.90 & 37.65 & 82.40 &  \\ 
\textbf{500} & -0.98 & -0.16 & -0.06 &  & 1.77 & 0.70 & 0.34 &  & 10.35 & 
6.75 & 5.70 &  & 78.30 & 99.85 & 100.00 &  \\ 
\textbf{1,000} & -0.93 & -0.17 & -0.06 &  & 1.42 & 0.51 & 0.24 &  & 15.90 & 
7.40 & 4.70 &  & 96.30 & 100.00 & 100.00 &  \\ 
\textbf{3,000} & -0.91 & -0.18 & -0.05 &  & 1.09 & 0.32 & 0.14 &  & 33.00 & 
10.50 & 5.60 &  & 100.00 & 100.00 & 100.00 &  \\ \hline
& \multicolumn{16}{l}{PME estimator with $q=4$ sub-samples} \\ \hline
\textbf{50} & -1.46 & -0.36 & -0.09 &  & 4.09 & 1.79 & 0.85 &  & 9.05 & 7.35
& 6.30 &  & 23.30 & 52.00 & 95.40 &  \\ 
\textbf{500} & -1.47 & -0.33 & -0.10 &  & 1.89 & 0.64 & 0.28 &  & 23.65 & 
9.65 & 5.45 &  & 97.50 & 100.00 & 100.00 &  \\ 
\textbf{1,000} & -1.41 & -0.35 & -0.10 &  & 1.65 & 0.53 & 0.21 &  & 39.95 & 
15.50 & 6.20 &  & 100.00 & 100.00 & 100.00 &  \\ 
\textbf{3,000} & -1.40 & -0.35 & -0.09 &  & 1.48 & 0.42 & 0.14 &  & 82.65 & 
36.05 & 12.65 &  & 100.00 & 100.00 & 100.00 &  \\ \hline
& \multicolumn{16}{l}{System Pooled Mean Group Estimator} \\ \hline
\textbf{50} & -0.33 & 0.05 & -0.01 &  & 7.10 & 1.87 & 0.81 &  & 58.50 & 24.95
& 15.00 &  & 62.80 & 67.95 & 99.05 &  \\ 
\textbf{500} & -0.05 & 0.03 & 0.00 &  & 2.21 & 0.58 & 0.25 &  & 59.45 & 25.10
& 13.20 &  & 83.15 & 100.00 & 100.00 &  \\ 
\textbf{1,000} & -0.01 & 0.01 & 0.00 &  & 1.59 & 0.41 & 0.18 &  & 59.25 & 
22.90 & 13.05 &  & 92.40 & 100.00 & 100.00 &  \\ 
\textbf{3,000} & -0.01 & 0.00 & 0.00 &  & 0.89 & 0.24 & 0.10 &  & 57.45 & 
24.30 & 13.65 &  & 99.75 & 100.00 & 100.00 &  \\ \hline
& \multicolumn{16}{l}{Breitung's 2-Step Estimator} \\ \hline
\textbf{50} & 5.25 & 1.28 & 0.36 &  & 6.45 & 2.03 & 0.84 &  & 53.95 & 23.30
& 12.30 &  & 28.55 & 33.05 & 96.15 &  \\ 
\textbf{500} & 5.25 & 1.26 & 0.36 &  & 5.39 & 1.35 & 0.43 &  & 99.70 & 84.00
& 41.90 &  & 70.55 & 97.95 & 100.00 &  \\ 
\textbf{1,000} & 5.27 & 1.25 & 0.35 &  & 5.34 & 1.29 & 0.39 &  & 99.90 & 
97.75 & 63.75 &  & 91.00 & 100.00 & 100.00 &  \\ 
\textbf{3,000} & 5.29 & 1.24 & 0.36 &  & 5.32 & 1.26 & 0.37 &  & 100.00 & 
100.00 & 98.20 &  & 99.90 & 100.00 & 100.00 &  \\ \hline
& \multicolumn{16}{l}{Pooled Mean Group Estimator} \\ \hline
\textbf{50} & 1.69 & 0.40 & 0.08 &  & 5.63 & 1.76 & 0.78 &  & 44.90 & 19.90
& 10.85 &  & 43.55 & 59.50 & 98.70 &  \\ 
\textbf{500} & 1.91 & 0.36 & 0.08 &  & 2.52 & 0.64 & 0.25 &  & 66.25 & 28.45
& 12.80 &  & 52.45 & 100.00 & 100.00 &  \\ 
\textbf{1,000} & 1.92 & 0.33 & 0.08 &  & 2.25 & 0.50 & 0.19 &  & 79.65 & 
33.55 & 14.35 &  & 61.00 & 100.00 & 100.00 &  \\ 
\textbf{3,000} & 1.90 & 0.32 & 0.08 &  & 2.02 & 0.39 & 0.13 &  & 98.30 & 
56.95 & 22.00 &  & 80.95 & 100.00 & 100.00 &  \\ \hline
& \multicolumn{16}{l}{Pooled Bewley Estimator} \\ \hline
\textbf{50} & 3.70 & 0.75 & 0.19 &  & 5.17 & 1.70 & 0.76 &  & 20.05 & 10.20
& 7.20 &  & 7.15 & 35.65 & 96.80 &  \\ 
\textbf{500} & 3.76 & 0.73 & 0.18 &  & 3.92 & 0.86 & 0.29 &  & 92.40 & 34.15
& 12.65 &  & 11.60 & 99.95 & 100.00 &  \\ 
\textbf{1,000} & 3.78 & 0.71 & 0.18 &  & 3.87 & 0.79 & 0.24 &  & 99.80 & 
58.55 & 19.80 &  & 18.45 & 100.00 & 100.00 &  \\ 
\textbf{3,000} & 3.80 & 0.71 & 0.18 &  & 3.82 & 0.73 & 0.21 &  & 100.00 & 
95.75 & 50.20 &  & 41.95 & 100.00 & 100.00 &  \\ \hline
& \multicolumn{16}{l}{Panel FMOLS estimator} \\ \hline
\textbf{50} & 10.27 & 4.27 & 2.01 &  & 11.08 & 4.69 & 2.24 &  & 96.10 & 90.85
& 85.70 &  & 85.30 & 47.05 & 52.85 &  \\ 
\textbf{500} & 10.30 & 4.27 & 2.03 &  & 10.38 & 4.32 & 2.05 &  & 100.00 & 
100.00 & 100.00 &  & 100.00 & 88.95 & 98.10 &  \\ 
\textbf{1,000} & 10.33 & 4.25 & 2.03 &  & 10.37 & 4.28 & 2.04 &  & 100.00 & 
100.00 & 100.00 &  & 100.00 & 97.90 & 99.95 &  \\ 
\textbf{3,000} & 10.36 & 4.25 & 2.03 &  & 10.37 & 4.26 & 2.04 &  & 100.00 & 
100.00 & 100.00 &  & 100.00 & 100.00 & 100.00 &  \\ \hline
& \multicolumn{16}{l}{Panel Dynamic OLS Estimator} \\ \hline
\textbf{50} & 4.15 & 1.15 & 0.36 &  & 6.51 & 2.15 & 0.91 &  & 23.45 & 15.10
& 9.40 &  & 12.85 & 24.85 & 90.25 &  \\ 
\textbf{500} & 4.13 & 1.12 & 0.35 &  & 4.44 & 1.25 & 0.43 &  & 72.85 & 58.95
& 30.50 &  & 16.05 & 94.50 & 100.00 &  \\ 
\textbf{1,000} & 4.19 & 1.11 & 0.35 &  & 4.35 & 1.18 & 0.39 &  & 87.45 & 
84.20 & 50.80 &  & 21.65 & 99.80 & 100.00 &  \\ 
\textbf{3,000} & 4.21 & 1.10 & 0.35 &  & 4.26 & 1.13 & 0.37 &  & 91.25 & 
99.45 & 93.10 &  & 39.50 & 99.80 & 100.00 &  \\ \hline\hline
\end{tabular}%
\vspace{-0.2in}
\end{center}

\begin{flushleft}
\scriptsize%
\singlespacing%
Notes: The long-run relation is given by $\mathbf{\beta }_{1,0}^{\prime }%
\mathbf{w}_{it}=w_{it,1}-w_{it,2}=\beta _{11,0}w_{it,1}+\beta
_{12,0}w_{it,2} $, and identified with $\beta _{11,0}=1$. Coefficient $\beta
_{12,0}=-1$ is estimated. Data generating process used for results reported
in this table is taken from 
\citeN{ChudikPesaranSmith2021PB}%
. Section 3.1 of 
\citeN{ChudikPesaranSmith2021PB}
provides full account of this design. Reported results are based on $R=2,000$
Monte Carlo replications. Size and Power are computed at 5 percent nominal
level. Simulated powers are computed under $H_{1}:\beta _{12}=-0.97$,
compared to null value of $-1$. \pagebreak
\end{flushleft}

\normalsize%
\onehalfspacing%

\subsubsection{Performance of PME in VARMA(1,1) designs}

Results when using VARMA(1,1) designs with $r_{0}=2$ are presented in Tables
4 for models without time effects, and in Table 5 for models with
interactive time effects. As before, these results are averages across eight
VARMA(1,1) experiments featuring $r_{0}=2$ long-run relations that differ in
terms of error distributions (Gaussian or chi-squared), fit (high or low),
and speed of convergence toward long run (moderate or low).\ Figures 1-2
give empirical power curves for tests on $\beta _{13,0}$ and $\beta _{23,0}$
using PME estimators computed with $q=2$ sub-samples, in VARMA(1,1) designs
with $r_{0}=2$ long-run relations, and for $n=50,$ $500,$ and $1000.$
Separate panels show $T=20,$ $T=50,$ $T=100.$ This experiment has a slow
speed of convergence, $PR_{nT}^{2}=0.2,$ Gaussian errors and without
interactive time effects (Figure 1) or with interactive time effects (Figure
2). In both cases, the power increases with $n$ and $T$ as expected.

Qualitatively, the VARMA results in Table 4 are similar to the VAR results
summarized in Table 3, with one important exception. Allowing for an MA
component in the DGP tends to reduce power of the tests. For example, in the
case where $n=500$, $T=20$, and $q=2$, the power of testing $\beta
_{13,0}=-1 $ against the alternative $\beta _{13}=-0.97$ is $67.73$ per cent
when using VAR design (Table 3) as compared to $29.62$ per cent when using
the VARMA design (Table 4). Otherwise, the results for bias, RMSE and size
are comparable across the two designs. Similarly, allowing for interactive
time effects (Table 5) does not alter these conclusions, and PME seems to be
quite robust to the inclusion of interactive time effects, in line with the
theoretical results of Section \ref{IntEffects}, so long as the time effects
are not trended (deterministic or stochastic).

\subsubsection{Comparisons with available estimators in the design with
single long-run relation and one-way long-run causality\label{SubS_MC_c}}

Last but not least, we investigate how PME compares with existing approaches
for the estimation of a single long-run relation, in a design that is
favorable to single equation estimators, some of which rely on the direction
of long-run causality to be one-way and known, and the lag order of the VAR
to be well specified. We report results for panel FMOLS estimator by 
\citeANP{Pedroni1996} (\citeyearNP{Pedroni1996}, \citeyearNP{Pedroni2001}, \citeyearNP{Pedroni2001ReStat})%
, the Pooled Mean Group (PMG) estimator by 
\citeN{PesaranShinSmith1999}%
, panel Dynamic OLS (PDOLS) by 
\citeN{MarkSul2003}%
, the two-step system estimator of 
\citeN{Breitung2005}%
, the system PMG estimator of 
\citeN{ChudikPesaranSmith2023}%
, and the pooled Bewley estimator by 
\citeN{ChudikPesaranSmith2021PB}%
.\ For large $T$ panels with moderate $n$, we would expect these single
equation techniques to perform better than the PME that allows for MA
components and does not assume long-run causality. This is confirmed by the
results summarized in Table 6. When $T=100$ and $n=50$, PME ($q=2$) has
higher RMSE than all other estimators reported in Table 6 except for the
FMOLS estimator. In contrast, PME estimator ($q=2$) is much more balanced in
terms of bias, RMSE and size of the tests, when $T$ is small ($=20$), and $n$
quite large ($=1000$). None of the alternative estimators to PME in the case
of $r_{0}=1$ (and known one-way long-run causality) work well in samples
where $T$ is not very large and $n$ much larger than $T$.

In terms of bias and size, the PME estimator with $q=2$ performs much better
than all the other single equation estimators under consideration, even if
we consider $T=100$ and $n=50$. Overall, Monte Carlo findings show that the
PME estimator with $q=2$ can have satisfactory performance for panels where $%
n$ is quite large relative to $T$, in particular for panels with $n$ and $T$
combinations similar to the ones we consider in our empirical application
discussed below.

\section{Empirical Applications\label{EA}}

We provide two empirical applications to illustrate wide applicability of
the PME approach to both micro and macro panels.

\subsection{Estimation of long-run financial relations}

The fist application considers micro panels of firms where $n$ is quite
large relative to the available time dimension. We use logarithms of six key
financial variables from CRSP/Compustat, available from Wharton Research
Data Services. The six variables (measured in logarithms) are book value ($%
BV_{it}$), market value ($MV_{it}$), short-term debt ($SD_{it}$), long-term
debt ($LD_{it}$), total assets ($TA_{it}$) and total debt outstanding ($%
DO_{it}$). These are variables among which one would expect some key
relations. We consider them in three sets of two or three variables, using
two unbalanced panels both begin in 1950, the shorter one ends in 2010 and
the longer one in 2021. The maximum $T$ is $71$. We set a minimum $T$ of $20$%
, and the average $T$ is around 30. The number of firms $n$ varies from
about $1,000$ to $2,500$ depending on the set of variables under
consideration. As well as estimating the number of long-run relations and
their parameters we test whether the coefficients in the linear combinations
of logarithms take the value -1, to relate our results to the ratios used in
corporate finance. In corporate finance accounting ratios, constructed from
balance sheet data, are commonly used to measure the profitability,
liquidity, and solvency of a firm. The rationales for the use of ratios
include correcting for size in the cross section dimension and eliminating
common trends in the time series dimension to render the ratios stationary.
These two objectives are not always compatible. In an early contribution,
that remains relevant, 
\citeN{LevSunder1979}
comment \textquotedblleft It appears that the extensive use of financial
ratios by both practitioners and researchers is often motivated by tradition
and convenience rather than resulting from theoretical considerations or
from a careful statistical analysis.\textquotedblright\ 
\shortciteN{Geelen_etal_2024}
provide a more recent study of the use of financial ratios in empirical
corporate finance literature.

Given that the theory is often not very specific, it is desirable to have a
statistical criteria to judge which are the appropriate long-run relations
among the set of variables considered and whether the logarithm of their
ratios is stationary. The time series stationarity of finance ratios like
the aggregate dividend price ratio have been studied, but the question has
not, to our knowledge, been addressed in corporate finance, where the
context is somewhat different. Corporate finance studies tend to use
unbalanced panels with large $n$ and relatively small $T$ and and the vector
of accounting variables, of the sort one gets from Compustat, may include
multiple long-run relations. Thus the PME estimator which is appropriate for
multiple long-run relations in a large $n$, moderate $T$ panel seems well
designed to determine whether there are long-run relations in accounting
data.

Using a multiplicative specification, the relation between $y_{it},$ $x_{it}$
and $z_{it}$ can be readily cast in terms of $\mathbf{w}_{it}=(\ln
y_{it},\ln x_{it},\ln z_{it})^{\prime }$ and the PME procedure can be used
to test (a) if $\mathbf{\beta }_{0}^{\prime }\mathbf{w}_{it}$ is stationary;
and (b) if $\beta _{11,0}=1,$ $\beta _{12,0}=-1,$ $\beta _{13,0}=0;$ to
validate the the use of $\ln \left( y_{it}/x_{it}\right) $ or $y_{it}/x_{it}$
in econometric analysis. If step (a) cannot be validated then $\mathbf{\beta 
}_{0}^{\prime }\mathbf{w}_{it}$ will not be stationary and its use in
econometric analysis could lead to spurious results. But if step (a) is
validated but not (b), whilst it would not be advisable to use $\ln
y_{it}-\ln x_{it}$, one can still consider using $\mathbf{\tilde{\beta}}%
^{\prime }\mathbf{w}_{it}$, where $\mathbf{\tilde{\beta}}$ could be the
exactly identified PME estimator of $\mathbf{\beta }_{0}$, in second stage
panel regressions that allow for short term dynamics as well as other
stationary variables.\footnote{%
To simplify the notations, we use the symbol tilde in this section to denote
PME estimates of the exactly identified long-run relations.} Such a two-step
procedure is justified, noting that $\mathbf{\tilde{\beta}}$ is super
consistent, converging to $\mathbf{\beta }_{0}$ at the rate of $T\sqrt{n}$.

Data sources, full definitions of the variables, summary statistics and
additional details on construction of the samples for each of the three
variable sets are provided in Section \ref{Sup_micro} of the supplement. The
following data filters are applied sequentially to each variable set and
sample period, separately. Firms are omitted if, for a given variable set
and sample: they do not have data for all variables in the set (without gaps
and covering at least 20 time periods); have nonpositive entries on any of
the variables, since we are using logarithms; and average value of key
ratios fall below the 1st or above the 99th percentiles estimated after the
application of the first two filters. This is similar to the filtering 
\shortciteN{Geelen_etal_2024}
use, except that we require a longer time series dimension.\footnote{%
\shortciteN{Geelen_etal_2024}
state: \textquotedblleft We winsorize all variables at the 1\% and 99\%
levels to mitigate the impact of outliers. We drop all observations with
missing values on one or more variables of interest. We remove observations
with a market-to-book ratio larger than 20, negative book equity or negative
EBITDA. Our final sample consists of 68,833 firm-year observations with
6,001 unique firms.\textquotedblright\ This gives $\bar{T}=11.5,$ though
some regressions use less.}

The variables are grouped into three sets, where we have prior expectations
about possible cointegration amongst them. To illustrate the procedure we
start with the simplest case where $m=2$ and $\mathbf{w}_{it}$ include the
logarithm of total debt outstanding and logarithm of total assets: \{$%
DO_{it} $, $TA_{it}$\}. The ratio of total debt to total assets is often
used as a measure of leverage, so there is a single hypothesized long-run
relation. To showcase the performance of the PME procedure with multiple
long-run relations, we consider two other sets of variables with $m=3$. They
are: the logarithms of short and long term debt and total assets, \{$SD_{it}$%
, $LD_{it}$, $TA_{it}$\}; and the logarithms of total debt outstanding, book
value and market value, \{$DO_{it}$, $BV_{it}$, $MV_{it}$\}. We expect two
hypothesized long-run relations in both of these sets. Since these variables
are often used to construct accounting ratios, whether the long-run relation
has a unit coefficient is also of interest. Because of missing firm
observations on some variables, the number of firms with minimum of $20$
data points on all the variables under consideration falls as we include
more variables in a set. For this reason we do not consider all six
variables together.

Table 7\textit{\ }gives the estimates of the number of long-run relations
for two and three variable models for the 1950-2021 and 1950-2010 unbalanced
samples using the eigenvalue thresholding\textit{\ }procedure, given by (\ref%
{r_est}). It also reports the eigenvalues of the correlation matrix, $%
\mathbf{R}_{ww}$ defined by (\ref{Rww}), together with the associated
threshold values, $T_{ave}^{-\delta }$, where $T_{ave}=n^{-1}%
\sum_{i=1}^{n}T_{i}$. Since the panel is unbalanced we base the thresholds
on $T_{ave}$ and provide $\tilde{r}$ for $\delta =1/2$ and $\delta =1/4$,
using $q=2$ sub-sample time averages.

The preferred threshold based on $\delta =1/4$ gives $\tilde{r}=1$ long-run
relation for the panel data models with $m=2,$ and $\tilde{r}=2$ long-run
relations for the two cases with $m=3$. The threshold with $\delta =1/2$
also yields $\tilde{r}=1$ in the case with $m=2$, but when used in the case
of panels with $m=3$ it selects $\tilde{r}=1$ rather than $\tilde{r}=2$.
Given the theoretical discussion above and the Monte Carlo results, we
proceed using the estimates of the number of long-run relations obtained
using the preferred value of $\delta =1/4$, namely $one$ long-run relation
when $m=2$ and $two$ long-run relations when $m=3$.\footnote{%
Table S10 in the supplement reports IPS panel unit root tests by 
\shortciteN{ImPesaranShin2003}
(using a 40-year balanced panel), which do not reject the null of unit root
in all cases except for short-term debt (SD). Given the $5$ per cent chance
that the IPS test could be in error, we proceed assuming that all the six
variables are $I(1)$.}

Tables 8-10 present PME estimates of the coefficients in the long-run
relations, their standard errors and $t$-statistics for testing the null
hypothesis that the long-run coefficient in question is equal to $-1$. The
first set of estimates, in Table 8, are for panels with $\mathbf{w}%
_{it}=(DO_{it},TA_{it})^{\prime }$. Recall that $DO_{it}$ and $TA_{it}$ are 
\textit{logarithms} of debt outstanding and total assets. There is a single
hypothesized long-run relation: $\beta _{11,0}DO_{it}+\beta _{12,0}TA_{it}$.
Panel A of Table 8 uses the exact identifying condition $\beta _{11,0}=1$
and provides PME estimates of $\beta _{12,0}$ and t-statistics for the null
value of $\beta _{12,0}=-1$. \ The PME estimates of $\beta _{12,0}$ at $%
-1.142$ and $-1.113$ for the two sample periods ending in 2021 and 2010,
respectively, are similar and close to but significantly different from $-1$%
. To illustrate that, unlike regression based methods, the PME estimator is
invariant to normalization, part B of Table 8 uses the exact identifying
condition $\beta _{12,0}=1$, and reports the PME estimate of $\beta _{11,0}$%
. It is confirmed that up to rounding error $\tilde{\beta}_{11}=1/\tilde{%
\beta}_{12}$.\pagebreak

\begin{center}
\singlespacing%
TABLE 7:\ Estimates of the number of long-run relations ($\tilde{r}$) by
eigenvalue thresholding using firm-level data and $q=2$ sub-sample time
averages\smallskip

\renewcommand{\arraystretch}{1.12}\setlength{\tabcolsep}{4pt}%
\scriptsize%
\begin{tabular}{rcccccrrrrrr}
\hline\hline
\multicolumn{2}{r}{Variables:} & \multicolumn{2}{c}{} & \multicolumn{2}{c}{$%
\mathbf{w}_{it}=\left( DO_{it},TA_{it}\right) ^{\prime }$} &  & 
\multicolumn{2}{c}{$\mathbf{w}_{it}=\left( SD_{it},LD_{it},TA_{it}\right)
^{\prime }$} &  & \multicolumn{2}{c}{$\mathbf{w}_{it}=\left(
DO_{it},BV_{it},MV_{it}\right) ^{\prime }$} \\ 
\cline{3-4}\cline{5-6}\cline{8-9}\cline{11-12}
\multicolumn{2}{r}{Sample end year:} &  &  & {\footnotesize 1950-2021} & 
{\footnotesize 1950-2010} &  & {\footnotesize 1950-2021} & {\footnotesize %
1950-2010} &  & {\footnotesize 1950-2021} & {\footnotesize 1950-2010} \\ 
\cline{3-12}
\multicolumn{11}{l}{\textbf{Estimated number of long-run relations (}$\tilde{%
r}$\textbf{)}} &  \\ 
$\tilde{r}$ ($\delta =1/2$) &  &  &  & 1 & 1 & \multicolumn{1}{c}{} & 
\multicolumn{1}{c}{1} & \multicolumn{1}{c}{1} & \multicolumn{1}{c}{} & 
\multicolumn{1}{c}{1} & \multicolumn{1}{c}{1} \\ 
$\tilde{r}$ ($\delta =1/4$) &  &  &  & 1 & 1 & \multicolumn{1}{c}{} & 
\multicolumn{1}{c}{2} & \multicolumn{1}{c}{2} & \multicolumn{1}{c}{} & 
\multicolumn{1}{c}{2} & \multicolumn{1}{c}{2} \\ \hline
\multicolumn{3}{l}{\textbf{Eigenvalues}} &  &  &  & \multicolumn{1}{c}{} & 
\multicolumn{1}{c}{} & \multicolumn{1}{c}{} & \multicolumn{1}{c}{} & 
\multicolumn{1}{c}{} & \multicolumn{1}{c}{} \\ \hline
$\tilde{\lambda}_{1}$ &  &  &  & 0.090 & 0.079 & \multicolumn{1}{c}{} & 
\multicolumn{1}{c}{0.106} & \multicolumn{1}{c}{0.100} & \multicolumn{1}{c}{}
& \multicolumn{1}{c}{0.058} & \multicolumn{1}{c}{0.046} \\ 
$\tilde{\lambda}_{2}$ &  &  &  & 1.910 & 1.921 & \multicolumn{1}{c}{} & 
\multicolumn{1}{c}{0.251} & \multicolumn{1}{c}{0.223} & \multicolumn{1}{c}{}
& \multicolumn{1}{c}{0.226} & \multicolumn{1}{c}{0.198} \\ 
$\tilde{\lambda}_{3}$ &  &  &  & - & - & \multicolumn{1}{c}{} & 
\multicolumn{1}{c}{2.644} & \multicolumn{1}{c}{2.678} & \multicolumn{1}{c}{}
& \multicolumn{1}{c}{2.717} & \multicolumn{1}{c}{2.756} \\ \hline
\multicolumn{3}{l}{\textbf{Threshold }$T_{ave}^{-\delta }$} & 
\multicolumn{1}{l}{} &  &  & \multicolumn{1}{c}{} & \multicolumn{1}{c}{} & 
\multicolumn{1}{c}{} & \multicolumn{1}{c}{} & \multicolumn{1}{c}{} & 
\multicolumn{1}{c}{} \\ \hline
$\delta =1/2$ &  &  &  & 0.176 & 0.178 & \multicolumn{1}{c}{} & 
\multicolumn{1}{c}{0.177} & \multicolumn{1}{c}{0.179} & \multicolumn{1}{c}{}
& \multicolumn{1}{c}{0.178} & \multicolumn{1}{c}{0.182} \\ 
$\delta =1/4$ &  &  &  & 0.419 & 0.422 & \multicolumn{1}{c}{} & 
\multicolumn{1}{c}{0.421} & \multicolumn{1}{c}{0.423} & \multicolumn{1}{c}{}
& \multicolumn{1}{c}{0.422} & \multicolumn{1}{c}{0.426} \\ \hline
\multicolumn{3}{l}{\textbf{Sample dimensions}} & \multicolumn{1}{l}{} &  & 
& \multicolumn{1}{c}{} & \multicolumn{1}{c}{} & \multicolumn{1}{c}{} & 
\multicolumn{1}{c}{} & \multicolumn{1}{c}{} & \multicolumn{1}{c}{} \\ \hline
$n$ &  &  &  & 2,555 & 1,901 & \multicolumn{1}{c}{} & \multicolumn{1}{c}{
1,373} & \multicolumn{1}{c}{1,101} & \multicolumn{1}{c}{} & 
\multicolumn{1}{c}{1,415} & \multicolumn{1}{c}{1,164} \\ 
$T_{ave}$ &  &  &  & 32.4 & 31.6 & \multicolumn{1}{c}{} & \multicolumn{1}{c}{
31.8} & \multicolumn{1}{c}{31.1} & \multicolumn{1}{c}{} & \multicolumn{1}{c}{
31.5} & \multicolumn{1}{c}{30.3} \\ 
$\max_{i}T_{i}$ &  &  &  & 73 & 61 & \multicolumn{1}{c}{} & 
\multicolumn{1}{c}{73} & \multicolumn{1}{c}{61} & \multicolumn{1}{c}{} & 
\multicolumn{1}{c}{61} & \multicolumn{1}{c}{49} \\ 
$\sum_{i=1}^{n}T_{i}$ &  &  &  & 82,837 & 60,118 & \multicolumn{1}{c}{} & 
\multicolumn{1}{c}{43,621} & \multicolumn{1}{c}{34,262} & \multicolumn{1}{c}{
} & \multicolumn{1}{c}{44,592} & \multicolumn{1}{c}{35,293} \\ \hline\hline
\end{tabular}%
\vspace{-0.25in}
\end{center}

\begin{flushleft}
\scriptsize%
\singlespacing%
Notes: This table reports eigenvalues $\tilde{\lambda}_{j}$ for $%
j=1,2...,m\, $\ (in ascending order) of $\mathbf{R}_{_{\bar{w}\bar{w}}}$
given by (\ref{Rww}) using $q=2$ sub-sample time averages and the
corresponding eigenvalue thresholding estimates of the number of long-run
relations given by $\tilde{r}=\sum_{j=1}^{m}\mathcal{I}\left( \tilde{\lambda}%
_{j}<T_{ave}^{-\delta }\right) $, for $\delta =1/4,1/2,$ where $%
T_{ave}=n^{-1}\sum_{i=1}^{n}T_{i}$, and $T_{i}\geq 20$ for all $i$. The
elements of $\mathbf{w}_{it}$ are logarithms of book value ($BV_{it}$),
market value ($MV_{it}$), short-term debt ($SD_{it}$), long-term debt ($%
LD_{it}$), total assets ($TA_{it}$) and total debt outstanding ($DO_{it}$).
See Section \ref{Sup_micro} of the supplement for variable definitions, data
sources and availability, and filters applied.
\end{flushleft}

\normalsize%
\onehalfspacing%

For panels with $m=3$ and $\tilde{r}=2$ we need two exactly identifying
conditions on each of the two long-run relations. In the case where $\mathbf{%
w}_{it}=(SD_{it},LD_{it},TA_{it})^{\prime }$, we use the conditions $\beta
_{11,0}=1$ and $\beta _{12,0}=0$ to exactly identify the first long-run
relation, and conditions $\beta _{21,0}=0$ and $\beta _{22,0}=1$ to identify
the second long-run relation. Hence the two identified long-run relations
are $SD_{it}+\beta _{13,0}TA_{it}$ and $LD_{it}+\beta _{23,0}TA_{it}$. PME
estimates for $\beta _{13,0}$ in Table 9 are -1.025 and -1.024 for the two
unbalanced samples, both very close to -1 and statistically not different
from -1. PME estimates for $\beta _{23,0}$ are -0.875 and -0.899, both
statistically different from -1 at the one percent level. Rotation of the
two identified long-run relations reported in Table 9 imply the long-run
relations $SD_{it}-1.156LD_{it}$ and $SD_{it}-1.130LD_{it}$ for the samples
ending in 2021 and 2010, respectively. Both of these PME estimates, $-1.156$
and $-1.130$, are statistically significantly different from $-1$ at the one
per cent level. Hence, in the case of the variable set $\left\{
SD_{it},LD_{it},TA_{it}\right\} $, the unit long-run elasticity hypothesis
could be accepted for short term debt to total assets only..\pagebreak

\begin{center}
\singlespacing%
TABLE 8:\textit{\ }PME estimates for the set $\left\{
DO_{it},TA_{it}\right\} $, using $q=2$\ sub-sample time averages and one\
long-run relation

\renewcommand{\arraystretch}{1.1}\setlength{\tabcolsep}{4pt}%
\scriptsize%
\begin{tabular}{rccc}
\multicolumn{4}{r}{\textbf{A. Exact identifying condition }$\beta _{11,0}=1$}
\\ \hline\hline
\multicolumn{4}{r}{Exactly identified long-run relation} \\ 
& \multicolumn{3}{c}{$\mathbf{\beta }_{1,0}^{\prime }\mathbf{w}%
_{it}=DO_{it}+\beta _{12,0}TA_{it}$} \\ \hline
Sample ends: &  & {\footnotesize 1950-2021} & {\footnotesize 1950-2010} \\ 
\hline
\multicolumn{1}{l}{$\tilde{\beta}_{12}$} &  & -1.142 & -1.113 \\ 
\multicolumn{1}{l}{s.e.} &  & (0.010) & (0.011) \\ 
\multicolumn{1}{l}{t($\beta _{12,0}=-1)$} &  & -14.6 & -10.4 \\ \hline
\multicolumn{1}{l}{$n$} &  & 2,555 & 1,901 \\ 
\multicolumn{1}{l}{$\sum_{i=1}^{n}T_{i}$} &  & 82,837 & 60,118 \\ 
\hline\hline
\end{tabular}
\qquad \qquad\ \ 
\begin{tabular}{rccc}
\multicolumn{4}{r}{\textbf{B. Exact identifying condition }$\beta _{12,0}=1$}
\\ \hline\hline
\multicolumn{4}{r}{Exactly identified long-run relation} \\ 
& \multicolumn{3}{c}{$\mathbf{\beta }_{1,0}^{\prime }\mathbf{w}_{it}=\beta
_{11,0}DO_{it}+TA_{it}$} \\ \hline
Sample ends: &  & {\footnotesize 1950-2021} & {\footnotesize 1950-2010} \\ 
\hline
\multicolumn{1}{l}{$\tilde{\beta}_{11}$} &  & -0.875 & -0.899 \\ 
\multicolumn{1}{l}{s.e.} &  & (0.010) & (0.012) \\ 
\multicolumn{1}{l}{t($\beta _{11,0}=-1)$} &  & -12.8 & -8.7 \\ \hline
\multicolumn{1}{l}{$n$} &  & 2,555 & 1,901 \\ 
\multicolumn{1}{l}{$\sum_{i=1}^{n}T_{i}$} &  & 82,837 & 60,118 \\ 
\hline\hline
\end{tabular}%
\vspace{-0.2in}
\end{center}

\begin{flushleft}
\scriptsize%
\singlespacing%
Notes: $TA_{it}$ and $DO_{it}$ are logarithms of total assets and total debt
outstanding. Left panel reports PME estimate of $\beta _{12,0}$ using the
exact identifying condition $\beta _{11,0}=1$. Right panel reports PME
estimate of $\beta _{11,0}$ using the exact identifying condition $\beta
_{12,0}=1$. PME estimator, given by (\ref{BETAdot_hat}), is computed using $%
q=2$\ sub-sample time averages, subject to $T_{i}\geq 20$, for all $i$. To
simplify the notations we use the tilde symbols to denote the exactly
identified PME estimates. The row labelled t($\beta _{12,0}=-1)$ and t($%
\beta _{11,0}=-1)$ gives the t statistic for testing $H_{0}:\beta _{12,0}=-1$
and $H_{0}:\beta _{11,0}=-1$, respectively. See Section \ref{Sup_micro} of
the supplement for variable definitions, data sources and availability, and
filters applied.\vspace{-0.3in}
\end{flushleft}

\begin{center}
\singlespacing%
TABLE 9:\textit{\ }PME estimates for the variable set $\left\{
SD_{it},LD_{it},TA_{it}\right\} $, using $q=2$\ sub-sample time averages and
two\ long-run relations

\renewcommand{\arraystretch}{1.1}\setlength{\tabcolsep}{5pt}%
\scriptsize%
\begin{tabular}{rccccccc}
\hline\hline
& \multicolumn{2}{c}{First exactly identified} &  &  &  & \multicolumn{2}{c}{
Second exactly identified} \\ 
& \multicolumn{2}{c}{long-run relation} &  &  &  & \multicolumn{2}{c}{
long-run relation} \\ 
\multicolumn{1}{l}{} & \multicolumn{2}{c}{$\mathbf{\beta }_{1,0}^{\prime }%
\mathbf{w}_{it}=SD_{it}+\beta _{13,0}TA_{it}$} &  &  &  & \multicolumn{2}{c}{%
$\mathbf{\beta }_{2,0}^{\prime }\mathbf{w}_{it}=LD_{it}+\beta _{23,0}TA_{it}$%
} \\ \hline
Sample end year: & {\footnotesize 1950-2021} & {\footnotesize 1950-2010} & 
&  & Sample end year: & {\footnotesize 1950-2021} & {\footnotesize 1950-2010}
\\ \hline
\multicolumn{1}{l}{$\tilde{\beta}_{13}$} & -1.025 & -1.024 &  &  & 
\multicolumn{1}{l}{$\tilde{\beta}_{23}$} & -1.185 & -1.158 \\ 
\multicolumn{1}{l}{s.e.} & (0.019) & (0.019) &  &  & \multicolumn{1}{l}{s.e.}
& (0.015) & (0.016) \\ 
\multicolumn{1}{l}{t($\beta _{13,0}=-1)$} & -1.4 & -1.3 &  &  & 
\multicolumn{1}{l}{t($\beta _{23,0}=-1)$} & -12.0 & -9.7 \\ \hline
\multicolumn{1}{l}{$n$} & 1,373 & 1,101 &  &  & \multicolumn{1}{l}{$n$} & 
1,373 & 1,101 \\ 
\multicolumn{1}{l}{$\sum_{i=1}^{n}T_{i}$} & 43,621 & 34,262 &  &  & 
\multicolumn{1}{l}{$\sum_{i=1}^{n}T_{i}$} & 43,621 & 34,262 \\ \hline\hline
\end{tabular}%
\vspace{-0.1in}\vspace{-0.1in}
\end{center}

\begin{flushleft}
\scriptsize%
\singlespacing%
Notes: Long run relations are $\mathbf{\beta }_{j,0}^{\prime }\mathbf{w}%
_{it}=\beta _{j1,0}SD_{it}+\beta _{j2,0}LD_{it}+\beta _{j3,0}TA_{it}$, for $%
j=1,2$, where $SD_{it}$, $LD_{it}$ and $TA_{it}$ are logarithms of
short-term and long-term debts, and total assets. The first long-run
relation $(j=1)$ is identified using the exact identifying conditions $\beta
_{11,0}=1$ and $\beta _{12,0}=0$. The second long-run relation ($j=2$) is
identified using the exact identifying conditions $\beta _{22,0}=1$ and $%
\beta _{21,0}=0$. This table reports estimates for $\beta _{13,0}$ and $%
\beta _{23,0}$ using PME estimator given by (\ref{BETAdot_hat}) with $q=2$
sub-sample time averages, subject to $T_{i}\geq 20$, for all $i$. To
simplify the notations we use tilde symbol to denote the exactly identified
PME estimates. See Section \ref{Sup_micro} of the supplement for variable
definitions, data sources and availability, and filters applied.\vspace{%
-0.3in}
\end{flushleft}

\begin{center}
\singlespacing%
TABLE 10:\textit{\ }PME estimation results for the variable set $\left\{
DO_{it},BV_{it},MV_{it}\right\} $, using $q=2$\ sub-sample time averages and
two\ long-run relations

\renewcommand{\arraystretch}{1.1}\setlength{\tabcolsep}{5pt}%
\scriptsize%
\begin{tabular}{rcccccc}
\hline\hline
& \multicolumn{2}{c}{First exactly identified} &  &  & \multicolumn{2}{c}{
Second exactly identified} \\ 
& \multicolumn{2}{c}{long-run relation} &  &  & \multicolumn{2}{c}{long-run
relation} \\ 
\multicolumn{1}{l}{} & \multicolumn{2}{c}{$\mathbf{\beta }_{1,0}^{\prime }%
\mathbf{w}_{it}=DO+\beta _{13,0}BV$} &  &  & \multicolumn{2}{c}{$\mathbf{%
\beta }_{2,0}^{\prime }\mathbf{w}_{it}=BV+\beta _{23,0}MV$} \\ \hline
Sample end year: & 2021 & 2010 &  & Sample end year: & 2021 & 2010 \\ \hline
\multicolumn{1}{l}{$\tilde{\beta}_{13}$} & -1.055 & -0.990 &  & 
\multicolumn{1}{l}{$\tilde{\beta}_{23}$} & -0.937 & -0.927 \\ 
\multicolumn{1}{l}{s.e.} & (0.019) & (0.020) &  & \multicolumn{1}{l}{s.e.} & 
(0.008) & (0.009) \\ 
\multicolumn{1}{l}{t($\beta _{13,0}=-1)$} & 2.8 & -0.5 &  & 
\multicolumn{1}{l}{t($\beta _{23,0}=-1)$} & -7.7 & -8.6 \\ \hline
\multicolumn{1}{l}{$n$} & 1,415 & 1,164 &  & $n$ & 1,415 & 1,164 \\ 
\multicolumn{1}{l}{$\sum_{i=1}^{n}T_{i}$} & 44,592 & 35,293 &  & $%
\sum_{i=1}^{n}T_{i}$ & 44,592 & 35,293 \\ \hline\hline
\end{tabular}%
\vspace{-0.1in}\vspace{-0.1in}
\end{center}

\begin{flushleft}
\scriptsize%
\singlespacing%
Notes: Long run relations are $\mathbf{\beta }_{j,0}^{\prime }\mathbf{w}%
_{it}=\beta _{j1,0}DO_{it}+\beta _{j2,0}BV_{it}+\beta _{j3,0}MV_{it}$, for $%
j=1,2$, where $DO_{it}$, $BV_{it}$ and $MV_{it}$ are logarithms of total
debt outstanding, book, and market values. The first long-run relation ($j=1$%
) is identified using the exact identifying conditions $\beta _{11,0}=1$ and 
$\beta _{12,0}=0$. The second long-run relation ($j=2$) is identified using
the exact identifying conditions $\beta _{22,0}=1$ and $\beta _{21,0}=0$.
See also the notes to Table 9.\pagebreak
\end{flushleft}

\normalsize%
\onehalfspacing%

Table 10 reports similar results for $\mathbf{w}_{it}=(DO_{it},$ $BV_{it},$ $%
MV_{it})^{\prime }$ and two exactly identified long-run relations associated
with logarithms total debt, market value, and book value. PME estimates are
close to -1, but the unit long-run elasticity between pairs of variables in
this set is rejected for all cases except the logarithms of total debt
outstanding and market value in the sample ending in 2010.\footnote{%
Rotation of the two identified long-run relationships reported in Table 10
imply the long-run relations $DO_{it}-0.888BV_{it}$ and $%
DO_{it}-0.937BV_{it} $ for the samples ending in 2021 and 2010,
respectively, both of these PME estimates are statistically significanlty
different from -1 at the 1 percent level.}

Overall, the results support the existence of long-run relations between the
logarithms of a number of key financial variables considered in the
corporate finance literature. But with the notable exception of the
logarithm of short term debt to total asset ratio, the use of other
financial ratios as stationary variables in financial analysis is not
supported by our empirical findings. Instead our study recommends using
estimated long-run relations such as $DO_{i,t-1}-1.14$ $TA_{i,t-1}$, $%
LD_{i,t-1}-1.19$ $TA_{i,t-1}$, and $BV_{i,t-1}-0.927MV_{i,t-1}$, as error
correction terms in dynamic panel regressors that allow for short term
dynamics and other stationary variables. This allows the analysis of short
term dynamics of financial variables to be coherently embedded in a long-run
equilibrating framework.

\subsection{International macro applications using Penn World Table}

Our second empirical application considers cross country macroeconomic time
series data from the Penn World Table\footnote{%
We use version 10.01 of PWT database, available at
https://www.rug.nl/ggdc/productivity/pwt/, see 
\shortciteN{FeenstraInklaarTimmer2015}%
. See also the supplement for a detailed description of data constructions.}
(PWT), where $n$ (the number of countries) is smaller and the average time
dimension is larger as compared to the corporate finance data. Given the
good small sample performance of the PME approach even for smaller values of 
$n$ in our Monte Carlo experiments, we have confidence in using the PME
estimation also in the case of the macro application. We focus on four key
macro variables, namely real merchandise exports per capita ($ex_{it}$),
real merchandise imports per capita ($im_{it}$), real productivity per hour
worked ($prod_{it}$) and real wages per hour worked ($wage_{it}$). The
choice of these variables was motivated by two widely maintained hypotheses.
Firstly, real wages and productivity should balance for steady state growth
to be feasible. Secondly export and imports should balance for international
solvency, though the constraint may not be binding for reserve-currency
countries such as the US. We first consider the two pairs that we expect to
cointegrate separately, then we consider all four variables together. As
with the corporate finance application, we are interested both in whether
they cointegrate and whether there is a unit coefficient. In both cases,
labour market balance and trade balance, there is no clear causal ordering
between the variables. Since the constraints that produce balance may be
somewhat different for advanced and emerging economies, we report estimates
for the country groupings separately as well as together.

The PME results for the analysis of a possible long-run relation between $%
ex_{it}$ and $im_{it}$ are summarized in Table 11. This table gives the two
eigenvalues of $\mathbf{R}_{_{\bar{w}\bar{w}}}$ for $\mathbf{w}_{it}=\left(
ex_{it},im_{it}\right) ^{\prime }$ and $q=2$. (See (\ref{Rww})) As can be
seen there is a clear separation between the first and second eigenvalues
supporting the existence of a long-run relation between $ex_{it}$ and $%
im_{it}$. This result holds for both choices of the threshold parameter $%
\delta =1/2$ and $1/4$, and for all three country groupings. Given this
result we then estimated the long-run relation $\mathbf{\beta }^{\prime }%
\mathbf{w}_{it}=$ $\beta _{11}ex_{it}+\beta _{12}im_{it}$, normalizing on $%
\beta _{12}=1$, for all three country groupings. There is a marked
difference between the estimates of $\beta _{11}$ across the advanced and
emerging economies. For the advanced economies we have $\hat{\beta}%
_{11}=-0.914$ ($0.029$), which strongly rejects the null hypothesis of $-1$
and suggests systematic differences can persist between merchandise imports
and exports for advanced economies. In contrast the estimate for the
emerging economies, namely $\hat{\beta}_{11}=-0.992$ ( $0.045$), is not
significantly different from $-1$. For all countries pooled, $\hat{\beta}%
_{11}=$ $-0.972$ ($0.034$) and does not reject the null of $\beta _{11}=-1$.%
\footnote{%
Normalizing on $\beta _{11}$ yields the same results for $\beta _{12}$ since
PME estimate of $\beta _{12}$ is exactly equal to $1/\hat{\beta}_{11}$ and
by construction $Var(\hat{\beta}_{12})=Var(1/\hat{\beta}_{11})$.} The
difference in the estimates obtained for advanced and emerging economies
could be due to greater ability of advanced economies in financing their
goods trade imbalances by increasing their export of services and having
easier access to international capital market.

Similar results are obtained when we consider the relation between wages and
labour productivity. For all country groupings and both choices of the
thresholds we find $\tilde{r}=1$. See Table 12. The PME estimates of the
long-run relation, $\mathbf{\beta }^{\prime }\mathbf{w}_{it}=$ $\beta
_{11}wage_{it}+prod_{it}$ for advanced and emerging economies are $\hat{\beta%
}_{11}=-0.953$ ($0.013$) and $\hat{\beta}_{11}=-0.984$ ($0.043$),
respectively. Once again estimates of $\beta _{11}$ are quite close to $-1$,
but as in the case of imports and exports the null of $\beta _{11}=-1$ is
rejected for advanced economies but not for emerging economies\textbf{.}

\begin{center}
\singlespacing%
TABLE 11:\textit{\ }PME estimates for $\mathbf{w}_{it}=\left(
ex_{it},im_{it}\right) ^{\prime }$, using $q=2$\ sub-sample time averages.

\renewcommand{\arraystretch}{1.1}\setlength{\tabcolsep}{5pt}%
\scriptsize%
\begin{tabular}{rccr}
\hline\hline
Sample: & Advanced & Emerging & All economies \\ \hline
\multicolumn{4}{l}{\textbf{Eigenvalues }of $\mathbf{R}_{_{\bar{w}\bar{w}}}$%
\textbf{\ given by (\ref{Rww}) }(in ascending order)} \\ \hline
$\tilde{\lambda}_{1}$ & 0.026 & 0.102 & \multicolumn{1}{c}{0.084} \\ 
$\tilde{\lambda}_{2}$ & 1.974 & 1.898 & \multicolumn{1}{c}{1.916} \\ \hline
\multicolumn{4}{l}{\textbf{Threshold }$\bar{T}^{-\delta }$} \\ \hline
$\delta =1/2$ & 0.128 & 0.134 & \multicolumn{1}{c}{0.132} \\ 
$\delta =1/4$ & 0.357 & 0.366 & \multicolumn{1}{c}{0.364} \\ \hline
\multicolumn{4}{l}{\textbf{Estimated number of long-run relations (}$\tilde{r%
}$\textbf{)}} \\ \hline
$\tilde{r}$ ($\delta =1/2$) & 1 & 1 & \multicolumn{1}{c}{1} \\ 
$\tilde{r}$ ($\delta =1/4$) & 1 & 1 & \multicolumn{1}{c}{1} \\ \hline
\multicolumn{4}{l}{\textbf{Exactly identified long-run relations}} \\ 
\multicolumn{4}{c}{$\mathbf{\beta }^{\prime }\mathbf{w}_{it}=\left( 
\begin{array}{cc}
\beta _{11} & 1%
\end{array}%
\right) \left( 
\begin{array}{c}
ex_{it} \\ 
im_{it}%
\end{array}%
\right) =\beta _{11}ex_{it}+im_{it}$} \\ \hline
$\hat{\beta}_{11}$ & -0.914 & -0.992 & \multicolumn{1}{c}{-0.972} \\ 
& (0.029) & (0.045) & \multicolumn{1}{c}{(0.034)} \\ \hline
\multicolumn{4}{l}{\textbf{Sample dimensions}} \\ \hline
$n$ & 38 & 139 & \multicolumn{1}{c}{177} \\ 
$\bar{T}=n^{-1}\sum_{i=1}^{n}T_{i}$ & 61.3 & 56.1 & \multicolumn{1}{c}{57.2}
\\ \hline\hline
\end{tabular}%
\vspace{-0.1in}\vspace{-0.1in}
\end{center}

\begin{flushleft}
\scriptsize%
\singlespacing%
Notes: Standard errors are reported in parentheses. See Section \ref%
{Sup_macro} of the supplement for variable definitions, data sources and
availability, and filters applied.
\end{flushleft}

\begin{center}
\normalsize%
\singlespacing%
TABLE 12:\textit{\ }PME estimates for $\mathbf{w}_{it}=\left(
wage_{it},prod_{it}\right) ^{\prime }$, using $q=2$\ sub-sample time
averages.

\renewcommand{\arraystretch}{1.1}\setlength{\tabcolsep}{5pt}%
\scriptsize%
\begin{tabular}{rrrr}
\hline\hline
Sample: & Advanced & Emerging & All economies \\ \hline
\multicolumn{4}{l}{\textbf{Eigenvalues }of $\mathbf{R}_{_{\bar{w}\bar{w}}}$%
\textbf{\ given by (\ref{Rww}) }(in ascending order)} \\ \hline
$\tilde{\lambda}_{1}$ & \multicolumn{1}{c}{0.004} & \multicolumn{1}{c}{0.042}
& \multicolumn{1}{c}{0.015} \\ 
$\tilde{\lambda}_{2}$ & \multicolumn{1}{c}{1.996} & \multicolumn{1}{c}{1.958}
& \multicolumn{1}{c}{1.985} \\ \hline
\multicolumn{4}{l}{\textbf{Threshold }$\bar{T}^{-\delta }$} \\ \hline
$\delta =1/2$ & \multicolumn{1}{c}{0.135} & \multicolumn{1}{c}{0.146} & 
\multicolumn{1}{c}{0.139} \\ 
$\delta =1/4$ & \multicolumn{1}{c}{0.367} & \multicolumn{1}{c}{0.382} & 
\multicolumn{1}{c}{0.373} \\ \hline
\multicolumn{4}{l}{\textbf{Estimated number of long-run relations (}$\tilde{r%
}$\textbf{)}} \\ \hline
$\tilde{r}$ ($\delta =1/2$) & \multicolumn{1}{c}{1} & \multicolumn{1}{c}{1}
& \multicolumn{1}{c}{1} \\ 
$\tilde{r}$ ($\delta =1/4$) & \multicolumn{1}{c}{1} & \multicolumn{1}{c}{1}
& \multicolumn{1}{c}{1} \\ \hline
\multicolumn{4}{l}{\textbf{Exactly identified long-run relations}} \\ 
\multicolumn{4}{c}{$\mathbf{\beta }^{\prime }\mathbf{w}_{it}=\left( 
\begin{array}{cc}
\beta _{11} & 1%
\end{array}%
\right) \left( 
\begin{array}{c}
prod_{it} \\ 
wage_{it}%
\end{array}%
\right) =\beta _{11}prod_{it}+wage_{it}$} \\ \hline
$\hat{\beta}_{11}$ & \multicolumn{1}{c}{-0.953} & \multicolumn{1}{c}{-0.984}
& \multicolumn{1}{c}{-0.962} \\ 
& \multicolumn{1}{c}{(0.013)} & \multicolumn{1}{c}{(0.043)} & 
\multicolumn{1}{c}{(0.016)} \\ \hline
\multicolumn{4}{l}{\textbf{Sample dimensions}} \\ \hline
$n$ & \multicolumn{1}{c}{35} & \multicolumn{1}{c}{24} & \multicolumn{1}{c}{59
} \\ 
$\bar{T}=n^{-1}\sum_{i=1}^{n}T_{i}$ & \multicolumn{1}{c}{55.5} & 
\multicolumn{1}{c}{47.4} & \multicolumn{1}{c}{52.2} \\ \hline\hline
\end{tabular}%
\vspace{-0.1in}\vspace{-0.1in}
\end{center}

\begin{flushleft}
\scriptsize%
\singlespacing%
Notes: See notes to Table 11.\vspace{-0.15in}
\end{flushleft}

\normalsize%
\onehalfspacing%

To illustrate how our proposed methods perform in the case of multiple
long-run relations we now consider all the four variables together and set $%
\mathbf{w}_{it}=\left( ex_{it},im_{it},prod_{it},wage_{it}\right) ^{\prime }$%
. Given the above pair-wise results we would expect at least two long-run
relations amongst these four variables. But the four eigenvalues of $\mathbf{%
R}_{_{\bar{w}\bar{w}}}$ reported in Table 13 clearly suggest that there are
three long-run relations among the four variables, irrespective whether we
use $\delta =1/4$ or $1/2$ as the threshold parameter. This result
highlights the advantage of considering the possibility of multiple long-run
relations and can help with discovery of hitherto unnoticed or overlooked
long relations. For the third long-run relation we consider a possible
long-run relation between exports and productivity. There is extensive
microeconomic evidence suggesting exporting firms tend to have higher
productivity. See, for example, 
\citeN{BernardJensen2004}%
. However, there is controversy about whether the relation arises because
high productivity firms export more or whether the competitive pressure of
exporting boosts productivity. See for example 
\shortciteN{Aghion_etal2018}%
.\textbf{\ }Also, while the firm level micro relation has been examined for
many countries, the country level macro relation has been less intensively
explored. Our application provides cross country evidence on the relation
between exports and productivity without making any assumption about the
direction of causality between these variables. Accordingly, we consider the
following exactly identified long-run relations assuming that $r_{0}=3$
amongst the four variables $\mathbf{w}_{it}$,\vspace{-0.15in}%
\begin{equation}
\mathbf{B}^{\prime }\mathbf{w}_{it}=\left( 
\begin{array}{cccc}
\beta _{11} & 1 & 0 & 0 \\ 
0 & 0 & \beta _{23} & 1 \\ 
\beta _{31} & 0 & 1 & 0%
\end{array}%
\right) \left( 
\begin{array}{c}
ex_{it} \\ 
im_{it} \\ 
prod_{it} \\ 
wage_{it}%
\end{array}%
\right) \text{.}  \label{id4}
\end{equation}%
PME estimates for $\beta _{11}$ and $\beta _{23}$ in Table 13 are in line
with the estimates in Tables 11 and 12. For the third long-run relation, we
obtain $\hat{\beta}_{31}=-0.509$ ($0.023$) for advanced economies and $\hat{%
\beta}_{31}=-0.426$ ( $0.032$) for emerging economies. Hence, we discovered
that exports and productivity are related with the expected sign. Unlike the
other two long-run relations, we do not have any \textit{a priori} reason to
believe the estimates of $\beta _{31}$ should be close to $-1$.

To corroborate the evidence in Table 13 regarding the third long-run
relation, Table 14 reports PME findings for $\mathbf{w}_{it}=\left(
prod_{it},ex_{it}\right) ^{\prime }$. There is a clear separation of
eigenvalues, indicating existence of a long-run relation, in line with
findings in Table 14 for the four-variable vector $\mathbf{w}_{it}$. In
addition, the estimated coefficients in Table 14 are very similar to the
corresponding coefficients reported in Table 13.\vspace{-0.2in}

\begin{center}
\singlespacing%
TABLE 13:\textit{\ }PME estimates for the variable set $\left\{
wage_{it},prod_{it},im_{it},ex_{it}\right\} $, using $q=2$\ sub-sample time
averages.

\renewcommand{\arraystretch}{1.1}\setlength{\tabcolsep}{5pt}%
\scriptsize%
\begin{tabular}{rrrr}
\hline\hline
Sample: & Advanced & Emerging & All economies \\ \hline
\multicolumn{4}{l}{\textbf{Eigenvalues }of $\mathbf{R}_{_{\bar{w}\bar{w}}}$%
\textbf{\ given by (\ref{Rww}) }(in ascending order)} \\ \hline
$\tilde{\lambda}_{1}$ & \multicolumn{1}{c}{0.003} & \multicolumn{1}{c}{0.012}
& \multicolumn{1}{c}{0.014} \\ 
$\tilde{\lambda}_{2}$ & \multicolumn{1}{c}{0.010} & \multicolumn{1}{c}{0.037}
& \multicolumn{1}{c}{0.015} \\ 
$\tilde{\lambda}_{3}$ & \multicolumn{1}{c}{0.064} & \multicolumn{1}{c}{0.111}
& \multicolumn{1}{c}{0.088} \\ 
$\tilde{\lambda}_{4}$ & \multicolumn{1}{c}{3.923} & \multicolumn{1}{c}{3.840}
& \multicolumn{1}{c}{3.883} \\ \hline
\multicolumn{4}{l}{\textbf{Threshold }$\bar{T}^{-\delta }$} \\ \hline
$\delta =1/2$ & \multicolumn{1}{c}{0.135} & \multicolumn{1}{c}{0.146} & 
\multicolumn{1}{c}{0.139} \\ 
$\delta =1/4$ & \multicolumn{1}{c}{0.367} & \multicolumn{1}{c}{0.382} & 
\multicolumn{1}{c}{0.373} \\ \hline
\multicolumn{4}{l}{\textbf{Estimated number of long-run relations (}$\tilde{r%
}$\textbf{)}} \\ \hline
$\tilde{r}$ ($\delta =1/2$) & \multicolumn{1}{c}{3} & \multicolumn{1}{c}{3}
& \multicolumn{1}{c}{3} \\ 
$\tilde{r}$ ($\delta =1/4$) & \multicolumn{1}{c}{3} & \multicolumn{1}{c}{3}
& \multicolumn{1}{c}{3} \\ \hline
\multicolumn{4}{l}{\textbf{Exactly identified long-run relations}} \\ 
\multicolumn{4}{c}{$\mathbf{B}^{\prime }\mathbf{w}_{it}=\left( 
\begin{array}{cccc}
\beta _{11} & 1 & 0 & 0 \\ 
0 & 0 & \beta _{23} & 1 \\ 
\beta _{31} & 0 & 1 & 0%
\end{array}%
\right) \left( 
\begin{array}{c}
ex_{it} \\ 
im_{it} \\ 
prod_{it} \\ 
wage_{it}%
\end{array}%
\right) $} \\ \hline
$\hat{\beta}_{11}$ & \multicolumn{1}{c}{-0.882} & \multicolumn{1}{c}{-1.005}
& \multicolumn{1}{c}{-0.928} \\ 
& \multicolumn{1}{c}{(0.027)} & \multicolumn{1}{c}{(0.031)} & 
\multicolumn{1}{c}{(0.023)} \\ 
$\hat{\beta}_{23}$ & \multicolumn{1}{c}{-0.952} & \multicolumn{1}{c}{-0.954}
& \multicolumn{1}{c}{-0.953} \\ 
& \multicolumn{1}{c}{(0.013)} & \multicolumn{1}{c}{(0.041)} & 
\multicolumn{1}{c}{(0.015)} \\ 
$\hat{\beta}_{31}$ & \multicolumn{1}{c}{-0.509} & \multicolumn{1}{c}{-0.426}
& \multicolumn{1}{c}{-0.478} \\ 
& \multicolumn{1}{c}{(0.023)} & \multicolumn{1}{c}{(0.032)} & 
\multicolumn{1}{c}{(0.021)} \\ \hline
\multicolumn{4}{l}{\textbf{Sample dimensions}} \\ \hline
$n$ & \multicolumn{1}{c}{35} & \multicolumn{1}{c}{24} & \multicolumn{1}{c}{59
} \\ 
$\bar{T}=n^{-1}\sum_{i=1}^{n}T_{i}$ & \multicolumn{1}{c}{55.5} & 
\multicolumn{1}{c}{47.4} & \multicolumn{1}{c}{52.2} \\ \hline\hline
\end{tabular}%
\vspace{-0.1in}\vspace{-0.1in}
\end{center}

\begin{flushleft}
\scriptsize%
\singlespacing%
Notes: See Section \ref{Sup_macro} of the supplement for variable
definitions, data sources and availability, and filters applied.\vspace{%
-0.15in}
\end{flushleft}

\begin{center}
\normalsize%
\singlespacing%
TABLE 14:\textit{\ }PME estimates for $\mathbf{w}_{it}=\left(
ex_{it},prod_{it}\right) ^{\prime }$, using $q=2$\ sub-sample time averages.

\renewcommand{\arraystretch}{1.1}\setlength{\tabcolsep}{5pt}%
\scriptsize%
\begin{tabular}{rrrr}
\hline\hline
Sample: & Advanced & Emerging & All economies \\ \hline
\multicolumn{4}{l}{\textbf{Eigenvalues }of $\mathbf{R}_{_{\bar{w}\bar{w}}}$%
\textbf{\ given by (\ref{Rww}) }(in ascending order)} \\ \hline
$\tilde{\lambda}_{1}$ & \multicolumn{1}{c}{0.025} & \multicolumn{1}{c}{0.080}
& \multicolumn{1}{c}{0.061} \\ 
$\tilde{\lambda}2$ & \multicolumn{1}{c}{1.975} & \multicolumn{1}{c}{1.920} & 
\multicolumn{1}{c}{1.939} \\ \hline
\multicolumn{4}{l}{\textbf{Threshold }$\bar{T}^{-\delta }$} \\ \hline
$\delta =1/2$ & \multicolumn{1}{c}{0.135} & \multicolumn{1}{c}{0.146} & 
\multicolumn{1}{c}{0.139} \\ 
$\delta =1/4$ & \multicolumn{1}{c}{0.367} & \multicolumn{1}{c}{0.382} & 
\multicolumn{1}{c}{0.373} \\ \hline
\multicolumn{4}{l}{\textbf{Estimated number of long-run relations (}$\tilde{r%
}$\textbf{)}} \\ \hline
$\tilde{r}$ ($\delta =1/2$) & \multicolumn{1}{c}{1} & \multicolumn{1}{c}{1}
& \multicolumn{1}{c}{1} \\ 
$\tilde{r}$ ($\delta =1/4$) & \multicolumn{1}{c}{1} & \multicolumn{1}{c}{1}
& \multicolumn{1}{c}{1} \\ \hline
\multicolumn{4}{l}{\textbf{Exactly identified long-run relations}} \\ 
\multicolumn{4}{c}{$\mathbf{B}^{\prime }\mathbf{w}_{it}=\left( 
\begin{array}{cc}
\beta _{11} & 1%
\end{array}%
\right) \left( 
\begin{array}{c}
ex_{it} \\ 
prod_{it}%
\end{array}%
\right) =\beta _{11}ex_{it}+prod_{it}$} \\ \hline
$\hat{\beta}_{11}$ & \multicolumn{1}{c}{-0.510} & \multicolumn{1}{c}{-0.357}
& \multicolumn{1}{c}{-0.432} \\ 
& \multicolumn{1}{c}{(0.023)} & \multicolumn{1}{c}{(0.044)} & 
\multicolumn{1}{c}{(0.036)} \\ \hline
\multicolumn{4}{l}{\textbf{Sample dimensions}} \\ \hline
$n$ & \multicolumn{1}{c}{35} & \multicolumn{1}{c}{29} & \multicolumn{1}{c}{64
} \\ 
$\bar{T}=n^{-1}\sum_{i=1}^{n}T_{i}$ & \multicolumn{1}{c}{55.5} & 
\multicolumn{1}{c}{47.0} & \multicolumn{1}{c}{51.7} \\ \hline\hline
\end{tabular}%
\vspace{-0.1in}\vspace{-0.1in}
\end{center}

\begin{flushleft}
\scriptsize%
\singlespacing%
Notes: See notes to Table 11.\vspace{-0.15in}
\end{flushleft}

\begin{landscape}%

\begin{center}
\singlespacing%
TABLE 15: Comparison of PME estimates with alternative estimates of single
long-run relation, $\mathbf{w}_{it}=\left( w_{1it},w_{2,t}\right) ^{\prime }$%
\bigskip

\renewcommand{\arraystretch}{1.2}\setlength{\tabcolsep}{2pt}%
\scriptsize%
\begin{tabular}{rcccrcccrcccrcccrcccrccc}
\hline\hline
$w_{1it}:$ & \multicolumn{7}{c}{$ex_{it}$} &  & \multicolumn{7}{c}{$%
prod_{it} $} &  & \multicolumn{7}{c}{$ex_{it}$} \\ 
$w_{2,t}:$ & \multicolumn{7}{c}{$im_{it}$} &  & \multicolumn{7}{c}{$%
wage_{it} $} &  & \multicolumn{7}{c}{$prod_{it}$} \\ 
\cline{2-8}\cline{10-16}\cline{18-24}
Coint. relation: & \multicolumn{3}{c}{$\beta _{11}w_{1it}+w_{2,t}$} &  & 
\multicolumn{3}{c}{$w_{1it}+\beta _{12}w_{2,t}$} &  & \multicolumn{3}{c}{$%
\beta _{11}w_{1it}+w_{2,t}$} &  & \multicolumn{3}{c}{$w_{1it}+\beta
_{12}w_{2,t}$} &  & \multicolumn{3}{c}{$\beta _{11}w_{1it}+w_{2,t}$} &  & 
\multicolumn{3}{c}{$w_{1it}+\beta _{12}w_{2,t}$} \\ 
& \multicolumn{3}{c}{$\hat{\beta}_{11}$} &  & \multicolumn{3}{c}{$\hat{\beta}%
_{12}$} &  & \multicolumn{3}{c}{$\hat{\beta}_{11}$} &  & \multicolumn{3}{c}{$%
\hat{\beta}_{12}$} &  & \multicolumn{3}{c}{$\hat{\beta}_{11}$} &  & 
\multicolumn{3}{c}{$\hat{\beta}_{12}$} \\ 
\cline{2-4}\cline{6-8}\cline{10-12}\cline{14-16}\cline{18-20}\cline{22-24}
Sample: & Adv. & Eme. & All & \multicolumn{1}{c}{} & Adv. & Eme. & All & 
\multicolumn{1}{c}{} & Adv. & Eme. & All & \multicolumn{1}{c}{} & Adv. & Eme.
& All & \multicolumn{1}{c}{} & Adv. & Eme. & All & \multicolumn{1}{c}{} & 
Adv. & Eme. & All \\ \hline
PME & -0.914 & -0.992 & -0.972 & \multicolumn{1}{c}{} & -1.094 & -1.008 & 
-1.029 & \multicolumn{1}{c}{} & -0.953 & -0.984 & -0.962 & 
\multicolumn{1}{c}{} & -1.049 & -1.016 & -1.039 & \multicolumn{1}{c}{} & 
-0.510 & -0.357 & -0.432 & \multicolumn{1}{c}{} & -1.962 & -2.803 & -2.315
\\ 
& (0.029) & (0.045) & (0.034) & \multicolumn{1}{c}{} & (0.035) & (0.045) & 
(0.036) & \multicolumn{1}{c}{} & (0.013) & (0.043) & (0.016) & 
\multicolumn{1}{c}{} & (0.015) & (0.056) & (0.021) & \multicolumn{1}{c}{} & 
(0.023) & (0.044) & (0.036) & \multicolumn{1}{c}{} & (0.075) & (0.251) & 
(0.119) \\ 
SPMG & -0.962 & -1.026 & -0.976 & \multicolumn{1}{c}{} & -1.040 & -0.974 & 
-1.025 & \multicolumn{1}{c}{} & -1.072 & -0.987 & -1.043 & 
\multicolumn{1}{c}{} & -0.933 & -1.013 & -0.959 & \multicolumn{1}{c}{} & 
-0.367 & -0.704 & -0.371 & \multicolumn{1}{c}{} & -2.729 & -1.421 & -2.697
\\ 
& (0.004) & (0.006) & (0.004) & \multicolumn{1}{c}{} & (0.004) & (0.006) & 
(0.004) & \multicolumn{1}{c}{} & (0.005) & (0.004) & (0.003) & 
\multicolumn{1}{c}{} & (0.005) & (0.004) & (0.003) & \multicolumn{1}{c}{} & 
(0.004) & (0.013) & (0.003) & \multicolumn{1}{c}{} & (0.027) & (0.033) & 
(0.024) \\ 
Breitung's 2-step & -0.842 & -0.924 & -0.902 & \multicolumn{1}{c}{} & -1.048
& -0.882 & -0.922 & \multicolumn{1}{c}{} & -1.144 & -1.010 & -1.066 & 
\multicolumn{1}{c}{} & -0.832 & -0.924 & -0.900 & \multicolumn{1}{c}{} & 
-0.522 & -0.371 & -0.455 & \multicolumn{1}{c}{} & -1.701 & -1.901 & -1.756
\\ 
& (0.010) & (0.021) & (0.016) & \multicolumn{1}{c}{} & (0.019) & (0.016) & 
(0.013) & \multicolumn{1}{c}{} & (0.008) & (0.020) & (0.005) & 
\multicolumn{1}{c}{} & (0.005) & (0.018) & (0.004) & \multicolumn{1}{c}{} & 
(0.013) & (0.013) & (0.010) & \multicolumn{1}{c}{} & (0.042) & (0.095) & 
(0.040) \\ 
PMG & -0.985 & -0.996 & -0.989 & \multicolumn{1}{c}{} & -0.973 & -0.949 & 
-0.960 & \multicolumn{1}{c}{} & -1.140 & -0.990 & -1.100 & 
\multicolumn{1}{c}{} & -0.837 & -0.962 & -0.886 & \multicolumn{1}{c}{} & 
-0.291 & -0.718 & -0.306 & \multicolumn{1}{c}{} & -1.568 & -1.350 & -1.527
\\ 
& (0.007) & (0.008) & (0.005) & \multicolumn{1}{c}{} & (0.009) & (0.009) & 
(0.006) & \multicolumn{1}{c}{} & (0.008) & (0.007) & (0.005) & 
\multicolumn{1}{c}{} & (0.006) & (0.009) & (0.004) & \multicolumn{1}{c}{} & 
(0.007) & (0.020) & (0.006) & \multicolumn{1}{c}{} & (0.026) & (0.077) & 
(0.024) \\ 
PB & -0.868 & -0.813 & -0.828 & \multicolumn{1}{c}{} & -0.993 & -0.869 & 
-0.891 & \multicolumn{1}{c}{} & -1.136 & -0.966 & -1.097 & 
\multicolumn{1}{c}{} & -0.844 & -0.961 & -0.888 & \multicolumn{1}{c}{} & 
-0.486 & -0.379 & -0.437 & \multicolumn{1}{c}{} & -1.688 & -1.967 & -1.771
\\ 
& (0.027) & (0.033) & (0.025) & \multicolumn{1}{c}{} & (0.039) & (0.037) & 
(0.031) & \multicolumn{1}{c}{} & (0.006) & (0.035) & (0.002) & 
\multicolumn{1}{c}{} & (0.005) & (0.031) & (0.002) & \multicolumn{1}{c}{} & 
(0.028) & (0.056) & (0.038) & \multicolumn{1}{c}{} & (0.066) & (0.293) & 
(0.103) \\ 
PFMOLS & -0.857 & -0.835 & -0.841 & \multicolumn{1}{c}{} & -1.083 & -0.886 & 
-0.934 & \multicolumn{1}{c}{} & -1.241 & -0.978 & -1.070 & 
\multicolumn{1}{c}{} & -0.776 & -0.944 & -0.896 & \multicolumn{1}{c}{} & 
-0.528 & -0.329 & -0.445 & \multicolumn{1}{c}{} & -1.723 & -2.051 & -1.822
\\ 
& (0.023) & (0.035) & (0.026) & \multicolumn{1}{c}{} & (0.050) & (0.043) & 
(0.032) & \multicolumn{1}{c}{} & (0.015) & (0.485) & (0.009) & 
\multicolumn{1}{c}{} & (0.010) & (0.163) & (0.008) & \multicolumn{1}{c}{} & 
(0.064) & (0.060) & (0.092) & \multicolumn{1}{c}{} & (0.059) & (0.191) & 
(0.081) \\ 
PDOLS & -0.912 & -0.841 & -0.856 & \multicolumn{1}{c}{} & -1.045 & -0.801 & 
-0.853 & \multicolumn{1}{c}{} & -1.093 & -1.029 & -1.089 & 
\multicolumn{1}{c}{} & -0.868 & -1.029 & -0.891 & \multicolumn{1}{c}{} & 
-0.520 & -0.409 & -0.470 & \multicolumn{1}{c}{} & -1.853 & -1.957 & -1.901
\\ 
& (0.004) & (0.006) & (0.004) & \multicolumn{1}{c}{} & (0.005) & (0.006) & 
(0.004) & \multicolumn{1}{c}{} & (0.006) & (0.004) & (0.003) & 
\multicolumn{1}{c}{} & (0.004) & (0.004) & (0.003) & \multicolumn{1}{c}{} & 
(0.004) & (0.008) & (0.004) & \multicolumn{1}{c}{} & (0.015) & (0.038) & 
(0.015) \\ \hline
\multicolumn{1}{l}{Sample size} &  &  &  & \multicolumn{1}{c}{} &  &  &  & 
\multicolumn{1}{c}{} &  &  &  & \multicolumn{1}{c}{} &  &  &  & 
\multicolumn{1}{c}{} &  &  &  & \multicolumn{1}{c}{} &  &  &  \\ \hline
$n$ & 38 & 139 & 177 & \multicolumn{1}{c}{} & 38 & 139 & 177 & 
\multicolumn{1}{c}{} & 35 & 24 & 59 & \multicolumn{1}{c}{} & 35 & 24 & 59 & 
\multicolumn{1}{c}{} & 35 & 29 & 64 & \multicolumn{1}{c}{} & 35 & 29 & 64 \\ 
$\bar{T}=n^{-1}\sum_{i=1}^{n}T_{i}$ & 61.3 & 56.1 & 57.2 & 
\multicolumn{1}{c}{} & 61.3 & 56.1 & 57.2 & \multicolumn{1}{c}{} & 55.5 & 
47.4 & 52.2 & \multicolumn{1}{c}{} & 55.5 & 47.4 & 52.2 & \multicolumn{1}{c}{
} & 55.5 & 47.0 & 51.7 & \multicolumn{1}{c}{} & 55.5 & 47.0 & 51.7 \\ 
\hline\hline
\end{tabular}%
\vspace{-0.1in}\vspace{-0.1in}
\end{center}

\begin{flushleft}
\scriptsize%
\singlespacing%
Notes: PME and SPMG are the only estimators where $\hat{\beta}_{11}=\hat{%
\beta}_{12}^{-1}$. This is not the case for the remaining estimators. PME is
the pooled minimum eigenvalue estimator using $q=2$ subsamples. SPMG\ is the
system PMG estimator of 
\citeN{ChudikPesaranSmith2023}%
, using $p=2$ lags in levels ($1$ lag in first differences). Breitung's
2-step is the two-step system estimator of 
\citeN{Breitung2005}%
, using $p=2$ lags in levels. PMG\ is the Pooled Mean Group (PMG) estimator
by 
\citeN{PesaranShinSmith1999}%
, using $p=2$ lags in levels. PB is the pooled Bewley estimator by 
\citeN{ChudikPesaranSmith2021PB}%
, using $p=2$ lags in levels. PFMOLS is the panel FMOLS estimator by 
\citeANP{Pedroni1996} (\citeyearNP{Pedroni1996}, \citeyearNP{Pedroni2001}, \citeyearNP{Pedroni2001ReStat})%
. PDOLS is the panel Dynamic OLS (PDOLS) by 
\citeN{MarkSul2003}%
, using one lead and one lag of first differences.
\end{flushleft}

\end{landscape}%

\normalsize%
\onehalfspacing%

\subsection{Comparison of PME and their estimates in the case of pair-wise
relations}

Here we provide comparative estimates for the pair-wise long-run relations
for which a number of alternative estimators are proposed in the literature.
Specifically, we compare PME estimates reported in Tables 11, 12, and 14
with the SPMG, Breitung's 2-step, PMG, PB, FMOLS, and PDOLS estimators
discussed in Sub-Section \ref{SubS_MC_c}.\footnote{%
The SPMG estimator is proposed by 
\citeN{ChudikPesaranSmith2023}%
, the Breitung's two-step system estimator is proposed by 
\citeN{Breitung2005}%
, the PMG\ estimator is proposed by 
\citeN{PesaranShinSmith1999}%
, the PB (Pooled Bewley) estimator is proposed by 
\citeN{ChudikPesaranSmith2021PB}%
, the panel FMOLS\ estimator is proposed by 
\citeANP{Pedroni1996} (\citeyearNP{Pedroni1996}, \citeyearNP{Pedroni2001}, \citeyearNP{Pedroni2001ReStat})%
, and the panel DOLS estimator is proposed by 
\citeN{MarkSul2003}%
.}\textbf{\ }The results are summarized in Table 15. PME and SPMG are the
only estimators that are invariant to the normalization imposed on $\mathbf{%
\beta }^{\prime }\mathbf{w}_{it}=\beta _{11}w_{1,it}+\beta _{12}w_{2,it}$.
As a result, the equality $\hat{\beta}_{11}\hat{\beta}_{12}=1$ holds only in
the case of PME and SPMG estimators.

The estimates of $\beta _{11}$ and $\beta _{12}$ are generally similar but
there are some large differences. For instance, the PME estimate of the
long-run coefficient in $\beta _{11}ex_{it}+prod_{it}$ in the case of
advanced economies is $-0.510$ ($0.023$) compared with SPMG estimate of $%
-0.367$ ($0.004$). Such difference could arise from the possibility that
some of the SPMG modeling assumptions regarding short-run dynamics are not
met, whereas the PME estimator is more general in that it does not require
modeling of short-run dynamics. We also observe that standard errors of the
estimates proposed in the literature are in some cases 2 to 10-fold smaller
compared to PME. As we have seen in Monte Carlo experiments all of the
estimators proposed in the literature severely over-reject the null, whilst
this is not the case if we consider the MC results for size reported in
Table 6.

\section{Concluding remarks\label{CON}}

This paper provides a new, pooled minimum eigenvalue (PME), methodology for
the analysis of multiple long-run relations in panel data models where the
cross section dimension, $n$, is large relative to the time series
dimension, $T$. \ It uses non-overlapping sub-sample time averages as
deviations from their full-sample counterpart and estimates the number of
long-run relations and their coefficients using eigenvalues and eigenvectors
of the pooled covariance matrix of these sub-sample deviations. It applies
to unbalanced panels generated from general linear processes with
interactive stationary time effects and does not require knowing long-run
causal linkages. The PME estimator is consistent and asymptotically normally
distributed as $n$ and $T$ $\rightarrow \infty $ jointly, such that $%
T\approx n^{d}$, with $d>0$ for consistency and $d>1/2$ for asymptotic
normality. Extensive Monte Carlo studies show that the number of long-run
relations can be estimated with high precision and the PME estimates of the
long-run coefficients show small bias and $RMSE$ and have good size and
power properties. The utility of our approach is illustrated with both micro
and macro applications. The micro application uncovers long-run relations
among key financial variables in an unbalanced panel of US firms from merged
CRSP-Compustat data set covering $2,000$ plus firms over the period $%
1950-2021$. The macro application uses cross country macroeconomic time
series data covering up to $177$ countries over slightly shorter time
period, $1950-2019.$

As well as application in other areas than corporate finance and
macroeconomics the procedure opens up a range of theoretical developments.
One question that we are considering investigating is whether it is possible
to develop a unit root test based on similar principles. Our method is
semi-parametric, in that it does not model the short run dynamics. Because
the estimates of the long-run relations are super consistent, then having
estimated them, the PME estimates of the long-run relations could be used as
inputs into second stage models using stationary variables. For instance,
they can provide estimates of the disequilibrium terms in error correction
models of the short run dynamics, to measure speeds of adjustment. Such
models could also be used for medium-term forecasting and counterfactual
analysis.

\newpage

\noindent 
\appendix%

\numberwithin{equation}{section}%

\onehalfspacing%

\section{\textbf{Appendix}}

This appendix provides proofs of propositions and theorems stated in the
paper. This appendix follows the same notations as in Section \ref{LongRR}.
Sub-sample time averages for a generic vector $\mathbf{x}_{it}$ are defined
as $\mathbf{\bar{x}}_{i\ell }=\frac{1}{T_{q}}\sum_{t=(\ell -1)T_{q}+1}^{\ell
T_{q}}\mathbf{x}_{it},$ for $\ell =1,2,...,q$, and the full-sample time
average by $\mathbf{\bar{x}}_{i\circ }=q^{-1}\sum_{\ell =1}^{q}\mathbf{\bar{x%
}}_{i\ell }$, where $q$ $(\geq 2)$, $m$ is fixed, and $T_{q}=T/q$ is an
integer. Small and large finite positive constants that do not depend on
sample sizes $n$ and $T$ are denoted by $\epsilon $ and $K$, respectively.
They can take different values at different instances. All lemmas referenced
in the proofs below are provided in the supplement.

\begin{proof}[Proof of Proposition \protect\ref{Qwbar}]
Consider the last three terms on the right side of (\ref{qwbd}). For its
second term we have $E\left\Vert n^{-1}\sum_{i=1}^{n}\mathbf{C}_{i}\mathbf{Q}%
_{\bar{s}_{i}\bar{v}_{i}}\right\Vert \leq n^{-1}\sum_{i=1}^{n}\left\Vert 
\mathbf{C}_{i}\right\Vert E\left\Vert \mathbf{Q}_{\bar{s}_{i}\bar{v}%
_{i}}\right\Vert \leq \sup_{i}\left\Vert \mathbf{C}_{i}\right\Vert
\sup_{i}E\left\Vert \mathbf{Q}_{\bar{s}_{i}\bar{v}_{i}}\right\Vert $. By
Assumption \ref{ASS2} $\sup_{i}\left\Vert \mathbf{C}_{i}\right\Vert <K$, and
using (\ref{bQsv}) in Lemma \ref{LQB} of the supplement we have $%
\sup_{i}E\left\Vert \mathbf{Q}_{\bar{s}_{i}\bar{v}_{i}}\right\Vert =O\left(
T^{-1}\right) $. Hence, it follows that $E\left\Vert n^{-1}\sum_{i=1}^{n}%
\mathbf{C}_{i}\mathbf{Q}_{\bar{s}_{i}\bar{v}_{i}}\right\Vert =O\left(
T^{-1}\right) $. Similarly, $\ E\left\Vert n^{-1}\sum_{i=1}^{n}\mathbf{Q}_{%
\bar{v}_{i}\bar{s}_{i}}^{\prime }\mathbf{C}_{i}^{\prime }\right\Vert
=O\left( T^{-1}\right) $, and, using (\ref{bQvv}) of Lemma \ref{LQB}, $%
E\left\Vert n^{-1}\sum_{i=1}^{n}\mathbf{Q}_{\bar{v}_{i}\bar{v}%
_{i}}\right\Vert =O\left( T^{-2}\right) $. Using these results in (\ref{qwbd}%
) now yields%
\begin{equation}
E\left\Vert \mathbf{Q}_{\bar{w}\bar{w}}-n^{-1}\sum_{i=1}^{n}\mathbf{C}_{i}%
\mathbf{Q}_{\bar{s}_{i}\bar{s}_{i}}\mathbf{C}_{i}^{\prime }\right\Vert
=O\left( T^{-1}\right) \text{.}  \label{p1pd}
\end{equation}%
Consider $n^{-1}\sum_{i=1}^{n}\mathbf{C}_{i}\mathbf{Q}_{\bar{s}_{i}\bar{s}%
_{i}}\mathbf{C}_{i}^{\prime }$ and note that it can be written as%
\begin{equation}
n^{-1}\sum_{i=1}^{n}\mathbf{C}_{i}\mathbf{Q}_{\bar{s}_{i}\bar{s}_{i}}\mathbf{%
C}_{i}^{\prime }=n^{-1}\sum_{i=1}^{n}\mathbf{C}_{i}E\left( \mathbf{Q}_{\bar{s%
}_{i}\bar{s}_{i}}\right) \mathbf{C}_{i}^{\prime }+n^{-1}\sum_{i=1}^{n}%
\mathbf{V}_{i}\text{,}  \label{p1pc}
\end{equation}%
where $\mathbf{V}_{i}=\mathbf{C}_{i}\left[ \mathbf{Q}_{\bar{s}_{i}\bar{s}%
_{i}}-E\left( \mathbf{Q}_{\bar{s}_{i}\bar{s}_{i}}\right) \right] \mathbf{C}%
_{i}^{\prime }$. Using result (\ref{EQss}) of Lemma \ref{LEQss} and noting
that $sup_{i}\left\Vert \mathbf{\Sigma }_{i}\right\Vert <K$ under Assumption %
\ref{ASS1}, we obtain%
\begin{equation}
n^{-1}\sum_{i=1}^{n}\mathbf{C}_{i}E\left( \mathbf{Q}_{\bar{s}_{i}\bar{s}%
_{i}}\right) \mathbf{C}_{i}^{\prime }=\frac{(q-1)}{6q}\mathbf{\Psi }%
_{n}+O\left( T^{-2}\right) \text{,}  \label{p1pb}
\end{equation}%
where $\mathbf{\Psi }_{n}=n^{-1}\sum_{i=1}^{n}\mathbf{C}_{i}\mathbf{\Sigma }%
_{i}\mathbf{C}_{i}^{\prime }$. In addition, uniformly bounded fourth moments
of individual elements of $\mathbf{u}_{it}$ ensure variances of individual
elements $\mathbf{Q}_{\bar{s}_{i}\bar{s}_{i}}$ are bounded. Since $%
\sup_{i}\left\Vert \mathbf{C}_{i}\right\Vert <K$ by Assumption \ref{ASS2},
it also follows that the variances of the individual elements of $\mathbf{V}%
_{i}$ are uniformly bounded. Noting that $E(\mathbf{V}_{i})=\mathbf{0}$ by
construction, and that $\mathbf{V}_{i}$ is independently distributed of $%
\mathbf{V}_{j}$ for all $i\neq j$ (by Assumption \ref{ASS1}), we have 
\begin{equation}
n^{-1}\sum_{i=1}^{n}\mathbf{V}_{i}=O_{p}\left( n^{-1/2}\right) \text{.}
\label{p1pa}
\end{equation}%
Using (\ref{p1pb}) and (\ref{p1pa}) in (\ref{p1pc}) yields%
\begin{equation}
n^{-1}\sum_{i=1}^{n}\mathbf{C}_{i}\mathbf{Q}_{\bar{s}_{i}\bar{s}_{i}}\mathbf{%
C}_{i}^{\prime }=\frac{(q-1)}{6q}\mathbf{\Psi }_{n}+O_{p}\left(
n^{-1/2}\right) +O\left( T^{-2}\right) \text{,}  \label{rq1}
\end{equation}%
and using this result in (\ref{p1pd}), we obtain $\mathbf{Q}_{\bar{w}\bar{w}%
}\rightarrow _{p}\frac{(q-1)}{6q}\mathbf{\Psi }$, as $n,T\rightarrow \infty $
(in no particular order), where $\mathbf{\Psi }=\lim_{n\rightarrow \infty }%
\mathbf{\Psi }_{n}$. This completes the proof of (\ref{Qbar2}). Result (\ref%
{QbarBeta}) follows from (\ref{Qbar2}) by noting that $\mathbf{C}%
_{i}^{\prime }\mathbf{\mathring{B}}_{0}\mathbf{=0}$ for all $i$ by
Assumption \ref{ASS3}, which in turn yields $\mathbf{\Psi }_{n}\mathbf{%
\mathring{B}}_{0}\mathbf{=\mathbf{\Psi }\mathring{B}}_{0}=\mathbf{0}$.
\end{proof}

\begin{proof}[Proof of Proposition \protect\ref{PropID}]
Under Assumption \ref{ASS3} $\left( n^{-1}\sum_{i=1}^{n}\mathbf{C}_{i}%
\mathbf{\Sigma }_{i}\mathbf{C}_{i}^{\prime }\right) \mathbf{\mathring{B}}_{0}%
\mathbf{=\Psi }_{n}\mathbf{\mathring{B}}_{0}\mathbf{=0,}$ for any $n$ and as 
$n\rightarrow \infty $. Then we have $\mathbf{\Psi \mathbf{\mathring{B}}%
_{0}=0=}\left( \mathbf{P}^{\prime }\mathbf{P}\right) \mathbf{\mathring{B}}%
_{0}\mathbf{=0}$. But by condition $rank\left( \mathbf{\Psi }\right)
=m-r_{0} $ of Assumption \ref{ASS3}, $\mathbf{P}$ is an $(m-r_{0})\times m\,$%
full row rank matrix. Then $\mathbf{PP}^{\prime }$ is non-singular, and we
must also have $\mathbf{P\mathbf{\mathring{B}}_{0}=0}$. Combing this result
with (\ref{ExactRes}) now yields 
\begin{equation}
\left( 
\begin{array}{c}
\mathbf{R} \\ 
\mathbf{P}%
\end{array}%
\right) \mathbf{\mathring{B}}_{0}\mathbf{=}\left( 
\begin{array}{c}
\mathbf{A} \\ 
\mathbf{0}%
\end{array}%
\right) .  \label{sys1}
\end{equation}%
Under Assumptions \ref{ASS3} and the exact $r_{0}^{2}$ restrictions given by
(\ref{ExactRes}) with $rank\left( \mathbf{A}\right) =rank\left( \mathbf{R}%
\right) =r_{0}<m$, condition $rank\left( 
\begin{array}{c}
\mathbf{P}_{n} \\ 
\mathbf{R}%
\end{array}%
\right) =rank\left( 
\begin{array}{c}
\mathbf{P} \\ 
\mathbf{R}%
\end{array}%
\right) =m$ holds, $\mathbf{R}^{\prime }\mathbf{R+P}^{\prime }\mathbf{P}$ is
a positive definite matrix, and\ $\mathbf{\beta \,}$is uniquely determined
by 
\begin{equation*}
\mathbf{\mathring{B}}_{0}=\left( \mathbf{R}^{\prime }\mathbf{R+P}^{\prime }%
\mathbf{P}\right) ^{-1}\left( 
\begin{array}{c}
\mathbf{A} \\ 
\mathbf{0}%
\end{array}%
\right) =\left( \mathbf{R}^{\prime }\mathbf{R+\Psi }\right) ^{-1}\left( 
\begin{array}{c}
\mathbf{A} \\ 
\mathbf{0}%
\end{array}%
\right) \text{.}
\end{equation*}%
Now using the normalizations $\mathbf{R=(I}_{r_{0}},\mathbf{0})^{\prime }$
and $\mathbf{A=I}_{r_{0}}$ and conformably partitioning $\mathbf{\Psi }$ we
have%
\begin{equation*}
\mathbf{R}^{\prime }\mathbf{R+\Psi =}\left( 
\begin{array}{cc}
\mathbf{I}_{r_{0}}+\mathbf{\Psi }_{11} & \mathbf{\Psi }_{21}^{\prime } \\ 
\mathbf{\Psi }_{21} & \mathbf{\Psi }_{22}%
\end{array}%
\right) .
\end{equation*}%
Since $\mathbf{R}^{\prime }\mathbf{R+\Psi }$ is a positive definite matrix
then the $r_{0}\times r_{0}$ and $(m-r_{0})\times \left( m-r_{0}\right) $
matrices $\mathbf{I}_{r_{0}}+\mathbf{\Psi }_{11}$ and $\mathbf{\Psi }_{22}$
are also positive definite. Solving the following system of equations now
yields 
\begin{equation*}
\left( 
\begin{array}{cc}
\mathbf{I}_{r_{0}}+\mathbf{\Psi }_{11} & \mathbf{\Psi }_{21}^{\prime } \\ 
\mathbf{\Psi }_{21} & \mathbf{\Psi }_{22}%
\end{array}%
\right) \left( 
\begin{array}{c}
\mathbf{\mathring{B}}_{0,1} \\ 
\mathbf{\Theta }%
\end{array}%
\right) =\left( 
\begin{array}{c}
\mathbf{I}_{r_{0}} \\ 
\mathbf{0}%
\end{array}%
\right) ,
\end{equation*}%
and we obtain $\left( \mathbf{I}_{r}+\mathbf{\Psi }_{11}\right) \mathbf{%
\mathring{B}}_{0,1}+\mathbf{\Psi }_{21}^{\prime }\mathbf{\Theta }=\mathbf{I}%
_{r_{0}},$ and $\mathbf{\Psi }_{21}\mathbf{\beta }_{1}+\mathbf{\Psi }_{22}%
\mathbf{\Theta =0}$. Also, since $\mathbf{\mathring{B}}_{0,1}=\mathbf{I}%
_{r_{0}}$ it follows that $\left( \mathbf{I}_{r_{0}}+\mathbf{\Psi }%
_{11}\right) +\mathbf{\Psi }_{21}^{\prime }\mathbf{\Theta }=\mathbf{I}%
_{r_{0}},$ and $\mathbf{\Theta =-\Psi }_{22}^{-1}\mathbf{\Psi }_{21},$which
implies that $\mathbf{\Psi }_{11}=\mathbf{\Psi }_{21}^{\prime }\mathbf{\Psi }%
_{22}^{-1}\mathbf{\Psi }_{21}$, and in turn ensures that $\mathbf{\Psi }$ is
rank deficient, as required under Assumption \ref{ASS3}).
\end{proof}

\begin{proof}[Proof of Theorem \protect\ref{Tcons}]
Multiplying (\ref{rq1}) by $\mathbf{\hat{B}}_{0}^{\prime }$ from the left
and by $\mathbf{\hat{B}}_{0}$ from the right, and noting eigenvectors $%
\mathbf{\hat{B}}_{0}$ are normalized so that $\mathbf{\hat{B}}_{0}^{\prime }%
\mathbf{\hat{B}}_{0}=\mathbf{I}_{r}$ yields 
\begin{equation}
\mathbf{\hat{B}}_{0}^{\prime }\mathbf{Q}_{\bar{w}\bar{w}}\mathbf{\hat{B}}%
_{0}=\frac{(q-1)}{6q}\mathbf{\hat{B}}_{0}^{\prime }\mathbf{\Psi }_{n}\mathbf{%
\hat{B}}_{0}+O_{p}\left( n^{-1/2}\right) +O_{p}\left( T^{-2}\right) \text{,}
\label{ca}
\end{equation}%
where $lim_{n\rightarrow \infty }\mathbf{\Psi }_{n}=\mathbf{\Psi }$ as $%
n\rightarrow \infty $. Under Assumption \ref{ASS3}, we have $rank\left( 
\mathbf{\Psi }\right) =m-r_{0}$, $rank\left( \mathbf{\mathring{B}}%
_{0}\right) =r$ and $\mathbf{\mathring{B}}_{0}^{\prime }\mathbf{\Psi }=%
\mathbf{0}$. Hence the space space spanned by the column vectors of $m\times
r_{0}$ matrix $\mathbf{\mathring{B}}_{0}$ is the same as the space spanned
by the first $r_{0}$ eigenvectors of $\mathbf{\Psi }$ associated with its $r$
smallest eigenvalues. It now follows from (\ref{ca}) that the space spanned
by the column vectors of $\mathbf{\hat{B}}_{0}$ converges to the space
spanned by the column vectors of $\mathbf{\mathring{B}}_{0}$, as $%
n,T\rightarrow \infty $ (in no particular order), and $\mathbf{\hat{B}}_{0}%
\mathbf{H}\rightarrow _{p}\mathbf{\mathring{B}}_{0}$, for a suitable choice
of $r_{0}\times r_{0}$ nonsingular matrix $\mathbf{H}$.\footnote{%
The rotation matrix is given by $\mathbf{H=}\left[ \left( \mathbf{\mathring{B%
}}_{0}^{\prime }\mathbf{\mathring{B}}_{0}\right) ^{-1}\mathbf{\mathring{B}}%
_{0}^{\prime }\mathbf{\hat{B}}\right] ^{-1}$, see Lemma 13.1 of 
\citeN{Johansen1995}%
.}
\end{proof}

\begin{proof}[Proof of Theorem \protect\ref{Tb}]
Multiplying both sides of (\ref{qwbd}) by $\mathbf{\mathring{B}}_{0}$ from
the right, and noting that $\mathbf{C}_{i}^{\prime }\mathbf{\mathring{B}}%
_{0}=\mathbf{0}$ under Assumption \ref{ASS3}, we have%
\begin{equation}
\mathbf{Q}_{\bar{w}\bar{w}}\mathbf{\mathring{B}}_{0}=\left(
n^{-1}\sum_{i=1}^{n}\mathbf{C}_{i}\mathbf{Q}_{\bar{s}_{i}\bar{v}_{i}}\right) 
\mathbf{\mathring{B}}_{0}+\mathbf{Q}_{\bar{v}\bar{v}}\mathbf{\mathring{B}}%
_{0}  \label{tcpa}
\end{equation}%
Lemma \ref{LQb} established $\left\Vert \mathbf{Q}_{\bar{w}\bar{w}}\mathbf{%
\hat{B}}_{0}\right\Vert =O_{p}\left( n^{-1/2}T^{-2}\right) $, and given that 
$\widehat{\mathbf{\mathring{B}}}_{0}$ is an $O_{p}\left( 1\right) $ rotation
of $\mathbf{\hat{B}}_{0}$, it follows%
\begin{equation}
\mathbf{Q}_{\bar{w}\bar{w}}\widehat{\mathbf{\mathring{B}}}_{0}=O_{p}\left(
n^{-1/2}T^{-2}\right) \text{.}  \label{tcpb}
\end{equation}%
Subtracting (\ref{tcpa}) from (\ref{tcpb}) yields 
\begin{equation}
\mathbf{Q}_{\bar{w}\bar{w}}\left( \widehat{\mathbf{\mathring{B}}}_{0}-%
\mathbf{\mathring{B}}_{0}\right) =-\left( n^{-1}\sum_{i=1}^{n}\mathbf{C}_{i}%
\mathbf{Q}_{\bar{s}_{i}\bar{v}_{i}}\right) \mathbf{\mathring{B}}_{0}-\mathbf{%
Q}_{\bar{v}\bar{v}}\mathbf{\mathring{B}}_{0}+O_{p}\left(
n^{-1/2}T^{-2}\right) \mathbf{.}  \label{QwwD}
\end{equation}%
Result (\ref{sqvvs}) of Lemma (\ref{L_sqvvs}) implies $\mathbf{Q}_{\bar{v}%
\bar{v}}=O_{p}\left( n^{-1/2}T^{-2}\right) $, and we have%
\begin{equation}
\mathbf{Q}_{\bar{w}\bar{w}}\sqrt{n}T\left( \widehat{\mathbf{\mathring{B}}}%
_{0}-\mathbf{\mathring{B}}_{0}\right) =-\left( n^{-1/2}\sum_{i=1}^{n}T%
\mathbf{C}_{i}\mathbf{Q}_{\bar{s}_{i}\bar{v}_{i}}\right) \mathbf{\mathring{B}%
}_{0}+O_{p}\left( T^{-1}\right) .  \label{t2a}
\end{equation}%
The first term on the right side of (\ref{t2a}) can be written as%
\begin{equation*}
\left( n^{-1/2}\sum_{i=1}^{n}T\mathbf{C}_{i}\mathbf{Q}_{\bar{s}_{i}\bar{v}%
_{i}}\right) \mathbf{\mathring{B}}_{0}=\left( n^{-1/2}\sum_{i=1}^{n}\mathbf{Z%
}_{i}\right) \mathbf{\mathring{B}}_{0}+\frac{\sqrt{n}}{T}\left[
n^{-1}\sum_{i=1}^{n}\mathbf{C}_{i}E\left( T^{2}\mathbf{Q}_{\bar{s}_{i}\bar{v}%
_{i}}\right) \right] \mathbf{\mathring{B}}_{0}\text{,}
\end{equation*}%
where $\mathbf{Z}_{i}=\mathbf{C}_{i}\left[ T\mathbf{Q}_{\bar{s}_{i}\bar{v}%
_{i}}-E\left( T\mathbf{Q}_{\bar{s}_{i}\bar{v}_{i}}\right) \right] \mathbf{%
\mathring{B}}_{0}$. Using result (\ref{EQsvb}) of Lemma \ref{LQB}, we obtain%
\begin{equation}
\left\Vert n^{-1}\sum_{i=1}^{n}\mathbf{C}_{i}E\left( T^{2}\mathbf{Q}_{\bar{s}%
_{i}\bar{v}_{i}}\right) \mathbf{\mathring{B}}_{0}\right\Vert \leq
T^{2}\sup_{i}\left\Vert \mathbf{C}_{i}\right\Vert \sup_{i}\left\Vert E\left( 
\mathbf{Q}_{\bar{s}_{i}\bar{v}_{i}}\right) \right\Vert \left\Vert \mathbf{%
\mathring{B}}_{0}\right\Vert <K\text{,}  \label{t2c}
\end{equation}%
where $\left\Vert \mathbf{\mathring{B}}_{0}\right\Vert <K$ and $%
\sup_{i}\left\Vert \mathbf{C}_{i}\right\Vert _{1}<K$ by Assumption \ref{ASS2}%
. Using the above result in (\ref{t2a}) it follows that 
\begin{equation*}
\mathbf{Q}_{\bar{w}\bar{w}}\sqrt{n}T\left( \widehat{\mathbf{\mathring{B}}}%
_{0}-\mathbf{\mathring{B}}_{0}\right) =-n^{-1/2}\sum_{i=1}^{n}\mathbf{Z}%
_{i}+O_{p}\left( \frac{\sqrt{n}}{T}\right) +O_{p}\left( T^{-1}\right) ,
\end{equation*}%
Vectorizing the above equation we have%
\begin{equation}
\left( \mathbf{I}_{r}\mathbf{\otimes Q}_{\bar{w}\bar{w}}\right) \sqrt{n}T%
\text{ $vec$}\left( \widehat{\mathbf{\mathring{B}}}_{0}-\mathbf{\mathring{B}}%
_{0}\right) =n^{-1/2}\sum_{i=1}^{n}\left( \mathbf{\mathring{B}}_{0}^{\prime }%
\mathbf{\otimes \mathbf{C}}_{i}\right) \mathbf{\tilde{\xi}}_{iq}+O_{p}\left( 
\frac{\sqrt{n}}{T}\right) +O_{p}\left( T^{-1}\right) \text{,}  \label{t2d}
\end{equation}%
where $\mathbf{\tilde{\xi}}_{iq}=\mathbf{\bar{\xi}}_{iq}-E\left( \mathbf{%
\bar{\xi}}_{iq}\right) $, and $\mathbf{\bar{\xi}}_{iq}$ is given by (recall $%
\mathbf{Z}_{i}=\mathbf{C}_{i}\left[ T\mathbf{Q}_{\bar{s}_{i}\bar{v}%
_{i}}-E\left( T\mathbf{Q}_{\bar{s}_{i}\bar{v}_{i}}\right) \right] \mathbf{%
\mathring{B}}_{0}$ and $\mathbf{Q}_{\bar{s}_{i}\bar{v}_{i}}=T^{-1}q^{-1}%
\sum_{\ell =1}^{q}\left( \mathbf{\bar{s}}_{i\ell }-\mathbf{\bar{s}}_{i\circ
}\right) \left( \mathbf{\bar{v}}_{i\ell }-\mathbf{\bar{v}}_{i\circ }\right)
^{\prime }$)%
\begin{eqnarray*}
\mathbf{\bar{\xi}}_{iq} &=&q^{-1}\sum_{\ell =1}^{q}vec\left[ \left( \mathbf{%
\bar{s}}_{i\ell }-\mathbf{\bar{s}}_{i\circ }\right) \left( \mathbf{\bar{v}}%
_{i\ell }-\mathbf{\bar{v}}_{i\circ }\right) ^{\prime }\right]
=q^{-1}\sum_{\ell =1}^{q}\left( \mathbf{\bar{v}}_{i\ell }-\mathbf{\bar{v}}%
_{i\circ }\right) \mathbf{\otimes }\left( \mathbf{\bar{s}}_{i\ell }-\mathbf{%
\bar{s}}_{i\circ }\right) \text{,} \\
&=&q^{-1}\sum_{\ell =1}^{q}\mathbf{\bar{v}}_{i\ell }\mathbf{\otimes \bar{s}}%
_{i\ell }-\left( q^{-1}\sum_{\ell =1}^{q}\mathbf{\bar{v}}_{i\ell }\right) 
\mathbf{\otimes \bar{s}}_{i\circ }-\mathbf{\bar{v}}_{i\circ }\mathbf{\otimes 
}\left( q^{-1}\sum_{\ell =1}^{q}\mathbf{\bar{s}}_{i\ell }\right) +\mathbf{%
\bar{v}}_{i\circ }\mathbf{\otimes \bar{s}}_{i\circ }\text{.}
\end{eqnarray*}%
Using $q^{-1}\sum_{\ell =1}^{q}\mathbf{\bar{v}}_{i\ell }=\mathbf{\bar{v}}%
_{i\circ }$ and $q^{-1}\sum_{\ell =1}^{q}\mathbf{\bar{s}}_{i\ell }=\mathbf{%
\bar{s}}_{i\circ }$, the expression for $\mathbf{\bar{\xi}}_{iq}$ simplifies
to $\mathbf{\bar{\xi}}_{iq}=q^{-1}\sum_{\ell =1}^{q}\left( \mathbf{\bar{v}}%
_{i\ell }-\mathbf{\bar{v}}_{i\circ }\right) \mathbf{\otimes }\left( \mathbf{%
\bar{s}}_{i\ell }-\mathbf{\bar{s}}_{i\circ }\right) =q^{-1}\sum_{\ell
=1}^{q}\left( \mathbf{\bar{v}}_{i\ell }\mathbf{\otimes \bar{s}}_{i\ell
}\right) -\mathbf{\bar{v}}_{i\circ }\mathbf{\otimes \bar{s}}_{i\circ }$.
Lemma \ref{Ld} established convergence in distribution for $%
n^{-1/2}\sum_{i=1}^{n}\left( \mathbf{\mathring{B}}^{\prime }\mathbf{\otimes 
\mathbf{C}}_{i}\right) \mathbf{\tilde{\xi}}_{iq}$. Using this Lemma in (\ref%
{t2d}), and noting that $d>1/2$ implies$\sqrt{n}/T\rightarrow 0$ as $%
n,T\rightarrow \infty $, we obtain (\ref{dist}), as required.
\end{proof}

\bibliographystyle{chicago}
\bibliography{LAref}

\pagebreak

\normalsize%

\setcounter{section}{0} \renewcommand{\thesection}{S\arabic{section}}

\setcounter{page}{1} \renewcommand{\thepage}{S.\arabic{page}}

\setcounter{equation}{0} \renewcommand{\theequation}{S.\arabic{equation}}

\onehalfspacing%

\ \quad \vspace{0.05in}

\begin{center}
\textbf{{\large Supplement} }\\[0pt]

\textbf{{\large {\textquotedblleft Analysis of Multiple Long-Run Relations
in Panel Data Models\textquotedblright \footnote{%
The views expressed in this paper are those of the authors and do not
necessarily reflect those of the Federal Reserve Bank of Dallas or the
Federal Reserve System. }}}}\bigskip \bigskip\ \\[0pt]

Alexander Chudik

Federal Reserve Bank of Dallas\bigskip

M. Hashem Pesaran

Trinity College, Cambridge, UK and University of Southern California,
USA\bigskip

Ron P. Smith

Birkbeck, University of London, United Kingdom\\[0pt]
\bigskip \bigskip

\today\bigskip
\end{center}

\noindent This supplement is organized in seven sections. Section \ref{A2}
presents lemmas and their proofs. Section \ref{Aunb} describes
implementation of the PME estimator for unbalanced panels. Section \ref{TSIE}
provides theorems and proofs for the consistency and asymptotic distribution
of the PME estimator for the model with interactive time effects. Section %
\ref{Sup_MCd} describes the Monte Carlo data generating processes and
provides a list of individual Monte Carlo experiments. Section \ref{Sup_MCs}
investigates sensitivity of the PME estimator of $r_{0}$, $\tilde{r}$ given
by (\ref{r_est}), to different scaling of the observations on $\mathbf{w}%
_{it}$. Section \ref{Sup_MCr} presents Monte Carlo findings for robustness
of PME estimators to GARCH and threshold autoregressive effects. Section \ref%
{Sup_micro} describes data sources and variable constructions for the micro
application, and Section \ref{Sup_macro} provides supplementary information
for the macro application.

\section{Lemmas\label{A2}}

\begin{lemma}
\label{Lacf}Let $\mathbf{v}_{it}=\mathbf{C}_{i}^{\ast }(L)\mathbf{u}%
_{it}=\sum_{j=0}^{\infty }\mathbf{C}_{ij}^{\ast }\mathbf{u}_{i,t-j}$ and
suppose Assumptions \ref{ASS1} and \ref{ASS2} hold. Then there exist finite
positive constants $K_{1}$, $K_{2}$, and $0<\rho <1$ such that%
\begin{equation}
\sup_{i}\left\Vert \mathbf{\Gamma }_{i}(h)\right\Vert <K_{1}\rho ^{h}\text{,}
\label{supGh}
\end{equation}%
and%
\begin{equation}
\sum_{h=0}^{\infty }\sup_{i}\left\Vert \mathbf{\Gamma }_{i}(h)\right\Vert
<K_{2}\text{,}  \label{sumG}
\end{equation}%
where $\mathbf{\Gamma }_{i}(h)=E\left( \mathbf{v}_{it}\mathbf{v}%
_{i,t-h}^{\prime }\right) ,$ for $h=0,1,2,...\,\ $is the autocovariance
function of $\mathbf{v}_{it}$.
\end{lemma}

\begin{proof}
The autocovariance function of $\mathbf{v}_{it}$ is%
\begin{eqnarray*}
\mathbf{\Gamma }_{i}(h) &=&E\left( \mathbf{v}_{it}\mathbf{v}_{i,t-h}^{\prime
}\right) =E\left( \sum_{j=0}^{\infty }\mathbf{C}_{ij}^{\ast }\mathbf{u}%
_{i,t-j}\right) \left( \sum_{j^{\prime }=0}^{\infty }\mathbf{u}%
_{i,t-j^{\prime }-h}^{\prime }\mathbf{C}_{ij^{\prime }}^{\ast \prime }\right)
\\
&=&\sum_{j=0}^{\infty }\sum_{j^{\prime }=0}^{\infty }\mathbf{C}_{ij}^{\ast
}E\left( \mathbf{u}_{i,t-j}\mathbf{u}_{i,t-j^{\prime }-h}^{\prime }\right) 
\mathbf{C}_{ij^{\prime }}^{\ast \prime }=\sum_{j=h}^{\infty }\mathbf{C}%
_{ij}^{\ast }\mathbf{\Sigma }_{i}\mathbf{C}_{i,j-h}^{\ast ^{\prime }}=%
\mathbf{\Gamma }_{i}^{\prime }(-h).
\end{eqnarray*}%
Taking spectral norm and supremum over $i$ yields%
\begin{equation*}
\sup_{i}\left\Vert \mathbf{\Gamma }_{i}(h)\right\Vert \leq
\sum_{j=h}^{\infty }\sup_{i}\left\Vert \mathbf{C}_{ij}^{\ast }\right\Vert
\sup_{i}\left\Vert \mathbf{\Sigma }_{i}\right\Vert \sup_{i}\left\Vert 
\mathbf{C}_{i,j-h}^{\ast }\right\Vert \leq K_{0}\sum_{j=h}^{\infty }\rho
^{j}\rho ^{j-h}=\frac{K_{0}\rho ^{h}}{1-\rho ^{2}}<K_{1}\rho ^{h},
\end{equation*}%
where $\sup_{i}\left\Vert \mathbf{\Sigma }_{i}\right\Vert <K$ by Assumption %
\ref{ASS1} and $\sup_{i}\left\Vert \mathbf{C}_{ij}^{\ast }\right\Vert <K\rho
^{j}$ with $0<\rho <1$ by Assumption \ref{ASS2}. It follows $%
\sum_{h=0}^{\infty }\sup_{i}\left\Vert \mathbf{\Gamma }_{i}(h)\right\Vert
<K_{2}$.
\end{proof}

\begin{lemma}
\label{LEQss} Consider the $m\times m$ matrix $\mathbf{Q}_{\bar{s}_{i}\bar{s}%
_{i}}=T^{-1}q^{-1}\sum_{\ell =1}^{q}\left( \mathbf{\bar{s}}_{i\ell }-\mathbf{%
\bar{s}}_{i\circ }\right) \left( \mathbf{\bar{s}}_{i\ell }-\mathbf{\bar{s}}%
_{i\circ }\right) ^{\prime }$, where $\mathbf{\bar{s}}_{i\ell }$ and $%
\mathbf{\bar{s}}_{i\circ }$ are sub-sample and full sample time averages of
the partial sum process $\mathbf{s}_{it}=\sum_{\ell =1}^{t}\mathbf{u}_{i\ell
}$. Suppose $E(\mathbf{u}_{it}\mathbf{u}_{it}^{\prime })=\mathbf{\Sigma }%
_{i} $ for $i=1,2,...,n$ and $t=1,2,...,T$, and $E(\mathbf{u}_{it}\mathbf{u}%
_{it^{\prime }}^{\prime })=\mathbf{0}$ for $i=1,2,...,n$, and all $t\neq
t^{\prime }$, $t,t^{\prime }=1,2,...,T$. Then 
\begin{equation}
E\left( \mathbf{Q}_{\bar{s}_{i}\bar{s}_{i}}\right) =\frac{(q-1)}{6}\left( 
\frac{1}{q}+\frac{1}{T^{2}}\right) \mathbf{\Sigma }_{i}\text{, }
\label{EQss}
\end{equation}%
for $q\geq 2$ and $i=1,2,...,n$.
\end{lemma}

\begin{proof}
We note that 
\begin{equation}
E\left( T\mathbf{Q}_{\bar{s}_{i}\bar{s}_{i}}\right) =E\left[
q^{-1}\sum_{\ell =1}^{q}\left( \mathbf{\bar{s}}_{i\ell }-\mathbf{\bar{s}}%
_{i\circ }\right) \left( \mathbf{\bar{s}}_{i\ell }-\mathbf{\bar{s}}_{i\circ
}\right) ^{\prime }\right] =q^{-1}\sum_{\ell =1}^{q}E\left( \mathbf{\bar{s}}%
_{i\ell }\mathbf{\bar{s}}_{i\ell }^{\prime }\right) -E\left( \mathbf{\bar{s}}%
_{i\circ }\mathbf{\bar{s}}_{i\circ }^{\prime }\right) \text{.}
\label{peqss1}
\end{equation}%
Also%
\begin{eqnarray*}
\mathbf{\bar{s}}_{i1} &=&\frac{1}{T_{q}}\left( \mathbf{s}_{i1}+\mathbf{s}%
_{i2}+....+\mathbf{s}_{i,T_{q}}\right) =\frac{1}{T_{q}}\left[ \mathbf{u}%
_{i1}+(\mathbf{u}_{i1}+\mathbf{u}_{i2})+....\left( \mathbf{u}_{i1}+\mathbf{u}%
_{i2}+...+\mathbf{u}_{i,T_{q}}\right) \right] \text{,} \\
&=&\frac{1}{T_{q}}\left[ T_{q}\mathbf{u}_{i1}+(T_{q}-1)\mathbf{u}%
_{i2}+(T_{q}-2)\mathbf{u}_{i3}+....+2\mathbf{u}_{i,T_{q}-1}+\mathbf{u}%
_{i,T_{q}}\right] \text{,}
\end{eqnarray*}%
which can be written more compactly as $\mathbf{\bar{s}}_{i1}=T_{q}^{-1}%
\sum_{t=1}^{T_{q}}\left( T_{q}-t+1\right) \mathbf{u}_{it}$. Similarly 
\begin{eqnarray*}
\mathbf{\bar{s}}_{i2} &=&\frac{1}{T_{q}}\left( \mathbf{s}_{i,T_{q}+1}+%
\mathbf{s}_{i,T_{q}+2}+....+\mathbf{s}_{i,2T_{q}}\right) \\
&=&\frac{1}{T_{q}}\left[ T_{q}\left( \mathbf{u}_{i1}+\mathbf{u}_{i2}+...+%
\mathbf{u}_{i,T_{q}}\right) +\sum_{t=T_{q}+1}^{2T_{q}}\left(
2T_{q}-t+1\right) \mathbf{u}_{it}\right] \text{,} \\
&=&\sum_{t=1}^{T_{q}}\mathbf{u}_{it}+\frac{1}{T_{q}}%
\sum_{t=T_{q}+1}^{2T_{q}}\left( 2T_{q}-t+1\right) \mathbf{u}_{it}\text{,}
\end{eqnarray*}%
and more generally%
\begin{equation}
\mathbf{\bar{s}}_{i\ell }=\sum_{t=1}^{(\ell -1)T_{q}}\mathbf{u}_{it}+\frac{1%
}{T_{q}}\sum_{t=(\ell -1)T_{q}+1}^{\ell T_{q}}\left( \ell T_{q}+1-t\right) 
\mathbf{u}_{it}\text{, for }\ell =1,2,...,q\text{.}  \label{siell}
\end{equation}%
It now readily follows that 
\begin{equation*}
E\left( \mathbf{\bar{s}}_{i\ell }\mathbf{\bar{s}}_{i\ell }^{\prime }\right) =%
\left[ \left( \ell -1\right) T_{q}+\left( \frac{1}{T_{q}}\right)
^{2}\sum_{t=\left( \ell -1\right) T_{q}+1}^{\ell T_{q}}\left( \ell
T_{q}-t+1\right) ^{2}\right] \mathbf{\Sigma }_{i}\text{.}
\end{equation*}%
But%
\begin{equation*}
\sum_{t=\left( \ell -1\right) T_{q}+1}^{\ell T_{q}}\left( \ell
T_{q}-t+1\right) ^{2}=T_{q}^{2}+(T_{q}-1)^{2}+...+1=\frac{%
T_{q}(T_{q}+1)(2T_{q}+1)}{6}\text{,}
\end{equation*}%
and we obtain 
\begin{equation*}
E\left( \mathbf{\bar{s}}_{i\ell }\mathbf{\bar{s}}_{i\ell }^{\prime }\right) =%
\left[ \left( \ell -1\right) T_{q}+\left( \frac{1}{T_{q}}\right) ^{2}\frac{%
T_{q}(T_{q}+1)(2T_{q}+1)}{6}\right] \mathbf{\Sigma }_{i}=\left[ \left( \ell
-1\right) T_{q}+\frac{(T_{q}+1)(2T_{q}+1)}{6T_{q}}\right] \mathbf{\Sigma }%
_{i}\text{,}
\end{equation*}%
and%
\begin{equation*}
q^{-1}\sum_{\ell =1}^{q}E\left( \mathbf{\bar{s}}_{i\ell }\mathbf{\bar{s}}%
_{i\ell }^{\prime }\right) =q^{-1}\sum_{\ell =1}^{q}\left( \left( \ell
-1\right) T_{q}+\frac{(T_{q}+1)(2T_{q}+1)}{6T_{q}}\right) \mathbf{\Sigma }%
_{i}\text{.}
\end{equation*}%
Noting that%
\begin{equation*}
q^{-1}\sum_{\ell =1}^{q}\left( \ell -1\right) T_{q}=q^{-1}\left[
T_{q}+2T_{q}+...+(q-1)T_{q}\right] =q^{-1}T_{q}\frac{q(q-1)}{2}=\frac{%
(q-1)T_{q}}{2}\text{,}
\end{equation*}%
we have 
\begin{equation*}
q^{-1}\sum_{\ell =1}^{q}E\left( \mathbf{\bar{s}}_{i\ell }\mathbf{\bar{s}}%
_{i\ell }^{\prime }\right) =\left[ \frac{(q-1)T_{q}}{2}+\frac{%
(T_{q}+1)(2T_{q}+1)}{6T_{q}}\right] \mathbf{\Sigma }_{i}\text{.}
\end{equation*}%
Recalling that $T_{q}=T/q$, we can write the expression above as%
\begin{equation}
q^{-1}\sum_{\ell =1}^{q}E\left( \mathbf{\bar{s}}_{i\ell }\mathbf{\bar{s}}%
_{i\ell }^{\prime }\right) =\left[ \frac{(q-1)T}{2q}+\frac{(T+q)(2T+q)}{6qT}%
\right] \mathbf{\Sigma }_{i}\text{.}  \label{sumsisi}
\end{equation}%
Similarly, for the full sample average we have $\mathbf{\bar{s}}_{i\circ
}=T^{-1}\sum_{t=1}^{T}\left( T-t+1\right) \mathbf{u}_{it}$, and%
\begin{equation}
E\left( \mathbf{\bar{s}}_{i\circ }\mathbf{\bar{s}}_{i\circ }^{\prime
}\right) =\frac{1}{T^{2}}\left[ T^{2}+(T-1)^{2}+....+1\right] \mathbf{\Sigma 
}_{i}=\frac{T(T+1)(2T+1)}{6T^{2}}\mathbf{\Sigma }_{i}\text{.}  \label{sumss}
\end{equation}%
Using (\ref{sumsisi}) and (\ref{sumss}) in (\ref{peqss1}), yields 
\begin{equation}
E\left( T\mathbf{Q}_{\bar{s}_{i}\bar{s}_{i}}\right) =\left[ \frac{T(q-1)}{2q}%
+\frac{(T+q)(2T+q)}{6Tq}\right] \mathbf{\Sigma }_{i}-\frac{(T+1)(2T+1)}{6T}%
\mathbf{\Sigma }_{i}=\frac{(q-1)}{6}\left( \frac{T}{q}+\frac{1}{T}\right) 
\mathbf{\Sigma }_{i}\text{,}  \label{Qsisi}
\end{equation}%
and result (\ref{EQss}) follows.
\end{proof}

\bigskip

\begin{lemma}
\label{LQB}Consider $m\times m$ matrices $\mathbf{Q}_{\bar{v}_{i}\bar{v}%
_{i}}=T^{-1}q^{-1}\sum_{\ell =1}^{q}\left( \mathbf{\bar{v}}_{i\ell }-\mathbf{%
\bar{v}}_{i\circ }\right) \left( \mathbf{\bar{v}}_{i\ell }-\mathbf{\bar{v}}%
_{i\circ }\right) ^{\prime }$ and \newline
$\mathbf{Q}_{\bar{s}_{i}\bar{v}_{i}}=T^{-1}q^{-1}\sum_{\ell =1}^{q}\left( 
\mathbf{\bar{s}}_{i\ell }-\mathbf{\bar{s}}_{i\circ }\right) \left( \mathbf{%
\bar{v}}_{i\ell }-\mathbf{\bar{v}}_{i\circ }\right) ^{\prime }$, where $%
\mathbf{\bar{s}}_{i\ell }$, $\mathbf{\bar{s}}_{i\circ }$ are the sub-sample
and full sample time averages of the partial sum process $\mathbf{s}%
_{it}=\sum_{\ell =1}^{t}\mathbf{u}_{it},$ where $m$ and $q$ ($\geq 2$) are
fixed. Also $\mathbf{\bar{v}}_{i\ell }$ and $\mathbf{\bar{v}}_{i\circ }$ are
the sub-sample and full sample time averages of $\mathbf{v}_{it}=\mathbf{C}%
_{i}^{\ast }(L)\mathbf{u}_{it}=\sum_{j=0}^{\infty }\mathbf{C}_{ij}^{\ast }%
\mathbf{u}_{i,t-j}$, and Assumptions \ref{ASS1} and \ref{ASS2} hold. Then%
\begin{equation}
\sup_{i}E\left\Vert \mathbf{Q}_{\bar{v}_{i}\bar{v}_{i}}\right\Vert =O\left(
T^{-2}\right) ,  \label{bQvv}
\end{equation}%
\begin{equation}
\sup_{i}\left\Vert T^{2}\text{ }E\left( \mathbf{Q}_{\bar{v}_{i}\bar{v}%
_{i}}\right) \right\Vert =O\left( T^{-1}\right) +O\left( \rho ^{T/q}\right) 
\text{, }  \label{EQvv}
\end{equation}%
\begin{equation}
\sup_{i}E\left\Vert \mathbf{Q}_{\bar{s}_{i}\bar{v}_{i}}\right\Vert =O\left(
T^{-1}\right) \text{,}  \label{bQsv}
\end{equation}%
and 
\begin{equation}
\sup_{i}\left\Vert E\left( \mathbf{Q}_{\bar{s}_{i}\bar{v}_{i}}\right)
\right\Vert =O\left( T^{-2}\right) \text{.}  \label{EQsvb}
\end{equation}
\end{lemma}

\begin{proof}
Consider 
\begin{equation*}
T\mathbf{Q}_{\bar{v}_{i}\bar{v}_{i}}=q^{-1}\sum_{\ell =1}^{q}\left( \mathbf{%
\bar{v}}_{i\ell }-\mathbf{\bar{v}}_{i\circ }\right) \left( \mathbf{\bar{v}}%
_{i\ell }-\mathbf{\bar{v}}_{i\circ }\right) ^{\prime }=q^{-1}\sum_{\ell
=1}^{q}\mathbf{\bar{v}}_{i\ell }\mathbf{\bar{v}}_{i\ell }^{\prime }-\mathbf{%
\bar{v}}_{i\circ }\mathbf{\bar{v}}_{i\circ }^{\prime }\text{,}
\end{equation*}%
where $\mathbf{\bar{v}}_{i\circ }=q^{-1}\sum_{\ell =1}^{q}\mathbf{\bar{v}}%
_{i\ell }$. Taking spectral norm and expectations yields%
\begin{equation*}
E\left\Vert T\mathbf{Q}_{\bar{v}_{i}\bar{v}_{i}}\right\Vert \leq
q^{-1}\sum_{\ell =1}^{q}E\left\Vert \mathbf{\bar{v}}_{i\ell }\right\Vert
^{2}+E\left\Vert \mathbf{\bar{v}}_{i\circ }\right\Vert ^{2}\text{\textbf{,}}
\end{equation*}%
and it follows%
\begin{equation}
\sup_{i}E\left\Vert T\mathbf{Q}_{\bar{v}_{i}\bar{v}_{i}}\right\Vert \leq
\sup_{i,\ell }E\left\Vert \mathbf{\bar{v}}_{i\ell }\right\Vert
^{2}+\sup_{i}E\left\Vert \mathbf{\bar{v}}_{i\circ }\right\Vert ^{2}\text{.}
\label{vb}
\end{equation}%
Term $E\left\Vert \mathbf{\bar{v}}_{i\ell }\right\Vert ^{2}$ can be written
as%
\begin{equation*}
E\left\Vert \mathbf{\bar{v}}_{i\ell }\right\Vert ^{2}=E\left( \mathbf{\bar{v}%
}_{i\ell }^{\prime }\mathbf{\bar{v}}_{i\ell }\right) =E\left[ tr\left( 
\mathbf{\bar{v}}_{i\ell }\mathbf{\bar{v}}_{i\ell }^{\prime }\right) \right]
=tr\left[ E\left( \mathbf{\bar{v}}_{i\ell }\mathbf{\bar{v}}_{i\ell }^{\prime
}\right) \right] \text{,}
\end{equation*}%
where, under Assumptions \ref{ASS1}-\ref{ASS2} process $\mathbf{v}_{it}=%
\mathbf{C}_{i}^{\ast }(L)\mathbf{u}_{it}$ is covariance-stationary with
uniformly (in $i$) absolutely summable autocovariances, and 
\begin{equation}
E\left( \mathbf{\bar{v}}_{i\ell }\mathbf{\bar{v}}_{i\ell }^{\prime }\right) =%
\frac{1}{T_{q}}\left\{ \Gamma _{i}(0)+\sum_{h=1}^{T_{q}-1}\left( 1-\frac{h}{%
T_{q}}\right) \left[ \Gamma _{i}(h)+\Gamma _{i}^{\prime }(h)\right] \right\} 
\text{,}  \label{Evl}
\end{equation}%
in which $\mathbf{\Gamma }_{i}(h)=E\left( \mathbf{v}_{it}\mathbf{v}%
_{i,t-h}^{\prime }\right) $. Using Lemma \ref{Lacf}, it follows $%
\sup_{i,\ell }E\left\Vert \mathbf{\bar{v}}_{i\ell }\right\Vert ^{2}=O\left(
T_{q}^{-1}\right) $, and given that $q$ is fixed (does not change with the
sample size), then $T_{q}\approx T$, and we in turn obtain 
\begin{equation}
\sup_{i,\ell }E\left\Vert \mathbf{\bar{v}}_{i\ell }\right\Vert ^{2}=O\left(
T^{-1}\right) \text{.}  \label{vil}
\end{equation}%
Similarly, $E\left\Vert \mathbf{\bar{v}}_{i\circ }\right\Vert ^{2}=tr\left[
E\left( \mathbf{\bar{v}}_{i\circ }\mathbf{\bar{v}}_{i\circ }^{\prime
}\right) \right] $, where 
\begin{equation}
E\left( \mathbf{\bar{v}}_{i\circ }\mathbf{\bar{v}}_{i\circ }^{\prime
}\right) =\frac{1}{T}\left\{ \Gamma _{i}(0)+\sum_{h=1}^{T-1}\left( 1-\frac{h%
}{T}\right) \left[ \Gamma _{i}(h)+\Gamma _{i}^{\prime }(h)\right] \right\} 
\text{,}  \label{Ev}
\end{equation}%
and using Lemma \ref{Lacf}, we obtain%
\begin{equation}
\sup_{i}E\left\Vert \mathbf{\bar{v}}_{i\circ }\right\Vert ^{2}=O\left(
T^{-1}\right) \text{.}  \label{vid}
\end{equation}%
Using (\ref{vil}) and (\ref{vid}) in (\ref{vb}) yields result (\ref{bQvv}).
Consider now (\ref{EQvv}) and note that 
\begin{equation*}
T\text{ }E\left( \mathbf{Q}_{\bar{v}_{i}\bar{v}_{i}}\right)
=q^{-1}\sum_{\ell =1}^{q}E\left( \mathbf{\bar{v}}_{i\ell }\mathbf{\bar{v}}%
_{i\ell }^{\prime }\right) -E\left( \mathbf{\bar{v}}_{i\circ }\mathbf{\bar{v}%
}_{i\circ }^{\prime }\right) ,
\end{equation*}%
and upon using (\ref{Evl}) we have (recall that $T_{q}=T/q)$%
\begin{equation*}
q^{-1}\sum_{\ell =1}^{q}E\left( \mathbf{\bar{v}}_{i\ell }\mathbf{\bar{v}}%
_{i\ell }^{\prime }\right) =\frac{1}{T}\left\{ \Gamma
_{i}(0)+\sum_{h=1}^{T_{q}-1}\left( 1-\frac{h}{T_{q}}\right) \left[ \Gamma
_{i}(h)+\Gamma _{i}^{\prime }(h)\right] \right\} .
\end{equation*}%
Also using (\ref{Ev}) we now have 
\begin{eqnarray*}
T\text{ }E\left( \mathbf{Q}_{\bar{v}_{i}\bar{v}_{i}}\right) &=&\frac{1}{T}%
\sum_{h=1}^{T_{q}-1}\left( 1-\frac{h}{T_{q}}\right) \left[ \Gamma
_{i}(h)+\Gamma _{i}^{\prime }(h)\right] -\frac{1}{T}\sum_{h=1}^{T-1}\left( 1-%
\frac{h}{T}\right) \left[ \Gamma _{i}(h)+\Gamma _{i}^{\prime }(h)\right] 
\text{,} \\
&=&\frac{1}{T}\sum_{h=1}^{T/q-1}\left[ \left( 1-\frac{qh}{T}\right) -\left(
1-\frac{h}{T}\right) \right] \left[ \Gamma _{i}(h)+\Gamma _{i}^{\prime }(h)%
\right] \\
&&-\frac{1}{T}\sum_{h=T/q+1}^{T-1}\left( 1-\frac{h}{T}\right) \left[ \Gamma
_{i}(h)+\Gamma _{i}^{\prime }(h)\right] \text{.}
\end{eqnarray*}%
Hence, as required we have%
\begin{equation*}
T^{2}\text{ }E\left( \mathbf{Q}_{\bar{v}_{i}\bar{v}_{i}}\right) =\frac{%
\left( q-1\right) }{T}\sum_{h=1}^{T/q-1}h\left[ \Gamma _{i}(h)+\Gamma
_{i}^{\prime }(h)\right] -\sum_{h=T/q+1}^{T-1}\left( 1-\frac{h}{T}\right) %
\left[ \Gamma _{i}(h)+\Gamma _{i}^{\prime }(h)\right] \text{, }
\end{equation*}%
and%
\begin{equation*}
T^{2}\text{ }\left\Vert E\left( \mathbf{Q}_{\bar{v}_{i}\bar{v}_{i}}\right)
\right\Vert \leq \frac{2\left( q-1\right) }{T}\sum_{h=1}^{T/q-1}h\left\Vert
\Gamma _{i}(h)\right\Vert +\sum_{h=T/q+1}^{T-1}\left\Vert \Gamma
_{i}(h)\right\Vert .
\end{equation*}%
Since $\sup_{i}\left\Vert \mathbf{\Gamma }_{i}(h)\right\Vert <K_{1}\rho ^{h}$%
, (see (\ref{supGh}) in Lemma \ref{Lacf}) it now follows that 
\begin{equation}
T^{2}\text{ sup}_{i}\left\Vert E\left( \mathbf{Q}_{\bar{v}_{i}\bar{v}%
_{i}}\right) \right\Vert \leq \frac{2\left( q-1\right) }{T}%
K_{1}\sum_{h=1}^{T/q-1}h\rho ^{h}+K_{1}\sum_{h=T/q+1}^{T-1}\rho ^{h}.
\label{EQvivi}
\end{equation}%
Also 
\begin{equation*}
\sum_{h=1}^{T/q-1}h\rho ^{h}=\rho \left( \sum_{h=1}^{T/q-1}h\rho
^{h-1}\right) =\rho \frac{d\left( \sum_{h=1}^{T/q-1}\rho ^{h}\right) }{d\rho 
}=\rho \frac{d}{d\rho }\left( \frac{\rho -\rho ^{T/q}}{1-\rho }\right) <K,
\end{equation*}%
and $\sum_{h=T/q+1}^{T-1}\rho ^{h}=\left( \rho ^{T/q+1}-\rho ^{T}\right)
/\left( 1-\rho \right) $. Using these results in (\ref{EQvivi})\ we have $%
T^{2}$ sup$_{i}\left\Vert E\left( \mathbf{Q}_{\bar{v}_{i}\bar{v}_{i}}\right)
\right\Vert =O\left( T^{-1}\right) +O\left( \rho ^{T/q}\right) ,$ as
required. We establish (\ref{bQsv}) next. 
\begin{equation*}
T\mathbf{Q}_{\bar{s}_{i}\bar{v}_{i}}=q^{-1}\sum_{\ell =1}^{q}\left( \mathbf{%
\bar{s}}_{i\ell }-\mathbf{\bar{s}}_{i\circ }\right) \left( \mathbf{\bar{v}}%
_{i\ell }-\mathbf{\bar{v}}_{i\circ }\right) ^{\prime }=q^{-1}\sum_{\ell
=1}^{q}\mathbf{\bar{s}}_{i\ell }\mathbf{\bar{v}}_{i\ell }^{\prime }-\mathbf{%
\bar{s}}_{i\circ }\mathbf{\bar{v}}_{i\circ }^{\prime }\text{,}
\end{equation*}%
where $\mathbf{\bar{v}}_{i\circ }=q^{-1}\sum_{\ell =1}^{q}\mathbf{\bar{v}}%
_{i\ell }$ and similarly $\mathbf{\bar{s}}_{i\circ }=q^{-1}\sum_{\ell =1}^{q}%
\mathbf{\bar{s}}_{i\ell }$. Taking spectral norm yields $\left\Vert T\mathbf{%
Q}_{\bar{s}_{i}\bar{v}_{i}}\right\Vert \leq q^{-1}\sum_{\ell
=1}^{q}\left\Vert \mathbf{\bar{s}}_{i\ell }\mathbf{\bar{v}}_{i\ell }^{\prime
}\right\Vert \mathbf{+}\left\Vert \mathbf{\bar{s}}_{i\circ }\mathbf{\bar{v}}%
_{i\circ }^{\prime }\right\Vert $\textbf{. }By Cauchy-Schwarz inequality,%
\begin{equation}
E\left\Vert T\mathbf{Q}_{\bar{s}_{i}\bar{v}_{i}}\right\Vert \leq
q^{-1}\sum_{\ell =1}^{q}\left( E\left\Vert \mathbf{\bar{s}}_{i\ell
}\right\Vert ^{2}\right) ^{1/2}\left( E\left\Vert \mathbf{\bar{v}}_{i\ell
}\right\Vert ^{2}\right) ^{1/2}+\left( E\left\Vert \mathbf{\bar{s}}_{i\circ
}\right\Vert ^{2}\right) ^{1/2}\left( E\left\Vert \mathbf{\bar{v}}_{i\circ
}\right\Vert ^{2}\right) ^{1/2}\text{. }  \label{pqsv}
\end{equation}%
Assumption \ref{ASS1} stipulates $\sup_{i}\left\Vert \mathbf{\Sigma }%
_{i}\right\Vert <K$. Using this bound in (\ref{sumsisi}) and (\ref{sumss})
yields%
\begin{equation}
\sup_{i}E\left\Vert \mathbf{\bar{s}}_{i\circ }\right\Vert ^{2}=\sup_{i}tr%
\left[ E\left( \mathbf{\bar{s}}_{i\circ }\mathbf{\bar{s}}_{i\circ }^{\prime
}\right) \right] =O\left( T\right) \text{, and similarly }\sup_{i,\ell
}E\left\Vert \mathbf{\bar{s}}_{i\ell }\right\Vert ^{2}=O\left( T\right) 
\text{.}  \label{so}
\end{equation}%
In addition, we have already established $\sup_{i,\ell }E\left\Vert \mathbf{%
\bar{v}}_{i\ell }\right\Vert ^{2}=O\left( T^{-1}\right) $ (see (\ref{vil}))
and $\sup_{i}E\left\Vert \mathbf{\bar{v}}_{i\circ }\right\Vert ^{2}=O\left(
T^{-1}\right) $ (see (\ref{vid})). Using these results in (\ref{pqsv})
yields $\sup_{i}E\left\Vert T\mathbf{Q}_{\bar{s}_{i}\bar{v}_{i}}\right\Vert
=O\left( 1\right) $, which in turn implies result (\ref{bQsv}).

We establish (\ref{EQsvb}) next. Taking expectation of $\mathbf{Q}_{\bar{s}%
_{i}\bar{v}_{i}}$, we have%
\begin{equation}
E\left( \mathbf{Q}_{\bar{s}_{i}\bar{v}_{i}}\right) =T^{-1}q^{-1}\sum_{\ell
=1}^{q}\left[ E\left( \mathbf{\bar{s}}_{i\ell }\mathbf{\bar{v}}_{i\ell
}^{\prime }\right) -E\left( \mathbf{\bar{s}}_{i\ell }\mathbf{\bar{v}}%
_{i\circ }^{\prime }\right) -E\left( \mathbf{\bar{s}}_{i\circ }\mathbf{\bar{v%
}}_{i\ell }^{\prime }\right) +E\left( \mathbf{\bar{s}}_{i\circ }\mathbf{\bar{%
v}}_{i\circ }^{\prime }\right) \right] \text{,}  \label{eqp1}
\end{equation}%
where $\mathbf{\bar{s}}_{i\ell }=\sum_{t=1}^{(\ell -1)T_{q}}\mathbf{u}%
_{it}+T_{q}^{-1}\sum_{t=(\ell -1)T_{q}+1}^{\ell T_{q}}\left( \ell
T_{q}+1-t\right) \mathbf{u}_{it}$, $\mathbf{\bar{s}}_{i\circ
}=T^{-1}\sum_{t=1}^{T}\left( T-t+1\right) \mathbf{u}_{it}$, \newline
$\mathbf{\bar{v}}_{i\ell }=T_{q}^{-1}\sum_{t=(\ell -1)T_{q}+1}^{\ell T_{q}}%
\mathbf{v}_{it}$, $\mathbf{\bar{v}}_{i\circ }=T^{-1}\sum_{t=1}^{T}\mathbf{v}%
_{it}$, and $\mathbf{v}_{it}=\mathbf{C}_{i}^{\ast }(L)\mathbf{u}_{it}$ with $%
\mathbf{C}_{i}^{\ast }(L)=\sum_{j=0}^{\infty }\mathbf{C}_{ij}^{\ast }L^{j}$.
We consider each of the four terms inside the summation operator on the
right side of (\ref{eqp1}) in turn. For $E\left( \mathbf{\bar{s}}_{i\ell }%
\mathbf{\bar{v}}_{i\ell }^{\prime }\right) $, we have%
\begin{equation}
E\left( \mathbf{\bar{s}}_{i\ell }\mathbf{\bar{v}}_{i\ell }^{\prime }\right) =%
\frac{1}{T_{q}}\sum_{t=1}^{(\ell -1)T_{q}}\sum_{t^{\prime }=(\ell
-1)T_{q}+1}^{\ell T_{q}}E\left( \mathbf{u}_{it}\mathbf{v}_{it^{\prime
}}^{\prime }\right) +\frac{1}{T_{q}^{2}}\sum_{t=(\ell -1)T_{q}+1}^{\ell
T_{q}}\sum_{t^{\prime }=(\ell -1)T_{q}+1}^{\ell T_{q}}\left( \ell
T_{q}+1-t\right) E\left( \mathbf{u}_{it}\mathbf{v}_{it^{\prime }}^{\prime
}\right) \text{.}  \label{esisvis}
\end{equation}%
Noting that 
\begin{equation}
E\left( \mathbf{u}_{it}\mathbf{v}_{it^{\prime }}^{\prime }\right) =\left\{ 
\begin{array}{l}
\mathbf{\Sigma }_{i}\mathbf{C}_{i,t^{\prime }-t}^{\ast \prime }\text{, for }%
t\leq t^{\prime }\text{,} \\ 
\mathbf{0}\text{, for }t>t^{\prime }\text{,}%
\end{array}%
\right.  \label{euv}
\end{equation}%
we obtain%
\begin{equation*}
\frac{1}{T_{q}}\sum_{t=1}^{(\ell -1)T_{q}}\sum_{t^{\prime }=(\ell
-1)T_{q}+1}^{\ell T_{q}}E\left( \mathbf{u}_{it}\mathbf{v}_{it^{\prime
}}^{\prime }\right) =\frac{1}{T_{q}}\sum_{t=1}^{(\ell
-1)T_{q}}\sum_{t^{\prime }=(\ell -1)T_{q}+1}^{\ell T_{q}}\mathbf{\Sigma }_{i}%
\mathbf{C}_{i,t^{\prime }-t}^{\ast \prime }\text{.}
\end{equation*}%
Recalling that under Assumptions \ref{ASS1} and \ref{ASS2}, $%
\sup_{i}\left\Vert \mathbf{\Sigma }_{i}\right\Vert <K$ and $%
\sup_{i}\left\Vert \mathbf{C}_{ij}^{\ast \prime }\right\Vert <K\rho ^{j},$
for $\rho <1$, then it follows that 
\begin{eqnarray*}
\sup_{i}\left\Vert \frac{1}{T_{q}}\sum_{t=1}^{(\ell -1)T_{q}}\sum_{t^{\prime
}=(\ell -1)T_{q}+1}^{\ell T_{q}}E\left( \mathbf{u}_{it}\mathbf{v}%
_{it^{\prime }}^{\prime }\right) \right\Vert &\leq &\frac{1}{T_{q}}%
\sum_{t=1}^{(\ell -1)T_{q}}\sum_{t^{\prime }=(\ell -1)T_{q}+1}^{\ell
T_{q}}\sup_{i}\left\Vert \mathbf{\Sigma }_{i}\right\Vert \sup_{i}\left\Vert 
\mathbf{C}_{i,t^{\prime }-t}^{\ast \prime }\right\Vert \text{,} \\
&\leq &\frac{1}{T_{q}}\sum_{t=1}^{(\ell -1)T_{q}}\sum_{t^{\prime }=(\ell
-1)T_{q}+1}^{\ell T_{q}}K\rho ^{t^{\prime }-t}\text{,} \\
&\leq &\frac{K}{T_{q}}\sum_{t=1}^{(\ell -1)T_{q}}\rho ^{(\ell
-1)T_{q}-t+1}\sum_{j=0}^{T_{q}}\rho ^{j}\text{.}
\end{eqnarray*}%
But both sums $\sum_{j=0}^{T_{q}}\rho ^{j}$ and $\sum_{t=1}^{(\ell
-1)T_{q}}\rho ^{(\ell -1)T_{q}-t+1}$ are bounded in $T$ (recall $T_{q}=T/q$
where $q$ does not change with sample size and is fixed). Hence%
\begin{equation}
\sup_{i}\left\Vert \frac{1}{T_{q}}\sum_{t=1}^{(\ell -1)T_{q}}\sum_{t^{\prime
}=(\ell -1)T_{q}+1}^{\ell T_{q}}E\left( \mathbf{u}_{it}\mathbf{v}%
_{it^{\prime }}^{\prime }\right) \right\Vert =O\left( T^{-1}\right) \text{.}
\label{pt2b}
\end{equation}%
For the second term on the right side of (\ref{esisvis}), we have%
\begin{eqnarray*}
&&\frac{1}{T_{q}^{2}}\sum_{t=(\ell -1)T_{q}+1}^{\ell T_{q}}\sum_{t^{\prime
}=(\ell -1)T_{q}+1}^{\ell T_{q}}\left( \ell T_{q}+1-t\right) E\left( \mathbf{%
u}_{it}\mathbf{v}_{it^{\prime }}^{\prime }\right) = \\
&=&\frac{1}{T_{q}^{2}}\sum_{t=(\ell -1)T_{q}+1}^{\ell T_{q}}\left( \ell
T_{q}+1-t\right) \left[ \sum_{t^{\prime }=(\ell -1)T_{q}+1}^{t-1}E\left( 
\mathbf{u}_{it}\mathbf{v}_{it^{\prime }}^{\prime }\right) +\sum_{t^{\prime
}=t}^{\ell T_{q}}E\left( \mathbf{u}_{it}\mathbf{v}_{it^{\prime }}^{\prime
}\right) \right] \text{,}
\end{eqnarray*}%
and substituting (\ref{euv}), we obtain%
\begin{eqnarray}
&&\frac{1}{T_{q}^{2}}\sum_{t=(\ell -1)T_{q}+1}^{\ell T_{q}}\sum_{t^{\prime
}=(\ell -1)T_{q}+1}^{\ell T_{q}}\left( \ell T_{q}+1-t\right) E\left( \mathbf{%
u}_{it}\mathbf{v}_{it^{\prime }}^{\prime }\right)  \notag \\
&=&\frac{1}{T_{q}^{2}}\sum_{t=(\ell -1)T_{q}+1}^{\ell T_{q}}\left( \ell
T_{q}+1-t\right) \sum_{t^{\prime }=t}^{\ell T_{q}}\mathbf{\Sigma }_{i}%
\mathbf{C}_{i,t^{\prime }-t}^{\ast \prime }=\frac{1}{T_{q}^{2}}%
\sum_{k=0}^{T_{q}-1}\left( \mathbf{\Sigma }_{i}\mathbf{C}_{ik}^{\ast \prime
}\sum_{j=k+1}^{T_{q}}j\right) \text{,}  \notag \\
&=&\frac{1}{T_{q}^{2}}\sum_{k=0}^{T_{q}-1}\frac{\left( T_{q}+k+1\right)
\left( T_{q}-k\right) }{2}\mathbf{\Sigma }_{i}\mathbf{C}_{ik}^{\ast \prime }%
\text{.}  \label{pt1b}
\end{eqnarray}%
Using (\ref{pt2b}) and (\ref{pt1b}) in (\ref{esisvis}) yields%
\begin{equation}
\sup_{i}\left\Vert E\left( \mathbf{\bar{s}}_{i\ell }\mathbf{\bar{v}}_{i\ell
}^{\prime }\right) -\frac{1}{T_{q}^{2}}\sum_{k=0}^{T_{q}-1}\frac{\left(
T_{q}+k+1\right) \left( T_{q}-k\right) }{2}\mathbf{\Sigma }_{i}\mathbf{C}%
_{ik}^{\ast \prime }\right\Vert =O\left( T^{-1}\right) \text{.}  \label{pe1}
\end{equation}%
Next consider $E\left( \mathbf{\bar{s}}_{i\ell }\mathbf{\bar{v}}_{i\circ
}^{\prime }\right) $, and note that 
\begin{equation}
E\left( \mathbf{\bar{s}}_{i\ell }\mathbf{\bar{v}}_{i\circ }^{\prime }\right)
=\frac{1}{T}\sum_{t=1}^{(\ell -1)T_{q}}\sum_{t^{\prime }=1}^{T}E\left( 
\mathbf{u}_{it}\mathbf{v}_{it^{\prime }}^{\prime }\right) +\frac{1}{T_{q}T}%
\sum_{t=(\ell -1)T_{q}+1}^{\ell T_{q}}\sum_{t^{\prime }=1}^{T}\left( \ell
T_{q}+1-t\right) E\left( \mathbf{u}_{it}\mathbf{v}_{it^{\prime }}^{\prime
}\right) \text{.}  \label{esbvbp1}
\end{equation}%
For the first term on the right side of (\ref{esbvbp1}), we obtain%
\begin{equation}
\frac{1}{T}\sum_{t=1}^{(\ell -1)T_{q}}\sum_{t^{\prime }=1}^{T}E\left( 
\mathbf{u}_{it}\mathbf{v}_{it^{\prime }}^{\prime }\right) =\frac{1}{T}%
\sum_{t=1}^{(\ell -1)T_{q}}\left[ \sum_{t^{\prime }=1}^{t-1}E\left( \mathbf{u%
}_{it}\mathbf{v}_{it^{\prime }}^{\prime }\right) +\sum_{t^{\prime
}=t}^{T}E\left( \mathbf{u}_{it}\mathbf{v}_{it^{\prime }}^{\prime }\right) %
\right] =\frac{1}{T}\sum_{t=1}^{(\ell -1)T_{q}}\sum_{t^{\prime }=t}^{T}%
\mathbf{\Sigma }_{i}\mathbf{C}_{i,t^{\prime }-t}^{\ast \prime }\text{,} 
\notag
\end{equation}%
and using $k=t^{\prime }-t$, we have%
\begin{equation*}
\frac{1}{T}\sum_{t=1}^{(\ell -1)T_{q}}\sum_{t^{\prime }=1}^{T}E\left( 
\mathbf{u}_{it}\mathbf{v}_{it^{\prime }}^{\prime }\right) =\frac{1}{T}%
\sum_{t=1}^{(\ell -1)T_{q}}\sum_{k=0}^{T-t}\mathbf{\Sigma }_{i}\mathbf{C}%
_{ik}^{\ast \prime }=\frac{1}{T}\sum_{t=1}^{(\ell
-1)T_{q}}\sum_{k=0}^{T_{q}-1}\mathbf{\Sigma }_{i}\mathbf{C}_{ik}^{\ast
\prime }+\frac{1}{T}\sum_{t=1}^{(\ell -1)T_{q}}\sum_{k=T_{q}-1}^{T-t}\mathbf{%
\Sigma }_{i}\mathbf{C}_{ik}^{\ast \prime }\text{,}
\end{equation*}%
where (noting that $T=qT_{q}$) $T^{-1}\sum_{t=1}^{(\ell
-1)T_{q}}\sum_{k=0}^{T_{q}-1}\mathbf{\Sigma }_{i}\mathbf{C}_{ik}^{\ast
\prime }=q^{-1}\left( \ell -1\right) \sum_{k=0}^{T_{q}-1}\mathbf{\Sigma }_{i}%
\mathbf{C}_{ik}^{\ast \prime }$, and (noting that $\left\Vert \mathbf{C}%
_{ij}^{\ast \prime }\right\Vert <K\rho ^{j}$ and $\left\Vert \mathbf{\Sigma }%
_{i}\right\Vert <K$)%
\begin{eqnarray*}
\sup_{i}\left\Vert \frac{1}{T}\sum_{t=1}^{(\ell
-1)T_{q}}\sum_{k=T_{q}-1}^{T-t}\mathbf{\Sigma }_{i}\mathbf{C}_{ik}^{\ast
\prime }\right\Vert &\leq &\frac{1}{T}\sum_{t=1}^{(\ell
-1)T_{q}}\sum_{k=T_{q}}^{T-t}\sup_{i}\left\Vert \mathbf{\Sigma }%
_{i}\right\Vert \sup_{i}\left\Vert \mathbf{C}_{ik}^{\ast \prime }\right\Vert
< \\
\frac{K}{T}\sum_{t=1}^{(\ell -1)T_{q}}\sum_{k=T_{q}}^{T-t}\rho ^{k} &<&K%
\frac{(\ell -1)T_{q}}{T}\rho ^{T_{q}}\sum_{j=0}^{T-t-T_{q}}\rho
^{j}<K_{2}\rho ^{T_{q}}\text{,}
\end{eqnarray*}%
where $\left\vert \rho \right\vert <1$. Hence%
\begin{equation}
\sup_{i}\left\Vert \frac{1}{T}\sum_{t=1}^{(\ell -1)T_{q}}\sum_{t^{\prime
}=1}^{T}E\left( \mathbf{u}_{it}\mathbf{v}_{it^{\prime }}^{\prime }\right) -%
\frac{\ell -1}{q}\sum_{k=0}^{T_{q}-1}\mathbf{\Sigma }_{i}\mathbf{C}%
_{ik}^{\ast \prime }\right\Vert =O\left( \rho ^{T_{q}}\right) \text{.}
\label{esbvp2}
\end{equation}%
For the second term on the right side of (\ref{esbvbp1}), we have%
\begin{eqnarray*}
&&\frac{1}{T_{q}T}\sum_{t=(\ell -1)T_{q}+1}^{\ell T_{q}}\sum_{t^{\prime
}=1}^{T}\left( \ell T_{q}+1-t\right) E\left( \mathbf{u}_{it}\mathbf{v}%
_{it^{\prime }}^{\prime }\right) = \\
&&\frac{1}{T_{q}T}\sum_{t=(\ell -1)T_{q}+1}^{\ell T_{q}}\left[
\sum_{t^{\prime }=1}^{t-1}\left( \ell T_{q}+1-t\right) E\left( \mathbf{u}%
_{it}\mathbf{v}_{it^{\prime }}^{\prime }\right) +\sum_{t^{\prime
}=t}^{T}\left( \ell T_{q}+1-t\right) E\left( \mathbf{u}_{it}\mathbf{v}%
_{it^{\prime }}^{\prime }\right) \right] \text{,} \\
&=&\frac{1}{T_{q}T}\sum_{t=(\ell -1)T_{q}+1}^{\ell T_{q}}\sum_{t^{\prime
}=t}^{T}\left( \ell T_{q}+1-t\right) \mathbf{\Sigma }_{i}\mathbf{C}%
_{i,t^{\prime }-t}^{\ast \prime }\text{.}
\end{eqnarray*}%
Using $k=t^{\prime }-t$ and $j=t-(\ell -1)T_{q}$, we obtain%
\begin{equation*}
\frac{1}{T_{q}T}\sum_{t=(\ell -1)T_{q}+1}^{\ell T_{q}}\sum_{t^{\prime
}=t}^{T}\left( \ell T_{q}+1-t\right) \mathbf{\Sigma }_{i}\mathbf{C}%
_{i,t^{\prime }-t}^{\ast \prime }=\frac{1}{T_{q}T}\sum_{j=1}^{T_{q}}%
\sum_{k=0}^{T-(\ell -1)T_{q}-j}\left( T_{q}+1-j\right) \mathbf{\Sigma }_{i}%
\mathbf{C}_{ik}^{\ast \prime }\text{.}
\end{equation*}%
Consider the case $\ell <q$ and $\ell =q$ in turn. For $\ell <q$, we can
write%
\begin{equation*}
\sum_{k=0}^{T-(\ell -1)T_{q}-j}\mathbf{C}_{ik}^{\ast \prime
}=\sum_{k=0}^{T_{q}-1}\mathbf{C}_{ik}^{\ast \prime }+\sum_{k=T_{q}}^{T-(\ell
-1)T_{q}-j}\mathbf{C}_{ik}^{\ast \prime }\text{,}
\end{equation*}%
where%
\begin{equation*}
\sup_{i}\left\Vert \sum_{k=T_{q}}^{T-(\ell -1)T_{q}-j}\mathbf{C}_{ik}^{\ast
\prime }\right\Vert \leq \sum_{k=T_{q}}^{T-(\ell
-1)T_{q}-j}\sup_{i}\left\Vert \mathbf{C}_{ik}^{\ast \prime }\right\Vert
<K\sum_{k=T_{q}}^{T-(\ell -1)T_{q}-j}\rho ^{k}<K_{3}\rho ^{T_{q}}\text{,}
\end{equation*}%
since $\left\vert \rho \right\vert <1$. Hence, for $\ell <q$, we have%
\begin{eqnarray*}
\frac{1}{T_{q}T}\sum_{j=1}^{T_{q}}\sum_{k=0}^{T-(\ell -1)T_{q}-j}\left(
T_{q}+1-j\right) \mathbf{\Sigma }_{i}\mathbf{C}_{ik}^{\ast \prime } &=&\frac{%
1}{T_{q}T}\mathbf{\Sigma }_{i}\sum_{j=1}^{T_{q}}\left[ \left(
T_{q}+1-j\right) \left( \sum_{k=0}^{T_{q}-1}\mathbf{C}_{ik}^{\ast \prime
}+O\left( \rho ^{T_{q}}\right) \right) \right] \text{,} \\
&=&\frac{1}{T_{q}T}\mathbf{\Sigma }_{i}\sum_{j=1}^{T_{q}}%
\sum_{k=0}^{T_{q}-1}\left( T_{q}+1-j\right) \mathbf{C}_{ik}^{\ast \prime
}+O\left( \rho ^{T_{q}}\right) \text{,} \\
&=&\frac{1}{T_{q}T}\frac{\left( T_{q}+1\right) T_{q}}{2}\mathbf{\Sigma }%
_{i}\sum_{k=0}^{T_{q}-1}\mathbf{C}_{ik}^{\ast \prime }+O\left( \rho
^{T_{q}}\right) \text{.}
\end{eqnarray*}%
where the $O\left( \rho ^{T_{q}}\right) $ term is uniform in $i$. Noting
that 
\begin{equation*}
\frac{1}{T_{q}T}\frac{\left( T_{q}+1\right) T_{q}}{2}=\frac{1}{2q}+O\left(
T^{-1}\right) \text{,}
\end{equation*}%
and using the fact that upper bound $O\left( \rho ^{T_{q}}\right) \,$is
dominated by $O\left( T^{-1}\right) $, we obtain (for $\ell <q$ )%
\begin{equation*}
\frac{1}{T_{q}T}\sum_{t=(\ell -1)T_{q}+1}^{\ell T_{q}}\sum_{t^{\prime
}=t}^{T}\left( \ell T_{q}+1-t\right) \mathbf{\Sigma }_{i}\mathbf{C}%
_{i,t^{\prime }-t}^{\ast \prime }=\frac{1}{2q}\sum_{k=0}^{T_{q}-1}\mathbf{%
\Sigma }_{i}\mathbf{C}_{ik}^{\ast \prime }+O\left( T^{-1}\right) \text{,}
\end{equation*}%
where the $O\left( T^{-1}\right) $ term is uniform in $i$. For $\ell =q$, we
have%
\begin{eqnarray*}
\frac{1}{T_{q}T}\sum_{j=1}^{T_{q}}\sum_{k=0}^{T-H_{q-1}-j}\left(
T_{q}+1-j\right) \mathbf{\Sigma }_{i}\mathbf{C}_{ik}^{\ast \prime } &=&\frac{%
1}{T_{q}T}\sum_{j=1}^{T_{q}}\sum_{k=0}^{T_{q}-j}\left( T_{q}+1-j\right) 
\mathbf{\Sigma }_{i}\mathbf{C}_{ik}^{\ast \prime }\text{,} \\
&=&\frac{1}{T_{q}T}\sum_{k=0}^{T_{q}-1}\frac{\left( T_{q}+k+1\right) \left(
T_{q}-k\right) }{2}\mathbf{\Sigma }_{i}\mathbf{C}_{ik}^{\ast \prime }\text{.}
\end{eqnarray*}%
But 
\begin{equation*}
\frac{1}{T_{q}T}\sum_{k=0}^{T_{q}-1}\frac{\left( T_{q}+k+1\right) \left(
T_{q}-k\right) }{2}\mathbf{\Sigma }_{i}\mathbf{C}_{ik}^{\ast \prime }=\frac{1%
}{T_{q}T}\sum_{k=0}^{T_{q}-1}\frac{\left( T_{q}+1\right) T_{q}}{2}\mathbf{%
\Sigma }_{i}\mathbf{C}_{ik}^{\ast \prime }-\frac{1}{T_{q}T}%
\sum_{k=0}^{T_{q}-1}\frac{\left( k+1\right) k}{2}\mathbf{\Sigma }_{i}\mathbf{%
C}_{ik}^{\ast \prime }\text{,}
\end{equation*}%
where the second term can be bounded by%
\begin{equation*}
\sup_{i}\left\Vert \frac{1}{T_{q}T}\sum_{k=1}^{T_{q}-j}\frac{\left(
k+1\right) k}{2}\mathbf{\Sigma }_{i}\mathbf{C}_{ik}^{\ast \prime
}\right\Vert <\frac{K}{T_{q}T}\sum_{k=1}^{T_{q}-j}\frac{\left( k+1\right) k}{%
2}\rho ^{k}=O\left( T^{-1}\right) \text{.}
\end{equation*}%
Hence regardless of $\ell =q$ or $\ell <q$, we have 
\begin{equation}
\sup_{i}\left\Vert \frac{1}{T_{q}T}\sum_{t=(\ell -1)T_{q}+1}^{\ell
T_{q}}\sum_{t^{\prime }=t}^{T}\left( \ell T_{q}+1-t\right) \mathbf{\Sigma }%
_{i}\mathbf{C}_{i,t^{\prime }-t}^{\ast \prime }-\frac{1}{2q}%
\sum_{k=0}^{T_{q}-1}\mathbf{\Sigma }_{i}\mathbf{C}_{ik}^{\ast \prime
}\right\Vert =O\left( T^{-1}\right)  \label{esbvp4}
\end{equation}%
Using (\ref{esbvp4}) and (\ref{esbvp2}) in (\ref{esbvbp1}) and noting that $%
O\left( \rho ^{T_{q}}\right) \,$is dominated by $O\left( T^{-1}\right) $, we
obtain%
\begin{equation}
\sup_{i}\left\Vert E\left( \mathbf{\bar{s}}_{i\ell }\mathbf{\bar{v}}_{i\circ
}\right) -\frac{\ell -1}{q}\sum_{k=0}^{T_{q}-1}\mathbf{\Sigma }_{i}\mathbf{C}%
_{ik}^{\ast \prime }-\frac{1}{2q}\sum_{k=0}^{T_{q}-1}\mathbf{\Sigma }_{i}%
\mathbf{C}_{ik}^{\ast \prime }\right\Vert =O\left( T^{-1}\right) \text{.}
\label{pe2}
\end{equation}

Consider the term $E\left( \mathbf{\bar{s}}_{i\circ }\mathbf{\bar{v}}_{i\ell
}^{\prime }\right) $ next.%
\begin{eqnarray*}
E\left( \mathbf{\bar{s}}_{i\circ }\mathbf{\bar{v}}_{i\ell }^{\prime }\right)
&=&\frac{1}{TT_{q}}\sum_{t=1}^{T}\sum_{t^{\prime }=(\ell -1)T_{q}+1}^{\ell
T_{q}}\left( T-t+1\right) E\left( \mathbf{u}_{it}\mathbf{v}_{it^{\prime
}}^{\prime }\right) \text{,} \\
&=&\frac{1}{TT_{q}}\sum_{t^{\prime }=(\ell -1)T_{q}+1}^{\ell T_{q}}\left[
\sum_{t=1}^{t^{\prime }}\left( T-t+1\right) E\left( \mathbf{u}_{it}\mathbf{v}%
_{it^{\prime }}^{\prime }\right) +\sum_{t=t^{\prime }+1}^{T}\left(
T-t+1\right) \left( \mathbf{u}_{it}\mathbf{v}_{it^{\prime }}^{\prime
}\right) \right] \text{,} \\
&=&\frac{1}{TT_{q}}\sum_{t^{\prime }=(\ell -1)T_{q}+1}^{\ell
T_{q}}\sum_{t=1}^{t^{\prime }}\left( T-t+1\right) \mathbf{\Sigma }_{i}%
\mathbf{C}_{i,t^{\prime }-t}^{\ast \prime }\text{.}
\end{eqnarray*}%
Using $k=t^{\prime }-t$ and $j=t^{\prime }-H_{s-1}-1$, we obtain%
\begin{eqnarray}
E\left( \mathbf{\bar{s}}_{i\circ }\mathbf{\bar{v}}_{i\ell }^{\prime }\right)
&=&\frac{1}{TT_{q}}\sum_{j=0}^{T_{q}-1}\sum_{k=0}^{j+(\ell -1)T_{q}}\left(
T-(\ell -1)T_{q}-j+k\right) \mathbf{\Sigma }_{i}\mathbf{C}_{ik}^{\ast \prime
}  \label{aab1} \\
&=&\frac{1}{TT_{q}}\sum_{j=0}^{T_{q}-1}\left( 
\begin{array}{c}
\sum_{k=0}^{(\ell -1)T_{q}}\left( T-(\ell -1)T_{q}-j+k\right) \mathbf{\Sigma 
}_{i}\mathbf{C}_{ik}^{\ast \prime } \\ 
+\sum_{k=(\ell -1)T_{q}+1}^{j+(\ell -1)T_{q}}\left[ T-(\ell -1)T_{q}-j+k%
\right] \mathbf{\Sigma }_{i}\mathbf{C}_{ik}^{\ast \prime }%
\end{array}%
\right) \text{.}  \notag
\end{eqnarray}%
But%
\begin{eqnarray}
&&\sup_{i}\left\Vert \sum_{k=(\ell -1)T_{q}+1}^{j+(\ell -1)T_{q}}\left(
T-(\ell -1)T_{q}-j+k\right) \mathbf{\Sigma }_{i}\mathbf{C}_{ik}^{\ast \prime
}\right\Vert  \label{aab2} \\
&\leq &\sum_{k=(\ell -1)T_{q}+1}^{j+(\ell -1)T_{q}}\left( T-(\ell
-1)T_{q}-j+k\right) \sup_{i}\left\Vert \mathbf{\Sigma }_{i}\right\Vert
\sup_{i}\left\Vert \mathbf{C}_{ik}^{\ast \prime }\right\Vert
<KT\sum_{k=(\ell -1)T_{q}+1}^{j+(\ell -1)T_{q}}\rho ^{k}<K_{1}\rho _{1}^{T}%
\text{,}  \notag
\end{eqnarray}%
where $\left\vert \rho \right\vert <\rho _{1}<1$. In addition,%
\begin{equation*}
\left\Vert \frac{1}{TT_{q}}\sum_{j=0}^{T_{q}-1}\sum_{k=0}^{(\ell -1)T_{q}}k%
\mathbf{\Sigma }_{i}\mathbf{C}_{ik}^{\ast \prime }\right\Vert \leq \frac{1}{%
TT_{q}}\sum_{j=0}^{T_{q}-1}\sum_{k=0}^{(\ell -1)T_{q}}k\left\Vert \mathbf{%
\Sigma }_{i}\right\Vert \left\Vert \mathbf{C}_{ik}^{\ast \prime }\right\Vert
<\frac{K}{TT_{q}}\sum_{j=0}^{T_{q}-1}\sum_{k=0}^{(\ell -1)T_{q}}k\rho ^{k}%
\text{.}
\end{equation*}%
Since $\sum_{k=0}^{(\ell -1)T_{q}}k\rho ^{k}<K$, we obtain%
\begin{equation}
\sup_{i}\left\Vert \frac{1}{TT_{q}}\sum_{j=0}^{T_{q}-1}\sum_{k=0}^{(\ell
-1)T_{q}}k\mathbf{\Sigma }_{i}\mathbf{C}_{ik}^{\ast \prime }\right\Vert
=O\left( T^{-1}\right) \text{.}  \label{aab3}
\end{equation}%
Using (\ref{aab2}) and (\ref{aab3}) in (\ref{aab1}), it follows%
\begin{equation*}
\sup_{i}\left\Vert E\left( \mathbf{\bar{s}}_{i\circ }\mathbf{\bar{v}}_{i\ell
}\right) -\frac{1}{TT_{q}}\sum_{j=0}^{T_{q}-1}\sum_{k=0}^{(\ell
-1)T_{q}}\left( T-(\ell -1)T_{q}-j\right) \mathbf{\Sigma }_{i}\mathbf{C}%
_{ik}^{\ast \prime }\right\Vert =O\left( T^{-1}\right) \text{.}
\end{equation*}%
Given that 
\begin{equation*}
\frac{1}{TT_{q}}\sum_{j=0}^{T_{q}-1}\left( T-(\ell -1)T_{q}-j\right) =\frac{%
\left( 2T-2(\ell -1)T_{q}-T_{q}+1\right) T_{q}}{2TT_{q}}=1-\frac{\ell -1}{q}-%
\frac{1}{2q}+O\left( T^{-1}\right) \text{,}
\end{equation*}%
it further follows%
\begin{equation}
\sup_{i}\left\Vert E\left( \mathbf{\bar{s}}_{i\circ }\mathbf{\bar{v}}_{i\ell
}\right) -\left( 1-\frac{\ell -1}{q}-\frac{1}{2q}\right) \sum_{k=0}^{(\ell
-1)T_{q}}\mathbf{\Sigma }_{i}\mathbf{C}_{ik}^{\ast \prime }\right\Vert
=O\left( T^{-1}\right) \text{.}  \label{pe3a}
\end{equation}%
For the purpose of this proof, it is convenient to decompose%
\begin{equation*}
\sum_{k=0}^{(\ell -1)T_{q}}\mathbf{\Sigma }_{i}\mathbf{C}_{ik}^{\ast \prime
}=\sum_{k=0}^{T_{q}-1}\mathbf{\Sigma }_{i}\mathbf{C}_{ik}^{\ast \prime
}+\sum_{k=T_{q}}^{(\ell -1)T_{q}}\mathbf{\Sigma }_{i}\mathbf{C}_{ik}^{\ast
\prime }\text{,}
\end{equation*}%
where, as before, 
\begin{equation*}
\sup_{i}\left\Vert \sum_{k=T_{q}}^{(\ell -1)T_{q}}\mathbf{\Sigma }_{i}%
\mathbf{C}_{ik}^{\ast \prime }\right\Vert \leq \sum_{k=T_{q}}^{(\ell
-1)T_{q}}\sup_{i}\left\Vert \mathbf{\Sigma }_{i}\right\Vert
\sup_{i}\left\Vert \mathbf{C}_{ik}^{\ast \prime }\right\Vert
<K\sum_{k=T_{q}}^{(\ell -1)T_{q}}\rho ^{k}<K_{1}\rho ^{T_{q}}\text{.}
\end{equation*}%
Using this result in (\ref{pe3a}) yields%
\begin{equation}
\sup_{i}\left\Vert E\left( \mathbf{\bar{s}}_{i\circ }\mathbf{\bar{v}}_{i\ell
}^{\prime }\right) -\left( 1-\frac{\ell -1}{q}-\frac{1}{2q}\right)
\sum_{k=0}^{T_{q}-1}\mathbf{\Sigma }_{i}\mathbf{C}_{ik}^{\ast \prime
}\right\Vert =O\left( T^{-1}\right) \text{.}  \label{pe3}
\end{equation}

Finally for the last term, $E\left( \mathbf{\bar{s}}_{i\circ }\mathbf{\bar{v}%
}_{i\circ }^{\prime }\right) $, we have%
\begin{eqnarray}
E\left( \mathbf{\bar{s}}_{i\circ }\mathbf{\bar{v}}_{i\circ }^{\prime
}\right) &=&\frac{1}{T^{2}}\sum_{t=1}^{T}\sum_{t^{\prime }=1}^{T}\left(
T-t+1\right) E\left( \mathbf{u}_{it}\mathbf{v}_{it^{\prime }}^{\prime
}\right) \text{,}  \notag \\
&=&\frac{1}{T^{2}}\sum_{t=1}^{T}\left( T-t+1\right) \left[ \sum_{t^{\prime
}=1}^{t-1}E\left( \mathbf{u}_{it}\mathbf{v}_{it^{\prime }}^{\prime }\right)
+\sum_{t^{\prime }=t}^{T}E\left( \mathbf{u}_{it}\mathbf{v}_{it^{\prime
}}^{\prime }\right) \right]  \notag \\
&=&\frac{1}{T^{2}}\sum_{t=1}^{T}\left( T-t+1\right) \sum_{t^{\prime }=t}^{T}%
\mathbf{\Sigma }_{i}\mathbf{C}_{i,t^{\prime }-t}^{\ast \prime }\text{,} 
\notag \\
&=&\frac{1}{T^{2}}\sum_{k=0}^{T-1}\left( \mathbf{\Sigma }_{i}\mathbf{C}%
_{ik}^{\ast \prime }\sum_{j=k+1}^{T}j\right) \text{,}  \notag \\
&=&\frac{1}{T^{2}}\sum_{k=0}^{T-1}\frac{\left( T+k+1\right) \left(
T-k\right) }{2}\mathbf{\Sigma }_{i}\mathbf{C}_{ik}^{\ast \prime }\text{.}
\label{pe4a}
\end{eqnarray}%
It is further useful to split the sum above into first $T_{q}$ terms and the
remainder, which can be appropriately bounded, namely%
\begin{eqnarray*}
&&\frac{1}{T^{2}}\sum_{k=0}^{T-1}\frac{\left( T+k+1\right) \left( T-k\right) 
}{2}\mathbf{\Sigma }_{i}\mathbf{C}_{ik}^{\ast \prime } \\
&=&\frac{1}{T^{2}}\sum_{k=0}^{T_{q}-1}\frac{\left( T+k+1\right) \left(
T-k\right) }{2}\mathbf{\Sigma }_{i}\mathbf{C}_{ik}^{\ast \prime }+\frac{1}{%
T^{2}}\sum_{k=T_{q}}^{T-1}\frac{\left( T+k+1\right) \left( T-k\right) }{2}%
\mathbf{\Sigma }_{i}\mathbf{C}_{ik}^{\ast \prime }\text{,}
\end{eqnarray*}%
where (noting that $\left( T+k+1\right) \left( T-k\right) /2T^{2}=O(1)$)%
\begin{eqnarray*}
\sup_{i}\left\Vert \frac{1}{T^{2}}\sum_{k=T_{q}}^{T-1}\frac{\left(
T+k+1\right) \left( T-k\right) }{2}\mathbf{\Sigma }_{i}\mathbf{C}_{ik}^{\ast
\prime }\right\Vert &\leq &\sum_{k=T_{q}}^{T-1}\frac{\left( T+k+1\right)
\left( T-k\right) }{2T^{2}}\sup_{i}\left\Vert \mathbf{\Sigma }%
_{i}\right\Vert \sup_{i}\left\Vert \mathbf{C}_{ik}^{\ast \prime }\right\Vert 
\text{,} \\
&<&K\sum_{k=T_{q}}^{T-1}\rho ^{k}<K_{1}\rho ^{T_{q}}\text{.}
\end{eqnarray*}%
Hence, we obtain for the last term $E\left( \mathbf{\bar{s}}_{i\circ }%
\mathbf{\bar{v}}_{i\circ }^{\prime }\right) $,%
\begin{equation}
\sup_{i}\left\Vert E\left( \mathbf{\bar{s}}_{i\circ }\mathbf{\bar{v}}%
_{i\circ }\right) -\frac{1}{T^{2}}\sum_{k=0}^{T_{q}-1}\frac{\left(
T+k+1\right) \left( T-k\right) }{2}\mathbf{\Sigma }_{i}\mathbf{C}_{ik}^{\ast
\prime }\right\Vert =O\left( \rho ^{T_{q}}\right) \text{.}  \label{pe4}
\end{equation}

Consider now $E\left( \mathbf{\bar{s}}_{i\ell }\mathbf{\bar{v}}_{i\ell
}^{\prime }\right) -E\left( \mathbf{\bar{s}}_{i\ell }\mathbf{\bar{v}}%
_{i\circ }^{\prime }\right) -E\left( \mathbf{\bar{s}}_{i\circ }\mathbf{\bar{v%
}}_{i\ell }^{\prime }\right) +E\left( \mathbf{\bar{s}}_{i\circ }\mathbf{\bar{%
v}}_{i\circ }^{\prime }\right) $. Using (\ref{pe1}), (\ref{pe2}), (\ref{pe3}%
), and (\ref{pe4}), we obtain%
\begin{equation*}
\sup_{i}\left\Vert E\left( \mathbf{\bar{s}}_{i\ell }\mathbf{\bar{v}}_{i\ell
}^{\prime }\right) -E\left( \mathbf{\bar{s}}_{i\ell }\mathbf{\bar{v}}%
_{i\circ }^{\prime }\right) -E\left( \mathbf{\bar{s}}_{i\circ }\mathbf{\bar{v%
}}_{i\ell }^{\prime }\right) +E\left( \mathbf{\bar{s}}_{i\circ }\mathbf{\bar{%
v}}_{i\circ }^{\prime }\right) \right\Vert _{1}<\sup_{i}\left\Vert
\sum_{k=0}^{T_{q}-1}a_{k}\mathbf{\Sigma }_{i}\mathbf{C}_{ik}^{\ast \prime
}\right\Vert +\frac{K}{T}\text{,}
\end{equation*}%
where $a_{k}$, for $k=0,1,...,T_{q}-1$, are given by%
\begin{eqnarray*}
a_{k} &=&\frac{\left( T_{q}+k+1\right) \left( T_{q}-k\right) }{2T_{q}^{2}}%
-\left( \frac{\ell -1}{q}+\frac{1}{2q}\right) -\left( 1-\frac{\ell -1}{q}-%
\frac{1}{2q}\right) +\frac{\left( T+k+1\right) \left( T-k\right) }{2T^{2}}%
\text{,} \\
&=&\frac{T_{q}-k-k^{2}}{2T_{q}^{2}}+\frac{T-k-k^{2}}{2T^{2}}=k\left( \frac{1%
}{2kT_{q}}-\frac{1}{2T_{q}^{2}}-\frac{k}{2T_{q}^{2}}+\frac{1}{2Tk}-\frac{1}{%
2T^{2}}-\frac{k}{2T^{2}}\right) \text{,} \\
&=&\frac{k}{T}\left( \frac{q}{2k}-\frac{q}{2T_{q}}-\frac{kq}{2T_{q}}+\frac{1%
}{2k}-\frac{1}{2T}-\frac{k}{2T}\right) \text{.}
\end{eqnarray*}%
It is easily seen that $\left\vert a_{k}\right\vert <\left( \frac{k}{T}%
\right) K$, and%
\begin{equation*}
\sup_{i}\left\Vert \sum_{k=0}^{T_{q}-1}a_{k}\mathbf{\Sigma }_{i}\mathbf{C}%
_{ik}^{\ast \prime }\right\Vert <\sum_{k=0}^{T_{q}-1}\left\vert
a_{k}\right\vert \sup_{i}\left\Vert \mathbf{\Sigma }_{i}\right\Vert
\sup_{i}\left\Vert \mathbf{C}_{ik}^{\ast \prime }\right\Vert <\frac{K}{T}%
\sum_{k=0}^{T_{q}-1}k\rho ^{k}=O\left( T^{-1}\right) \text{.}
\end{equation*}%
Hence%
\begin{equation*}
\sup_{i}\left\Vert E\left( \mathbf{\bar{s}}_{i\ell }\mathbf{\bar{v}}_{i\ell
}^{\prime }\right) -E\left( \mathbf{\bar{s}}_{i\ell }\mathbf{\bar{v}}%
_{i\circ }^{\prime }\right) -E\left( \mathbf{\bar{s}}_{i\circ }\mathbf{\bar{v%
}}_{i\ell }^{\prime }\right) +E\left( \mathbf{\bar{s}}_{i\circ }\mathbf{\bar{%
v}}_{i\circ }^{\prime }\right) \right\Vert =O\left( T^{-1}\right) \text{,}
\end{equation*}%
and using this bound in (\ref{eqp1}) yields $\sup_{i}\left\Vert E\left( 
\mathbf{Q}_{\bar{s}_{i}\bar{v}_{i}}\right) \right\Vert _{1}=O\left(
T^{-2}\right) $, as required. This completes the proof of (\ref{EQsvb}).

\bigskip
\end{proof}

\begin{lemma}
\label{Llc}Suppose Assumptions \ref{ASS1} and \ref{ASS2} hold, $m$ and $q$ $%
\left( \geq 2\right) $ are fixed, and consider $\mathbf{\tilde{\xi}}_{iq}=%
\mathbf{\bar{\xi}}_{iq}-E\left( \mathbf{\bar{\xi}}_{iq}\right) $, where $%
\mathbf{\bar{\xi}}_{iq}=q^{-1}\sum_{\ell =1}^{q}\left( \mathbf{\bar{v}}%
_{i\ell }\mathbf{\otimes \bar{s}}_{i\ell }\right) -\mathbf{\bar{v}}_{i\circ }%
\mathbf{\otimes \bar{s}}_{i\circ }$, $\mathbf{\bar{s}}_{i\ell }$ and $%
\mathbf{\bar{s}}_{i\circ }$ are, respectively, the sub-sample and full
sample time averages of the partial sum process $\mathbf{s}_{it}=\sum_{\ell
=1}^{t}\mathbf{u}_{it}$, and $\mathbf{\bar{v}}_{i\ell }$ and $\mathbf{\bar{v}%
}_{i\circ }$ are, respectively, the sub-sample and full sample time averages
of $\mathbf{v}_{it}=\mathbf{C}_{i}^{\ast }(L)\mathbf{u}_{it}=\sum_{j=0}^{%
\infty }\mathbf{C}_{ij}^{\ast }\mathbf{u}_{i,t-j}$. Let $\kappa =\left(
4+\epsilon \right) /2>2$, where $\epsilon >0$ is the constant from
Assumption \ref{ASS1}. Then%
\begin{equation}
\sup_{i,k,T}E\left\vert \tilde{\xi}_{iqk}\right\vert ^{\kappa }<K\text{,}
\label{ex3}
\end{equation}%
\begin{equation}
\sup_{i,\ell ,T}E\left\Vert \sqrt{T}\left( \mathbf{\bar{v}}_{i\ell }-\mathbf{%
\bar{v}}_{i\circ }\right) \right\Vert ^{2\kappa }<K\text{,}  \label{lyv}
\end{equation}%
and%
\begin{equation}
\sup_{i,\ell ,T}\left\Vert T^{-1/2}\mathbf{\bar{s}}_{i\ell }\right\Vert
^{2\kappa }<K\text{,}  \label{ls}
\end{equation}%
where $\tilde{\xi}_{iqk}$ is $k$-th element of $\mathbf{\tilde{\xi}}_{iq}$.
\end{lemma}

\begin{proof}
Denote the individual elements of $\mathbf{\bar{v}}_{i\ell }$ and $\mathbf{%
\bar{s}}_{i\ell }$ by $\bar{v}_{i\ell k}$ and $\bar{s}_{i\ell k}$, for $%
k=1,2,...,m$,$\,$respectively. These vectors depend on $T_{q}$ but the
subscript $T_{q}$ is suppressed to simplify notations. Given that $m$ and $q$
are fixed, sufficient condition for (\ref{ex3}) is 
\begin{equation}
\sup_{i,\ell ,\ell ^{\prime },k,k^{\prime },T}E\left\vert \bar{v}_{i\ell k}%
\bar{s}_{i\ell ^{\prime }k^{\prime }}\right\vert ^{\kappa }<K\text{,}
\label{c2}
\end{equation}%
where $\kappa =\left( 4+\epsilon \right) /2>2$ since $\epsilon >0$ by
Assumption \ref{ASS1}. It is convenient to write $\bar{v}_{i\ell k}\bar{s}%
_{i\ell ^{\prime }k^{\prime }}$ as $\left( T^{1/2}\bar{v}_{i\ell k}\right)
\left( T^{-1/2}\bar{s}_{i\ell ^{\prime }k^{\prime }}\right) $. Using
Cauchy-Schwarz inequality, we obtain%
\begin{equation}
E\left\vert \left( \sqrt{T}\bar{v}_{i\ell k}\right) \frac{\bar{s}_{i\ell
^{\prime }k^{\prime }}}{\sqrt{T}}\right\vert ^{\kappa }\leq \left(
E\left\vert \sqrt{T}\bar{v}_{i\ell k}\right\vert ^{2\kappa }\right)
^{1/2}\left( E\left\vert \frac{\bar{s}_{i\ell ^{\prime }k^{\prime }}}{\sqrt{T%
}}\right\vert ^{2\kappa }\right) ^{1/2}\text{.}  \label{c1}
\end{equation}%
Consider $\sqrt{T}\bar{v}_{i\ell k}$ first, and note that it can be written
as%
\begin{eqnarray}
\sqrt{T}\bar{v}_{i\ell k} &=&\sqrt{T}T_{q}^{-1}\sum_{t=(\ell
-1)T_{q}+1}^{\ell T_{q}}\mathbf{e}_{k}^{\prime }\mathbf{v}%
_{it}=qT^{-1/2}\sum_{t=(\ell -1)T_{q}+1}^{\ell T_{q}}\sum_{h=0}^{\infty }%
\mathbf{e}_{k}^{\prime }\mathbf{C}_{ih}^{\ast }\mathbf{u}_{i,t-h}\text{,} 
\notag \\
&=&q\sum_{h=0}^{\infty }\mathbf{e}_{k}^{\prime }\mathbf{C}_{ih}^{\ast
}\left( T^{-1/2}\sum_{t=(\ell -1)T_{q}+1}^{\ell T_{q}}\mathbf{u}%
_{i,t-h}\right) =q\sum_{h=0}^{\infty }\mathbf{e}_{k}^{\prime }\mathbf{C}%
_{ih}^{\ast }\left( T^{-1/2}\mathbf{\vartheta }_{i\ell h,T}\right) \text{,}
\label{tvba}
\end{eqnarray}%
where $\mathbf{e}_{k}$ is $m\times 1$ selection vector for $k$-th element,
and $\mathbf{\vartheta }_{i\ell h,T}=\sum_{t=(\ell -1)T/q+1}^{\ell T/q}%
\mathbf{u}_{i,t-h}$ is sum of $T_{q}=T/q$ independent random vectors
distributed with zero means. Under Assumption \ref{ASS1}, $%
\sup_{it}E\left\Vert \mathbf{u}_{it}\right\Vert ^{2\kappa
}=\sup_{it}E\left\Vert \mathbf{u}_{it}\right\Vert ^{4+\epsilon }<K$. Hence,
using the result on moments of the sums of independent random variables
given by equation (2) of 
\citeN{Petrov1992}%
,\footnote{%
Let $X_{1},X_{2},...,X_{n}$ be independent random variables and let $%
S_{n}=\sum_{\ell =1}^{n}X_{\ell }$. If $E\left( X_{\ell }\right) =0$ for $%
\ell =1,2,...,n$, and $p\geq 2$ then $E\left\vert S_{n}\right\vert ^{p}\leq
c\left( p\right) n^{p/2-1}\sum_{\ell =1}^{n}E\left\vert X_{\ell }\right\vert
^{p}$, where $c\left( p\right) >0$ depends only on $p$. See 
\citeN{Petrov1992}%
.} we have (for a fixed $q$) $\sup_{i\ell hT}E\left\Vert \mathbf{\vartheta }%
_{i\ell h,T}\right\Vert ^{2\kappa }=O\left( T^{\kappa }\right) $, and it
follows%
\begin{equation}
\sup_{i\ell hT}E\left\Vert T^{-1/2}\mathbf{\vartheta }_{i\ell
h,T}\right\Vert ^{2\kappa }<K\text{,}  \label{sva}
\end{equation}%
Consider $E\left\Vert \sqrt{T}\bar{v}_{i\ell k}\right\Vert ^{2\kappa }$
next. Using (\ref{tvba}) and the Minkowski inequality, we obtain the
following upper bound%
\begin{eqnarray*}
\left( E\left\Vert \sqrt{T}\bar{v}_{i\ell k}\right\Vert ^{2\kappa }\right) ^{%
\frac{1}{2\kappa }} &\leq &q\sum_{h=0}^{\infty }\left[ E\left\Vert \mathbf{e}%
_{k}^{\prime }\mathbf{C}_{ih}^{\ast }\left( T^{-1/2}\mathbf{\vartheta }%
_{i\ell h,T}\right) \right\Vert ^{2\kappa }\right] ^{\frac{1}{2\kappa }}%
\text{,} \\
&\leq &q\sum_{h=0}^{\infty }\left\Vert \mathbf{C}_{ih}^{\ast }\right\Vert
\left( E\left\Vert T^{-1/2}\mathbf{\vartheta }_{i\ell h,T}\right\Vert
^{2\kappa }\right) ^{\frac{1}{2\kappa }}\text{.}
\end{eqnarray*}%
Using (\ref{sva}) and noting that under Assumption \ref{ASS2} $%
\sup_{ih}\sum_{h=0}^{\infty }\left\Vert \mathbf{C}_{ih}^{\ast }\right\Vert
<K $, we obtain%
\begin{equation}
\sup_{i\ell kT}E\left\Vert \sqrt{T}\bar{v}_{i\ell k}\right\Vert ^{2\kappa }<K%
\text{.}  \label{vbo}
\end{equation}%
Existence of a uniform bound for $E\left\vert T^{-1/2}\bar{s}_{i\ell
k}\right\vert ^{2\kappa }$ is established next in a similar way. Using (\ref%
{siell}), $s_{i\ell k}$ is given by%
\begin{equation*}
s_{i\ell k}=\sum_{t=1}^{(\ell -1)T_{q}}u_{ikt}+\frac{1}{T_{q}}\sum_{t=(\ell
-1)T_{q}+1}^{\ell T_{q}}\left( \ell T_{q}+1-t\right)
u_{ikt}=\sum_{t=1}^{\ell T_{q}}\zeta _{i\ell kt,T}\text{,}
\end{equation*}%
where $\zeta _{i\ell kt,T}=a_{t\ell ,T}u_{ikt}$%
\begin{equation*}
a_{t\ell ,T}=\left\{ 
\begin{array}{cc}
1 & \text{for }t=1,2,...,\left( \ell -1\right) T_{q} \\ 
T_{q}^{-1}\left( \ell T_{q}+1-t\right) & \text{for }t=(\ell
-1)T_{q}+1,...,\ell T_{q}%
\end{array}%
\right. \text{,}
\end{equation*}%
and we have $\left\vert a_{t\ell ,T}\right\vert \leq 1$. Hence $s_{i\ell k}$
is sum of independent random variables distributed with zero means. In
addition, Assumption \ref{ASS1} and $\left\vert a_{t\ell ,T}\right\vert \leq
1$ imply $\sup_{i\ell ktT}E\left\vert \zeta _{i\ell kt,T}\right\vert
^{2\kappa }$. Using equation (2) of 
\citeN{Petrov1992}
we obtain $\sup_{i\ell kT}E\left\vert \bar{s}_{i\ell k}\right\vert ^{2\kappa
}=O\left( T^{\kappa }\right) $, and it follows%
\begin{equation}
\sup_{i\ell kT}E\left\vert \frac{\bar{s}_{i\ell k}}{\sqrt{T}}\right\vert
^{2\kappa }<K\text{.}  \label{sbo}
\end{equation}%
Using (\ref{vbo}) and (\ref{sbo}) in (\ref{c1}) establishes (\ref{c2}). This
completes the proof of (\ref{ex3}). Consider (\ref{lyv}) next. Given that $m$
and $q$ are fixed, sufficient condition for (\ref{lyv}) to hold is (\ref{vbo}%
), which was already established. Similarly, the last result of this lemma, (%
\ref{ls}), follows from (\ref{sbo}). This completes the proof.
\end{proof}

\bigskip

\begin{lemma}
\label{L_sqvvs}Consider $\mathbf{Q}_{\bar{v}\bar{v}}=T^{-1}n^{-1}q^{-1}%
\sum_{i=1}^{n}\sum_{\ell =1}^{q}\left( \mathbf{\bar{v}}_{i\ell }-\mathbf{%
\bar{v}}_{i\circ }\right) \left( \mathbf{\bar{v}}_{i\ell }-\mathbf{\bar{v}}%
_{i\circ }\right) ^{\prime }$ and suppose Assumptions \ref{ASS1} and \ref%
{ASS2} hold, $q\left( \geq 2\right) $ is fixed and $m$ is fixed. Let $\kappa
=\left( 4+\epsilon \right) /2>2$, where $\epsilon >0$ is the constant from
Assumption \ref{ASS1}. Then for any $m\times 1$ vector $\mathbf{b}$ of unit
length, $\left\Vert \mathbf{b}\right\Vert =1$, we have 
\begin{equation}
E\left( \left\vert \mathbf{b}^{\prime }\mathbf{Q}_{\bar{v}\bar{v}}\mathbf{b}%
\right\vert ^{\kappa }\right) =O\left( n^{-\kappa /2}T^{-2\kappa }\right) 
\text{.}  \label{sqvvs}
\end{equation}
\end{lemma}

\begin{proof}
Denote $\left( k,\ell \right) $-th element of the scaled $m\times m$ matrix $%
\sqrt{n}T^{2}\mathbf{Q}_{\bar{v}\bar{v}}$ as $\tilde{q}_{vv,k\ell }$, which
can be written as 
\begin{equation*}
\tilde{q}_{vv,k\ell }=n^{-1/2}\sum_{i=1}^{n}\nu _{ik}\nu _{i\ell }\text{,}
\end{equation*}%
where $\nu _{ik}$ is $k$-th element of $q^{-1}\sum_{\ell =1}^{q}\sqrt{T}%
\left( \mathbf{\bar{v}}_{i\ell }-\mathbf{\bar{v}}_{i\circ }\right) $.
Consider $\nu _{ik}\nu _{i\ell }$. Using Cauchy-Schwarz inequality,%
\begin{equation*}
E\left\vert \nu _{ik}\nu _{i\ell }\right\vert ^{\kappa }\leq \left(
E\left\vert \nu _{ik}\right\vert ^{2\kappa }\right) ^{1/2}\left( E\left\vert
\nu _{i\ell }\right\vert ^{2\kappa }\right) ^{1/2}\text{,}
\end{equation*}%
and by result (\ref{lyv}) of Lemma \ref{Llc}, we obtain $E\left\vert \nu
_{ik}\nu _{i\ell }\right\vert ^{\kappa }$ is uniformly bounded, namely%
\begin{equation*}
\sup_{k,\ell }E\left\vert \nu _{ik}\nu _{i\ell }\right\vert ^{\kappa }<K%
\text{.}
\end{equation*}%
Since $\nu _{ik}\nu _{i\ell }$ is independently distributed over $i$, we can
use the result on moments of the sums of independent random variables given
by equation (2) of 
\citeN{Petrov1992}%
, and obtain $E\left\vert \tilde{q}_{vv,k\ell }\right\vert ^{\kappa
}=O\left( 1\right) $. This implies $E\left( \left\vert \mathbf{b}^{\prime
}\left( \sqrt{n}T^{2}\mathbf{Q}_{\bar{v}\bar{v}}\right) \mathbf{b}%
\right\vert ^{\kappa }\right) =O\left( 1\right) $ for any $m\times 1$ vector 
$\mathbf{b}$ of unit length. Result (\ref{sqvvs}) now readily follows.
\end{proof}

\bigskip

\begin{lemma}
\label{L_bqb}Consider $\mathbf{Q}_{\bar{w}\bar{w}}=T^{-1}n^{-1}q^{-1}%
\sum_{i=1}^{n}\sum_{\ell =1}^{q}\left( \mathbf{\bar{w}}_{i\ell }-\mathbf{%
\bar{w}}_{i\circ }\right) \left( \mathbf{\bar{w}}_{i\ell }-\mathbf{\bar{w}}%
_{i\circ }\right) ^{\prime }$ and suppose the $m\times 1$ vector $\mathbf{w}%
_{it}$ is given by (\ref{Grep}) without interactive time effects ($\mathbf{G}%
_{i}=\mathbf{0}$), Assumptions \ref{ASS1} to \ref{ASS3} hold, $q\left( \geq
2\right) $ is fixed and $m$ is fixed. Then 
\begin{equation}
E\left\vert \mathbf{\beta }_{j0}^{\prime }\mathbf{Q}_{\bar{w}\bar{w}}\mathbf{%
\beta }_{j0}\right\vert =O\left( n^{-1/2}T^{-2}\right) \text{, for }%
j=1,2,...,r_{0}\text{,}  \label{order1}
\end{equation}%
where $\mathbf{\beta }_{j0}$ are eigenvectors of $\left( 6q\right)
^{-1}\left( q-1\right) \mathbf{\Psi }_{n}$ defined in Assumption \ref{ASS3}.
\end{lemma}

\begin{proof}
Premultiplying (\ref{qwbd}) by $\mathbf{B}_{0}^{\prime }$, postmultiplying
by $\mathbf{B}_{0}$, and noting $\mathbf{B}_{0}^{\prime }\mathbf{C}_{i}=%
\mathbf{0}$ for $i=1,2,...,n$ under Assumption \ref{ASS3}, we obtain%
\begin{equation*}
\mathbf{B}_{0}^{\prime }\mathbf{Q}_{\bar{w}\bar{w}}\mathbf{B}_{0}=\mathbf{B}%
_{0}^{\prime }\mathbf{Q}_{\bar{v}\bar{v}}\mathbf{B}_{0}\text{,}
\end{equation*}%
where by the orthonormality requirement $\mathbf{B}_{0}^{\prime }\mathbf{B}%
_{0}=\mathbf{I}_{r_{0}}$. Using Lemma \ref{L_sqvvs}, there exists a positive
constant $K$ such that for any $j=1,2,...,r$,%
\begin{equation*}
E\left( \mathbf{\beta }_{j0}\mathbf{Q}_{\bar{v}\bar{v}}\mathbf{\beta }%
_{j0}\right) ^{2}<Kn^{-1}T^{-4}\text{.}
\end{equation*}%
Result (\ref{order1}) directly follows.
\end{proof}

\begin{lemma}
\label{LQb}Consider $\mathbf{Q}_{\bar{w}\bar{w}}=T^{-1}n^{-1}q^{-1}%
\sum_{i=1}^{n}\sum_{\ell =1}^{q}\left( \mathbf{\bar{w}}_{i\ell }-\mathbf{%
\bar{w}}_{i\circ }\right) \left( \mathbf{\bar{w}}_{i\ell }-\mathbf{\bar{w}}%
_{i\circ }\right) ^{\prime }$ and the associated $m\times r$ matrix $\mathbf{%
\hat{B}}_{0}$ given by orthonormal eigenvectors of $\mathbf{Q}_{\bar{w}\bar{w%
}}$ corresponding to its $r_{0}$ smallest eigenvalues collected in the $%
r_{0}\times r_{0}$ diagonal matrix $\mathbf{\hat{\Lambda}}_{0}=diag\left( 
\hat{\lambda}_{1},\hat{\lambda}_{2},...,\hat{\lambda}_{r_{0}}\right) $.
Suppose the $m\times 1$ vector $\mathbf{w}_{it}$ is given by (\ref{Grep})
without interactive time effects ($\mathbf{G}_{i}=\mathbf{0}$), Assumptions %
\ref{ASS1} to \ref{ASS3} hold, $m$ and $q$ $\left( \geq 2\right) $ are fixed
integers. Then%
\begin{equation}
E\left\Vert \mathbf{\hat{\Lambda}}_{0}\right\Vert =O\left(
n^{-1/2}T^{-2}\right) \text{.}  \label{el}
\end{equation}
\begin{equation}
\mathbf{Q}_{\bar{w}\bar{w}}\mathbf{\hat{B}}_{0}=O_{p}\left(
n^{-1/2}T^{-2}\right) \text{.}  \label{qbt}
\end{equation}
\end{lemma}

\begin{proof}
We have $\mathbf{Q}_{\bar{w}\bar{w}}\mathbf{\hat{B}}_{0}=\mathbf{\mathbf{%
\hat{B}}}_{0}\mathbf{\hat{\Lambda}}_{0}$,\textbf{\ }where $\mathbf{\hat{%
\Lambda}}_{0}=diag\left( \hat{\lambda}_{1},\hat{\lambda}_{2},...,\hat{\lambda%
}_{r_{0}}\right) $ is consisting of $r_{0}$ smallest eigenvalues of $\mathbf{%
Q}_{\bar{w}\bar{w}}$, denoted as $0\leq \hat{\lambda}_{1}\leq \hat{\lambda}%
_{2}\leq ...\leq \hat{\lambda}_{r_{0}}$, and $\mathbf{\hat{B}}_{0}^{\prime }%
\mathbf{\hat{B}}_{0}=\mathbf{I}_{r}$. Noting $\mathbf{B}_{0}^{\prime }%
\mathbf{B}_{0}=\mathbf{I}_{r}$ then 
\begin{equation*}
0\leq tr\left( \mathbf{\hat{\Lambda}}_{0}\right) =tr\left( \mathbf{\hat{B}}%
_{0}^{\prime }\mathbf{Q}_{\bar{w}\bar{w}}\mathbf{\hat{B}}_{0}\right) \leq
tr\left( \mathbf{B}_{0}^{\prime }\mathbf{Q}_{\bar{w}\bar{w}}\mathbf{B}%
_{0}\right) \text{.}
\end{equation*}%
But result (\ref{order1}) of Lemma \ref{L_bqb} implies $E\left[ tr\left( 
\mathbf{B}_{0}^{\prime }\mathbf{Q}_{\bar{w}\bar{w}}\mathbf{B}_{0}\right) %
\right] =O\left( n^{-1/2}T^{-2}\right) $ and it follows that $E\left\Vert 
\mathbf{\hat{\Lambda}}_{0}\right\Vert =O\left( n^{-1/2}T^{-2}\right) $, as
required. Taking norm of $\mathbf{Q}_{\bar{w}\bar{w}}\mathbf{\hat{B}}_{0}=%
\mathbf{\hat{B}}_{0}\mathbf{\hat{\Lambda}}_{0}$, and noting that $\left\Vert 
\mathbf{\hat{B}}_{0}\right\Vert =1$, we obtain $\left\Vert \mathbf{Q}_{\bar{w%
}\bar{w}}\mathbf{\hat{B}}_{0}\right\Vert \leq \left\Vert \mathbf{\hat{\Lambda%
}}_{0}\right\Vert $. Taking expectations and using (\ref{el}) yields $%
E\left\Vert \mathbf{Q}_{\bar{w}\bar{w}}\mathbf{\hat{B}}_{0}\right\Vert
=O\left( n^{-1/2}T^{-2}\right) $, and result (\ref{qbt}) follows.
\end{proof}

\begin{lemma}
\label{Ld}Suppose Assumptions \ref{ASS1} to \ref{ASS3} hold, $q\left( \geq
2\right) $ is fixed and $m$ is fixed. Consider $\mathbf{\bar{\xi}}%
_{iq}=q^{-1}\sum_{\ell =1}^{q}\left( \mathbf{\bar{v}}_{i\ell }\mathbf{%
\otimes \bar{s}}_{i\ell }\right) -\mathbf{\bar{v}}_{i\circ }\mathbf{\otimes 
\bar{s}}_{i\circ }$, where $\mathbf{\bar{s}}_{i\ell }$, $\mathbf{\bar{s}}%
_{i\circ }$ are the sub-sample and full sample time averages of the partial
sum process $\mathbf{s}_{it}=\sum_{\ell =1}^{t}\mathbf{u}_{it}$, and $%
\mathbf{\bar{v}}_{i\ell }$,$\mathbf{\bar{v}}_{i\circ }$ are the sub-sample
and full sample time averages of $\mathbf{v}_{it}=\mathbf{C}_{i}^{\ast }(L)%
\mathbf{u}_{it}=\sum_{j=0}^{\infty }\mathbf{C}_{ij}^{\ast }\mathbf{u}%
_{i,t-j} $. Let $\mathbf{\Omega }_{\xi _{qi}}=Var\left( \mathbf{\bar{\xi}}%
_{iq}\right) $ and suppose $\mathbf{\Omega }_{q}=\lim_{n,T\rightarrow \infty
}\left( \mathbf{\mathring{B}}_{0}^{\prime }\mathbf{\otimes \mathbf{C}}%
_{i}\right) \mathbf{\Omega }_{\xi _{qi}}\left( \mathbf{\mathring{B}}_{0}%
\mathbf{\otimes \mathbf{C}}_{i}^{\prime }\right) $ is positive definite. Then%
\begin{equation}
n^{-1/2}\sum_{i=1}^{n}\left( \mathbf{\mathring{B}}_{0}^{\prime }\mathbf{%
\otimes \mathbf{C}}_{i}\right) \mathbf{\tilde{\xi}}_{iq}\rightarrow
_{d}N\left( \mathbf{0,\Omega }_{q}\right) ,  \label{dr1}
\end{equation}%
for $n,T\rightarrow \infty \,$jointly (in no particular order), where $%
\mathbf{\tilde{\xi}}_{iq}=\mathbf{\bar{\xi}}_{iq}-E\left( \mathbf{\bar{\xi}}%
_{iq}\right) $.
\end{lemma}

\begin{proof}
Let $n_{T}=n\left( T\right) $ be any non-decreasing integer-valued function
of $T$ such that $\lim_{T\rightarrow \infty }n_{T}=\infty $, and let $%
\mathbf{\omega }$ be any $m^{2}\times 1$ vector of unit length, namely $%
\left\Vert \mathbf{\omega }\right\Vert =1$. Define the following triangular
array%
\begin{equation*}
a_{n_{T},i}=\frac{1}{\sqrt{n_{T}}}\mathbf{\omega }^{\prime }\mathbf{\Omega }%
_{q}^{-1/2}\left( \mathbf{\mathring{B}}_{0}^{\prime }\mathbf{\otimes \mathbf{%
C}}_{i}\right) \left[ \mathbf{\bar{\xi}}_{iq}-E\left( \mathbf{\bar{\xi}}%
_{iq}\right) \right] \text{,}
\end{equation*}%
where $m^{2}\times 1$ vector $\mathbf{\bar{\xi}}_{iq}$ depends on $T$, but
this subscript is suppressed for notational simplicity. By construction
array $\left\{ a_{n_{T},i}\right\} $ is zero mean. By assumption \ref{ASS1}, 
$\mathbf{u}_{it}$ is independently distributed over $i$ and $t$, and
therefore $a_{n_{T},i}$ is independent of $a_{n_{T},j}$ for $i\neq j$. In
addition, 
\begin{equation*}
\sum_{i=1}^{n_{T}}Var\left( a_{n_{T},i}\right) =1\text{.}
\end{equation*}%
We establish next the Liapunov's condition, given by%
\begin{equation}
\lim_{n\rightarrow \infty }\sum_{i=1}^{n_{T}}E\left\vert
a_{n_{T},i}\right\vert ^{2+\epsilon }=0\text{ for some }\epsilon >0\text{,}
\label{lia}
\end{equation}%
is met. Since $\mathbf{\Omega }_{\xi _{q}}$ is positive definite then $%
\left\Vert \mathbf{\omega }^{\prime }\mathbf{\Omega }_{\xi
_{q}}^{-1/2}\left( \mathbf{\mathring{B}}_{0}^{\prime }\mathbf{\otimes 
\mathbf{C}}_{i}\right) \right\Vert <K$, and (\ref{lia}) will hold if 
\begin{equation}
\sup_{i,k,T}E\left\vert \tilde{\xi}_{iqk}\right\vert ^{2+\epsilon }<K\text{
for some }\epsilon >0\text{,}  \label{exb}
\end{equation}%
where $\tilde{\xi}_{iqk}$ is $k$-th element of $\mathbf{\tilde{\xi}}_{iq}$.
Lemma \ref{Llc} establish (\ref{exb}) holds, and therefore (\ref{lia}) is
met, as required. Then, by Liapunov's theorem (Theorem 23.11 in 
\citeN{Davidson1994}%
), the Lindeberg condition is also met, and (\ref{dr1}) follows by Theorem
23.6 in 
\citeN{Davidson1994}%
.
\end{proof}

\begin{lemma}
\label{Lcqb}Consider $m\times m$ matrices $\mathbf{Q}_{\bar{s}_{i}\bar{v}%
_{i}}=T^{-1}q^{-1}\sum_{\ell =1}^{q}\left( \mathbf{\bar{s}}_{i\ell }-\mathbf{%
\bar{s}}_{i\circ }\right) \left( \mathbf{\bar{v}}_{i\ell }-\mathbf{\bar{v}}%
_{i\circ }\right) ^{\prime }$, where $\mathbf{\bar{s}}_{i\ell }$, $\mathbf{%
\bar{s}}_{i\circ }$ are the sub-sample and full sample time averages of the
partial sum process $\mathbf{s}_{it}=\sum_{\ell =1}^{t}\mathbf{u}_{it}$, and 
$\mathbf{\bar{v}}_{i\ell }$,$\mathbf{\bar{v}}_{i\circ }$ are the sub-sample
and full sample time averages of $\mathbf{v}_{it}=\mathbf{C}_{i}^{\ast }(L)%
\mathbf{u}_{it}=\sum_{j=0}^{\infty }\mathbf{C}_{ij}^{\ast }\mathbf{u}%
_{i,t-j} $. Suppose Assumptions \ref{ASS1} and \ref{ASS2} hold, $q$($\geq 2$%
) is fixed and $m$ is fixed. Then 
\begin{equation}
n^{-1}\sum_{i=1}^{n}\mathbf{C}_{i}\mathbf{Q}_{\bar{s}_{i}\bar{v}%
_{i}}=O_{p}\left( n^{-1/2}T^{-1}\right) +O\left( T^{-2}\right)  \label{cqb}
\end{equation}
\end{lemma}

\begin{proof}
For the purpose of this proof, define $m^{2}\times 1$ vector $\mathbf{b}%
_{nT} $ obtained by vectorization of $n^{-1}\sum_{i=1}^{n}\mathbf{C}_{i}%
\mathbf{Q}_{\bar{s}_{i}\bar{v}_{i}}$, 
\begin{equation*}
\mathbf{b}_{nT}=vec\left( n^{-1}\sum_{i=1}^{n}\mathbf{C}_{i}\mathbf{Q}_{\bar{%
s}_{i}\bar{v}_{i}}\right) =T^{-1}n^{-1}\sum_{i=1}^{n}\left( \mathbf{I}%
_{m}\otimes \mathbf{C}_{i}\right) \mathbf{\bar{\xi}}_{iq}\text{,}
\end{equation*}%
where $\mathbf{\bar{\xi}}_{iq}=q^{-1}\sum_{\ell =1}^{q}\left( \mathbf{\bar{v}%
}_{i\ell }\mathbf{\otimes \bar{s}}_{i\ell }\right) -\mathbf{\bar{v}}_{i\circ
}\mathbf{\otimes \bar{s}}_{i\circ }$. Denote the $k$-$th$ element of $%
\mathbf{b}_{nT}$ as $b_{k,nT}=\mathbf{e}_{k}^{\prime }\mathbf{b}_{nT}$,
where $\mathbf{e}_{k}$ is $m^{2}\times 1$ selection vector for the $k$-th
element, $k=1,2,...,m^{2}$. Define $\tilde{b}_{k,nT}=b_{k,nT}-E\left(
b_{k,nT}\right) $. Using result (\ref{EQsvb}) of Lemma \ref{LQB} and noting $%
\sup_{i}\left\Vert \mathbf{C}_{i}\right\Vert <K$ by Assumption \ref{ASS2},
we obtain%
\begin{equation}
\sup_{k}E\left( b_{k,nT}\right) =O\left( T^{-2}\right) \text{.}
\label{lcbp1}
\end{equation}%
Consider $E\left( \tilde{b}_{k,nT}\right) ^{2}$ next. We have 
\begin{equation*}
E\left( \tilde{b}_{k,nT}\right) ^{2}=E\left\{ \left[ T^{-1}n^{-1}%
\sum_{i=1}^{n}\mathbf{e}_{k}^{\prime }\left( I_{m}\otimes \mathbf{C}%
_{i}\right) \mathbf{\tilde{\xi}}_{iq}\right] ^{2}\right\} \text{,}
\end{equation*}%
where $\mathbf{\tilde{\xi}}_{iq}=\mathbf{\bar{\xi}}_{iq}-E\left( \mathbf{%
\bar{\xi}}_{iq}\right) $. By construction $E\left( \mathbf{\tilde{\xi}}%
_{iq}\right) =0$. In addition, $\mathbf{\tilde{\xi}}_{iq}$ is independent of 
$\mathbf{\tilde{\xi}}_{jq}$ for any $i\neq j$. It follows%
\begin{eqnarray*}
E\left( \tilde{b}_{k,nT}\right) ^{2} &=&n^{-2}\sum_{i=1}^{n}E\left( \mathbf{e%
}_{k}^{\prime }\left( \mathbf{I}_{m}\otimes \mathbf{C}_{i}\right) \mathbf{%
\tilde{\xi}}_{iq}\right) ^{2}\text{,} \\
&\leq &T^{-2}n^{-2}\sum_{i=1}^{n}\left\Vert \mathbf{e}_{k}^{\prime }\left( 
\mathbf{I}_{m}\otimes \mathbf{C}_{i}\right) \right\Vert ^{2}E\left(
\left\Vert \mathbf{\tilde{\xi}}_{iq}\right\Vert ^{2}\right) \text{.}
\end{eqnarray*}%
Using result (\ref{ex3}) of Lemma \ref{Llc} and $\sup_{i}\left\Vert \mathbf{C%
}_{i}\right\Vert <K$ (by Assumption \ref{ASS2}), then 
\begin{equation}
\sup_{k}E\left( \tilde{b}_{k,nT}\right) ^{2}=O\left( n^{-1}T^{-2}\right) 
\text{.}  \label{lcbp2}
\end{equation}%
Using (\ref{lcbp1}) and (\ref{lcbp2}) in $b_{k,nT}=\tilde{b}_{k,nT}+E\left(
b_{k,nT}\right) $, we obtain (\ref{cqb}).
\end{proof}

\begin{lemma}
\label{LQB_ie}Consider $m\times m$ matrices $\mathbf{Q}_{\bar{f}_{i}\bar{f}%
_{i}}=T^{-1}q^{-1}\sum_{\ell =1}^{q}\mathbf{G}_{i}\left( \mathbf{\bar{f}}%
_{\ell }-\mathbf{\bar{f}}_{\circ }\right) \left( \mathbf{\bar{f}}_{\ell }-%
\mathbf{\bar{f}}_{\circ }\right) ^{\prime }\mathbf{G}_{i}^{\prime }$, 
\newline
$\mathbf{Q}_{\bar{f}_{i}\bar{v}_{i}}=T^{-1}q^{-1}\sum_{\ell =1}^{q}\mathbf{G}%
_{i}\left( \mathbf{\bar{f}}_{\ell }-\mathbf{\bar{f}}_{\circ }\right) \left( 
\mathbf{\bar{v}}_{i\ell }-\mathbf{\bar{v}}_{i\circ }\right) ^{\prime }$ and $%
\mathbf{Q}_{\bar{f}_{i}\bar{s}_{i}}=T^{-1}q^{-1}\sum_{\ell =1}^{q}\mathbf{G}%
_{i}\left( \mathbf{\bar{f}}_{\ell }-\mathbf{\bar{f}}_{\circ }\right) \left( 
\mathbf{\bar{s}}_{i\ell }-\mathbf{\bar{s}}_{i\circ }\right) ^{\prime }$,
where $\mathbf{\bar{f}}_{\ell }$ and $\mathbf{\bar{f}}_{\circ }$ are the
sub-sample and full sample time averages of $\mathbf{f}_{t}=\sum_{j=0}^{%
\infty }\mathbf{\Phi }_{f\ell }L^{\ell }\mathbf{\varepsilon }_{f,t-\ell }$, $%
\mathbf{\bar{s}}_{i\ell }$ and $\mathbf{\bar{s}}_{i\circ }$ are the
sub-sample and full sample time averages of the partial sum process $\mathbf{%
s}_{it}=\sum_{\ell =1}^{t}\mathbf{u}_{it}$, and $\mathbf{\bar{v}}_{i\ell }$
and $\mathbf{\bar{v}}_{i\circ }$ are the sub-sample and full sample time
averages of $\mathbf{v}_{it}=\mathbf{C}_{i}^{\ast }(L)\mathbf{u}%
_{it}=\sum_{j=0}^{\infty }\mathbf{C}_{ij}^{\ast }\mathbf{u}_{i,t-j}$.
Suppose Assumptions \ref{ASS1}, \ref{ASS2} and \ref{ASSfactors} hold, and $q$%
($\geq 2$), $m$ and $m_{f}$ are fixed. Then,%
\begin{equation}
\sup_{i}E\left\Vert \mathbf{Q}_{\bar{f}_{i}\bar{f}_{i}}\right\Vert =O\left(
T^{-2}\right) \text{,}  \label{bQff}
\end{equation}%
\begin{equation}
\sup_{i}E\left\Vert \mathbf{Q}_{\bar{f}_{i}\bar{v}_{i}}\right\Vert =O\left(
T^{-2}\right) \text{,}  \label{bQfv}
\end{equation}%
and 
\begin{equation}
\sup_{i}E\left\Vert \mathbf{Q}_{\bar{f}_{i}\bar{s}_{i}}\right\Vert =O\left(
T^{-1}\right) \text{.}  \label{bQfs}
\end{equation}
\end{lemma}

\begin{proof}
Consider (\ref{bQff}) first. We can write $\mathbf{Q}_{\bar{f}_{i}\bar{f}%
_{i}}\,$as%
\begin{equation*}
\mathbf{Q}_{\bar{f}_{i}\bar{f}_{i}}=\mathbf{G}_{i}\mathbf{Q}_{\bar{f}\bar{f}}%
\mathbf{G}_{i}^{\prime }\text{,}
\end{equation*}%
where%
\begin{equation*}
\mathbf{Q}_{\bar{f}\bar{f}}=T^{-1}q^{-1}\sum_{\ell =1}^{q}\left( \mathbf{%
\bar{f}}_{\ell }-\mathbf{\bar{f}}_{\circ }\right) \left( \mathbf{\bar{f}}%
_{\ell }-\mathbf{\bar{f}}_{\circ }\right) ^{\prime }\text{.}
\end{equation*}%
Taking norm, expectation and $\sup $ over $i$, we have%
\begin{equation}
\sup_{i}E\left\Vert T\mathbf{Q}_{\bar{f}_{i}\bar{f}_{i}}\right\Vert \leq
\left( \sup_{i}\left\Vert \mathbf{G}_{i}\right\Vert \right) ^{2}E\left\Vert 
\mathbf{Q}_{\bar{f}\bar{f}}\right\Vert \leq K\cdot E\left\Vert T\mathbf{Q}_{%
\bar{f}\bar{f}}\right\Vert \text{,}  \label{ppf1}
\end{equation}%
where $\sup_{i}\left\Vert \mathbf{G}_{i}\right\Vert <K$ by Assumption \ref%
{ASSfactors}. In addition, $\mathbf{f}_{t}$ is a covariance stationary
process that satisfies essentially the same assumptions as the covariance
stationary process $\mathbf{v}_{it}$. Hence, using the same arguments as in
the proof of result (\ref{bQvv}) in Lemma \ref{LQB} in Appendix, we obtain%
\begin{equation}
E\left\Vert T\mathbf{Q}_{\bar{f}\bar{f}}\right\Vert \leq \sup_{\ell
}E\left\Vert \mathbf{\bar{f}}_{\ell }\right\Vert ^{2}+\sup_{i}E\left\Vert 
\mathbf{\bar{f}}_{\circ }\right\Vert ^{2}=O\left( T^{-1}\right) \text{,}
\label{ppf2}
\end{equation}%
where 
\begin{equation}
\sup_{\ell }E\left\Vert \mathbf{\bar{f}}_{\ell }\right\Vert ^{2}=O\left(
T^{-1}\right) \text{ and }\sup_{i}E\left\Vert \mathbf{\bar{f}}_{\circ
}\right\Vert ^{2}=O\left( T^{-1}\right) \text{.}  \label{ppf3}
\end{equation}%
Using (\ref{ppf2}) in (\ref{ppf1}), we obtain (\ref{bQff}).

Consider (\ref{bQfv}) next. We can write $T\mathbf{Q}_{\bar{f}_{i}\bar{v}%
_{i}}\,$as%
\begin{equation*}
T\mathbf{Q}_{\bar{f}_{i}\bar{v}_{i}}=q^{-1}\sum_{\ell =1}^{q}\mathbf{G}%
_{i}\left( \mathbf{\bar{f}}_{\ell }-\mathbf{\bar{f}}_{\circ }\right) \left( 
\mathbf{\bar{v}}_{i\ell }-\mathbf{\bar{v}}_{i\circ }\right) ^{\prime
}=q^{-1}\sum_{\ell =1}^{q}\mathbf{G}_{i}\mathbf{\bar{f}}_{\ell }\mathbf{\bar{%
v}}_{i\ell }^{\prime }-\mathbf{G}_{i}\mathbf{\bar{f}}_{\circ }\mathbf{\bar{v}%
}_{i\circ }^{\prime }\text{.}
\end{equation*}%
Hence%
\begin{equation}
E\left\Vert T\mathbf{Q}_{\bar{f}_{i}\bar{v}_{i}}\right\Vert \leq
q^{-1}\sum_{\ell =1}^{q}\left\Vert \mathbf{G}_{i}\right\Vert \left(
E\left\Vert \mathbf{\bar{f}}_{\ell }\right\Vert ^{2}\right) ^{1/2}\left(
E\left\Vert \mathbf{\bar{v}}_{i\ell }\right\Vert ^{2}\right)
^{1/2}+\left\Vert \mathbf{G}_{i}\right\Vert \left( E\left\Vert \mathbf{\bar{f%
}}_{\circ }\right\Vert ^{2}\right) ^{1/2}\left( E\left\Vert \mathbf{\bar{v}}%
_{i\circ }\right\Vert ^{2}\right) ^{1/2}\text{.}  \label{ppf4}
\end{equation}%
Using (\ref{vil}), (\ref{vid}), (\ref{ppf3}) and $\sup_{i}\left\Vert \mathbf{%
G}_{i}\right\Vert <K$ (by Assumption \ref{ASSfactors}) in (\ref{ppf4}),
result (\ref{bQfv}) follows.

Consider the last result, (\ref{bQfs}). Similarly to (\ref{ppf4}), we have%
\begin{equation}
E\left\Vert T\mathbf{Q}_{\bar{f}_{i}\bar{s}_{i}}\right\Vert \leq
q^{-1}\sum_{\ell =1}^{q}\left\Vert \mathbf{G}_{i}\right\Vert \left(
E\left\Vert \mathbf{\bar{f}}_{\ell }\right\Vert ^{2}\right) ^{1/2}\left(
E\left\Vert \mathbf{\bar{s}}_{i\ell }\right\Vert ^{2}\right)
^{1/2}+\left\Vert \mathbf{G}_{i}\right\Vert \left( E\left\Vert \mathbf{\bar{f%
}}_{\circ }\right\Vert ^{2}\right) ^{1/2}\left( E\left\Vert \mathbf{\bar{s}}%
_{i\circ }\right\Vert ^{2}\right) ^{1/2}\text{.}  \label{ppf5}
\end{equation}%
Using (\ref{so}), (\ref{ppf3}) and $\sup_{i}\left\Vert \mathbf{G}%
_{i}\right\Vert <K$ (by Assumption \ref{ASSfactors}) in (\ref{ppf5}), result
(\ref{bQfs}) follows.
\end{proof}

\begin{lemma}
\label{L_bqb_ie}Consider $\mathbf{Q}_{\bar{w}\bar{w}}=T^{-1}n^{-1}q^{-1}%
\sum_{i=1}^{n}\sum_{\ell =1}^{q}\left( \mathbf{\bar{w}}_{i\ell }-\mathbf{%
\bar{w}}_{i\circ }\right) \left( \mathbf{\bar{w}}_{i\ell }-\mathbf{\bar{w}}%
_{i\circ }\right) ^{\prime }$ and suppose the $m\times 1$ vector $\mathbf{w}%
_{it}$ is given by (\ref{gmf}) featuring interactive time effects,
Assumptions \ref{ASS1} to \ref{ASSfactors} hold, and $q\left( \geq 2\right) $%
, $m$ and $m_{f}$ are fixed. Then%
\begin{equation}
E\left\vert \mathbf{\beta }_{j0}^{\prime }\mathbf{Q}_{\bar{w}\bar{w}}\mathbf{%
\beta }_{j0}^{\prime }\right\vert =O\left( T^{-2}\right) \text{, for }%
j=1,2,...,r_{0}\text{,}  \label{order1f}
\end{equation}%
where $\mathbf{\beta }_{j0}$ (for $j=1,2,...,r_{0}$) are defined in
Assumption \ref{ASS3}.
\end{lemma}

\begin{proof}
Premultiplying (\ref{qwbd}) by $\mathbf{B}_{0}^{\prime }$, postmultiplying
by $\mathbf{B}_{0}$, and noting $\mathbf{B}_{0}^{\prime }\mathbf{C}_{i}=%
\mathbf{0}$ for $i=1,2,...,n$ under Assumption \ref{ASS3}, we obtain%
\begin{equation*}
\mathbf{B}_{0}^{\prime }\mathbf{Q}_{\bar{w}\bar{w}}\mathbf{B}_{0}=\mathbf{B}%
_{0}^{\prime }\mathbf{Q}_{\bar{f}\bar{f}}\mathbf{B}_{0}+\mathbf{B}%
_{0}^{\prime }\left( \mathbf{Q}_{\bar{f}\bar{v}}+\mathbf{Q}_{\bar{v}\bar{f}%
}\right) \mathbf{B}_{0}+\mathbf{B}_{0}^{\prime }\mathbf{Q}_{\bar{v}\bar{v}}%
\mathbf{B}_{0}\text{,}
\end{equation*}%
where by the orthonormality requirement $\mathbf{B}_{0}^{\prime }\mathbf{B}%
_{0}=\mathbf{I}_{r_{0}}$. By Lemma \ref{L_sqvvs}, $E\left\Vert \mathbf{B}%
_{0}^{\prime }\mathbf{Q}_{\bar{v}\bar{v}}\mathbf{B}_{0}\right\Vert =O\left(
n^{-1/2}T^{-2}\right) $. Using results (\ref{bQff}) and (\ref{bQfv}) of
Lemma \ref{LQB_ie}, we obtain $E\left\Vert \mathbf{B}_{0}^{\prime }\mathbf{Q}%
_{\bar{f}\bar{f}}\mathbf{B}_{0}\right\Vert =O\left( T^{-2}\right) $ and 
\newline
$E\left\Vert \mathbf{B}_{0}^{\prime }\left( \mathbf{Q}_{\bar{f}\bar{v}}+%
\mathbf{Q}_{\bar{v}\bar{f}}\right) \mathbf{B}_{0}\right\Vert \mathbf{=}%
O\left( T^{-2}\right) $. Hence, the dominant term is $O\left( T^{-2}\right) $
and result (\ref{order1f}) follows.
\end{proof}

\begin{lemma}
\label{LQb_ie}Consider $\mathbf{Q}_{\bar{w}\bar{w}}=T^{-1}n^{-1}q^{-1}%
\sum_{i=1}^{n}\sum_{\ell =1}^{q}\left( \mathbf{\bar{w}}_{i\ell }-\mathbf{%
\bar{w}}_{i\circ }\right) \left( \mathbf{\bar{w}}_{i\ell }-\mathbf{\bar{w}}%
_{i\circ }\right) ^{\prime }$ and the associated $m\times r$ matrix $\mathbf{%
\hat{B}}_{0}$ given by orthonormal eigenvectors of $\mathbf{Q}_{\bar{w}\bar{w%
}}$ corresponding to its $r_{0}$ smallest eigenvalues. Suppose the $m\times
1 $ vector $\mathbf{w}_{it}$ is given by (\ref{gmf}) featuring interactive
time effects, Assumptions \ref{ASS1} to \ref{ASSfactors} hold, and $q\left(
\geq 2\right) $, $m$ and $m_{f}$ are fixed. Then 
\begin{equation}
\mathbf{Q}_{\bar{w}\bar{w}}\mathbf{\hat{B}}_{0}=O_{p}\left( T^{-2}\right) 
\text{.}  \label{qbt1}
\end{equation}
\end{lemma}

\begin{proof}
Similarly to proof of result (\ref{qbt}) in Lemma \ref{LQb} in Appendix,
when $\mathbf{w}_{it}$ is given by (\ref{gmf}), we continue to have $0\leq
tr\left( \mathbf{\hat{\Lambda}}\right) =tr\left( \mathbf{\hat{B}}%
_{0}^{\prime }\mathbf{Q}_{\bar{w}\bar{w}}\mathbf{\hat{B}}_{0}\right) \leq
tr\left( \mathbf{B}_{0}^{\prime }\mathbf{Q}_{\bar{w}\bar{w}}\mathbf{B}%
_{0}\right) $. But result (\ref{order1f}) of Lemma \ref{L_bqb_ie} implies $E%
\left[ tr\left( \mathbf{B}_{0}^{\prime }\mathbf{Q}_{\bar{w}\bar{w}}\mathbf{B}%
_{0}\right) \right] =O\left( T^{-2}\right) $ and it follows $E\left\Vert 
\mathbf{\hat{\Lambda}}\right\Vert =O\left( T^{-2}\right) $. Taking norm of $%
\mathbf{Q}_{\bar{w}\bar{w}}\mathbf{\hat{B}}_{0}=\mathbf{\hat{B}}_{0}\mathbf{%
\Lambda }$ and noting $\left\Vert \mathbf{\hat{B}}_{0}\right\Vert =1$, we
obtain $\left\Vert \mathbf{Q}_{\bar{w}\bar{w}}\mathbf{\hat{B}}%
_{0}\right\Vert \leq \left\Vert \mathbf{\hat{\Lambda}}\right\Vert $. Taking
expectations and using $E\left\Vert \mathbf{\hat{\Lambda}}\right\Vert
=O\left( T^{-2}\right) $ yields $E\left\Vert \mathbf{Q}_{\bar{w}\bar{w}}%
\mathbf{\hat{B}}_{0}\right\Vert =O\left( T^{-2}\right) $, which is
sufficient for (\ref{qbt1}). This completes the proof.
\end{proof}

\begin{lemma}
\label{Ldf}Suppose Assumptions \ref{ASS1} to \ref{ASSfactors} hold, and $%
q\left( \geq 2\right) $, $m$ and $m_{f}$ are fixed. Consider $\mathbf{\xi }%
_{iq}^{\ast }=q^{-1}\sum_{\ell =1}^{q}\left( \mathbf{\bar{\omega}}_{i\ell }-%
\mathbf{\bar{\omega}}_{i}\right) \mathbf{\otimes }\left( \mathbf{\bar{s}}%
_{i\ell }-\mathbf{\bar{s}}_{i}\right) $, where $\mathbf{\bar{s}}_{i\ell }$, $%
\mathbf{\bar{s}}_{i\circ }$ are the sub-sample and full sample time averages
of the partial sum process $\mathbf{s}_{it}=\sum_{\ell =1}^{t}\mathbf{u}%
_{it} $, and $\mathbf{\bar{\omega}}_{i\ell }$,$\mathbf{\bar{\omega}}_{i}$
are the sub-sample and full sample time averages of $\mathbf{\omega }_{it}=%
\mathbf{f}_{it}+\mathbf{v}_{it}$, $\mathbf{f}_{it}=\sum_{j=0}^{\infty }%
\mathbf{\Phi }_{fi\ell }L^{\ell }\mathbf{\varepsilon }_{f,t-\ell }$, $%
\mathbf{v}_{it}=\sum_{j=0}^{\infty }\mathbf{C}_{ij}^{\ast }\mathbf{u}%
_{i,t-j} $. Let $\mathbf{\Omega }_{\xi _{qi}^{\ast }}=Var\left( \mathbf{\bar{%
\xi}}_{iq}^{\ast }\right) $ and suppose $\mathbf{\Omega }_{q}^{\ast
}=\lim_{n,T\rightarrow \infty }\left( \mathbf{\mathring{B}}_{0}^{\prime }%
\mathbf{\otimes \mathbf{C}}_{i}\right) \mathbf{\Omega }_{\xi _{qi}^{\ast
}}\left( \mathbf{\mathring{B}}_{0}\mathbf{\otimes \mathbf{C}}_{i}^{\prime
}\right) $ is positive definite. Then%
\begin{equation}
n^{-1/2}\sum_{i=1}^{n}\left( \mathbf{\mathring{B}}_{0}^{\prime }\mathbf{%
\otimes \mathbf{C}}_{i}\right) \mathbf{\tilde{\xi}}_{iq}^{\ast }\rightarrow
_{d}N\left( \mathbf{0,\Omega }_{q}^{\ast }\right) ,  \label{dr1f}
\end{equation}%
for $n,T\rightarrow \infty \,$jointly (in no particular order), where $%
\mathbf{\tilde{\xi}}_{iq}^{\ast }=\mathbf{\bar{\xi}}_{iq}^{\ast }-E\left( 
\mathbf{\bar{\xi}}_{iq}^{\ast }\right) $.
\end{lemma}

\begin{proof}
(\ref{dr1f}) can be established in a similar way as the proof of (\ref{dr1}%
), but we rely on a central limit theorems for martingales (Theorem 24.3 in 
\citeN{Davidson1994}%
), since $\mathbf{\tilde{\xi}}_{iq}^{\ast }$ is no longer independently
distributed over $i$ in the presence of unobserved common factors.
\end{proof}

\section{PME estimator for unbalanced panels\label{Aunb}}

We assume there are $n$ cross section units, $i=1,2,...,n$, with each cross
section unit having $T_{i}$ consecutive observations. We do not allow for
gaps. Let $T_{m}=\max T_{i}$. We also assume all $T_{i}$'s expand at the
same rate with $n$. Specifically, there exists a positive constant $%
\varkappa _{u}>0$, which does not depend on the sample size, such that $%
T_{i}>\varkappa _{u}\cdot T_{m}$, for $i=1,2....,n$.

In practice, these assumption may require excluding some observations to
avoid gaps, and excluding individual cross section units where $T_{i}$ is
very small compared with the rest of the panel (see Subsection \ref{iu}). In
the exposition below, we assume that the unbalanced panel satisfies the
assumptions above.

To accommodate unbalanced panels, we consider the following generalization
of sub-sample time averages defined in Section \ref{LongRR} (see (\ref{la})),%
\begin{equation}
\mathbf{\bar{w}}_{i\ell }=\frac{1}{T_{i\ell }}\sum_{t=H_{i,\ell
-1}+1}^{H_{i\ell }}\mathbf{w}_{it},\text{ for }\ell =1,2,...,q_{i}\text{,}
\label{lau}
\end{equation}%
where $T_{i\ell }=H_{i\ell }-H_{i,\ell -1}$ is the sample size for computing
sub-sample time average $\ell $ for unit $i$. We continue to assume $%
H_{i\ell }>H_{i,\ell -1}$, $T_{i}=\sum_{\ell =1}^{q_{i}}T_{i\ell }$, with $%
T_{i\ell }$ being of the same order as $T_{i}$, namely there exists positive
constants $K>0$ that do not depend on the sample size such that $T_{i\ell
}/\max_{\ell =1,2,...,q_{i}}\left\{ T_{i\ell }\right\} >K_{0}$ for all $i$
and all $\ell $. The number of sub-sample time averages, $q_{i}$, in (\ref%
{lau}) are allowed to vary across units with $2\leq q_{i}\leq q_{m}$, where $%
q_{m}$ does not depend on the sample size. The time average $\mathbf{\bar{w}}%
_{i\circ }$ is now defined by 
\begin{equation*}
\mathbf{\bar{w}}_{i\circ }=q_{i}^{-1}\sum_{\ell =1}^{q_{i}}\mathbf{\bar{w}}%
_{i\ell }\text{,}
\end{equation*}%
and 
\begin{equation*}
\mathbf{Q}_{\bar{w}_{i}\bar{w}_{i}}=T_{i}^{-1}q_{i}^{-1}\sum_{\ell
=1}^{q_{i}}\left( \mathbf{\bar{w}}_{i\ell }-\mathbf{\bar{w}}_{i\circ
}\right) \left( \mathbf{\bar{w}}_{i\ell }-\mathbf{\bar{w}}_{i\circ }\right)
^{\prime }\text{.}
\end{equation*}%
with the associated pooled version given by%
\begin{equation*}
\mathbf{Q}_{\bar{w}\bar{w}}=n^{-1}\sum_{i=1}^{n}\mathbf{Q}_{\bar{w}_{i}\bar{w%
}_{i}}=T_{ave}^{-1}n^{-1}\sum_{i=1}^{n}\left( \phi _{i}q_{i}^{-1}\sum_{\ell
=1}^{q_{i}}\left( \mathbf{\bar{w}}_{i\ell }-\mathbf{\bar{w}}_{i\circ
}\right) \left( \mathbf{\bar{w}}_{i\ell }-\mathbf{\bar{w}}_{i\circ }\right)
^{\prime }\right) \text{,}
\end{equation*}%
where%
\begin{equation*}
\phi _{i}=\left( \frac{T_{i}}{T_{ave}}\right) ^{-1}\text{ and }%
T_{ave}=\left( n^{-1}\sum_{i=1}^{n}T_{i}^{-1}\right) ^{-1}\text{.}
\end{equation*}

Following the same arguments as in the case of balanced panels, it can be
established that $E\left\Vert \mathbf{Q}_{\bar{v}_{i}\bar{v}_{i}}\right\Vert
_{1}<K/T_{i}^{2}$, $E\left\Vert \mathbf{Q}_{\bar{s}_{i}\bar{v}%
_{i}}\right\Vert _{1}<K/T_{i}$, and $E\left\Vert \mathbf{Q}_{\bar{s}_{i}\bar{%
s}_{i}}\right\Vert <K$. These results imply $\mathbf{Q}_{\bar{w}\bar{w}}%
\mathbf{\beta }=O_{p}\left( T_{ave}^{-1}\right) $ and we can use the
eigenvectors corresponding to the $r_{0}$ smallest eigenvalues of $\mathbf{Q}%
_{\bar{w}\bar{w}}$ to estimate long-run relations subject to the exact
identifying assumptions.

To derive asymptotic distribution, let 
\begin{equation}
\mathbf{Z}_{i}=\mathbf{C}_{i}\left[ \phi _{i}q_{i}^{-1}\sum_{\ell
=1}^{q_{i}}\left( \mathbf{\bar{s}}_{i\ell }-\mathbf{\bar{s}}_{i\circ
}\right) \left( \mathbf{\bar{v}}_{i\ell }-\mathbf{\bar{v}}_{i\circ }\right)
^{\prime }\right] \mathbf{\mathring{B}}_{0}-\mathbf{C}_{i}\left[ \phi
_{i}q_{i}^{-1}\sum_{\ell =1}^{q_{i}}E\left( \mathbf{\bar{s}}_{i\ell }-%
\mathbf{\bar{s}}_{i\circ }\right) \left( \mathbf{\bar{v}}_{i\ell }-\mathbf{%
\bar{v}}_{i\circ }\right) ^{\prime }\right] \mathbf{\mathring{B}}_{0}\text{,}
\end{equation}%
and following the same arguments as in Section \ref{LongRR}, we obtain%
\begin{equation}
\mathbf{Q}_{\bar{w}\bar{w}}\sqrt{n}T_{ave}\left( \widehat{\mathbf{\mathring{B%
}}_{0}}-\mathbf{\mathring{B}}_{0}\right) =-n^{-1/2}\sum_{i=1}^{n}\mathbf{Z}%
_{i}+O_{p}\left( \frac{\sqrt{n}}{T_{ave}}\right) ,
\end{equation}%
Consider the exact identifying restrictions $\mathbf{\mathring{B}}%
_{0}=\left( \mathbf{I}_{r_{0}},\mathbf{\mathring{B}}_{0,2}^{\prime }\right)
^{\prime }$, where $\mathbf{I}_{r_{0}}$\ is an identity matrix of order $%
r_{0}$. Let $\mathbf{\hat{\Theta}}$ and $\mathbf{\Theta }_{0}$ be the lower $%
\left( m-r_{0}\right) \times r_{0}$ block of $\widehat{\mathbf{\mathring{B}}%
_{0}}$ and $\mathbf{\mathring{B}}_{0}$, respectively. Then, $\sqrt{n}%
T_{ave}vec\left( \mathbf{\hat{\Theta}}-\mathbf{\Theta }_{0}\right) $
converges to a normal distribution and the variance of $\mathbf{\hat{\Theta}}
$ can be estimated as%
\begin{equation}
\widehat{Var}\left( \mathbf{\hat{\Theta}}\right) =\frac{1}{nT_{ave}^{2}}%
\mathbf{Q}_{22,\bar{w}\bar{w}}^{-1}\mathbf{\hat{\Omega}}_{z,22}\mathbf{Q}%
_{22,\bar{w}\bar{w}}^{-1}\text{,}
\end{equation}%
where 
\begin{equation}
\mathbf{\hat{\Omega}}_{z}=n^{-1}\sum_{i=1}^{n}\phi
_{i}^{2}q_{i}^{-2}\sum_{\ell =1}^{q_{i}}\sum_{\ell ^{\prime
}=1}^{q_{i}}\left( \mathbf{\bar{E}}_{i\ell }\mathbf{\otimes I}_{m}\right)
\left( \mathbf{\bar{w}}_{i\ell }-\mathbf{\bar{w}}_{i\circ }\right) \left( 
\mathbf{\bar{w}}_{i\ell ^{\prime }}-\mathbf{\bar{w}}_{i\circ }\right)
^{\prime }\left( \mathbf{\mathbf{\bar{E}}}_{i\ell ^{\prime }}^{\prime }%
\mathbf{\otimes I}_{m}\right) .
\end{equation}%
and, as before, $\mathbf{Q}_{22,\bar{w}\bar{w}}$ and $\mathbf{\hat{\Omega}}%
_{z,22}$ are the $\left( m-r_{0}\right) \times \left( m-r_{0}\right) $ lower
blocks of matrices $\mathbf{Q}_{\bar{w}\bar{w}}$ and $\mathbf{\hat{\Omega}}%
_{z}$, respectively.

\subsection{Implementation of unbalanced PME estimator in empirical
application\label{iu}}

Let $T_{m}=\max T_{i}$. In empirical application in Section \ref{EA} we
exclude cross section units with gaps in data and we exclude units with $%
T_{i}<20$ years. The remaining units are used for estimation, and we denote
the number of these units as $n$.

Definition of sub-sample sample time averages in (\ref{lau}) is quite
general, and allows for a variety of asymptotically justified options to
implement the PME estimator for unbalanced panels. In view of inference in
small samples being adversely affected by a larger choices of $q$, we set $%
q_{i}=2$ for all $i$, and we divide the $T_{i}$ consecutive observations
into two halves when $T_{i}$ is even. If $T_{i}$ is odd, we compute the
first sub-sample time average based on $\left( T_{i}+1\right) /2$
observations, and the second sub-sample time average based on the remaining $%
\left( T_{i}-1\right) /2$ observations. This version of unbalanced PME
estimator is considered in the empirical application.

\section{Consistency and asymptotic distribution of PME estimator in the
model with interactive time effects\label{TSIE}}

Theorems \ref{Tconsf}-\ref{Tbf} extends Theorems \ref{Tcons}-\ref{Tb} to
model with interactive time effects.

\begin{theorem}
\label{Tconsf}Consider the panel data model for the $m\times 1$ vector $%
\mathbf{w}_{it}$ given by (\ref{gmf}) and suppose that Assumptions \ref{ASS1}
to \ref{ASSfactors} hold, and the number of long-run relations, $r_{0}$, is
known. Let $\mathbf{\hat{B}}_{0}$ be formed from the first $r_{0}$
orthonormalized eigenvectors of $\mathbf{Q}_{\bar{w}\bar{w}}$ given by (\ref%
{Qbar}), associated with its $r_{0}$ smallest eigenvalues. Then for a fixed $%
m,$ $m_{f}$ and $q\left( \geq 2\right) $, $\mathbf{\hat{B}}_{0}\mathbf{%
H\rightarrow }_{p}\mathbf{\mathring{B}}_{0}$ as $n,T\rightarrow \infty $
jointly such that $T_{n}\approx n^{d}$ and $d>0$, for some $r_{0}\times
r_{0} $ non-singular matrix $\mathbf{H}$.
\end{theorem}

\begin{proof}[Proof of Theorem \protect\ref{Tconsf}]
Using (\ref{CQss}) in (\ref{Qwwfo}), we have 
\begin{equation*}
\mathbf{Q}_{\bar{w}\bar{w}}=\mathbf{\Psi }_{n}+O_{p}\left( n^{-1/2}\right)
+O_{p}\left( T^{-1}\right) \text{.}
\end{equation*}%
Multiplying this expression by $\mathbf{\hat{B}}_{0}^{\prime }$ from the
left and by $\mathbf{\hat{B}}_{0}$ from the right, and noting eigenvectors $%
\mathbf{\hat{B}}_{0}$ are normalized so that $\mathbf{\hat{B}}_{0}^{\prime }%
\mathbf{\hat{B}}_{0}=\mathbf{I}_{r}$ yields 
\begin{equation}
\mathbf{\hat{B}}_{0}^{\prime }\mathbf{Q}_{\bar{w}\bar{w}}\mathbf{\hat{B}}%
_{0}=\frac{(q-1)}{6q}\mathbf{\hat{B}}_{0}^{\prime }\mathbf{\Psi }_{n}\mathbf{%
\hat{B}}_{0}+O_{p}\left( n^{-1/2}\right) +O_{p}\left( T^{-1}\right) \text{,}
\label{caf}
\end{equation}%
where $lim_{n\rightarrow \infty }\mathbf{\Psi }_{n}=\mathbf{\Psi }$ as $%
n\rightarrow \infty $. (\ref{caf}) is the same as (\ref{ca}) with the
exception of the $O_{p}\left( T^{-1}\right) $ term (as opposed to $%
O_{p}\left( T^{-2}\right) $). Consistency of $\mathbf{\hat{B}}_{0}$ now
follows using the same arguments as in the proof of Theorem \ref{Tcons}.
\end{proof}

\begin{theorem}
\label{Tbf}Consider the panel data model for the $m\times 1$ vector $\mathbf{%
w}_{it}$ given by (\ref{gmf}), and suppose that Assumptions \ref{ASS1} to %
\ref{ASSfactors} hold, $m$, $m_{f}$ and $q(\geq 2)$ are fixed, and the
number of long-run relations, $r_{0}$ ($m>r_{0}>0$) is known. Suppose
further that the long-run relations, $\mathbf{\mathring{B}}_{0}$, of
interest are subject to the exact identifying restrictions, $\mathbf{R}%
\mathbf{\mathring{B}}_{0}\mathbf{=A}$\textbf{, }given by (\ref{ExactRes}),
and consider the PME estimator of $\mathbf{\mathring{B}}_{0}$\textbf{\ }%
given by $\widehat{\mathbf{\mathbf{\mathring{B}}}}_{0}\mathbf{=\hat{B}}%
_{0}\left( \mathbf{R\hat{B}}_{0}\right) ^{-1}\mathbf{A}$\textbf{, }where $%
\mathbf{\hat{B}}_{0}=\left( \mathbf{\hat{\beta}}_{10},\mathbf{\hat{\beta}}%
_{20},...,\mathbf{\hat{\beta}}_{r_{0},0}\right) $ are the first $r_{0}$
orthonormalized eigenvectors of $\mathbf{Q}_{\bar{w}\bar{w}}$ defined by (%
\ref{Qbar}). Then%
\begin{equation}
\sqrt{n}T\left( \mathbf{I}_{r}\mathbf{\otimes Q}_{\bar{w}\bar{w}}\right) 
\text{$vec$}\left( \widehat{\mathbf{\mathring{B}}}_{0}-\mathbf{\mathring{B}}%
_{0}\right) \rightarrow _{d}N\left( \mathbf{0,\Omega }_{q}^{\ast }\right) ,
\label{distf}
\end{equation}%
as $n,T\rightarrow \infty $, jointly such that $T\approx n^{d}$ for $d>1/2$,
where 
\begin{equation}
\mathbf{\Omega }_{q}^{\ast }=lim_{n,T\rightarrow \infty }\left[
n^{-1}\sum_{i=1}^{n}\left( \mathbf{\mathring{B}}_{0}^{\prime }\mathbf{%
\otimes \mathbf{C}}_{i}\right) \mathbf{\Omega }_{\xi _{qi}^{\ast }}\left( 
\mathbf{\mathring{B}}_{0}\mathbf{\otimes \mathbf{C}}_{i}^{\prime }\right) %
\right] ,
\end{equation}%
and $\mathbf{\Omega }_{\xi _{qi}^{\ast }}=Var\left( \mathbf{\xi }_{iq}^{\ast
}\right) $, and $\mathbf{Q}_{\bar{w}\bar{w}}\rightarrow _{p}\frac{(q-1)}{6q}%
\mathbf{\Psi }$.
\end{theorem}

\begin{proof}[Proof of Theorem \protect\ref{Tbf}]
Averaging (\ref{Qwiwi_f}) over $i$, we have%
\begin{eqnarray}
\mathbf{Q}_{\bar{w}\bar{w}} &=&n^{-1}\sum_{i=1}^{n}\mathbf{C}_{i}\mathbf{Q}_{%
\bar{s}_{i}\bar{s}_{i}}\mathbf{C}_{i}^{\prime }+n^{-1}\sum_{i=1}^{n}\mathbf{C%
}_{i}\mathbf{Q}_{\bar{s}_{i}\bar{v}_{i}}+n^{-1}\sum_{i=1}^{n}\mathbf{C}_{i}%
\mathbf{Q}_{\bar{s}_{i}\bar{f}_{i}}+  \notag \\
&&+n^{-1}\sum_{i=1}^{n}\mathbf{Q}_{\bar{v}_{i}\bar{s}_{i}}^{\prime }\mathbf{C%
}_{i}^{\prime }+n^{-1}\sum_{i=1}^{n}\mathbf{Q}_{\bar{f}_{i}\bar{s}%
_{i}}^{\prime }\mathbf{C}_{i}+\mathbf{Q}_{\bar{f}\bar{f}}+\mathbf{Q}_{\bar{f}%
\bar{v}}+\mathbf{Q}_{\bar{v}\bar{f}}+\mathbf{Q}_{\bar{v}\bar{v}}\text{.}
\label{qwbdf}
\end{eqnarray}%
Multiplying both sides of (\ref{qwbdf}) by $\mathbf{\mathring{B}}_{0}$ from
the right, and noting that $\mathbf{C}_{i}^{\prime }\mathbf{\mathring{B}}%
_{0}=\mathbf{0}$ under Assumption \ref{ASS3}, we have%
\begin{equation}
\mathbf{Q}_{\bar{w}\bar{w}}\mathbf{\mathring{B}}_{0}=\left(
n^{-1}\sum_{i=1}^{n}\mathbf{C}_{i}\mathbf{Q}_{\bar{s}_{i}\bar{v}_{i}}\right) 
\mathbf{\mathring{B}}_{0}+\left( n^{-1}\sum_{i=1}^{n}\mathbf{C}_{i}\mathbf{Q}%
_{\bar{s}_{i}\bar{f}_{i}}\right) \mathbf{\mathring{B}}_{0}+\mathbf{Q}_{\bar{f%
}\bar{f}}\mathbf{\mathring{B}}_{0}+\mathbf{Q}_{\bar{f}\bar{v}}\mathbf{%
\mathring{B}}_{0}+\mathbf{Q}_{\bar{v}\bar{f}}\mathbf{\mathring{B}}_{0}+%
\mathbf{Q}_{\bar{v}\bar{v}}\mathbf{\mathring{B}}_{0}\text{,}  \label{tcpaf}
\end{equation}%
where by results (\ref{bQff})-(\ref{bQfv}) of Lemma \ref{LQB}, $\mathbf{Q}_{%
\bar{f}\bar{f}}=O_{p}\left( T^{-2}\right) $, $\mathbf{Q}_{\bar{f}\bar{v}}=%
\mathbf{Q}_{\bar{v}\bar{f}}^{\prime }=O_{p}\left( T^{-2}\right) $, and by
result (\ref{sqvvs}) of Lemma (\ref{L_sqvvs}) $\mathbf{Q}_{\bar{v}\bar{v}%
}=O_{p}\left( n^{-1/2}T^{-2}\right) $. Using result (\ref{qbt1}) of Lemma %
\ref{LQb_ie}, we have $\left\Vert \mathbf{Q}_{\bar{w}\bar{w}}\mathbf{\hat{B}}%
_{0}\right\Vert =O_{p}\left( T^{-2}\right) $, and given that $\widehat{%
\mathbf{\mathring{B}}}_{0}$ is an $O_{p}\left( 1\right) $ rotation of $%
\mathbf{\hat{B}}_{0}$, it follows%
\begin{equation}
\mathbf{Q}_{\bar{w}\bar{w}}\widehat{\mathbf{\mathring{B}}}_{0}=O_{p}\left(
T^{-2}\right) \text{.}  \label{tcpbf}
\end{equation}%
Subtracting (\ref{tcpaf}) from (\ref{tcpbf}) yields (noting $O_{p}\left(
T^{-2}\right) $ dominates $O_{p}\left( n^{-1/2}T^{-2}\right) $) 
\begin{equation*}
\mathbf{Q}_{\bar{w}\bar{w}}\left( \widehat{\mathbf{\mathring{B}}}-\mathbf{%
\mathring{B}}_{0}\right) =-\left( n^{-1}\sum_{i=1}^{n}\mathbf{C}_{i}\mathbf{Q%
}_{\bar{s}_{i}\bar{v}_{i}}\right) \mathbf{\mathring{B}}_{0}+\left(
n^{-1}\sum_{i=1}^{n}\mathbf{C}_{i}\mathbf{Q}_{\bar{s}_{i}\bar{f}_{i}}\right) 
\mathbf{\mathring{B}}_{0}+O_{p}\left( T^{-2}\right) \mathbf{.}
\end{equation*}%
Defining $\mathbf{Q}_{\bar{s}_{i}\bar{\omega}_{i}}\mathbf{=Q}_{\bar{s}_{i}%
\bar{v}_{i}}+\mathbf{Q}_{\bar{s}_{i}\bar{f}_{i}}$ and multiplying above
equation by $\sqrt{n}T$ we obtain,%
\begin{equation}
\mathbf{Q}_{\bar{w}\bar{w}}\sqrt{n}T\left( \widehat{\mathbf{\mathring{B}}_{0}%
}-\mathbf{\mathring{B}}_{0}\right) =-\left( n^{-1/2}\sum_{i=1}^{n}T\mathbf{C}%
_{i}\mathbf{Q}_{\bar{s}_{i}\bar{\omega}_{i}}\right) \mathbf{\mathring{B}}%
_{0}+O_{p}\left( \frac{\sqrt{n}}{T}\right) \mathbf{.}  \label{t2af}
\end{equation}%
Noting $\mathbf{s}_{it}$ is independently distributed of $\mathbf{f}%
_{it^{\prime }}$ for all $t,t^{\prime }$, the first term on the right side
of (\ref{t2af}) can be written as%
\begin{equation*}
\left( n^{-1/2}\sum_{i=1}^{n}T\mathbf{C}_{i}\mathbf{Q}_{\bar{s}_{i}\bar{%
\omega}_{i}}\right) \mathbf{\mathring{B}}_{0}=\left( n^{-1/2}\sum_{i=1}^{n}%
\mathbf{Z}_{i}^{\ast }\right) \mathbf{\mathring{B}}_{0}+\frac{\sqrt{n}}{T}%
\left[ n^{-1}\sum_{i=1}^{n}\mathbf{C}_{i}E\left( T^{2}\mathbf{Q}_{\bar{s}_{i}%
\bar{v}_{i}}\right) \right] \mathbf{\mathring{B}}_{0}\text{,}
\end{equation*}%
where $\mathbf{Z}_{i}^{\ast }=\mathbf{C}_{i}\left[ T\mathbf{Q}_{\bar{s}_{i}%
\bar{\omega}_{i}}-E\left( T\mathbf{Q}_{\bar{s}_{i}\bar{\omega}_{i}}\right) %
\right] \mathbf{\mathring{B}}_{0}$, and $E\left( T^{2}\mathbf{Q}_{\bar{s}_{i}%
\bar{\omega}_{i}}\right) =E\left( T^{2}\mathbf{Q}_{\bar{s}_{i}\bar{v}%
_{i}}\right) $. Using (\ref{t2c}), it follows that 
\begin{equation*}
\mathbf{Q}_{\bar{w}\bar{w}}\sqrt{n}T\left( \widehat{\mathbf{\mathring{B}}}%
_{0}-\mathbf{\mathring{B}}_{0}\right) =-n^{-1/2}\sum_{i=1}^{n}\mathbf{Z}%
_{i}^{\ast }+O_{p}\left( \frac{\sqrt{n}}{T}\right) ,
\end{equation*}%
Vectorizing the above equation we have%
\begin{equation}
\left( \mathbf{I}_{r}\mathbf{\otimes Q}_{\bar{w}\bar{w}}\right) \sqrt{n}T%
\text{ $vec$}\left( \widehat{\mathbf{\mathring{B}}}_{0}-\mathbf{\mathring{B}}%
_{0}\right) =n^{-1/2}\sum_{i=1}^{n}\left( \mathbf{\mathring{B}}_{0}^{\prime }%
\mathbf{\otimes \mathbf{C}}_{i}\right) \mathbf{\tilde{\xi}}_{iq}^{\ast
}+O_{p}\left( \frac{\sqrt{n}}{T}\right) \text{,}  \label{t2df}
\end{equation}%
where $\mathbf{\tilde{\xi}}_{iq}^{\ast }=\mathbf{\bar{\xi}}_{iq}^{\ast
}-E\left( \mathbf{\bar{\xi}}_{iq}^{\ast }\right) $, and $\mathbf{\bar{\xi}}%
_{iq}^{\ast }$ is given by (recall $\mathbf{Z}_{i}^{\ast }=\mathbf{C}_{i}%
\left[ T\mathbf{Q}_{\bar{s}_{i}\bar{\omega}_{i}}-E\left( T\mathbf{Q}_{\bar{s}%
_{i}\bar{\omega}_{i}}\right) \right] \mathbf{\mathring{B}}_{0}$ and $\mathbf{%
Q}_{\bar{s}_{i}\bar{\omega}_{i}}=T^{-1}q^{-1}\sum_{\ell =1}^{q}\left( 
\mathbf{\bar{s}}_{i\ell }-\mathbf{\bar{s}}_{i\circ }\right) \left( \mathbf{%
\bar{\omega}}_{i\ell }-\mathbf{\bar{\omega}}_{i\circ }\right) ^{\prime }$)%
\begin{equation*}
\mathbf{\bar{\xi}}_{iq}^{\ast }=q^{-1}\sum_{\ell =1}^{q}vec\left[ \left( 
\mathbf{\bar{s}}_{i\ell }-\mathbf{\bar{s}}_{i\circ }\right) \left( \mathbf{%
\bar{\omega}}_{i\ell }-\mathbf{\bar{\omega}}_{i\circ }\right) ^{\prime }%
\right] =q^{-1}\sum_{\ell =1}^{q}\left( \mathbf{\bar{\omega}}_{i\ell }%
\mathbf{\otimes \bar{s}}_{i\ell }\right) -\mathbf{\bar{\omega}}_{i\circ }%
\mathbf{\otimes \bar{s}}_{i\circ }\text{.}
\end{equation*}%
Lemma \ref{Ldf} established convergence in distribution for $%
n^{-1/2}\sum_{i=1}^{n}\left( \mathbf{\mathring{B}}_{0}^{\prime }\mathbf{%
\otimes \mathbf{C}}_{i}\right) \mathbf{\tilde{\xi}}_{iq}^{\ast }$. Using
this Lemma in (\ref{t2df}), and noting that $d>1/2$ implies$\sqrt{n}%
/T\rightarrow 0$ as $n,T\rightarrow \infty $, we obtain (\ref{distf}), as
required.
\end{proof}

\section{Description of Data Generating Processes\label{Sup_MCd}}

We have experiments with $r_{0}=0,1,2$ long-run relations, and with and
without interactive time effects. Overview of all experiments is provided in
the paper. Subsection \ref{SL} below provides full details of the DGPs with
long-run relations and without interactive time effects. Subsection \ref{SNL}
provides details of the DGPs without long-run relations and without
interactive time effects. Subsection \ref{SIFE} provides details on
augmentation of each of these DGPs with interactive time effects.

\subsection{Experiments with $r_{0}=1$ and $2$ long-run relations and $m=3$
variables\label{SL}}

We consider the following data generating process for $\mathbf{w}_{it}$,%
\begin{equation}
\Delta \mathbf{w}_{it}=\mathbf{d}_{i}-\mathbf{\Pi }_{i}\mathbf{w}_{i,t-1}+%
\mathbf{u}_{it}-\mathbf{\Theta }_{i}\mathbf{u}_{i,t-1},  \label{VECMA}
\end{equation}%
where%
\begin{equation*}
\mathbf{\Pi }_{i}=\mathbf{A}_{i}\mathbf{B}_{0}^{\prime },
\end{equation*}%
$\Delta \mathbf{w}_{it}$ is $m\times 1,$ $\mathbf{A}_{i}$ is $m\times r_{0},$
$\mathbf{B}_{0}$ is $m\times r_{0}.$ To ensure that $\mathbf{w}_{it}$ is not
trended we impose the restriction%
\begin{equation}
\mathbf{d}_{i}=\mathbf{\Pi }_{i}\mathbf{\mu }_{iw}=\mathbf{A}_{i}\mathbf{B}%
_{0}^{\prime }\mathbf{\mu }_{iw}.  \label{ai}
\end{equation}%
We set $m=3$ and consider two cases, one and two long-run relations, $%
r_{0}=1 $ and $r_{0}=2.$

For $r_{0}=1$, $\mathbf{B}_{0}\mathbf{=\beta }_{0}=(1,0,-1)^{\prime }$, and
for $r_{0}=2$%
\begin{equation*}
\mathbf{B}_{0}=\left( 
\begin{array}{cc}
\mathbf{\beta }_{10} & \mathbf{\beta }_{20}%
\end{array}%
\right) \mathbf{=}\left( 
\begin{array}{cc}
1 & 0 \\ 
0 & 1 \\ 
-1 & -1%
\end{array}%
\right) .
\end{equation*}

The VARMA representation of (\ref{VECMA}) is given by%
\begin{equation}
\mathbf{w}_{it}=\mathbf{d}_{i}+\left( \mathbf{I}_{m}-\mathbf{A}_{i}\mathbf{B}%
_{0}^{\prime }\right) \mathbf{w}_{i,t-1}+\mathbf{u}_{it}-\mathbf{\Theta }_{i}%
\mathbf{u}_{i,t-1}  \label{VARMAa}
\end{equation}%
which can be written more compactly as%
\begin{equation}
\mathbf{\Psi }_{i}(L)\mathbf{w}_{it}=\mathbf{d}_{i}+(\mathbf{I}_{m}-\mathbf{%
\Theta }_{i}L)\mathbf{u}_{it},  \label{VARMAb}
\end{equation}%
where $\mathbf{\Psi }_{i}(L)=\mathbf{I}_{m}-\mathbf{\Psi }_{i}L,$ and $%
\mathbf{\Psi }_{i}=\mathbf{I}_{m}-\mathbf{A}_{i}\mathbf{B}_{0}^{\prime }$.

\subsubsection{Relation to Granger Representation}

$\Delta \mathbf{w}_{it}$ can be written as%
\begin{equation}
\Delta \mathbf{w}_{it}=\left[ \mathbf{C}_{i}+(1-L)\mathbf{C}_{i}^{\ast }(L)%
\right] \mathbf{u}_{it}  \label{GrepS}
\end{equation}%
where $\mathbf{C}_{i}^{\ast }\left( L\right) =\sum_{\ell =0}^{\infty }%
\mathbf{C}_{i\ell }^{\ast }L^{\ell }$. First differencing (\ref{VARMAb}),%
\begin{equation*}
\mathbf{\Psi }_{i}(L)\Delta \mathbf{w}_{it}=(\mathbf{I}_{m}-\mathbf{\Theta }%
_{i}L)\Delta \mathbf{u}_{it},
\end{equation*}
and using (\ref{GrepS}), we obtain%
\begin{equation}
\mathbf{\Psi }_{i}(L)\left[ \mathbf{C}_{i}+(1-L)\mathbf{C}_{i}^{\ast }(L)%
\right] =(\mathbf{I}_{m}-\mathbf{\Theta }_{i}L)(1-L).  \label{Cil}
\end{equation}%
Using the above we obtain\ $\mathbf{C}_{i}$ and $\mathbf{C}_{i}^{\ast },$ $%
\ell =1,2,...$ in terms of $\mathbf{\Psi }_{i}$ and $\mathbf{\Theta }_{i}$.
Hence we can relate the parameters of (\ref{VECMA}) to the general linear
representation%
\begin{equation}
\mathbf{w}_{it}=\mathbf{a}_{i}+\mathbf{C}_{i}\mathbf{s}_{it}+\mathbf{C}%
_{i}^{\ast }(L)\mathbf{u}_{it}.  \label{GLrep}
\end{equation}%
where $\mathbf{s}_{it}=\mathbf{u}_{i1}+\mathbf{u}_{i2}+....+\mathbf{u}_{it}$%
. Pre-multiplying by $\mathbf{B}_{0}^{\prime }$, we obtain%
\begin{equation*}
\mathbf{B}_{0}^{\prime }\mathbf{w}_{it}=\mathbf{B}_{0}^{\prime }\mathbf{a}%
_{i}+\mathbf{B_{0}^{\prime }C}_{i}^{\ast }(L)\mathbf{u}_{it}.
\end{equation*}

\subsubsection{Parameterization}

We would expect $\mathbf{B}_{0}$ to be more difficult to estimate the more
persistent is $\mathbf{B}_{0}^{\prime }\mathbf{w}_{it}.$ In the extreme case
where $\mathbf{A}_{i}=\mathbf{0},$ $\mathbf{B}_{0}$ is not identified.
Pre-multiplying both sides of (\ref{VECMA}) by $\mathbf{B}_{0}^{\prime }$
yields%
\begin{equation}
\mathbf{B}_{0}^{\prime }\Delta \mathbf{w}_{it}=\mathbf{B}_{0}^{\prime }%
\mathbf{d}_{i}-\mathbf{B}_{0}^{\prime }\mathbf{A}_{i}\mathbf{B}_{0}^{\prime }%
\mathbf{w}_{i,t-1}+\mathbf{B}_{0}^{\prime }\mathbf{u}_{it}-\mathbf{B}%
_{0}^{\prime }\mathbf{\Theta }_{i}\mathbf{u}_{i,t-1}.  \label{ECMS}
\end{equation}%
Let $\mathbf{\xi }_{it}=\mathbf{B}_{0}^{\prime }\mathbf{w}_{it}$ then 
\begin{equation}
\mathbf{\xi }_{it}=\mathbf{B}_{0}^{\prime }\mathbf{d}_{i}+(\mathbf{I}_{r}-%
\mathbf{B}_{0}^{\prime }\mathbf{A}_{i})\mathbf{\xi }_{i,t-1}+\mathbf{B}%
_{0}^{\prime }\mathbf{u}_{it}-\mathbf{B}_{0}^{\prime }\mathbf{\Theta }_{i}%
\mathbf{u}_{i,t-1}.  \label{EC}
\end{equation}%
In the VAR(1) case where $\mathbf{\Theta }_{i}=\mathbf{0},$ the persistence
of $\mathbf{\xi }_{it}=\mathbf{B}_{0}^{\prime }\mathbf{w}_{it}$ is fully
determined by the largest eigenvalue of $(\mathbf{I}_{r}-\mathbf{B}%
_{0}^{\prime }\mathbf{A}_{i})$. Allowing for the MA component complicates
the analysis of persistence of $\mathbf{\xi }_{it}$, but it will still
depend on the size of the eigenvalues of $(\mathbf{I}_{r}-\mathbf{B}%
_{0}^{\prime }\mathbf{A}_{i})$. As an approximation it is reasonable to
control the eigenvalues of $(\mathbf{I}_{r}-\mathbf{B}_{0}^{\prime }\mathbf{A%
}_{i})$.

\subsubsection{Long run relations}

When $r_{0}=1,$ we set $\mathbf{B}_{0}^{\prime }=\mathbf{\beta }_{0}^{\prime
}=(1,0,-1)$ and $\mathbf{A}_{i}=\left( a_{i,11},a_{i,21},a_{i,31}\right)
^{\prime }$, and hence%
\begin{equation}
\mathbf{B}_{0}^{\prime }\mathbf{A}_{i}=a_{i,11}-a_{i,31}=\rho _{i}.
\label{rhoi}
\end{equation}%
In the case of VAR(1) the persistence of $\mathbf{\xi }_{it}$ does not
depend on $a_{i,21}$, which we set to $0$, for all $i$. We consider two sets
of values for $\rho _{i}$ and generate them as (for $i=1,2,...,n$)%
\begin{eqnarray*}
\text{slow }\rho _{i} &\thicksim &IIDU\left[ 0.1,0.2\right] \text{, and} \\
\text{moderate }\rho _{i} &\thicksim &IIDU\left[ 0.1,0.3\right] \text{.}
\end{eqnarray*}%
Slow speed of convergence corresponds to a median half-life of 4.3 periods
(years) and moderate speed of convergence corresponds to a median half-life
of 3.1 periods (years). Since both $a_{i,11}$ and $a_{i,31}$ are nonzero,
long-run causality runs from $w_{it,1}$ to $\left( w_{it,2},w_{it,3}\right) $
as well as from $w_{it,3}$ to $w_{it,1}$. (\ref{rhoi}) leaves us with one
free parameter to determine $\mathbf{A}_{i}$ which we choose to control a
system measure of fit, as described below.

When $r_{0}=2,$ we set the long-run relations as 
\begin{equation*}
\mathbf{B}_{0}=\left( \mathbf{\beta }_{10},\mathbf{\beta }_{20}\right)
=\left( 
\begin{array}{cc}
1 & 0 \\ 
0 & 1 \\ 
-1 & -1%
\end{array}%
\right) \text{.}
\end{equation*}%
Let%
\begin{equation*}
\mathbf{A}_{i}\mathbf{=}\left( 
\begin{array}{cc}
a_{i,11} & a_{i,12} \\ 
a_{i,21} & a_{i,22} \\ 
a_{i,31} & a_{i,32}%
\end{array}%
\right) \text{,}
\end{equation*}%
then%
\begin{eqnarray}
\mathbf{I}_{2}-\mathbf{B}_{0}^{\prime }\mathbf{A}_{i} &=&\mathbf{I}%
_{2}-\left( 
\begin{array}{ccc}
1 & 0 & -1 \\ 
0 & 1 & -1%
\end{array}%
\right) \left( 
\begin{array}{cc}
a_{i,11} & a_{i,12} \\ 
a_{i,21} & a_{i,22} \\ 
a_{i,31} & a_{i,32}%
\end{array}%
\right) \text{,}  \notag \\
&=&\left( 
\begin{array}{cc}
1-\left( a_{i,11}-a_{i,31}\right) & -\left( a_{i,12}-a_{i,32}\right) \\ 
-\left( a_{i,21}-a_{i,31}\right) & 1-\left( a_{i,22}-a_{i,32}\right)%
\end{array}%
\right) \text{.}  \notag
\end{eqnarray}%
Hence%
\begin{equation}
\mathbf{I}_{2}-\mathbf{B}_{0}^{\prime }\mathbf{A}_{i}=\left( 
\begin{array}{cc}
1-\rho _{i,11} & -\rho _{i,12} \\ 
-\rho _{i,21} & 1-\rho _{i,22}%
\end{array}%
\right) \text{.}  \label{Aix}
\end{equation}%
The eigenvalues of this matrix are given by 
\begin{eqnarray*}
\lambda _{1} &=&1-\frac{1}{2}\rho _{i,11}-\frac{1}{2}\rho _{i,22}-\frac{1}{2}%
\sqrt{\rho _{i,11}^{2}-2\rho _{i,11}\rho _{i,22}+\rho _{i,22}^{2}+4\rho
_{i,12}\rho _{i,21}}, \\
\lambda _{2} &=&1-\frac{1}{2}\rho _{i,11}-\frac{1}{2}\rho _{i,22}+\frac{1}{2}%
\sqrt{\rho _{i,11}^{2}-2\rho _{i,11}\rho _{i,22}+\rho _{i,22}^{2}+4\rho
_{i,12}\rho _{i,21}}.
\end{eqnarray*}%
To simplify the design we set $\rho _{i,12}=\left( a_{i,12}-a_{i,32}\right)
=0=\rho _{i,21}=\left( a_{i,21}-a_{i,31}\right) =0$. then%
\begin{equation*}
\rho _{i,11}=\left( a_{i,11}-a_{i,31}\right) \text{ and }\rho _{i,22}=\left(
a_{i,22}-a_{i,32}\right) ,
\end{equation*}%
and the eigenvalues are $\lambda _{1}=1-\rho _{i,11}$, and $\lambda
_{2}=1-\rho _{i,22}$. A stable solution arises so long as $%
sup_{i,j}\left\vert 1-\rho _{i,jj}\right\vert <1$. We consider two sets of
values for $\rho _{i,11}$ and $\rho _{i,22}$ and generate them as (for $%
i=1,2,...,n$)%
\begin{equation*}
\text{Slow }\rho _{i,11}\thicksim IIDU\left[ 0.1,0.2\right] \text{ ;
Moderate }\rho _{i,11}\thicksim IIDU\left[ 0.1,0.3\right] ,
\end{equation*}%
\begin{equation*}
\text{Slow }\rho _{i,22}\thicksim IIDU\left[ 0.1,0.2\right] \text{ ;
Moderate }\rho _{i,22}\thicksim IIDU\left[ 0.1,0.3\right] .
\end{equation*}%
We set $a_{i,j\ell }$ values below, to achieve a balance between the speed
of convergence to equilibrium, $\rho _{ij}$, and the fit of the error
correction equations.

\subsubsection{Deterministic terms, errors and initial values}

We generate $\mathbf{\mu }_{iw}=(\mu _{ij,w})$ as $\mu _{ij,w}\thicksim
IIDN(0,1)$, for $j=1,2,3;$ and $i=1,2,...,n$. These parameters do not enter
the distribution of the estimators of $\mathbf{B}_{0}$.

\textbf{Errors: }$\mathbf{u}_{it}$ are generated as%
\begin{equation}
\mathbf{u}_{it}=\mathbf{P}_{ui}\mathbf{\varepsilon }_{it}\text{,}
\end{equation}%
for $i=1,2,...,n$, and $t=1,2,...,T$, where two cases for considered for the
distribution of $\mathbf{\varepsilon }_{it}$: Gaussian case with elements of 
$\mathbf{\varepsilon }_{it}$ generated as $IIDN\left( 0,1\right) $, and
non-Gaussian case with individual elements of $\mathbf{\varepsilon }_{it}$
generated as $IID$ from a chi-squared distribution with 4 degrees of freedom
normalized to mean zero and unit variance, namely $\left[ \chi ^{2}\left(
4\right) -4\right] /\sqrt{8}$. Matrix $\mathbf{P}_{ui}\,$is a
lower-triangular matrix computed as Choleski factorization of $\mathbf{%
\Sigma }_{i}=\mathbf{P}_{ui}\mathbf{P}_{ui}^{\prime }$, where individual
elements of $\mathbf{\Sigma }_{i}$, denoted as $\sigma _{i,pq}$, for $%
p,q=1,2,3$ are generated as follows. We set the diagonal elements to one,
namely $\sigma _{i,pp}=1$ for $p=1,2,3$, and $i=1,2,...,n$. We generate the
off-diagonal elements as $\sigma _{i,pq}=\sigma _{i,qp}\sim IIDU\left(
0,0.5\right) $ for $\left( p,q\right) \in \left\{ \left( 1,2\right) ,\left(
1,3\right) ,\left( 2,3\right) \right\} $, and $i=1,2,...,n$. This ensures $%
\mathbf{\Sigma }_{i}$ is symmetric and positive definite with a probability
one.

For the MA part we consider two cases $\mathbf{\Theta }_{i}=0$ (VAR design),
and $\mathbf{\Theta }_{i}=diag(\theta _{ij},j=1,2,3)$, with $\theta _{ij},$ $%
j=1,2,3$ $\sim U(0,0.50)$. \ 

\textbf{Initial values: }We first generate $\Delta \mathbf{w}_{it}$, for $%
t=1,2,...,T$, using the representation, (\ref{VECMA}), and then cumulate
these differences to obtain $\mathbf{w}_{it}$. To this end we require the
initial values $\mathbf{B}^{\prime }\mathbf{w}_{i0}$ and $\Delta \mathbf{w}%
_{i0}$. Using the Granger representation let $\Upsilon _{i}\left( L\right) =%
\left[ \mathbf{C}_{i}+(1-L)\mathbf{C}_{i}^{\ast }(L)\right] =\sum_{\ell
=0}^{\infty }\Upsilon _{i\ell }L^{\ell }$, and 
\begin{equation}
\Delta \mathbf{w}_{i0}=\sum_{\ell =0}^{\infty }\Upsilon _{i,-\ell }\mathbf{u}%
_{i,-\ell }\text{,}  \label{dwi0}
\end{equation}%
and $\mathbf{B}_{0}^{\prime }\mathbf{w}_{i0}=\mathbf{B}_{0}^{\prime }\mathbf{%
a}_{i}+\sum_{\ell =0}^{\infty }\mathbf{B}_{0}^{\prime }\mathbf{C}_{i,-\ell
}^{\ast }\mathbf{u}_{i,-\ell }$. Hence, $E\left( \mathbf{B}_{0}^{\prime }%
\mathbf{w}_{i0}\left\vert \mathbf{a}_{i}\right. \right) =E\left( \mathbf{\xi 
}_{i0}\left\vert \mathbf{a}_{i}\right. \right) =\mathbf{B}_{0}^{\prime }%
\mathbf{a}_{i}$. Also using (\ref{EC}), and noting that $E\left( \Delta 
\mathbf{w}_{i,t-1}\right) =\mathbf{0}$, we have%
\begin{equation*}
E\left( \mathbf{\xi }_{it}\left\vert \mathbf{d}_{i},\mathbf{\alpha }%
_{i}\right. \right) =\mathbf{B}^{\prime }\mathbf{d}_{i}+(\mathbf{I}_{r}-%
\mathbf{B}_{0}^{\prime }\mathbf{A}_{i})E\left( \mathbf{\xi }%
_{i,t-1}\left\vert \mathbf{\alpha }_{i}\right. \right) ,
\end{equation*}%
and assuming $\left\vert \lambda _{\min }(\mathbf{I}_{r}-\mathbf{B}%
_{0}^{\prime }\mathbf{A}_{i})\right\vert <1$ we obtain $E\left( \mathbf{\xi }%
_{i0}\right) =(\mathbf{B}_{0}^{\prime }\mathbf{A}_{i})^{-1}\mathbf{B}%
_{0}^{\prime }\mathbf{d}_{i}$. Therefore, recalling that $\mathbf{d}_{i}=%
\mathbf{\Pi }_{i}\mathbf{\mu }_{iw}$ (see (\ref{ai}))%
\begin{equation*}
\mathbf{B}_{0}^{\prime }\mathbf{w}_{i0}=\mathbf{B}_{0}^{\prime }\mathbf{\mu }%
_{iw}+\sum_{\ell =0}^{\infty }\mathbf{B}_{0}^{\prime }\mathbf{C}_{i,-\ell
}^{\ast }\mathbf{u}_{i,-\ell }\text{.}
\end{equation*}%
Using (\ref{GLrep}) we also have%
\begin{equation}
Var\left( \mathbf{B}_{0}^{\prime }\mathbf{w}_{it}\right) =Var\left( \mathbf{B%
}_{0}^{\prime }\mathbf{w}_{i0}\right) =\mathbf{B_{0}^{\prime }}\left(
\sum_{s=0}^{\infty }\mathbf{C}_{is}^{\ast \prime }\mathbf{V}_{i}\mathbf{C}%
_{is}^{\ast }\right) \mathbf{B}_{0}.  \label{Varegzi}
\end{equation}%
We generate $\Delta \mathbf{w}_{i0}$ and $\mathbf{B}_{0}^{\prime }\mathbf{w}%
_{i0}$ according to 
\begin{equation*}
\Delta \mathbf{w}_{i0}=\sum_{\ell =0}^{M}\Upsilon _{i,-\ell }\mathbf{u}%
_{i,-\ell }\text{, and }\mathbf{B}_{0}^{\prime }\mathbf{w}_{i0}=\mathbf{B}%
_{0}^{\prime }\mathbf{\mu }_{iw}+\sum_{\ell =0}^{M}\mathbf{B}_{0}^{\prime }%
\mathbf{C}_{i,-\ell }^{\ast }\mathbf{u}_{i,-\ell }\text{,}
\end{equation*}%
where we set $M=50$.

\subsubsection{Fit of error correction equations and setting the error
correction coefficients}

An average measure of fit for the $m\times 1$ system of equations for $%
\Delta \mathbf{w}_{it}=(\Delta \mathring{w}_{it,1},\Delta \mathring{w}%
_{it,2},...,\Delta \mathring{w}_{it,m})^{\prime }$ is given by 
\begin{equation}
PR_{nT}^{2}=1-\frac{\sum_{j=1}^{m}\sum_{t=1}^{T}\sum_{i=1}^{n}u_{it,j}^{2}}{%
\sum_{j=1}^{m}\sum_{t=1}^{T}\sum_{i=1}^{n}(\Delta w_{it,j}-\Delta \bar{w}%
_{iT,j})^{2}},  \label{SysPR2}
\end{equation}%
where $\Delta \bar{w}_{iT,j}=T^{-1}\sum_{t=1}^{T}\Delta w_{it,j}$. This
system measure places equal weights on the fit of the $m$ different
error-correcting (EC) equations. We can control for $PR_{nT}^{2}$ by
generating remaining free parameters in $\mathbf{A}_{i}$.

\subsubsection{Case of $r_{0}=1$ long-run relation in VAR(1) design}

Using (\ref{SysPR2}), and for $\mathbf{\Theta }_{i}=\mathbf{0}$, under
stationarity and for large $n$, ($T$ does not need to be large given the
cross-sectional independence of $\mathbf{u}_{it}$) we have%
\begin{eqnarray}
PR_{n}^{2} &=&1-\frac{\sum_{j=1}^{m}\sum_{i=1}^{n}E\left(
u_{it,j}^{2}\right) }{\sum_{j=1}^{m}\sum_{i=1}^{n}E(\Delta w_{it,j}-\Delta 
\bar{w}_{iT,j})^{2}}  \notag \\
&=&\frac{\sum_{i=1}^{n}tr\left( \mathbf{A}_{i}\mathbf{\Omega }_{i}\mathbf{A}%
_{i}^{\prime }\right) }{\sum_{i=1}^{n}tr\left( \mathbf{A}_{i}\mathbf{\Omega }%
_{i}\mathbf{A}_{i}^{\prime }\right) +\sum_{i=1}^{n}tr(\mathbf{V}_{i})},
\label{PRn}
\end{eqnarray}%
where $\mathbf{\Omega }_{i}=Var\left( \mathbf{B}_{0}^{\prime }\mathbf{w}%
_{i,t-1}\right) $. Using (\ref{EC}) (when $\mathbf{\Theta }_{i}=\mathbf{0}$%
)we note that%
\begin{equation*}
\mathbf{\xi }_{it}=\mathbf{B}_{0}^{\prime }\mathbf{w}_{i,t-1}=\mathbf{B}%
_{0}^{\prime }\mathbf{d}_{i}+(\mathbf{I}_{r}-\mathbf{B}_{0}^{\prime }\mathbf{%
A}_{i})\mathbf{\xi }_{i,t-1}+\mathbf{B}_{0}^{\prime }\mathbf{u}_{it},
\end{equation*}%
and $\mathbf{\Omega }_{i}$ is given by 
\begin{equation}
\mathbf{\Omega }_{i}=(\mathbf{I}_{r}-\mathbf{B}_{0}^{\prime }\mathbf{A}_{i})%
\mathbf{\Omega }_{i}(\mathbf{I}_{r}-\mathbf{B}_{0}^{\prime }\mathbf{A}%
_{i})^{\prime }+\mathbf{B}_{0}^{\prime }\mathbf{V}_{i}\mathbf{B}_{0}.
\label{omegai}
\end{equation}%
In the case where $r_{0}=1$, $\mathbf{\Omega }_{i}$ is a scalar, $\mathbf{B}%
_{0}=\mathbf{\beta }_{1,0}$, and we have%
\begin{equation*}
\mathbf{\Omega }_{i}=\frac{\mathbf{\beta }_{1,0}^{\prime }\mathbf{V}_{i}%
\mathbf{\beta }_{1,0}.}{1-(1-\rho _{i})^{2}},
\end{equation*}%
Hence%
\begin{equation*}
PR_{n}^{2}=\frac{\sum_{i=1}^{n}\frac{\left( \mathbf{A}_{i}^{\prime }\mathbf{A%
}_{i}\right) \mathbf{\beta }_{1,0}^{\prime }\mathbf{V}_{i}\mathbf{\beta }%
_{1,0}}{1-(1-\rho _{i})^{2}}}{\sum_{i=1}^{n}\frac{\left( \mathbf{A}%
_{i}^{\prime }\mathbf{A}_{i}\right) \mathbf{\beta }_{1,0}^{\prime }\mathbf{V}%
_{i}\mathbf{\beta }_{1,0}}{1-(1-\rho _{i})^{2}}+\sum_{i=1}^{n}tr(\mathbf{V}%
_{i})}.
\end{equation*}%
For a given value of $\rho _{i}$, it is now possible to use $\mathbf{A}%
_{i}^{\prime }\mathbf{A}_{i}=a_{i,11}^{2}+\alpha _{i,21}^{2}+\alpha
_{i,31}^{2}$ as a scaling factor to achieve a desired value of $PR_{n}^{2}$.
But we need to take account of the fact that $\rho _{i}=a_{i,11}-a_{i,31}$
and both $a_{i,11}$ and $a_{i,31}$ can not be scaled up. Given $\rho _{i}$
we set 
\begin{equation*}
\mathbf{A}_{i}^{\prime }\mathbf{A}_{i}=\left( \rho _{i}+a_{i,31}\right)
^{2}+a_{i,21}^{2}+a_{i,31}^{2}=\varkappa ^{2},
\end{equation*}%
and then derive $a_{i,21}$ and $a_{i,31}$ that satisfy the above equation.
Then $\varkappa ^{2}$ can be set in terms of $PR_{n}^{2}:$ 
\begin{equation}
\varkappa ^{2}=\left( \frac{1-PR_{n}^{2}}{PR_{n}^{2}}\right) \left( \frac{%
\sum_{i=1}^{n}tr(\mathbf{V}_{i})}{\sum_{i=1}^{n}\frac{\mathbf{\beta }%
_{1,0}^{\prime }\mathbf{V}_{i}\mathbf{\beta }_{1,0}}{1-(1-\rho _{i})^{2}}}%
\right) .  \label{kappa}
\end{equation}%
We can allow variations across $a_{i,31}$ and $a_{i,21}$, so long as $\left(
\rho _{i}+a_{i,31}\right) ^{2}+a_{i,21}^{2}+a_{i,31}^{2}=\varkappa ^{2}$. \
To check the feasibility of the above procedure (results in real-valued $%
a_{i,j\ell }$) we set $a_{i,21}=0$, and note that $a_{i,31}$ must now
satisfy the quadratic equation $a_{i,31}^{2}+\rho _{i}a_{i,31}+\frac{1}{2}%
\left( \rho _{i}^{2}-\varkappa ^{2}\right) =0$. For this equation to have
real solutions we must have $\rho _{i}^{2}-2(\rho _{i}^{2}-\varkappa ^{2})>0$%
, or $\varkappa ^{2}>\sup_{i}\rho _{i}^{2}/2$. We consider two values of $%
PR_{n}^{2}=0.20$ and $0.30$.

\subsubsection{Case of $r_{0}=2$ long-run relations in VAR(1) design}

Setting $\mathbf{Q}_{i}=(\mathbf{I}_{r_{0}}-\mathbf{B}_{0}^{\prime }\mathbf{A%
}_{i})$\textbf{, }and using (\ref{omegai}), we have\footnote{%
Note that $vec\left( \mathbf{ABC}\right) =\left( \mathbf{C}^{\prime }\otimes 
\mathbf{A}\right) vec\left( \mathbf{B}\right) $.} 
\begin{equation*}
vec\left( \mathbf{\Omega }_{i}\right) =\left( \mathbf{Q}_{i}\otimes \mathbf{Q%
}_{i}\right) vec\left( \mathbf{\Omega }_{i}\right) +vec\left( \mathbf{B}%
_{0}^{\prime }\mathbf{V}_{i}\mathbf{B}_{0}\right)
\end{equation*}%
Then%
\begin{equation*}
vec\left( \mathbf{\Omega }_{i}\right) =\left[ \mathbf{I}_{r_{0}^{2}}-\left( 
\mathbf{Q}_{i}\otimes \mathbf{Q}_{i}\right) \right] ^{-1}vec\left( \mathbf{B}%
_{0}^{\prime }\mathbf{V}_{i}\mathbf{B}_{0}\right) .
\end{equation*}%
Also since%
\begin{equation*}
tr\left( \mathbf{A}_{i}\mathbf{\Omega }_{i}\mathbf{A}_{i}^{\prime }\right)
=tr\left( \mathbf{\Omega }_{i}\mathbf{A}_{i}^{\prime }\mathbf{A}_{i}\right)
=vec\left( \mathbf{A}_{i}^{\prime }\mathbf{A}_{i}\right) ^{\prime }vec\left( 
\mathbf{\Omega }_{i}\right) ,
\end{equation*}%
then 
\begin{equation*}
tr\left( \mathbf{A}_{i}\mathbf{\Omega }_{i}\mathbf{A}_{i}^{\prime }\right)
=vec\left( \mathbf{A}_{i}^{\prime }\mathbf{A}_{i}\right) ^{\prime }\left[ 
\mathbf{I}_{r_{0}^{2}}-\left( \mathbf{Q}_{i}\otimes \mathbf{Q}_{i}\right) %
\right] ^{-1}vec\left( \mathbf{B}_{0}^{\prime }\mathbf{V}_{i}\mathbf{B}%
_{0}\right) ,
\end{equation*}%
\begin{equation*}
\mathbf{I}_{r_{0}^{2}}-\left( \mathbf{Q}_{i}\otimes \mathbf{Q}_{i}\right)
=diag(1-\rho _{i,11}^{2},1-\rho _{i,11}\rho _{i,22},1-\rho _{i,11}\rho
_{i,22},1-\rho _{i,22}^{2})\text{.}
\end{equation*}%
As before $\mathbf{A}_{i}^{\prime }\mathbf{A}_{i}$ can be scaled up/down
fixing the value of $\mathbf{Q}_{i}$. Recall from (\ref{Aix}) that (since $%
\rho _{i,12}=\left( a_{i,12}-a_{i,32}\right) =0$, and $\rho _{i,21}=\left(
a_{i,21}-a_{i,31}\right) =0)$, 
\begin{equation*}
\mathbf{\Upsilon }_{i}=\left( 
\begin{array}{cc}
\rho _{i,11} & 0 \\ 
0 & \rho _{i,22}%
\end{array}%
\right)
\end{equation*}%
where $\rho _{i,11}=\left( a_{i,11}-a_{i,31}\right) $ and $\rho
_{i,22}=\left( a_{i,32}-a_{i,22}\right) $. Noting that 
\begin{equation*}
PR_{n}^{2}=\frac{\sum_{i=1}^{n}tr\left( \mathbf{A}_{i}\mathbf{\Omega }_{i}%
\mathbf{A}_{i}^{\prime }\right) }{\sum_{i=1}^{n}tr\left( \mathbf{A}_{i}%
\mathbf{\Omega }_{i}\mathbf{A}_{i}^{\prime }\right) +\sum_{i=1}^{n}tr(%
\mathbf{V}_{i})},
\end{equation*}%
then%
\begin{equation*}
\frac{1-PR_{n}^{2}}{PR_{n}^{2}}=\frac{\sum_{i=1}^{n}tr(\mathbf{V}_{i})}{%
\sum_{i=1}^{n}vec\left( \mathbf{A}_{i}^{\prime }\mathbf{A}_{i}\right)
^{\prime }\left[ \mathbf{I}_{r^{2}}-\left( \mathbf{Q}_{i}\otimes \mathbf{Q}%
_{i}\right) \right] ^{-1}vec\left( \mathbf{B}_{0}^{\prime }\mathbf{V}_{i}%
\mathbf{B}_{0}\right) }\text{,}
\end{equation*}%
where%
\begin{eqnarray*}
\mathbf{A}_{i}^{\prime }\mathbf{A}_{i} &=&\left( 
\begin{array}{ccc}
a_{i,11} & a_{i,21} & a_{i,31} \\ 
a_{i,12} & a_{i,22} & a_{i,32}%
\end{array}%
\right) \left( 
\begin{array}{cc}
a_{i,11} & a_{i,12} \\ 
a_{i,21} & a_{i,22} \\ 
a_{i,31} & a_{i,32}%
\end{array}%
\right) \\
&=&\left( 
\begin{array}{cc}
\sum_{s=1}^{3}a_{i,s1}^{2} & \sum_{s=1}^{3}a_{i,s1}a_{i,s2} \\ 
\sum_{s=1}^{3}a_{i,s1}a_{i,s2} & \sum_{s=1}^{3}a_{i,s2}^{2}%
\end{array}%
\right) .
\end{eqnarray*}%
As before $\rho _{i,11}$ and $\rho _{i,22}$ are pre-set: 
\begin{equation*}
\text{Slow }\rho _{i,11}\thicksim IIDU\left[ 0.1,0.2\right] \text{ ;
Moderate }\rho _{i,11}\thicksim IIDU\left[ 0.1,0.3\right] ,
\end{equation*}%
\begin{equation*}
\text{Slow }\rho _{i,22}\thicksim IIDU\left[ 0.1,0.2\right] \text{ ;
Moderate }\rho _{i,22}\thicksim IIDU\left[ 0.1,0.3\right] .
\end{equation*}%
and $\rho _{i,12}=\left( a_{i,12}-a_{i,32}\right) =0=\rho _{i,21}=\left(
a_{i,21}-a_{i,31}\right) =0$. Overall, we have the following restrictions%
\begin{eqnarray*}
a_{i,11} &=&a_{i,31}+\rho _{i,11}\text{, and }a_{i,22}=a_{i,32}+\rho _{i,22}
\\
a_{i,12} &=&a_{i,32}\text{ and }a_{i,21}=a_{i,31}
\end{eqnarray*}%
We are left with two free parameters to control the fit of the model. To
this end note that 
\begin{eqnarray*}
&&\mathbf{A}_{i}^{\prime }\mathbf{A}_{i}= \\
&&\hspace*{-0.5in}\left( 
\begin{array}{cc}
{\small (a}_{i,31}{\small +\rho }_{i,11}{\small )}^{2}{\small +2a}_{i,31}^{2}
& \left( a_{i,31}+\rho _{i,11}\right) {\small a}_{i,32}{\small +a}%
_{i,31}\left( a_{i,32}+\rho _{i,22}\right) {\small +a}_{i,31}{\small a}%
_{i,32} \\ 
\left( a_{i,31}+\rho _{i,11}\right) {\small a}_{i,32}{\small +a}%
_{i,31}\left( a_{i,32}+\rho _{i,22}\right) {\small +a}_{i,31}{\small a}%
_{i,32} & {\small (a}_{i,32}{\small +\rho }_{i,22}{\small )}^{2}{\small +2a}%
_{i,32}^{2}%
\end{array}%
\right) ,
\end{eqnarray*}%
and $\mathbf{A}_{i}^{\prime }\mathbf{A}_{i}$ can scaled up by scaling on $%
a_{i,31}$ and $a_{i,32}.$ We set $a_{i,31}=a_{i,32}=\varkappa $ which yields%
\begin{equation*}
\mathbf{A}_{i}^{\prime }\mathbf{A}_{i}=\left( 
\begin{array}{cc}
{\small \rho }_{i,11}^{2}{\small +2\varkappa \rho }_{i,11}{\small %
+3\varkappa }^{2} & \left( \rho _{i,11}+\rho _{i,22}\right) {\small %
\varkappa +3\varkappa }^{2} \\ 
\left( \rho _{i,11}+\rho _{i,22}\right) {\small \varkappa +3\varkappa }^{2}
& {\small \rho }_{i,22}^{2}{\small +2\varkappa \rho }_{i,22}{\small %
+3\varkappa }^{2}%
\end{array}%
\right) .
\end{equation*}%
Then we solve the following equation for $\varkappa $ that yields a desired
value of $PR_{n}^{2}$ 
\begin{equation*}
\frac{1-PR_{n}^{2}}{PR_{n}^{2}}=\frac{\sum_{i=1}^{n}tr(\mathbf{V}_{i})}{%
\sum_{i=1}^{n}vec\left( \mathbf{A}_{i}^{\prime }\mathbf{A}_{i}\right) \left[ 
\mathbf{I}_{r^{2}}-\left( \mathbf{Q}_{i}\otimes \mathbf{Q}_{i}\right) \right]
^{-1}vec\left( \mathbf{B}_{0}^{\prime }\mathbf{V}_{i}\mathbf{B}_{0}\right) },
\end{equation*}%
where%
\begin{equation*}
\left[ \mathbf{I}_{r^{2}}-\left( \mathbf{Q}_{i}\otimes \mathbf{Q}_{i}\right) %
\right] ^{-1}=diag(\frac{1}{1-\rho _{i,11}^{2}},\frac{1}{1-\rho _{i,11}\rho
_{i,22}},\frac{1}{1-\rho _{i,11}\rho _{i,22}},\frac{1}{1-\rho _{i,22}^{2}})%
\text{.}
\end{equation*}

We consider the same two values of $PR_{n}^{2}=0.20$ and $0.30$.

\subsubsection{Case of VARMA(1,1) design}

When $\mathbf{\Theta }_{i}\neq \mathbf{0}$, it is cumbersome to solve for $%
PR_{n}^{2}$ analytically. We proceed by computing $\varkappa $ using
stochastic simulations to ensure desired value of $PR_{RnT}^{2}\left(
\varkappa \right) $, where%
\begin{equation*}
PR_{RnT}^{2}\left( \varkappa \right) =\frac{1}{R}\sum_{rep=1}^{R}PR_{nT}^{2}%
\left( rep,\varkappa \right) \text{,}
\end{equation*}%
and%
\begin{equation*}
PR_{nT}^{2}\left( rep,\varkappa \right) =1-\frac{\sum_{j=1}^{m}%
\sum_{t=1}^{T}\sum_{i=1}^{n}\left( u_{it,j}^{\left( rep\right) }\right) ^{2}%
}{\sum_{j=1}^{m}\sum_{t=1}^{T}\sum_{i=1}^{n}(\Delta w_{it,j}^{\left(
rep\right) }-\Delta \bar{w}_{iT,j}^{\left( rep\right) })^{2}}\text{,}
\end{equation*}%
in which we use $rep=1,2,...,R$ to denote individual MC replications ($%
R=2000 $). We solve $PR_{RnT}^{2}\left( \varkappa \right) =0.2$ or $0.3$
using grid search method.

\subsection{Experiments with no long-run relations\label{SNL}}

Experiment with $I\left( 1\right) $ variables and $r_{0}=0$ are based on the
DGP given by%
\begin{equation*}
\Delta \mathbf{w}_{it}=\mathbf{\Phi }_{i}\Delta \mathbf{w}_{i,t-1}+\mathbf{u}%
_{it},
\end{equation*}%
for $i=1,2,...,n$, $t=1,2,...,T$, where $\mathbf{u}_{it}=\mathbf{P}_{i}%
\mathbf{\varepsilon }_{it}$, $\mathbf{\varepsilon }_{it}\thicksim IIDN(%
\mathbf{0},\mathbf{I}_{m})$, the error covariance matrix $\mathbf{P}_{i}%
\mathbf{P}_{i}^{\prime }\equiv \mathbf{\Sigma }_{ui}=\left[ \sigma _{i,\ell
q}\right] $ , is generated using $\sigma _{i,\ell \ell }=1$ for $i=1,2,...,n$
and $\ell =1,2,3$, and $\sigma _{i,\ell q}\thicksim IIDU(0,0.5)$, for $\ell
\neq q\,$, and $i=1,2,...,n$. Matrix $\mathbf{\Phi }_{i}$ is diagonal with $%
\phi _{ij}$ elements on the diagonal, for $j=1,2,...,m$. We consider three
options for $\phi _{ij}$: ($i$) low values $\phi _{ij}\sim U[0,0.8]$, ($ii$)
moderate values $\phi _{ij}\sim U[0.7,0.9]$, and ($iii$) high values $\phi
_{ij}\sim U[0.80,0.95]$. Initial values are generated as $\Delta
w_{i0,j}\sim IIDN\left[ 0,\left( 1-\phi _{ij}^{2}\right) ^{-1}\right] $ for $%
j=1,2,...,m$, and $i=1,2,...,n$, and $\mathbf{w}_{i,-1}$ is set to $\mathbf{0%
}$. This DGP is also a special case of (\ref{Grep}). Specifically, it leads
to $\mathbf{w}_{it}=\mathbf{w}_{i0}+\mathbf{G}_{i}\mathbf{f}_{t}+\mathbf{C}%
_{i}\mathbf{s}_{it}+\mathbf{C}_{i}^{\ast }(L)\mathbf{u}_{it},$ where $%
\mathbf{C}_{i}=\left( \mathbf{I}_{m}-\mathbf{\Phi }_{i}\right) ^{-1}$, and $%
\mathbf{C}_{i}^{\ast }(L)=-\mathbf{\Phi }_{i}\left( \mathbf{I}_{m}-\mathbf{%
\Phi }_{i}\right) ^{-1}\left( \mathbf{I}_{m}-\mathbf{\Phi }_{i}L\right)
^{-1} $.

\subsection{Experiments with interactive time effects\label{SIFE}}

We augment the general linear process versions of the above VARMA and VAR
specifications with $\mathbf{G}_{i}\mathbf{f}_{t}$\thinspace , namely%
\begin{equation}
\mathbf{w}_{it}=\mathbf{w}_{i0}+\mathbf{G}_{i}\mathbf{f}_{t}+\mathbf{C}_{i}%
\mathbf{s}_{it}+\mathbf{C}_{i}^{\ast }(L)\mathbf{u}_{it},  \label{s_dgps}
\end{equation}%
where $\mathbf{C}_{i}$ and $\mathbf{C}_{i}^{\ast }(L)$ are obtained from the
VARMA and VAR models above. $\mathbf{f}_{t}$ is an $m_{f}\times 1\,\ $vector
of unobserved common factors, $\mathbf{G}_{i}$ is an $m\times m_{f}$ matrix
of factor loadings. We set $m_{f}=4$ and generate latent factors, $\mathbf{f}%
_{t}$, to be serially correlated with a break,%
\begin{eqnarray*}
\mathbf{f}_{t} &=&\rho _{f_{1}}\mathbf{f}_{t-1}+\sqrt{1-\rho _{f1}^{2}}%
\mathbf{v}_{ft}\text{, for }t=1,2,...,[T/2]-1\text{,} \\
\mathbf{f}_{t} &=&\rho _{f2}\mathbf{f}_{t-1}+\sqrt{1-\rho _{f2}^{2}}\mathbf{v%
}_{t}\text{, for }t=[T/2],[T/2]+1,...,T-1\text{,}
\end{eqnarray*}%
where $\rho _{f1}=0.6$, $\rho _{f2}=0.4$, and $\mathbf{v}_{t}\sim IIDN\left( 
\mathbf{0},\mathbf{I}_{m_{f}}\right) $. Individual elements of $\mathbf{G}%
_{i}$ are generated as $IIDU\left[ 0.0.4\right] $.

\subsection{Design from Chudik, Pesaran and Smith (2021) \label{PBD}}

This design features $m=2$ variables in $\mathbf{w}_{it}=\left(
w_{1,it},w_{2,it}\right) ^{\prime }$, generated according to the
cross-sectionally independent DGP described in Section 3.1 of 
\citeN{ChudikPesaranSmith2021PB}%
. In this design, $\mathbf{w}_{it}$ is generated as 
\begin{eqnarray}
\Delta w_{1,it} &=&c_{i}-a_{i}\left( w_{1,i,t-1}-w_{2,i,t-1}\right) +u_{1,it}%
\text{,}  \label{y} \\
\Delta w_{2,it} &=&u_{2,it}\text{,}  \label{x}
\end{eqnarray}%
where $a_{i}\sim IIDU\left[ 0.2,0.3\right] $, $u_{1,it}=\sigma _{1i}e_{1,it}$%
, $u_{2,it}=\sigma _{2i}e_{2,it}$, $\sigma _{1,i}^{2},\sigma _{2,i}^{2}\sim
IIDU\left[ 0.8,1.2\right] $,%
\begin{equation*}
\left( 
\begin{array}{c}
e_{1,it} \\ 
e_{2,it}%
\end{array}%
\right) \sim IIDN\left( \mathbf{0}_{2},\mathbf{\Sigma }_{e}\right) \text{, }%
\mathbf{\Sigma }_{e}\sim \left( 
\begin{array}{cc}
1 & \rho _{ei} \\ 
\rho _{ei} & 1%
\end{array}%
\right) \text{, and }\rho _{ei}\sim IIDU\left[ 0.3,0.7\right] \text{.}
\end{equation*}%
Full details of this design are provided in Section 3.1 of 
\citeN{ChudikPesaranSmith2021PB}%
.

\subsection{List of individual experiments\label{Sup_MCf}}

Overall, we conducted 71 experiments for the estimation of $r_{0}$ and 65
experiments for the estimation of $\mathbf{B}_{0}$. Table provide a summary
of these experiments. Monte Carlo findings for individual experiments are
available from authors upon request. \bigskip \pagebreak

\begin{center}
\onehalfspacing%
TABLE S1: List of Monte Carlo experiments\bigskip

\renewcommand{\arraystretch}{1.0}\setlength{\tabcolsep}{2pt}%
\scriptsize%
\begin{tabular}{cccccccrr}
\hline\hline
\multicolumn{9}{l}{\textit{A. Experiments with no long-run relations used
for estimation of }$r_{0}$ only} \\ \hline\hline
Number of &  & Interactive &  & \multicolumn{3}{c}{} &  &  \\ 
experiments &  & effects &  & \multicolumn{3}{c}{$\phi _{i,\ell \ell }$} & 
&  \\ \hline
6 &  & yes, no &  & \multicolumn{3}{c}{low, moderate, high} &  &  \\ 
\hline\hline
&  &  &  & \multicolumn{3}{c}{} &  &  \\ 
\multicolumn{9}{l}{\textit{B. Experiments with long-run relations and
two-way long-run causality}} \\ \hline\hline
Number of &  & Interactive &  &  &  & Error & \multicolumn{1}{c}{} & 
\multicolumn{1}{c}{} \\ 
experiments & $r_{0}$ & effects &  & Model &  & distribution & 
\multicolumn{1}{c}{$PR^{2}$} & \multicolumn{1}{c}{Speed of convergence} \\ 
\hline
64 & 1,2 & yes, no &  & VAR(1), VARMA(1,1) &  & Gaussian, chi-squared & 
\multicolumn{1}{c}{$0.2$, $0.3$} & \multicolumn{1}{c}{moderate, slow} \\ 
\hline\hline
&  &  &  &  &  &  &  &  \\ 
\multicolumn{9}{l}{C. \textit{Experiments with single long-run relation and
one-way long-run causality}} \\ \hline\hline
Number of &  &  &  &  &  &  &  &  \\ 
experiments &  &  &  &  &  &  &  &  \\ \hline
1 &  & \multicolumn{7}{l}{Design from Chudik, Pesaran and Smith (2021)} \\ 
\hline\hline
\end{tabular}%
\vspace{-0.2in}
\end{center}

\begin{flushleft}
\scriptsize%
\singlespacing%
Notes: This table lists 71 Monte Carlo experiments.%
\normalsize%
\end{flushleft}

\section{Sensitivity of $\tilde{r}$ to scaling\label{Sup_MCs}}

It is clear that eigenvalues of $\mathbf{Q}_{\bar{w}\bar{w}}$ depend on the
scale of the observations, $\mathbf{w}_{it}=\left(
w_{it,1},w_{it,2},...,w_{it,m}\right) ^{\prime }$, and some form of scaling
of data is required to reduce or eliminate the sensitivity of the estimator
of $r_{0}$ to scaling. In Section \ref{qCTchoices} in the body of the paper,
we proposed using eigenvalues of the correlation matrix $\mathbf{R}_{_{\bar{w%
}\bar{w}}}$ given by (\ref{Rww}), which, for convenience, we reproduce here:%
\begin{equation*}
\mathbf{R}_{_{\bar{w}\bar{w}}}=\left[ diag\left( \mathbf{Q}_{\bar{w}\bar{w}%
}\right) \right] ^{-1/2}\mathbf{Q}_{\bar{w}\bar{w}}\left[ diag\left( \mathbf{%
Q}_{\bar{w}\bar{w}}\right) \right] ^{-1/2}.
\end{equation*}%
Accordingly, we defined in equation (\ref{r_est}) the following estimator of 
$r_{0}$,%
\begin{equation*}
\tilde{r}=\sum_{j=1}^{m}\mathcal{I}\left( \tilde{\lambda}_{j}<T^{-\delta
}\right) ,
\end{equation*}%
where $\tilde{\lambda}_{j}$, for $j=1,2,...,m$ are the eigenvalues of $%
\mathbf{R}_{_{\bar{w}\bar{w}}}$.

Consider the scaled vector $\mathbf{\dot{w}}_{it}=\mathbf{Dw}_{it}$, where $%
\mathbf{D}=diag\left( d_{11},d_{22},...,d_{mm}\right) $ is a diagonal $%
m\times m$ scaling matrix with $d_{kk}>0$ for all $k=1,2,...,m$. Then matrix 
$\mathbf{\dot{R}}_{_{\bar{w}\bar{w}}}$ computed based on scaled variables $%
\mathbf{\dot{w}}_{it}$ remains unaffected by the scaling matrix $\mathbf{D}$%
, namely $\mathbf{R}_{_{\bar{w}\bar{w}}}=\mathbf{\dot{R}}_{_{\bar{w}\bar{w}%
}} $. However, this no longer true when differential scaling is considered,
given by $\mathbf{\ddot{w}}_{it}=\mathbf{D}_{i}\mathbf{w}_{it}$, where $%
\mathbf{D}_{i}=diag\left( d_{i,11},d_{i,22},...,d_{i,mm}\right) $, for $%
i=1,2,...,n$, are unit-specific diagonal scaling matrices with $d_{i,kk}>0$
for all $k=1,2,...,m$ and $i=1,2,...,n$. In this case, matrix $\mathbf{\ddot{%
R}}_{_{\bar{w}\bar{w}}}$, computed based on $\mathbf{\ddot{w}}_{it}$, is
affected by scaling, namely $\mathbf{\ddot{R}}_{_{\bar{w}\bar{w}}}\neq 
\mathbf{R}_{_{\bar{w}\bar{w}}}$ in the general case where $\mathbf{D}_{i}$
differs over $i$.

To investigate impact of differential scaling given by $\mathbf{\ddot{w}}%
_{it}=\mathbf{D}_{i}\mathbf{w}_{it}$ on the small sample performance of $%
\tilde{r}$, we consider additional Monte Carlo experiments below. We
generate $\mathbf{D}_{i}=\varkappa _{i}\mathbf{I}_{3}$, for $i=1,2,...,n$,
where $\varkappa _{i}\sim IIDU\left[ 1,2\right] $. We compare the small
sample performance of $\tilde{r}$ based on the original data $\mathbf{w}%
_{it} $ and the scaled data $\mathbf{\ddot{w}}_{it}=\mathbf{D}_{i}\mathbf{w}%
_{it}=\varkappa _{i}\mathbf{w}_{it}$, for three experiments. Table S2
reports selection frequencies for the estimation of $r_{0}$ using $\tilde{r}$
based on original and scaled data as well as two choices of $q=2$ and $4$
and two choices of exponent $\delta =1/4$ and $1/2$, in experiments with $%
r_{0}=0$ (no long-run relations), given by the first-differenced VAR(1)
design (\ref{VARdif}) with high persistence $\phi _{i,\ell \ell }\sim
U[0.80,0.95]$. Table S3 reports the same set of results in the case of
VAR(1) experiments with $r_{0}=1$ long-run relation, interactive effects,
low speed of convergence, low value of $PR^{2}=0.2$ and chi-squared
distributed errors. Table S4 below reports findings for VAR(1) experiments
with $r_{0}=2$ long-run relations, interactive effects, low speed of
convergence, low value of $PR^{2}=0.2$ and chi-squared distributed errors.
Overall, these Monte Carlo findings show a very small dependence of $\tilde{r%
}$ on the adopted differential scaling. In addition, these findings also
show that performance of $\tilde{r}$ is almost identical for the two choices
of $q=2$ and $4$.

\pagebreak

\begin{center}
\onehalfspacing%
TABLE S2: Selection frequencies for the estimation of $r_{0}$ by eigenvalue
thresholding estimator $\tilde{r}$ with $\delta =1/4$ and $1/2$, $q=2$ and $%
4 $, and using original and scaled data, in experiments with no long-run
relation ($r_{0}=0$), $\phi _{ij}\sim U[0.8,0.95],$ and with interactive
time effects\bigskip

\renewcommand{\arraystretch}{0.9}\setlength{\tabcolsep}{5pt}%
\scriptsize%
\begin{tabular}{rrrrrrrrrrrrrrrrr}
\hline\hline
&  & \multicolumn{3}{c}{Frequency $\tilde{r}=0$} &  & \multicolumn{3}{c}{
Frequency $\tilde{r}=1$} &  & \multicolumn{3}{c}{Frequency $\tilde{r}=2$} & 
& \multicolumn{3}{c}{Frequency $\tilde{r}=3$} \\ 
\cline{3-5}\cline{7-9}\cline{11-13}\cline{15-17}
$n$ $\backslash $ $T$ &  & \textbf{20} & \textbf{50} & \textbf{100} &  & 
\textbf{20} & \textbf{50} & \textbf{100} &  & \textbf{20} & \textbf{50} & 
\textbf{100} &  & \textbf{20} & \textbf{50} & \textbf{100} \\ \hline
\multicolumn{10}{l}{\textbf{A. }$\tilde{r}$ is computed using original data $%
w_{it}$} &  &  &  &  &  &  &  \\ \hline
&  & \multicolumn{15}{l}{Estimator $\tilde{r}$ with $\delta =1/4$ and $q=2$}
\\ \hline
\textbf{50} &  & 0.95 & 1.00 & 1.00 &  & 0.05 & 0.00 & 0.00 &  & 0.00 & 0.00
& 0.00 &  & 0.00 & 0.00 & 0.00 \\ 
\textbf{500} &  & 1.00 & 1.00 & 1.00 &  & 0.00 & 0.00 & 0.00 &  & 0.00 & 0.00
& 0.00 &  & 0.00 & 0.00 & 0.00 \\ 
\textbf{1,000} &  & 1.00 & 1.00 & 1.00 &  & 0.00 & 0.00 & 0.00 &  & 0.00 & 
0.00 & 0.00 &  & 0.00 & 0.00 & 0.00 \\ 
\textbf{3,000} &  & 1.00 & 1.00 & 1.00 &  & 0.00 & 0.00 & 0.00 &  & 0.00 & 
0.00 & 0.00 &  & 0.00 & 0.00 & 0.00 \\ \hline
&  & \multicolumn{15}{l}{Estimator $\tilde{r}$ with $\delta =1/4$ and $q=4$}
\\ \hline
\textbf{50} &  & 0.97 & 1.00 & 1.00 &  & 0.03 & 0.00 & 0.00 &  & 0.00 & 0.00
& 0.00 &  & 0.00 & 0.00 & 0.00 \\ 
\textbf{500} &  & 1.00 & 1.00 & 1.00 &  & 0.00 & 0.00 & 0.00 &  & 0.00 & 0.00
& 0.00 &  & 0.00 & 0.00 & 0.00 \\ 
\textbf{1,000} &  & 1.00 & 1.00 & 1.00 &  & 0.00 & 0.00 & 0.00 &  & 0.00 & 
0.00 & 0.00 &  & 0.00 & 0.00 & 0.00 \\ 
\textbf{3,000} &  & 1.00 & 1.00 & 1.00 &  & 0.00 & 0.00 & 0.00 &  & 0.00 & 
0.00 & 0.00 &  & 0.00 & 0.00 & 0.00 \\ \hline
&  & \multicolumn{15}{l}{Estimator $\tilde{r}$ with $\delta =1/2$ and $q=2$}
\\ \hline
\textbf{50} &  & 1.00 & 1.00 & 1.00 &  & 0.00 & 0.00 & 0.00 &  & 0.00 & 0.00
& 0.00 &  & 0.00 & 0.00 & 0.00 \\ 
\textbf{500} &  & 1.00 & 1.00 & 1.00 &  & 0.00 & 0.00 & 0.00 &  & 0.00 & 0.00
& 0.00 &  & 0.00 & 0.00 & 0.00 \\ 
\textbf{1,000} &  & 1.00 & 1.00 & 1.00 &  & 0.00 & 0.00 & 0.00 &  & 0.00 & 
0.00 & 0.00 &  & 0.00 & 0.00 & 0.00 \\ 
\textbf{3,000} &  & 1.00 & 1.00 & 1.00 &  & 0.00 & 0.00 & 0.00 &  & 0.00 & 
0.00 & 0.00 &  & 0.00 & 0.00 & 0.00 \\ \hline
&  & \multicolumn{15}{l}{Estimator $\tilde{r}$ with $\delta =1/2$ and $q=4$}
\\ \hline
\textbf{50} &  & 1.00 & 1.00 & 1.00 &  & 0.00 & 0.00 & 0.00 &  & 0.00 & 0.00
& 0.00 &  & 0.00 & 0.00 & 0.00 \\ 
\textbf{500} &  & 1.00 & 1.00 & 1.00 &  & 0.00 & 0.00 & 0.00 &  & 0.00 & 0.00
& 0.00 &  & 0.00 & 0.00 & 0.00 \\ 
\textbf{1,000} &  & 1.00 & 1.00 & 1.00 &  & 0.00 & 0.00 & 0.00 &  & 0.00 & 
0.00 & 0.00 &  & 0.00 & 0.00 & 0.00 \\ 
\textbf{3,000} &  & 1.00 & 1.00 & 1.00 &  & 0.00 & 0.00 & 0.00 &  & 0.00 & 
0.00 & 0.00 &  & 0.00 & 0.00 & 0.00 \\ \hline
\multicolumn{15}{l}{\textbf{B. }$\tilde{r}$ is computed using scaled data $%
\varkappa _{i}\mathbf{w}_{it}$, $\varkappa _{i}\sim IIDU\left[ 1,2\right] $}
&  &  \\ \hline
&  & \multicolumn{15}{l}{Estimator $\tilde{r}$ with $\delta =1/4$ and $q=2$}
\\ \hline
\textbf{50} &  & 0.92 & 1.00 & 1.00 &  & 0.08 & 0.00 & 0.00 &  & 0.00 & 0.00
& 0.00 &  & 0.00 & 0.00 & 0.00 \\ 
\textbf{500} &  & 1.00 & 1.00 & 1.00 &  & 0.00 & 0.00 & 0.00 &  & 0.00 & 0.00
& 0.00 &  & 0.00 & 0.00 & 0.00 \\ 
\textbf{1,000} &  & 1.00 & 1.00 & 1.00 &  & 0.00 & 0.00 & 0.00 &  & 0.00 & 
0.00 & 0.00 &  & 0.00 & 0.00 & 0.00 \\ 
\textbf{3,000} &  & 1.00 & 1.00 & 1.00 &  & 0.00 & 0.00 & 0.00 &  & 0.00 & 
0.00 & 0.00 &  & 0.00 & 0.00 & 0.00 \\ \hline
&  & \multicolumn{15}{l}{Estimator $\tilde{r}$ with $\delta =1/4$ and $q=4$}
\\ \hline
\textbf{50} &  & 0.95 & 1.00 & 1.00 &  & 0.05 & 0.00 & 0.00 &  & 0.00 & 0.00
& 0.00 &  & 0.00 & 0.00 & 0.00 \\ 
\textbf{500} &  & 1.00 & 1.00 & 1.00 &  & 0.00 & 0.00 & 0.00 &  & 0.00 & 0.00
& 0.00 &  & 0.00 & 0.00 & 0.00 \\ 
\textbf{1,000} &  & 1.00 & 1.00 & 1.00 &  & 0.00 & 0.00 & 0.00 &  & 0.00 & 
0.00 & 0.00 &  & 0.00 & 0.00 & 0.00 \\ 
\textbf{3,000} &  & 1.00 & 1.00 & 1.00 &  & 0.00 & 0.00 & 0.00 &  & 0.00 & 
0.00 & 0.00 &  & 0.00 & 0.00 & 0.00 \\ \hline
&  & \multicolumn{15}{l}{Estimator $\tilde{r}$ with $\delta =1/2$ and $q=2$}
\\ \hline
\textbf{50} &  & 1.00 & 1.00 & 1.00 &  & 0.00 & 0.00 & 0.00 &  & 0.00 & 0.00
& 0.00 &  & 0.00 & 0.00 & 0.00 \\ 
\textbf{500} &  & 1.00 & 1.00 & 1.00 &  & 0.00 & 0.00 & 0.00 &  & 0.00 & 0.00
& 0.00 &  & 0.00 & 0.00 & 0.00 \\ 
\textbf{1,000} &  & 1.00 & 1.00 & 1.00 &  & 0.00 & 0.00 & 0.00 &  & 0.00 & 
0.00 & 0.00 &  & 0.00 & 0.00 & 0.00 \\ 
\textbf{3,000} &  & 1.00 & 1.00 & 1.00 &  & 0.00 & 0.00 & 0.00 &  & 0.00 & 
0.00 & 0.00 &  & 0.00 & 0.00 & 0.00 \\ \hline
&  & \multicolumn{15}{l}{Estimator $\tilde{r}$ with $\delta =1/2$ and $q=4$}
\\ \hline
\textbf{50} &  & 1.00 & 1.00 & 1.00 &  & 0.00 & 0.00 & 0.00 &  & 0.00 & 0.00
& 0.00 &  & 0.00 & 0.00 & 0.00 \\ 
\textbf{500} &  & 1.00 & 1.00 & 1.00 &  & 0.00 & 0.00 & 0.00 &  & 0.00 & 0.00
& 0.00 &  & 0.00 & 0.00 & 0.00 \\ 
\textbf{1,000} &  & 1.00 & 1.00 & 1.00 &  & 0.00 & 0.00 & 0.00 &  & 0.00 & 
0.00 & 0.00 &  & 0.00 & 0.00 & 0.00 \\ 
\textbf{3,000} &  & 1.00 & 1.00 & 1.00 &  & 0.00 & 0.00 & 0.00 &  & 0.00 & 
0.00 & 0.00 &  & 0.00 & 0.00 & 0.00 \\ \hline\hline
\end{tabular}%
\vspace{-0.2in}
\end{center}

\begin{flushleft}
\scriptsize%
\singlespacing%
Notes: This table reports selection frequencies for the number of estimated
long-run relations. $r_{0}$ denotes the true number of long-run relations. $%
\tilde{r}$ is given by (\ref{r_est}), namely $\tilde{r}=\sum_{j=1}^{m}%
\mathcal{I}\left( \tilde{\lambda}_{j}<T^{-\delta }\right) $, with $\tilde{%
\lambda}_{j}$, $j=1,2,...,m$, being the eigenvalues of $\mathbf{R}_{_{\bar{w}%
\bar{w}}}$ defined by (\ref{Rww}). See Subsection \ref{DGPsubsection} in the
paper for a summary of the design and Section \ref{Sup_MCd} in the
supplement for the full description. Reported results are based on $R=2000$
MC replications. \pagebreak
\end{flushleft}

\begin{center}
\normalsize%
\onehalfspacing%
TABLE S3: Selection frequencies for the estimation of $r_{0}$ by eigenvalue
thresholding estimator $\tilde{r}$ with $\delta =1/4$ and $1/2$, $q=2$ and $%
4 $, and using original and scaled data, in VAR(1)\ experiments with $%
r_{0}=1 $ long-run relation, slow speed of convergence toward long run,
chi-squared distributed errors, $PR_{nT}^{2}=0.2$, and with interactive time
effects\bigskip

\renewcommand{\arraystretch}{0.9}\setlength{\tabcolsep}{5pt}%
\scriptsize%
\begin{tabular}{rrrrrrrrrrrrrrrrr}
\hline\hline
&  & \multicolumn{3}{c}{Frequency $\tilde{r}=0$} &  & \multicolumn{3}{c}{
Frequency $\tilde{r}=1$} &  & \multicolumn{3}{c}{Frequency $\tilde{r}=2$} & 
& \multicolumn{3}{c}{Frequency $\tilde{r}=3$} \\ 
\cline{3-5}\cline{7-9}\cline{11-13}\cline{15-17}
$n$ $\backslash $ $T$ &  & \textbf{20} & \textbf{50} & \textbf{100} &  & 
\textbf{20} & \textbf{50} & \textbf{100} &  & \textbf{20} & \textbf{50} & 
\textbf{100} &  & \textbf{20} & \textbf{50} & \textbf{100} \\ \hline
\multicolumn{10}{l}{\textbf{A. }$\tilde{r}$ is computed using original data $%
w_{it}$} &  &  &  &  &  &  &  \\ \hline
&  & \multicolumn{15}{l}{Estimator $\tilde{r}$ with $\delta =1/4$ and $q=2$}
\\ \hline
\textbf{50} &  & 0.00 & 0.00 & 0.00 &  & 1.00 & 1.00 & 1.00 &  & 0.00 & 0.00
& 0.00 &  & 0.00 & 0.00 & 0.00 \\ 
\textbf{500} &  & 0.00 & 0.00 & 0.00 &  & 1.00 & 1.00 & 1.00 &  & 0.00 & 0.00
& 0.00 &  & 0.00 & 0.00 & 0.00 \\ 
\textbf{1,000} &  & 0.00 & 0.00 & 0.00 &  & 1.00 & 1.00 & 1.00 &  & 0.00 & 
0.00 & 0.00 &  & 0.00 & 0.00 & 0.00 \\ 
\textbf{3,000} &  & 0.00 & 0.00 & 0.00 &  & 1.00 & 1.00 & 1.00 &  & 0.00 & 
0.00 & 0.00 &  & 0.00 & 0.00 & 0.00 \\ \hline
&  & \multicolumn{15}{l}{Estimator $\tilde{r}$ with $\delta =1/4$ and $q=4$}
\\ \hline
\textbf{50} &  & 0.00 & 0.00 & 0.00 &  & 1.00 & 1.00 & 1.00 &  & 0.00 & 0.00
& 0.00 &  & 0.00 & 0.00 & 0.00 \\ 
\textbf{500} &  & 0.00 & 0.00 & 0.00 &  & 1.00 & 1.00 & 1.00 &  & 0.00 & 0.00
& 0.00 &  & 0.00 & 0.00 & 0.00 \\ 
\textbf{1,000} &  & 0.00 & 0.00 & 0.00 &  & 1.00 & 1.00 & 1.00 &  & 0.00 & 
0.00 & 0.00 &  & 0.00 & 0.00 & 0.00 \\ 
\textbf{3,000} &  & 0.00 & 0.00 & 0.00 &  & 1.00 & 1.00 & 1.00 &  & 0.00 & 
0.00 & 0.00 &  & 0.00 & 0.00 & 0.00 \\ \hline
&  & \multicolumn{15}{l}{Estimator $\tilde{r}$ with $\delta =1/2$ and $q=2$}
\\ \hline
\textbf{50} &  & 0.00 & 0.00 & 0.00 &  & 1.00 & 1.00 & 1.00 &  & 0.00 & 0.00
& 0.00 &  & 0.00 & 0.00 & 0.00 \\ 
\textbf{500} &  & 0.00 & 0.00 & 0.00 &  & 1.00 & 1.00 & 1.00 &  & 0.00 & 0.00
& 0.00 &  & 0.00 & 0.00 & 0.00 \\ 
\textbf{1,000} &  & 0.00 & 0.00 & 0.00 &  & 1.00 & 1.00 & 1.00 &  & 0.00 & 
0.00 & 0.00 &  & 0.00 & 0.00 & 0.00 \\ 
\textbf{3,000} &  & 0.00 & 0.00 & 0.00 &  & 1.00 & 1.00 & 1.00 &  & 0.00 & 
0.00 & 0.00 &  & 0.00 & 0.00 & 0.00 \\ \hline
&  & \multicolumn{15}{l}{Estimator $\tilde{r}$ with $\delta =1/2$ and $q=4$}
\\ \hline
\textbf{50} &  & 0.00 & 0.00 & 0.00 &  & 1.00 & 1.00 & 1.00 &  & 0.00 & 0.00
& 0.00 &  & 0.00 & 0.00 & 0.00 \\ 
\textbf{500} &  & 0.00 & 0.00 & 0.00 &  & 1.00 & 1.00 & 1.00 &  & 0.00 & 0.00
& 0.00 &  & 0.00 & 0.00 & 0.00 \\ 
\textbf{1,000} &  & 0.00 & 0.00 & 0.00 &  & 1.00 & 1.00 & 1.00 &  & 0.00 & 
0.00 & 0.00 &  & 0.00 & 0.00 & 0.00 \\ 
\textbf{3,000} &  & 0.00 & 0.00 & 0.00 &  & 1.00 & 1.00 & 1.00 &  & 0.00 & 
0.00 & 0.00 &  & 0.00 & 0.00 & 0.00 \\ \hline
\multicolumn{15}{l}{\textbf{B. }$\tilde{r}$ is computed using scaled data $%
\varkappa _{i}\mathbf{w}_{it}$, $\varkappa _{i}\sim IIDU\left[ 1,2\right] $}
&  &  \\ \hline
&  & \multicolumn{15}{l}{Estimator $\tilde{r}$ with $\delta =1/4$ and $q=2$}
\\ \hline
\textbf{50} &  & 0.00 & 0.00 & 0.00 &  & 1.00 & 1.00 & 1.00 &  & 0.00 & 0.00
& 0.00 &  & 0.00 & 0.00 & 0.00 \\ 
\textbf{500} &  & 0.00 & 0.00 & 0.00 &  & 1.00 & 1.00 & 1.00 &  & 0.00 & 0.00
& 0.00 &  & 0.00 & 0.00 & 0.00 \\ 
\textbf{1,000} &  & 0.00 & 0.00 & 0.00 &  & 1.00 & 1.00 & 1.00 &  & 0.00 & 
0.00 & 0.00 &  & 0.00 & 0.00 & 0.00 \\ 
\textbf{3,000} &  & 0.00 & 0.00 & 0.00 &  & 1.00 & 1.00 & 1.00 &  & 0.00 & 
0.00 & 0.00 &  & 0.00 & 0.00 & 0.00 \\ \hline
&  & \multicolumn{15}{l}{Estimator $\tilde{r}$ with $\delta =1/4$ and $q=4$}
\\ \hline
\textbf{50} &  & 0.00 & 0.00 & 0.00 &  & 1.00 & 1.00 & 1.00 &  & 0.00 & 0.00
& 0.00 &  & 0.00 & 0.00 & 0.00 \\ 
\textbf{500} &  & 0.00 & 0.00 & 0.00 &  & 1.00 & 1.00 & 1.00 &  & 0.00 & 0.00
& 0.00 &  & 0.00 & 0.00 & 0.00 \\ 
\textbf{1,000} &  & 0.00 & 0.00 & 0.00 &  & 1.00 & 1.00 & 1.00 &  & 0.00 & 
0.00 & 0.00 &  & 0.00 & 0.00 & 0.00 \\ 
\textbf{3,000} &  & 0.00 & 0.00 & 0.00 &  & 1.00 & 1.00 & 1.00 &  & 0.00 & 
0.00 & 0.00 &  & 0.00 & 0.00 & 0.00 \\ \hline
&  & \multicolumn{15}{l}{Estimator $\tilde{r}$ with $\delta =1/2$ and $q=2$}
\\ \hline
\textbf{50} &  & 0.00 & 0.00 & 0.00 &  & 1.00 & 1.00 & 1.00 &  & 0.00 & 0.00
& 0.00 &  & 0.00 & 0.00 & 0.00 \\ 
\textbf{500} &  & 0.00 & 0.00 & 0.00 &  & 1.00 & 1.00 & 1.00 &  & 0.00 & 0.00
& 0.00 &  & 0.00 & 0.00 & 0.00 \\ 
\textbf{1,000} &  & 0.00 & 0.00 & 0.00 &  & 1.00 & 1.00 & 1.00 &  & 0.00 & 
0.00 & 0.00 &  & 0.00 & 0.00 & 0.00 \\ 
\textbf{3,000} &  & 0.00 & 0.00 & 0.00 &  & 1.00 & 1.00 & 1.00 &  & 0.00 & 
0.00 & 0.00 &  & 0.00 & 0.00 & 0.00 \\ \hline
&  & \multicolumn{15}{l}{Estimator $\tilde{r}$ with $\delta =1/2$ and $q=4$}
\\ \hline
\textbf{50} &  & 0.00 & 0.00 & 0.00 &  & 1.00 & 1.00 & 1.00 &  & 0.00 & 0.00
& 0.00 &  & 0.00 & 0.00 & 0.00 \\ 
\textbf{500} &  & 0.00 & 0.00 & 0.00 &  & 1.00 & 1.00 & 1.00 &  & 0.00 & 0.00
& 0.00 &  & 0.00 & 0.00 & 0.00 \\ 
\textbf{1,000} &  & 0.00 & 0.00 & 0.00 &  & 1.00 & 1.00 & 1.00 &  & 0.00 & 
0.00 & 0.00 &  & 0.00 & 0.00 & 0.00 \\ 
\textbf{3,000} &  & 0.00 & 0.00 & 0.00 &  & 1.00 & 1.00 & 1.00 &  & 0.00 & 
0.00 & 0.00 &  & 0.00 & 0.00 & 0.00 \\ \hline\hline
\end{tabular}%
\vspace{-0.2in}
\end{center}

\begin{flushleft}
\scriptsize%
\singlespacing%
Notes: See notes to Table S2. \pagebreak
\end{flushleft}

\begin{center}
\normalsize%
\onehalfspacing%
TABLE S4: Selection frequencies for the estimation of $r_{0}$ by eigenvalue
thresholding estimator $\tilde{r}$ with $\delta =1/4$ and $1/2$, $q=2$ and $%
4 $, and using original and scaled data, in VAR(1)\ experiments with $%
r_{0}=2 $ long-run relations, slow speed of convergence toward long run,
chi-squared distributed errors, $PR_{nT}^{2}=0.2$, and with interactive time
effects\bigskip

\renewcommand{\arraystretch}{0.9}\setlength{\tabcolsep}{5pt}%
\scriptsize%
\begin{tabular}{rrrrrrrrrrrrrrrrr}
\hline\hline
&  & \multicolumn{3}{c}{Frequency $\tilde{r}=0$} &  & \multicolumn{3}{c}{
Frequency $\tilde{r}=1$} &  & \multicolumn{3}{c}{Frequency $\tilde{r}=2$} & 
& \multicolumn{3}{c}{Frequency $\tilde{r}=3$} \\ 
\cline{3-5}\cline{7-9}\cline{11-13}\cline{15-17}
$n$ $\backslash $ $T$ &  & \textbf{20} & \textbf{50} & \textbf{100} &  & 
\textbf{20} & \textbf{50} & \textbf{100} &  & \textbf{20} & \textbf{50} & 
\textbf{100} &  & \textbf{20} & \textbf{50} & \textbf{100} \\ \hline
\multicolumn{10}{l}{\textbf{A. }$\tilde{r}$ is computed using original data $%
w_{it}$} &  &  &  &  &  &  &  \\ \hline
&  & \multicolumn{15}{l}{Estimator $\tilde{r}$ with $\delta =1/4$ and $q=2$}
\\ \hline
\textbf{50} &  & 0.00 & 0.00 & 0.00 &  & 0.00 & 0.00 & 0.00 &  & 1.00 & 1.00
& 1.00 &  & 0.00 & 0.00 & 0.00 \\ 
\textbf{500} &  & 0.00 & 0.00 & 0.00 &  & 0.00 & 0.00 & 0.00 &  & 1.00 & 1.00
& 1.00 &  & 0.00 & 0.00 & 0.00 \\ 
\textbf{1,000} &  & 0.00 & 0.00 & 0.00 &  & 0.00 & 0.00 & 0.00 &  & 1.00 & 
1.00 & 1.00 &  & 0.00 & 0.00 & 0.00 \\ 
\textbf{3,000} &  & 0.00 & 0.00 & 0.00 &  & 0.00 & 0.00 & 0.00 &  & 1.00 & 
1.00 & 1.00 &  & 0.00 & 0.00 & 0.00 \\ \hline
&  & \multicolumn{15}{l}{Estimator $\tilde{r}$ with $\delta =1/4$ and $q=4$}
\\ \hline
\textbf{50} &  & 0.00 & 0.00 & 0.00 &  & 0.00 & 0.00 & 0.00 &  & 1.00 & 1.00
& 1.00 &  & 0.00 & 0.00 & 0.00 \\ 
\textbf{500} &  & 0.00 & 0.00 & 0.00 &  & 0.00 & 0.00 & 0.00 &  & 1.00 & 1.00
& 1.00 &  & 0.00 & 0.00 & 0.00 \\ 
\textbf{1,000} &  & 0.00 & 0.00 & 0.00 &  & 0.00 & 0.00 & 0.00 &  & 1.00 & 
1.00 & 1.00 &  & 0.00 & 0.00 & 0.00 \\ 
\textbf{3,000} &  & 0.00 & 0.00 & 0.00 &  & 0.00 & 0.00 & 0.00 &  & 1.00 & 
1.00 & 1.00 &  & 0.00 & 0.00 & 0.00 \\ \hline
&  & \multicolumn{15}{l}{Estimator $\tilde{r}$ with $\delta =1/2$ and $q=2$}
\\ \hline
\textbf{50} &  & 0.00 & 0.00 & 0.00 &  & 0.00 & 0.00 & 0.00 &  & 1.00 & 1.00
& 1.00 &  & 0.00 & 0.00 & 0.00 \\ 
\textbf{500} &  & 0.00 & 0.00 & 0.00 &  & 0.00 & 0.00 & 0.00 &  & 1.00 & 1.00
& 1.00 &  & 0.00 & 0.00 & 0.00 \\ 
\textbf{1,000} &  & 0.00 & 0.00 & 0.00 &  & 0.00 & 0.00 & 0.00 &  & 1.00 & 
1.00 & 1.00 &  & 0.00 & 0.00 & 0.00 \\ 
\textbf{3,000} &  & 0.00 & 0.00 & 0.00 &  & 0.00 & 0.00 & 0.00 &  & 1.00 & 
1.00 & 1.00 &  & 0.00 & 0.00 & 0.00 \\ \hline
&  & \multicolumn{15}{l}{Estimator $\tilde{r}$ with $\delta =1/2$ and $q=4$}
\\ \hline
\textbf{50} &  & 0.00 & 0.00 & 0.00 &  & 0.00 & 0.00 & 0.00 &  & 1.00 & 1.00
& 1.00 &  & 0.00 & 0.00 & 0.00 \\ 
\textbf{500} &  & 0.00 & 0.00 & 0.00 &  & 0.00 & 0.00 & 0.00 &  & 1.00 & 1.00
& 1.00 &  & 0.00 & 0.00 & 0.00 \\ 
\textbf{1,000} &  & 0.00 & 0.00 & 0.00 &  & 0.00 & 0.00 & 0.00 &  & 1.00 & 
1.00 & 1.00 &  & 0.00 & 0.00 & 0.00 \\ 
\textbf{3,000} &  & 0.00 & 0.00 & 0.00 &  & 0.00 & 0.00 & 0.00 &  & 1.00 & 
1.00 & 1.00 &  & 0.00 & 0.00 & 0.00 \\ \hline
\multicolumn{15}{l}{\textbf{B. }$\tilde{r}$ is computed using scaled data $%
\varkappa _{i}\mathbf{w}_{it}$, $\varkappa _{i}\sim IIDU\left[ 1,2\right] $}
&  &  \\ \hline
&  & \multicolumn{15}{l}{Estimator $\tilde{r}$ with $\delta =1/4$ and $q=2$}
\\ \hline
\textbf{50} &  & 0.00 & 0.00 & 0.00 &  & 0.00 & 0.00 & 0.00 &  & 1.00 & 1.00
& 1.00 &  & 0.00 & 0.00 & 0.00 \\ 
\textbf{500} &  & 0.00 & 0.00 & 0.00 &  & 0.00 & 0.00 & 0.00 &  & 1.00 & 1.00
& 1.00 &  & 0.00 & 0.00 & 0.00 \\ 
\textbf{1,000} &  & 0.00 & 0.00 & 0.00 &  & 0.00 & 0.00 & 0.00 &  & 1.00 & 
1.00 & 1.00 &  & 0.00 & 0.00 & 0.00 \\ 
\textbf{3,000} &  & 0.00 & 0.00 & 0.00 &  & 0.00 & 0.00 & 0.00 &  & 1.00 & 
1.00 & 1.00 &  & 0.00 & 0.00 & 0.00 \\ \hline
&  & \multicolumn{15}{l}{Estimator $\tilde{r}$ with $\delta =1/4$ and $q=4$}
\\ \hline
\textbf{50} &  & 0.00 & 0.00 & 0.00 &  & 0.00 & 0.00 & 0.00 &  & 1.00 & 1.00
& 1.00 &  & 0.00 & 0.00 & 0.00 \\ 
\textbf{500} &  & 0.00 & 0.00 & 0.00 &  & 0.00 & 0.00 & 0.00 &  & 1.00 & 1.00
& 1.00 &  & 0.00 & 0.00 & 0.00 \\ 
\textbf{1,000} &  & 0.00 & 0.00 & 0.00 &  & 0.00 & 0.00 & 0.00 &  & 1.00 & 
1.00 & 1.00 &  & 0.00 & 0.00 & 0.00 \\ 
\textbf{3,000} &  & 0.00 & 0.00 & 0.00 &  & 0.00 & 0.00 & 0.00 &  & 1.00 & 
1.00 & 1.00 &  & 0.00 & 0.00 & 0.00 \\ \hline
&  & \multicolumn{15}{l}{Estimator $\tilde{r}$ with $\delta =1/2$ and $q=2$}
\\ \hline
\textbf{50} &  & 0.00 & 0.00 & 0.00 &  & 0.00 & 0.00 & 0.00 &  & 1.00 & 1.00
& 1.00 &  & 0.00 & 0.00 & 0.00 \\ 
\textbf{500} &  & 0.00 & 0.00 & 0.00 &  & 0.00 & 0.00 & 0.00 &  & 1.00 & 1.00
& 1.00 &  & 0.00 & 0.00 & 0.00 \\ 
\textbf{1,000} &  & 0.00 & 0.00 & 0.00 &  & 0.00 & 0.00 & 0.00 &  & 1.00 & 
1.00 & 1.00 &  & 0.00 & 0.00 & 0.00 \\ 
\textbf{3,000} &  & 0.00 & 0.00 & 0.00 &  & 0.00 & 0.00 & 0.00 &  & 1.00 & 
1.00 & 1.00 &  & 0.00 & 0.00 & 0.00 \\ \hline
&  & \multicolumn{15}{l}{Estimator $\tilde{r}$ with $\delta =1/2$ and $q=4$}
\\ \hline
\textbf{50} &  & 0.00 & 0.00 & 0.00 &  & 0.00 & 0.00 & 0.00 &  & 1.00 & 1.00
& 1.00 &  & 0.00 & 0.00 & 0.00 \\ 
\textbf{500} &  & 0.00 & 0.00 & 0.00 &  & 0.00 & 0.00 & 0.00 &  & 1.00 & 1.00
& 1.00 &  & 0.00 & 0.00 & 0.00 \\ 
\textbf{1,000} &  & 0.00 & 0.00 & 0.00 &  & 0.00 & 0.00 & 0.00 &  & 1.00 & 
1.00 & 1.00 &  & 0.00 & 0.00 & 0.00 \\ 
\textbf{3,000} &  & 0.00 & 0.00 & 0.00 &  & 0.00 & 0.00 & 0.00 &  & 1.00 & 
1.00 & 1.00 &  & 0.00 & 0.00 & 0.00 \\ \hline\hline
\end{tabular}%
\vspace{-0.2in}
\end{center}

\begin{flushleft}
\scriptsize%
\singlespacing%
Notes: See notes to Table S2. 
\normalsize%
\pagebreak
\end{flushleft}

\section{Robustness of PME estimators to GARCH and threshold autoregressive
effects\label{Sup_MCr}}

We also conduct Monte Carlo experiments to investigate robustness of PME
estimators to GARCH and threshold autoregressive (TAR) effects. To this end,
we replace $\mathbf{u}_{it}$ with 
\begin{equation*}
\mathbf{\xi }_{it}=\Phi ^{+}\mathbf{I}\left( \mathbf{\xi }_{i,t-1}>0\right) 
\mathbf{\xi }_{i,t-1}+\Phi ^{-}\mathbf{I}\left( \mathbf{\xi }_{i,t-1}\leq
0\right) \mathbf{\xi }_{i,t-1}+\mathbf{v}_{it}\text{,}
\end{equation*}%
for $t=-50,-49,...,0,1,2,...,T$, with initial values $\mathbf{\xi }%
_{i,-51}=0 $, 
\begin{equation*}
\mathbf{I}\left( \mathbf{\xi }_{i,t-1}>0\right) =\left( 
\begin{tabular}{lll}
$I\left( \xi _{it,1}>0\right) $ & $0$ & $0$ \\ 
$0$ & $I\left( \xi _{it,2}>0\right) $ & $0$ \\ 
$0$ & $0$ & $I\left( \xi _{it,3}>0\right) $%
\end{tabular}%
\right) \text{,}
\end{equation*}%
$I\left( .\right) $ is an indicator function, 
\begin{equation*}
\Phi ^{-}=\phi ^{-}\mathbf{I}_{3}\text{, and }\Phi ^{+}=\phi ^{+}\mathbf{I}%
_{3}\text{.}
\end{equation*}%
$\mathbf{v}_{it}=\left( v_{it,1},v_{it,2},v_{it,3}\right) ^{\prime }$ is
generated according to $\mathbf{v}_{it}=\mathbf{D}_{it}\mathbf{P}_{i}\mathbf{%
\varepsilon }_{it}$, where $\mathbf{\varepsilon }_{it}$ is generated as
before, namely individual elements are either $IIDN\left( 0,1\right) $, or
generated as $IID$ from a chi-squared distribution with 4 degrees of freedom
normalized to mean zero and unit variance, namely $\left[ \chi ^{2}\left(
4\right) -4\right] /\sqrt{8}$. $\mathbf{P}_{i}$ is Choleski factor of $%
\mathbf{R}_{i}=\mathbf{P}_{i}\mathbf{P}_{i}^{\prime }$, and%
\begin{equation*}
\mathbf{R}_{i}\sim \left( 
\begin{array}{ccc}
1 & \rho _{\varepsilon i,12} & \rho _{\varepsilon i,13} \\ 
\rho _{\varepsilon i,12} & 1 & \rho _{\varepsilon i,23} \\ 
\rho _{\varepsilon i,13} & \rho _{\varepsilon i,23} & 1%
\end{array}%
\right) \text{, }\rho _{\varepsilon i,pq}\sim IIDU\left[ 0,0.5\right] \text{,%
}
\end{equation*}%
for $\left( p,q\right) \in \left\{ \left( 1,2\right) ,\left( 1,3\right)
,\left( 2,3\right) \right\} $, and $i=1,2,...,n$. We generate $\mathbf{D}%
_{it}$, for $i=1,2,...,n$, and $t=-50,-49,...,0,1,2,...,T$, as 
\begin{equation*}
\mathbf{D}_{it}=\left( 
\begin{array}{ccc}
\sigma _{it,1} & 0 & 0 \\ 
0 & \sigma _{it,2} & 0 \\ 
0 & 0 & \sigma _{it,3}%
\end{array}%
\right) \text{,}
\end{equation*}%
where%
\begin{equation*}
\sigma _{it,q}^{2}=\left( 1-\psi _{i1,q}-\psi _{i2,q}\right) +\psi
_{i1,q}\varepsilon _{i,t-1,q}^{2}+\psi _{i1,q}\sigma _{i,t-1,q}^{2}\text{,}
\end{equation*}%
for $q=1,2,3$,$\,$\ and $t=-50,-49,...,0,1,2,...,T$, with initial values $%
\sigma _{i,-51,q}^{2}=1$.

Experiments with GARCH effects are given by 
\begin{equation*}
\psi _{i1,q}\sim IIDU\left( 0.2,0.3\right) \text{, }\psi _{i2,q}\sim
IIDU\left( 0.3,0.6\right) \text{, }
\end{equation*}%
for $q=1,2,3$ and $i=1,2,...,n$. Experiments with TAR\ effects are given by $%
\phi ^{-}=0.2$ and $\phi ^{+}=0.6$.

Overall, we consider baseline experiments without GARCH and TAR effects ($%
\psi _{i1,q}=\psi _{i2,q}=\phi ^{-}=\phi ^{+}=0$), experiments with GARCH
effects, experiments with TAR effects, and experiments featuring both GARCH
and TAR effects. Tables S5 and S6 report Monte Carlo findings for
three-variable VARMA(1,1) as the DGP with $r_{0}=2$ long-run rlations,
chi-squared distributed errors, $PR_{nT}^{2}=0.2$, moderate speed of
convergence toward long run, and with interactive time effects.\pagebreak

\begin{center}
\normalsize%
\onehalfspacing%
TABLE S5: Selection frequencies for the estimation of $r_{0}=2$ by
eigenvalue thresholding estimator, $\tilde{r}$ with $\delta =1/4$ and $1/2$
using three-varaible $VARMA(1,1)$ as the DGP with $r_{0}=2$ long-run
relations, chi-squared distributed errors, $PR_{nT}^{2}=0.2$, moderate speed
of convergence toward long run, and with interactive time effects.\bigskip

\renewcommand{\arraystretch}{0.85}\setlength{\tabcolsep}{10pt}%
\scriptsize%
\begin{tabular}{rrrrrrrrrrrrr}
\hline\hline
&  & \multicolumn{3}{c}{Frequency $\tilde{r}=0$} &  & \multicolumn{3}{c}{
Frequency $\tilde{r}=1$} &  & \multicolumn{3}{c}{Frequency $\tilde{r}=2$} \\ 
\cline{3-5}\cline{7-9}\cline{11-13}\cline{11-13}
$n$ $\backslash $ $T$ &  & \textbf{20} & \textbf{50} & \textbf{100} &  & 
\textbf{20} & \textbf{50} & \textbf{100} &  & \textbf{20} & \textbf{50} & 
\textbf{100} \\ \hline
\multicolumn{13}{l}{\textbf{A. Baseline experiments without GARCH and
without TAR effects}} \\ \hline
&  & \multicolumn{11}{l}{Correlation matrix eigenvalue thresholding
estimator $\tilde{r}$, with $\delta =1/4$} \\ \hline
\textbf{50} &  & 0.00 & 0.00 & 0.00 &  & 0.00 & 0.00 & 0.00 &  & 1.00 & 1.00
& 1.00 \\ 
\textbf{500} &  & 0.00 & 0.00 & 0.00 &  & 0.00 & 0.00 & 0.00 &  & 1.00 & 1.00
& 1.00 \\ 
\textbf{1,000} &  & 0.00 & 0.00 & 0.00 &  & 0.00 & 0.00 & 0.00 &  & 1.00 & 
1.00 & 1.00 \\ 
\textbf{3,000} &  & 0.00 & 0.00 & 0.00 &  & 0.00 & 0.00 & 0.00 &  & 1.00 & 
1.00 & 1.00 \\ \hline
&  & \multicolumn{11}{l}{Correlation matrix eigenvalue thresholding
estimator $\tilde{r}$, with $\delta =1/2$} \\ \hline
\textbf{50} &  & 0.00 & 0.00 & 0.00 &  & 0.05 & 0.00 & 0.00 &  & 0.95 & 1.00
& 1.00 \\ 
\textbf{500} &  & 0.00 & 0.00 & 0.00 &  & 0.00 & 0.00 & 0.00 &  & 1.00 & 1.00
& 1.00 \\ 
\textbf{1,000} &  & 0.00 & 0.00 & 0.00 &  & 0.00 & 0.00 & 0.00 &  & 1.00 & 
1.00 & 1.00 \\ 
\textbf{3,000} &  & 0.00 & 0.00 & 0.00 &  & 0.00 & 0.00 & 0.00 &  & 1.00 & 
1.00 & 1.00 \\ \hline
\multicolumn{13}{l}{\textbf{B. Experiments with GARCH\ effects}} \\ \hline
&  & \multicolumn{11}{l}{Correlation matrix eigenvalue thresholding
estimator $\tilde{r}$, with $\delta =1/4$} \\ \hline
\textbf{50} &  & 0.00 & 0.00 & 0.00 &  & 0.00 & 0.00 & 0.00 &  & 1.00 & 1.00
& 1.00 \\ 
\textbf{500} &  & 0.00 & 0.00 & 0.00 &  & 0.00 & 0.00 & 0.00 &  & 1.00 & 1.00
& 1.00 \\ 
\textbf{1,000} &  & 0.00 & 0.00 & 0.00 &  & 0.00 & 0.00 & 0.00 &  & 1.00 & 
1.00 & 1.00 \\ 
\textbf{3,000} &  & 0.00 & 0.00 & 0.00 &  & 0.00 & 0.00 & 0.00 &  & 1.00 & 
1.00 & 1.00 \\ \hline
&  & \multicolumn{11}{l}{Correlation matrix eigenvalue thresholding
estimator $\tilde{r}$, with $\delta =1/2$} \\ \hline
\textbf{50} &  & 0.00 & 0.00 & 0.00 &  & 0.07 & 0.00 & 0.00 &  & 0.93 & 1.00
& 1.00 \\ 
\textbf{500} &  & 0.00 & 0.00 & 0.00 &  & 0.00 & 0.00 & 0.00 &  & 1.00 & 1.00
& 1.00 \\ 
\textbf{1,000} &  & 0.00 & 0.00 & 0.00 &  & 0.00 & 0.00 & 0.00 &  & 1.00 & 
1.00 & 1.00 \\ 
\textbf{3,000} &  & 0.00 & 0.00 & 0.00 &  & 0.00 & 0.00 & 0.00 &  & 1.00 & 
1.00 & 1.00 \\ \hline
\multicolumn{13}{l}{\textbf{C. Experiments with TAR\ effects}} \\ \hline
&  & \multicolumn{11}{l}{Correlation matrix eigenvalue thresholding
estimator $\tilde{r}$, with $\delta =1/4$} \\ \hline
\textbf{50} &  & 0.00 & 0.00 & 0.00 &  & 0.00 & 0.00 & 0.00 &  & 1.00 & 1.00
& 1.00 \\ 
\textbf{500} &  & 0.00 & 0.00 & 0.00 &  & 0.00 & 0.00 & 0.00 &  & 1.00 & 1.00
& 1.00 \\ 
\textbf{1,000} &  & 0.00 & 0.00 & 0.00 &  & 0.00 & 0.00 & 0.00 &  & 1.00 & 
1.00 & 1.00 \\ 
\textbf{3,000} &  & 0.00 & 0.00 & 0.00 &  & 0.00 & 0.00 & 0.00 &  & 1.00 & 
1.00 & 1.00 \\ \hline
&  & \multicolumn{11}{l}{Correlation matrix eigenvalue thresholding
estimator $\tilde{r}$, with $\delta =1/2$} \\ \hline
\textbf{50} &  & 0.00 & 0.00 & 0.00 &  & 0.00 & 0.00 & 0.00 &  & 1.00 & 1.00
& 1.00 \\ 
\textbf{500} &  & 0.00 & 0.00 & 0.00 &  & 0.00 & 0.00 & 0.00 &  & 1.00 & 1.00
& 1.00 \\ 
\textbf{1,000} &  & 0.00 & 0.00 & 0.00 &  & 0.00 & 0.00 & 0.00 &  & 1.00 & 
1.00 & 1.00 \\ 
\textbf{3,000} &  & 0.00 & 0.00 & 0.00 &  & 0.00 & 0.00 & 0.00 &  & 1.00 & 
1.00 & 1.00 \\ \hline
\multicolumn{13}{l}{\textbf{D. Experiments with GARCH and TAR\ effects}} \\ 
\hline
&  & \multicolumn{11}{l}{Correlation matrix eigenvalue thresholding
estimator $\tilde{r}$, with $\delta =1/4$} \\ \hline
\textbf{50} &  & 0.00 & 0.00 & 0.00 &  & 0.00 & 0.00 & 0.00 &  & 1.00 & 1.00
& 1.00 \\ 
\textbf{500} &  & 0.00 & 0.00 & 0.00 &  & 0.00 & 0.00 & 0.00 &  & 1.00 & 1.00
& 1.00 \\ 
\textbf{1,000} &  & 0.00 & 0.00 & 0.00 &  & 0.00 & 0.00 & 0.00 &  & 1.00 & 
1.00 & 1.00 \\ 
\textbf{3,000} &  & 0.00 & 0.00 & 0.00 &  & 0.00 & 0.00 & 0.00 &  & 1.00 & 
1.00 & 1.00 \\ \hline
&  & \multicolumn{11}{l}{Correlation matrix eigenvalue thresholding
estimator $\tilde{r}$, with $\delta =1/2$} \\ \hline
\textbf{50} &  & 0.00 & 0.00 & 0.00 &  & 0.00 & 0.00 & 0.00 &  & 1.00 & 1.00
& 1.00 \\ 
\textbf{500} &  & 0.00 & 0.00 & 0.00 &  & 0.00 & 0.00 & 0.00 &  & 1.00 & 1.00
& 1.00 \\ 
\textbf{1,000} &  & 0.00 & 0.00 & 0.00 &  & 0.00 & 0.00 & 0.00 &  & 1.00 & 
1.00 & 1.00 \\ 
\textbf{3,000} &  & 0.00 & 0.00 & 0.00 &  & 0.00 & 0.00 & 0.00 &  & 1.00 & 
1.00 & 1.00 \\ \hline\hline
\end{tabular}%
\vspace{-0.2in}
\end{center}

\begin{flushleft}
\scriptsize%
\singlespacing%
Notes: This table reports selection frequencies for the number of estimated
long-run relations. $r_{0}$ denotes the true number of long-run relations. $%
\tilde{r}$ is given by (\ref{r_est}), namely $\tilde{r}=\sum_{j=1}^{m}%
\mathcal{I}\left( \tilde{\lambda}_{j}<T^{-\delta }\right) $, with $\tilde{%
\lambda}_{j}$, $j=1,2,...,m$, being the eigenvalues of $\mathbf{R}_{_{\bar{w}%
\bar{w}}}$ defined by (\ref{Rww}). See Sections \ref{Sup_MCd} and \ref%
{Sup_MCr} in the supplement for the full description of the design. Reported
results are based on $R=2000$ MC replications.\pagebreak
\end{flushleft}

\begin{center}
\normalsize%
\onehalfspacing%
TABLE S6: Simulated bias, RMSE, size and power for the PME estimation of $%
\beta _{13,0}$ in three variable VARMA(1,1) experiments with $r_{0}=2$ long
run relations, chi-squared distributed errors, $PR_{nT}^{2}=0.2$, moderate
speed of convergence toward long run, and with interactive time
effects\bigskip

\renewcommand{\arraystretch}{0.85}\setlength{\tabcolsep}{6pt}%
\scriptsize%
\begin{tabular}{rrrrrrrrrrrrrrrrr}
\hline\hline
& \multicolumn{4}{c}{\textbf{Bias (}$\times 100$\textbf{)}} & 
\multicolumn{4}{|c}{\textbf{RMSE (}$\times 100$\textbf{)}} & 
\multicolumn{4}{|c}{\textbf{Size }($\times 100$)} & \multicolumn{4}{|c}{%
\textbf{Power (}$\times 100$\textbf{)}} \\ \cline{2-17}
$n\backslash T$ & \textbf{20} & \textbf{50} & \textbf{100} &  & 
\multicolumn{1}{|r}{\textbf{20}} & \textbf{50} & \textbf{100} &  & 
\multicolumn{1}{|r}{\textbf{20}} & \textbf{50} & \textbf{100} &  & 
\multicolumn{1}{|r}{\textbf{20}} & \textbf{50} & \textbf{100} &  \\ \hline
\multicolumn{17}{l}{\textbf{A. Baseline experiments without GARCH and
without TAR effects}} \\ \hline
& \multicolumn{16}{l}{PME estimator with $q=2$ sub-samples} \\ \hline
\textbf{50} & -1.12 & -0.38 & -0.20 &  & 8.60 & 4.85 & 2.91 &  & 7.95 & 7.20
& 7.65 &  & 10.50 & 13.40 & 25.95 &  \\ 
\textbf{500} & -0.59 & -0.34 & -0.15 &  & 2.76 & 1.58 & 0.90 &  & 6.85 & 5.65
& 5.70 &  & 28.10 & 59.50 & 94.20 &  \\ 
\textbf{1,000} & -0.44 & -0.28 & -0.14 &  & 1.95 & 1.13 & 0.66 &  & 6.35 & 
6.05 & 6.00 &  & 46.60 & 85.55 & 99.70 &  \\ 
\textbf{3,000} & -0.59 & -0.29 & -0.11 &  & 1.22 & 0.70 & 0.38 &  & 9.05 & 
7.15 & 5.35 &  & 92.10 & 99.95 & 100.00 &  \\ \hline
& \multicolumn{16}{l}{PME estimator with $q=4$ sub-samples} \\ \hline
\textbf{50} & -1.86 & -0.90 & -0.30 &  & 7.00 & 3.97 & 2.18 &  & 6.50 & 6.50
& 7.00 &  & 12.55 & 19.30 & 38.25 &  \\ 
\textbf{500} & -1.48 & -0.81 & -0.23 &  & 2.62 & 1.47 & 0.70 &  & 10.50 & 
9.55 & 5.90 &  & 52.95 & 86.95 & 99.60 &  \\ 
\textbf{1,000} & -1.39 & -0.75 & -0.22 &  & 2.07 & 1.14 & 0.53 &  & 14.70 & 
12.20 & 7.55 &  & 81.00 & 98.90 & 100.00 &  \\ 
\textbf{3,000} & -1.51 & -0.76 & -0.21 &  & 1.74 & 0.91 & 0.34 &  & 38.25 & 
29.70 & 11.60 &  & 99.95 & 100.00 & 100.00 &  \\ \hline
\multicolumn{17}{l}{\textbf{B. Experiments with GARCH\ effects}} \\ \hline
& \multicolumn{16}{l}{PME estimator with $q=2$ sub-samples} \\ \hline
\textbf{50} & -1.55 & -0.45 & -0.25 &  & 9.95 & 5.29 & 3.07 &  & 9.05 & 8.80
& 8.15 &  & 11.70 & 14.45 & 26.90 &  \\ 
\textbf{500} & -0.54 & -0.32 & -0.15 &  & 3.28 & 1.76 & 0.94 &  & 6.70 & 6.35
& 4.65 &  & 22.85 & 51.05 & 90.55 &  \\ 
\textbf{1,000} & -0.43 & -0.32 & -0.13 &  & 2.33 & 1.27 & 0.71 &  & 6.90 & 
6.45 & 6.65 &  & 34.50 & 77.15 & 99.50 &  \\ 
\textbf{3,000} & -0.58 & -0.31 & -0.11 &  & 1.46 & 0.79 & 0.41 &  & 7.60 & 
7.50 & 5.90 &  & 76.10 & 99.00 & 100.00 &  \\ \hline
& \multicolumn{16}{l}{PME estimator with $q=4$ sub-samples} \\ \hline
\textbf{50} & -2.23 & -0.97 & -0.34 &  & 8.22 & 4.37 & 2.30 &  & 7.80 & 7.65
& 7.85 &  & 14.10 & 19.30 & 36.95 &  \\ 
\textbf{500} & -1.45 & -0.82 & -0.25 &  & 2.97 & 1.61 & 0.75 &  & 8.95 & 9.70
& 6.90 &  & 41.60 & 77.35 & 98.90 &  \\ 
\textbf{1,000} & -1.35 & -0.78 & -0.22 &  & 2.30 & 1.25 & 0.56 &  & 11.20 & 
12.20 & 6.85 &  & 63.25 & 96.00 & 100.00 &  \\ 
\textbf{3,000} & -1.50 & -0.77 & -0.21 &  & 1.83 & 0.96 & 0.36 &  & 27.50 & 
26.40 & 10.20 &  & 97.50 & 100.00 & 100.00 &  \\ \hline
\multicolumn{17}{l}{\textbf{C. Experiments with TAR\ effects}} \\ \hline
& \multicolumn{16}{l}{PME estimator with $q=2$ sub-samples} \\ \hline
\textbf{50} & 1.56 & -0.12 & -0.08 &  & 6.54 & 3.50 & 1.71 &  & 9.60 & 7.90
& 8.55 &  & 7.90 & 17.70 & 50.25 &  \\ 
\textbf{500} & 2.78 & 0.30 & 0.05 &  & 3.44 & 1.11 & 0.53 &  & 30.40 & 5.90
& 4.80 &  & 6.30 & 71.10 & 100.00 &  \\ 
\textbf{1,000} & 2.84 & 0.33 & 0.08 &  & 3.18 & 0.82 & 0.39 &  & 52.15 & 6.90
& 5.70 &  & 5.95 & 94.25 & 100.00 &  \\ 
\textbf{3,000} & 2.74 & 0.31 & 0.07 &  & 2.86 & 0.55 & 0.23 &  & 92.00 & 
12.10 & 6.15 &  & 6.65 & 100.00 & 100.00 &  \\ \hline
\multicolumn{1}{l}{} & \multicolumn{16}{l}{PME estimator with $q=4$
sub-samples} \\ \hline
\textbf{50} & 1.48 & -0.36 & -0.12 &  & 5.38 & 3.02 & 1.43 &  & 9.50 & 7.40
& 8.45 &  & 7.10 & 22.95 & 64.40 & \multicolumn{1}{c}{} \\ 
\textbf{500} & 2.64 & -0.03 & 0.06 &  & 3.12 & 0.91 & 0.43 &  & 36.70 & 3.95
& 4.80 &  & 6.00 & 90.80 & 100.00 & \multicolumn{1}{c}{} \\ 
\textbf{1,000} & 2.68 & 0.01 & 0.07 &  & 2.93 & 0.65 & 0.32 &  & 63.40 & 4.95
& 5.60 &  & 6.15 & 99.35 & 100.00 & \multicolumn{1}{c}{} \\ 
\textbf{3,000} & 2.58 & 0.01 & 0.07 &  & 2.67 & 0.38 & 0.19 &  & 96.75 & 5.25
& 6.10 &  & 9.50 & 100.00 & 100.00 & \multicolumn{1}{c}{} \\ \hline
\multicolumn{17}{l}{\textbf{D. Experiments with GARCH and TAR\ effects}} \\ 
\hline
& \multicolumn{16}{l}{PME estimator with $q=2$ sub-samples} \\ \hline
\textbf{50} & 1.32 & -0.09 & -0.06 &  & 6.80 & 3.45 & 1.66 &  & 8.35 & 8.20
& 6.85 &  & 7.80 & 18.55 & 51.60 &  \\ 
\textbf{500} & 2.64 & 0.34 & 0.07 &  & 3.41 & 1.12 & 0.52 &  & 23.95 & 6.25
& 4.70 &  & 6.55 & 70.65 & 100.00 &  \\ 
\textbf{1,000} & 2.66 & 0.34 & 0.09 &  & 3.04 & 0.82 & 0.39 &  & 42.45 & 7.30
& 6.70 &  & 5.55 & 94.05 & 100.00 &  \\ 
\textbf{3,000} & 2.54 & 0.31 & 0.07 &  & 2.69 & 0.55 & 0.22 &  & 82.10 & 
10.85 & 5.90 &  & 8.35 & 100.00 & 100.00 &  \\ \hline
& \multicolumn{16}{l}{PME estimator with $q=4$ sub-samples} \\ \hline
\textbf{50} & 1.23 & -0.32 & -0.12 &  & 5.52 & 2.97 & 1.37 &  & 7.25 & 7.00
& 7.20 &  & 7.00 & 22.90 & 66.10 &  \\ 
\textbf{500} & 2.46 & 0.01 & 0.07 &  & 3.02 & 0.92 & 0.43 &  & 28.00 & 4.90
& 5.40 &  & 5.50 & 89.20 & 100.00 &  \\ 
\textbf{1,000} & 2.48 & 0.04 & 0.08 &  & 2.76 & 0.65 & 0.32 &  & 50.15 & 4.90
& 7.05 &  & 5.70 & 99.40 & 100.00 &  \\ 
\textbf{3,000} & 2.36 & 0.02 & 0.07 &  & 2.47 & 0.38 & 0.19 &  & 91.05 & 5.05
& 6.40 &  & 13.00 & 100.00 & 100.00 &  \\ \hline\hline
\end{tabular}%
\vspace{-0.2in}
\end{center}

\begin{flushleft}
\scriptsize%
\singlespacing%
Notes: The long-run relations are given by $\mathbf{\beta }_{1,0}^{\prime }%
\mathbf{w}_{it}=w_{it,1}-w_{it,3}$ and $\mathbf{\beta }_{2,0}^{\prime }%
\mathbf{w}_{it}=w_{it,2}-w_{it,3}$, and identified using $\beta
_{11,0}=\beta _{22,0}=1$, and $\beta _{12,0}=\beta _{21,0}=0$. Reported
results are based on $R=2,000$ Monte Carlo replications. Simulated power are
computed under $H_{1}:$ $\beta _{13}=-0.97$, as alternatives to $-1$ under
the null. A detailed account of the data generating processes provided in
Section \ref{Sup_MCd} and \ref{Sup_MCr} of this supplement. Size and Power
are computed at 5 percent nominal level.\pagebreak
\end{flushleft}

\normalsize%
\onehalfspacing%

\section{Supplementary information for micro application\label{Sup_micro}}

\subsection{Data sources\label{data sources}}

All data is from Wharton Research Data Services, WRDS, available at
wrds.wharton.upenn.edu, accessed on 2023-04-20. Specifically, our variables
come from annual \textquotedblleft CRSP/Compustat Merged\textquotedblright\
database available through WRDS. We follow literature (%
\shortciteN{Geelen_etal_2024}%
) in constructing the variables listed in Table S5.

\begin{center}
TABLE S5: \textit{Variable description*}\bigskip

\renewcommand{\arraystretch}{1.0}\setlength{\tabcolsep}{5pt}%
\small%
\begin{tabular}{lll}
\hline\hline
Variable & Definition & Compustat item \\ \hline
MV & Market value & CSHO $\times $ PRCC\_F \\ 
BV & Book value & See Note 1 \\ 
DO & Total debt outstanding, DO=SD+LT & DLC+DLTT \\ 
SD & Short-term debt & DLC \\ 
LD & Long-term debt & DLTT \\ 
TA & Total assets & AT \\ \hline\hline
\end{tabular}%
\vspace{-0.1in}\vspace{-0.1in}
\end{center}

\begin{flushleft}
\scriptsize%
\singlespacing%
Notes: (*) The source for all variables is Wharton Research Data Services,
wrds.wharton.upenn.edu, accessed on 2023-04-20. Variables come from annual
\textquotedblleft CRSP/Compustat Merged\textquotedblright\ database
available through WRDS.

(1) Book value = the book equity of shareholders + compustat item $TXDITC$ -
book value of preferred stocks. The book equity of shareholders is given by
(in sequential order, depending on availability): ($i$) compustat item SEQ,
or ($ii$), common equity (compustat item CEQ) + par value of preferred stock
(compustat item PSTK), or ($iii$) total assets (compustat item AT) - total
liabilities (compustat item LT). The book value of preferred stocks is
computed as (in sequential order, based on availability): ($i$) the
redemption value (compustat item PSTKRV), or ($ii$) the liquidation value
(compustat item PSTKL), or ($iii$) the par value (compustat item \ PSTK).
\end{flushleft}

\normalsize%

\subsection{Data availability, variable construction and summary statistics 
\label{das}}

Data availability is summarized in Table S6, for the three variable sets we
considered: \{DO, TA\}, \{SD, LD, TA\},\{DO, BV, MV\}. We apply following
data filters (in sequential order) to each of the variable set and data
sample separately.

\begin{itemize}
\item[Filter 1.] We omit firms with gaps in the data for the given variable
set and sample, and firms with $T_{i}<20$ (unbalanced samples).

\item[Filter 2.] We omit firms with nonpositive entries on any of the
variable in the given set and sample.

\item[Filter 3.] We omit firms where, for a given sample and variable set,
average value of key ratios fall below 1 or above 99 percentiles estimated
after the application of the first two filters.
\end{itemize}

Filter 1 ensures that sufficient data exists, Filter 2 excludes firms with
negative or zero values, and Filter 3 is the outlier filter based on
percentiles of key ratios. Tables S3-S5 report summary statistics for each
variable set after all filters were applied.

\begin{center}
TABLE S6: Number of firms with available data\bigskip

\renewcommand{\arraystretch}{1.2}\setlength{\tabcolsep}{5pt}%
\small%
\begin{tabular}{rrr}
\hline\hline
& 1950-2021 & 1950-2010 \\ \hline
\multicolumn{3}{l}{\textbf{Variable set DO, TA}} \\ \hline
\multicolumn{1}{l}{Filter 1} & 4193 & 3010 \\ 
\multicolumn{1}{l}{Filter 1, 2} & 2546 & 1909 \\ 
\multicolumn{1}{l}{Filter 1, 2, 3} & 2531 & 1901 \\ \hline
\multicolumn{3}{l}{\textbf{Variable set SD, LD, TA}} \\ \hline
\multicolumn{1}{l}{Filter 1} & 4193 & 3010 \\ 
\multicolumn{1}{l}{Filter 1, 2} & 1379 & 1110 \\ 
\multicolumn{1}{l}{Filter 1, 2, 3} & 1365 & 1101 \\ \hline
\multicolumn{3}{l}{\textbf{Variable set DO, BV, MV}} \\ \hline
\multicolumn{1}{l}{Filter 1} & 2907 & 2196 \\ 
\multicolumn{1}{l}{Filter 1, 2} & 1419 & 1172 \\ 
\multicolumn{1}{l}{Filter 1, 2, 3} & 1403 & 1164 \\ \hline\hline
\end{tabular}%
\vspace{-0.2in}
\end{center}

\begin{flushleft}
\footnotesize%
\singlespacing%
Notes: Variable definitions are provided in Table S5. Filter 1 omits firms
with gaps in the data for the given variable set and sample, and firms with
fewer than 20 years. Filter 2 omits firms with nonpositive entries. Filter
3, omits firms where average value of key ratios fall below 1 or above 99
percentiles after the application of the first two filters%
\normalsize%
\pagebreak
\end{flushleft}

\begin{center}
TABLE S7: Summary Statistics (min, mean, median, max) for individual
variables and ratios in the variable set \{DO, TA\} after the application of
all filters.\bigskip

\renewcommand{\arraystretch}{1}\setlength{\tabcolsep}{5pt}%
\scriptsize%
\begin{tabular}{rrr}
\hline\hline
& 1950-2021 & 1950-2010 \\ \hline
\textbf{Variable DO} &  &  \\ \hline
min & 0.001 & 0.001 \\ 
median & 160.4 & 94.77 \\ 
mean & 3,405 & 1,837 \\ 
max & 889,300 & 889,300 \\ \hline
\textbf{Variable TA} &  &  \\ \hline
min & 0.283 & 0.126 \\ 
median & 689.8 & 371.7 \\ 
mean & 13,809 & 6,621 \\ 
max & 3,743,567 & 3,001,251 \\ \hline
\textbf{Ratio DO/TA} &  &  \\ \hline
min & 0.000 & 0.000 \\ 
median & 0.270 & 0.276 \\ 
mean & 0.290 & 0.294 \\ 
max & 4.434 & 6.789 \\ \hline\hline
\end{tabular}%
\vspace{-0.1in}\vspace{-0.1in}
\end{center}

\begin{flushleft}
\footnotesize%
\singlespacing%
Notes: Unit for the top part of this table reporting the summary statistics
for individual variables is million U.S. dollars.%
\normalsize%
\end{flushleft}

\pagebreak

\begin{center}
TABLE S8: Summary Statistics (min, mean, median, max) for individual
variables and ratios in the variable set \{SD, LD, TA\} after the
application of all filters.\bigskip

\renewcommand{\arraystretch}{1}\setlength{\tabcolsep}{5pt}%
\scriptsize%
\begin{tabular}{rrr}
\hline\hline
& 1950-2021 & 1950-2010 \\ \hline
\textbf{Variable SD} &  &  \\ \hline
min & 0.001 & 0.001 \\ 
median & 21.77 & 12.98 \\ 
mean & 2,081 & 1,071 \\ 
max & 614,237 & 562,857 \\ \hline
\textbf{Variable LD} &  &  \\ \hline
min & 0.001 & 0.001 \\ 
median & 133.4 & 78.77 \\ 
mean & 2,882 & 1,453 \\ 
max & 486,876 & 486,876 \\ \hline
\textbf{Variable TA} &  &  \\ \hline
min & 0.652 & 0.652 \\ 
median & 795.7 & 419.2 \\ 
mean & 19,510 & 8,482 \\ 
max & 3,743,567 & 2,264,909 \\ \hline
\textbf{Ratio SD/TA} &  &  \\ \hline
min & 0.000 & 0.000 \\ 
median & 0.039 & 0.041 \\ 
mean & 0.066 & 0.067 \\ 
max & 1.585 & 1.585 \\ \hline
\textbf{Ratio LD/TA} &  &  \\ \hline
min & 0.000 & 0.000 \\ 
median & 0.215 & 0.217 \\ 
mean & 0.233 & 0.233 \\ 
max & 1.752 & 1.752 \\ \hline\hline
\end{tabular}%
\vspace{-0.1in}\vspace{-0.1in}
\end{center}

\begin{flushleft}
\footnotesize%
\singlespacing%
Notes: Unit for the top part of this table reporting the summary statistics
for individual variables is million U.S. dollars.%
\normalsize%
\end{flushleft}

\pagebreak

\begin{center}
TABLE S9: Summary Statistics (min, mean, median, max) for individual
variables and ratios in the variable set \{DO, BV, MV\} after the
application of all filters.\bigskip

\renewcommand{\arraystretch}{1}\setlength{\tabcolsep}{5pt}%
\scriptsize%
\begin{tabular}{rrr}
\hline\hline
& 1950-2021 & 1950-2010 \\ \hline
\textbf{Variable DO} &  &  \\ \hline
min & 0.002 & 0.002 \\ 
median & 138.9 & 91.61 \\ 
mean & 2,178 & 1,212 \\ 
max & 889,300 & 889,300 \\ \hline
\textbf{Variable BV} &  &  \\ \hline
min & 0.079 & 0.079 \\ 
median & 269.2 & 178.7 \\ 
mean & 2,752 & 1,594 \\ 
max & 595,878 & 212,294 \\ \hline
\textbf{Variable MV} &  &  \\ \hline
min & 0.177 & 0.177 \\ 
median & 377.0 & 227.4 \\ 
mean & 5,007 & 2,908 \\ 
max & 662,627 & 504,240 \\ \hline
\textbf{Ratio DO/MV} &  &  \\ \hline
min & 0.000 & 0.000 \\ 
median & 0.455 & 0.486 \\ 
mean & 0.832 & 0.869 \\ 
max & 63.26 & 63.26 \\ \hline
\textbf{Ratio BV/MV} &  &  \\ \hline
min & 0.002 & 0.003 \\ 
median & 0.773 & 0.833 \\ 
mean & 0.997 & 1.066 \\ 
max & 27.85 & 31.89 \\ \hline\hline
\end{tabular}%
\vspace{-0.1in}\vspace{-0.1in}
\end{center}

\begin{flushleft}
\footnotesize%
\singlespacing%
Notes: Unit for the top part of this table reporting the summary statistics
for individual variables is million U.S. dollars.%
\normalsize%
\pagebreak
\end{flushleft}

\begin{center}
\singlespacing%
TABLE S10:\textit{\ IPS unit root test results for 40-year balanced sample
ending 2021}\bigskip

\renewcommand{\arraystretch}{1.1}\setlength{\tabcolsep}{4pt}%
\footnotesize%
\vspace{-0.1in}%
\begin{tabular}{rrccrccccc}
\hline\hline
\multicolumn{10}{l}{Panel unit root test results} \\ \hline
&  & \multicolumn{2}{c}{} &  & \multicolumn{2}{c}{} &  & \multicolumn{2}{c}{}
\\ \cline{3-4}\cline{6-7}\cline{9-10}
& Lag order: & $p=1$ & $p=2$ & \multicolumn{1}{c}{} & $p=1$ & $p=2$ &  & $%
p=1 $ & $p=2$ \\ 
\multicolumn{10}{l}{A. Panel unit root test results for the variable set DO,
TA ($n=336$)} \\ \hline
&  & \multicolumn{2}{c}{DO} &  & \multicolumn{2}{c}{TA} & \multicolumn{1}{r}{
} & \multicolumn{1}{r}{} & \multicolumn{1}{r}{} \\ \cline{3-4}\cline{6-7}
& IPS stat.: & -1.534 & -1.504 & \multicolumn{1}{c}{} & -1.45 & -1.457 & 
\multicolumn{1}{r}{} & \multicolumn{1}{r}{} & \multicolumn{1}{r}{} \\ 
&  & \multicolumn{1}{r}{} & \multicolumn{1}{r}{} &  & \multicolumn{1}{r}{} & 
\multicolumn{1}{r}{} & \multicolumn{1}{r}{} & \multicolumn{1}{r}{} & 
\multicolumn{1}{r}{} \\ 
\multicolumn{10}{l}{B. Panel unit root test results for the variable set SD,
LD, TA ($n=176$)} \\ \hline
&  & \multicolumn{2}{c}{SD} &  & \multicolumn{2}{c}{LD} & \multicolumn{1}{r}{
} & \multicolumn{2}{c}{TA} \\ \cline{3-4}\cline{6-7}\cline{9-10}
& IPS stat.: & \multicolumn{1}{r}{-2.344**} & \multicolumn{1}{r}{-2.067**} & 
& \multicolumn{1}{r}{-1.492} & \multicolumn{1}{r}{-1.419} & 
\multicolumn{1}{r}{} & \multicolumn{1}{r}{-1.561} & \multicolumn{1}{r}{-1.585
} \\ 
&  & \multicolumn{1}{r}{} & \multicolumn{1}{r}{} &  & \multicolumn{1}{r}{} & 
\multicolumn{1}{r}{} & \multicolumn{1}{r}{} & \multicolumn{1}{r}{} & 
\multicolumn{1}{r}{} \\ 
\multicolumn{10}{l}{C. Panel unit root test results for the variable set DO,
BV, MV ($n=175$)} \\ \hline
&  & \multicolumn{2}{c}{DO} &  & \multicolumn{2}{c}{BV} & \multicolumn{1}{r}{
} & \multicolumn{2}{c}{MV} \\ \cline{3-4}\cline{6-7}\cline{9-10}
& IPS stat.: & \multicolumn{1}{r}{-1.328} & \multicolumn{1}{r}{-1.215} &  & 
\multicolumn{1}{r}{-1.153} & \multicolumn{1}{r}{-1.165} & \multicolumn{1}{r}{
} & \multicolumn{1}{r}{-1.328} & \multicolumn{1}{r}{-1.351} \\ 
&  & \multicolumn{1}{r}{} & \multicolumn{1}{r}{} &  & \multicolumn{1}{r}{} & 
\multicolumn{1}{r}{} & \multicolumn{1}{r}{} & \multicolumn{1}{r}{} & 
\multicolumn{1}{r}{} \\ \hline\hline
\end{tabular}%
\vspace{-0.2in}
\end{center}

\begin{flushleft}
\scriptsize%
\singlespacing%
Notes: This table reports IPS panel unit root test statistics by 
\citeN{ImPesaranShin2003}
for balanced sample using 40 years ending 2021. Rejections at 5 and 1
percent nominal level are highlighted by * and **, respectively.
\end{flushleft}

\section{Supplementary information for macro applications\label{Sup_macro}}

We use Penn World Table database, version 10.01, available at \newline
https://www.rug.nl/ggdc/productivity/pwt/. This dataset contains unbalanced
annual macroeconomic data covering the period $1950-2019$. The following
variables are constructed (variable names correspond to the variable
identifiers in PWT database)

\begin{enumerate}
\item (Exports per-capita and imports per-capita), \textit{exppc=csh\_x}$%
\mathit{\times }$\textit{rgdpna/pop and imppc=-csh\_m}$\mathit{\times }$%
\textit{rgdpna/pop}

\item (Real wages and productivity per hour worked) \textit{ewageph=labsh}$%
\mathit{\times }$\textit{rgdpna/(emp}$\mathit{\times }$\textit{avh)} and 
\textit{prodph=rgdpna/(emp}$\mathit{\times }$\textit{avh)}

\item (Output and Capital in efficiency units) \textit{connapc=(csh\_c}$%
\times $\textit{rgdpna)/pop}, \textit{invnapc=(csh\_i}$\times $\textit{%
rgdpna)/pop}, \textit{rgdpnapc=rgdpna/pop,}
\end{enumerate}

where: \textit{rgdpna} is Real GDP at constant 2017 national prices (in mil.
2017US\$), \textit{pop} is population (in millions), \textit{csh\_x }is
share of merchandise exports at current PPPs, \textit{csh\_m} is share of
merchandise imports at current PPPs\textit{, labsh} is share of labour
compensation in GDP at current national prices\textit{, emp} is number of
persons engaged (in millions),\textit{\ avh }is average annual hours worked
by persons engaged\textit{, csh\_c }is share of household consumption at
current PPPs, and \textit{csh\_i }is Share of gross capital formation at
current PPPs.

We use the following variables in our analysis: $ex_{it}=\ln \left( \mathit{%
exppc}_{it}\right) $, $im_{it}=\ln \left( \mathit{imppc}_{it}\right) $, $%
wage_{it}=ln\left( \mathit{ewageph}_{it}\right) $, $prod_{it}=\ln \left( 
\mathit{prodph}_{it}\right) $, $inv_{it}=\ln \left( \mathit{invnapc}%
_{it}\right) $, $cons_{it}=\ln \left( \mathit{connapc}\right) $, and $%
gdp_{it}=\ln \left( \mathit{rgdpnapc}_{it}\right) $.

For for each variable-pair, we apply the following filters : (A) Incldue
countries with $T_{i}\geq 20$. (B) Drop countries with annual per capita
observations that are below $\epsilon =0.01$.

\end{document}